\documentclass[letterpaper,fleqn,usenatbib,useAMS]{mnras}

\usepackage{newtxtext,newtxmath}
\usepackage[T1]{fontenc}
\usepackage{ae,aecompl}

\usepackage{graphicx,subfig}	% Including figure files
\usepackage{float}
\usepackage{amsmath}	% Advanced maths commands
\usepackage{amssymb}	% Extra maths symbols
\usepackage{natbib}
\usepackage{upgreek}
\usepackage{mathrsfs}
\usepackage{calrsfs}
\usepackage{hyperref}
\hypersetup{colorlinks=false}
\usepackage{pdflscape}
\usepackage{rotating}
\usepackage{longtable}
\usepackage{tabularx}
\usepackage{array}
\usepackage{supertabular}
\usepackage{booktabs}

\title[LOFAR Faraday rotation measures towards pulsars]{Low-frequency Faraday rotation measures towards pulsars using LOFAR: probing the 3-D Galactic halo magnetic field}

\author[C. Sobey et al.]{
C.~Sobey$^{1, 2, 3}$\thanks{E-mail: Charlotte.Sobey@csiro.au (CS)},
A.~V.~Bilous$^{4}$, %confirmed
J-M.~Grie{\ss}meier$^{5,6}$, %confirmed
J.~W.~T.~Hessels$^{3,4}$, \newauthor%confirmed
A.~Karastergiou$^{7,8,9}$, %confirmed
E.~F.~Keane$^{10}$, %confirmed
V.~I.~Kondratiev$^{3,11}$, %confirmed
M.~Kramer$^{12}$, %confirmed
D.~Michilli$^{4,3,13,14}$, \newauthor %confirmed
A.~Noutsos$^{12}$, %confirmed
M.~Pilia$^{15}$, %confirmed
E.~J.~Polzin$^{16}$, % confirmed
B.~W.~Stappers$^{16}$,  %confirmed
C.~M.~Tan$^{16}$, \newauthor%confirmed
J.~van~Leeuwen$^{3}$, % confirmed
J.~P.~W.~Verbiest$^{17,12}$, %confirmed
P.~Weltevrede$^{16}$, %confirmed
G.~Heald$^{2}$, % confirmed
M.~I.~R.~Alves$^{18}$, \newauthor%confirmed 
E.~Carretti$^{19}$, % INAF % confirmed
T.~En{\ss}lin$^{20}$, %confirmed
M.~Haverkorn$^{18}$, %confirmed
M.~Iacobelli$^{3}$, %confirmed
W.~Reich$^{12}$, %confirmed
C.~Van Eck$^{21}$\\ % confirmed
\\
$^{1}$International Centre for Radio Astronomy Research - Curtin University, GPO Box U1987, Perth, WA 6845, Australia\\
$^{2}$CSIRO Astronomy and Space Science, PO Box 1130 Bentley, WA 6102, Australia\\
$^{3}$ASTRON, Netherlands Institute for Radio Astronomy, Oude Hoogeveensedijk 4, 7991 PD, Dwingeloo, The Netherlands \\ %Jason,Danielle,Marco
$^{4}$Anton Pannekoek Institute for Astronomy, University of Amsterdam, Science Park 904, 1098 XH, Amsterdam, The Netherlands\\ %Jason, Danielle, Anya
$^{5}$LPC2E - Universit\'{e} d'Orl\'{e}ans /  CNRS, 45071 Orl\'{e}ans cedex 2, France\\ %J.M
$^{6}$Station de Radioastronomie de Nan\c{c}ay, Observatoire de Paris, PSL Research University, CNRS, Univ. Orl\'{e}ans, OSUC, 18330 Nan\c{c}ay, France\\ % J.M
$^{7}$Oxford Astrophysics, Denys Wilkinson Building, Keble Road, Oxford OX1 3RH, UK\\ %ArisK
$^{8}$Department of Physics \& Astronomy, University of the Western Cape, Private Bag X17, Bellville 7535, South Africa\\ %ArisK
$^{9}$Department of Physics and Electronics, Rhodes University, PO Box 94, Grahamstown 6140, South Africa \\%Aris K
$^{10}$SKA Organisation, Jodrell Bank Observatory, SK11 9DL, UK\\ %Evan
$^{11}$Astro Space Centre, Lebedev Physical Institute, Russian Academy of Sciences, Profsoyuznaya Str. 84/32, Moscow 117997, Russia\\ %Vlad
$^{12}$Max Planck Institute for Radio Astronomy, Auf dem H\"{u}gel 69, 53121 Bonn, Germany\\ %% Wolfgang
$^{13}$Department of Physics, McGill University, 3600 University Street, Montréal, QC H3A 2T8, Canada\\ %Danielle
$^{14}$McGill Space Institute, McGill University, 3550 University Street, Montréal, QC H3A 2A7, Canada\\ %Danielle
$^{15}$INAF - Osservatorio Astronomico di Cagliari, via della Scienza 5, I-09047 Selargius (CA), Italy\\ %Maura
$^{16}$Jodrell Bank Centre for Astrophysics, School of Physics and Astronomy, The University of Manchester, Manchester, M13 9PL, UK\\ %Ben, Elliott
$^{17}$Fakult\"{a}t f\"{u}r Physik, Universit\"{a}t Bielefeld, Postfach 100131, 33501 Bielefeld, Germany\\ %Joris
$^{18}$Department of Astrophysics/IMAPP, Radboud University Nijmegen, PO Box 9010, 6500 GL Nijmegen, The Netherlands\\ %Marta, Marijke
$^{19}$INAF - Istituto di Radioastronomia, Via Gobetti 101, 40129 Bologna, Italy\\ %%% Ettore
$^{20}$Max Planck Institut f\"{u}r Astrophysik, Karl-Schwarzschild-Str 1, D-85740 Garching, Germany\\ % Torsten
$^{21}$Dunlap Institute for Astronomy and Astrophysics, University of Toronto, 50 St. George Street, Toronto, ON M5S 3H4, Canada %%% Cameron
}
\date{Accepted 2019 January 17. Received 2019 January 17; in original form 2018 November 9}
\pubyear{2018}

\hypersetup{draft}

\begin{document}
\label{firstpage}
\pagerange{\pageref{firstpage}--\pageref{lastpage}}
\maketitle

%%%%%%%%%%%%%%%%% ABSTRACT %%%%%%%%%%%%%%%%%%

%This is a simple template for authors to write new MNRAS papers.
%The abstract should briefly describe the aims, methods, and main results of the paper.
%It should be a single paragraph not more than 250 words (200 words for Letters).
%No references should appear in the abstract.
% context heading (optional)  % aims heading (mandatory)  % methods heading (mandatory) % results heading (mandatory)  % conclusions heading (optional)
% LOCATE FOCUS REPORT ARGUE 

\begin{abstract} 
 % We determined Faraday rotation measurements (RMs) towards 137 pulsars in the northern sky, using previously published Low-Frequency Array (LOFAR) High-Band Antenna (HBA) census observations at 110--190\,MHz. This is the largest low-frequency RM catalogue to date. The RMs towards 25 of the pulsars have not been previously published.  For the pulsars with RMs in the literature, our measurements are 20-times more precise, on average, owing to the low frequency and large bandwidth provided by LOFAR, and aided by the use of RM-synthesis. The RMs were corrected for ionospheric Faraday rotation using a previously tested and verified method, also increasing the accuracy of our catalogue, compared to un-corrected measurements. In addition, we used \textsc{RM CLEAN} to find that the Faraday dispersion functions towards pulsars are extremely Faraday thin -- mostly less than 0.001\,rad m$^{\rm -2}$. We use these new precise RM measurements, in combination with previously published values towards pulsars (along with dispersion measures and distance estimates using a Galactic electron density model) to determine the Galactic halo magnetic scale height.  This sample represents an initial low-frequency catalogue that can provide valuable information about the three-dimensional structure of the Galactic magnetic field. 
    
We determined Faraday rotation measures (RMs) towards 137 pulsars in the northern sky, using Low-Frequency Array (LOFAR) observations at 110--190\,MHz.
This low-frequency RM catalogue, the largest to date, improves the precision of existing RM measurements on average by a factor of 20 -- due to the low frequency and wide bandwidth of the data, aided by the RM synthesis method. We report RMs towards 25 pulsars for the first time.
The RMs were corrected for ionospheric Faraday rotation to increase the accuracy of our catalogue to $\approx$0.1\,rad m$^{\rm -2}$. 
The ionospheric RM correction is currently the largest contributor to the measurement uncertainty.
In addition, we find that the Faraday dispersion functions towards pulsars are extremely Faraday thin -- mostly less than 0.001\,rad m$^{\rm -2}$. 
We use these new precise RM measurements (in combination with existing RMs, dispersion measures, and distance estimates) to estimate the scale height of the Galactic halo magnetic field:
2.0$\pm$0.3\,kpc for Galactic quadrants \textsc{I} and \textsc{II} above and below the Galactic plane (we also evaluate the scale height for these regions individually).
Overall, our initial low-frequency catalogue provides valuable information about the 3-D structure of the Galactic magnetic field.

\end{abstract}

%%%%%%%%%%%%%%%%% KEY WORDS %%%%%%%%%%%%%%%%%%



%%%%%%%%%%%%%%%%% INTRODUCTION %%%%%%%%%%%%%%%%%%

%This is a simple template for authors to write new MNRAS papers.
%See \texttt{mnras\_sample.tex} for a more complex example, and \texttt{mnras\_guide.tex} for a full user guide.
%All papers should start with an Introduction section, which sets the work in context, cites relevant earlier studies in the field by \citet{Others2013}, and describes the problem the authors aim to solve \citep[e.g.][]{Author2012}.

\section{Introduction}\label{sec:int}

Magnetic fields are ubiquitous in the Universe. They play a role in numerous astrophysical processes across a range of physical scales and field strengths.
Within galaxies, magnetic fields pervade the multi-phase diffuse interstellar medium (ISM). 

Galactic magnetic fields can be modelled as a combination of a large, kiloparsec-scale, coherent component; a component with a power spectrum of small-scale (1--100 parsec), random fluctuations; and an intermediate-scale component that is ordered overall but with small-scale field direction reversals \citep[e.g.][]{2010MNRAS.401.1013J}. The coherent component observed, for example, in `magnetic spiral arms' \citep[e.g.][]{2009IAUS..259....3B} accelerates and confines cosmic rays \citep[e.g.][and references therein]{2012SSRv..166...97A} and may be maintained by a turbulent dynamo \citep[e.g.][]{2016MNRAS.461.4482D}. The random component, caused by turbulence, supernovae and shocks, and other localised phenomena \citep[e.g.][]{2015aska.confE..96H} plays a role in, for example, the formation of molecular clouds and stars \citep[e.g.][]{2012ARA&A..50...29C}. The intermediate ordered component is thought to result from larger-scale effects, including shearing dynamics and compression, on the random component \citep[e.g.][]{2010MNRAS.401.1013J} and plays a role in, for example, the transport of heat and angular momentum \citep[e.g.][]{2015ASSL..407..483H}. %% citation needed %%
In the Milky Way, the random and ordered components can be between 0.5- and 5-times the magnitude of the coherent component, depending on the observables (see below), regions studied, and assumptions used \citep[e.g.][]{1989ApJ...343..760R,2004A&A...427..169H,2006ApJ...637L..33H,2007A&A...471L..21S,2010MNRAS.401.1013J,2011MNRAS.416.1152J,2016JCAP...05..056B}.

The (magnetic) structure of our Galaxy is challenging to study because we are embedded within it.  The Galactic magnetic field (GMF) was first measured over 65 years ago using polarisation of starlight \citep{1949Sci...109..166H,1949Sci...109..165H}. There are several other observables that provide information about the GMF, including Zeeman splitting of spectral lines \citep[e.g.][]{2010ApJ...725..466C}; polarised thermal emission from dust grains \citep[e.g.][]{2016A&A...596A.105P}; total intensity and polarisation of Galactic synchrotron emission \citep[e.g.][]{2008A&A...477..573S}; Faraday rotation of polarised sources \citep[e.g.][]{2011ApJ...728...97V}; and ultra-high energy cosmic rays \citep[e.g.][]{2013JCAP...01..023F}.  These observables (mostly 2-D tracers requiring ancillary data) usually provide information about the strength/direction of at least one component of the magnetic field parallel/perpendicular to the line-of-sight (LoS) within a particular phase of the ISM \citep[e.g.][]{2011MmSAI..82..824F,2015ASSL..407..483H}.  Several of these observables will be required to accurately reconstruct the multiple-scale components of the GMF \citep[e.g.][]{2010MNRAS.401.1013J,2018JCAP...08..049B}.

% Lyne & Smith (1968),
Our current picture of the large-scale GMF is that the field strength is on the order of $\sim$2--10\,$\upmu$G; $\approx$2\,$\upmu$G at the solar position and increasing towards the Galactic centre and decreasing towards the anticentre \citep[e.g.][]{2001SSRv...99..243B,2003ApJ...592L..29B,2006ApJ...642..868H}. Recent analyses of the large-scale disc component often favour an axisymmetric spiral with an overall clockwise direction (as viewed from the north Galactic pole) that seems to somewhat follow the spiral-arm structure, with one field direction reversal near the Sun in the direction of the Galactic centre along the Scutum-Centaurus arm \citep[e.g.][]{2011ApJ...728...97V,2012ApJ...761L..11J}. However, the total number and locations of the large-scale GMF reversals are still under debate \citep[e.g.][]{2006ApJ...642..868H}, and many open questions concerning the 3-D structure of the aggregate GMF (along with its generation and evolution) remain \citep[e.g.][]{2011MmSAI..82..824F}.  Furthermore, such reversals have not yet been observed in any other galaxies \citep[e.g.][]{2015aska.confE..94B}.
Another component of the large-scale GMF often considered separately to the disc, is the halo \citep[absolute Galactic latitude greater than a few degrees, e.g.,][]{2018ApJS..234...11H}. The 3-D GMF structure in the halo is less well understood, with several proposed geometries \citep[see, e.g.,][and references therein]{2015ASSL..407..483H,2017A&A...600A..29T}, including a north--south asymmetry across the Galactic disc \citep[][]{2012ApJ...755...21M}. Despite the challenges, our Galaxy also provides a unique opportunity to study the magnetic field structure at much smaller scales, essential for comparison with nearby galaxies \citep[e.g.][]{2010A&A...513A..65N}.

%The magnetic field strength and direction in the solar neighbourhood are relatively well known (e.g. $B_{\rm{reg}}\sim1.5\,\upmu$G in the clockwise direction with pitch angle $-8^{\circ}$). 
% There is a small vertical large-scale magnetic field component towards the south ($\sim0.3\,\upmu$G), while a small vertical magnetic field component towards the north could be attributed to the North Polar Spur.
%All interstellar matter but the densest, coldest clouds is sufficiently ionized (even with an ionization degree of only 10?4 ?10?3) for the neutral gas component to remain coupled to the ionized gas, and therefore be efficiently frozen into the magnetic field
% \citep[e.g.][]{2015ASSL..407..483H}
% (dust grain physics, cosmic ray or thermal electron densities, cosmic ray origin and charge)

Faraday rotation measurements of polarised radio sources have been used to measure the magnetic field strength and direction in the intervening (warm) ISM parallel to the LoS. This method has been used to investigate the structure of the GMF by using a large sample of sources, including pulsars \citep[e.g.][]{1972ApJ...172...43M,1974ApJ...188..637M,1994MNRAS.268..497R,1999MNRAS.306..371H,2006ApJ...642..868H,2008MNRAS.386.1881N}, extragalactic sources \citep[e.g.][]{2010ApJ...714.1170M,2012A&A...542A..93O,2015A&A...575A.118O}, and a combination of both \citep[e.g.][]{2007ApJ...663..258B,2011ApJ...728...97V}. Polarisation observations of a large set of pulsars can be used to probe the 3-D structure of the GMF efficiently. The ratio between the Faraday rotation measure (RM) and the dispersion measure (DM) towards a pulsar at distance $d$ (pc) provides an estimate of the electron density-weighted average magnetic field strength and net direction parallel to the LoS:
\begin{equation} %%%%WORK ON EQUATION%%%%%%%%%
\langle B_{\parallel}\rangle=
\frac{\int^{{d}}_{0} n_{\mathrm e} B_{\parallel}\mathrm{d}l}{\int^{{d}}_{0} n_{\mathrm e} \mathrm{d}l}=
1.232~\upmu\mathrm{G} \left( \frac{\mathrm {RM}}{\mathrm {rad~m^{\mathrm -2}}} \right)\left( \frac{\mathrm {DM}}{\mathrm{pc~cm^{\mathrm {-3}}}} \right)^{\mathrm {-1}},
  \label{eq:1}
\end{equation}
where ${n_{\mathrm e}}$ is the electron density and ${\rm{d}}l$ is the differential distance element. By definition, positive (negative) RMs indicate that the net direction of $\langle B_{\parallel}\rangle$ is towards (away from) the observer. Equation \ref{eq:1} assumes that the electron density and magnetic field are uncorrelated \citep[e.g.][]{2003A&A...411...99B}.
Furthermore, pulsar emission is often highly (linearly) polarised \citep[e.g.][]{2018MNRAS.474.4629J}, and Faraday rotation internal to the pulsar magnetosphere is negligible \citep[e.g.][]{2011MNRAS.417.1183W}, facilitating measurement of the RM due to the foreground ISM alone. Thus, it is also expected that pulsars are `Faraday thin' point sources, where the polarised flux detected lies at a single Faraday depth (i.e. the RM dispersion is negligible; in contrast to a Faraday thick source with polarised flux distributed over a range of Faraday depths).
There are 2659 known pulsars\footnote{Current version of pulsar catalogue (1.59) retrieved from \href{http://www.atnf.csiro.au/people/pulsar/psrcat/}{http://www.atnf.csiro.au/people/pulsar/psrcat/}} distributed throughout the Galaxy, particularly near the disc \citep{2005AJ....129.1993M}. Currently, 1133 pulsars (43 per cent) have published RMs  -- two-thirds of which are located in the Southern sky, near the Galactic plane, and are mostly concentrated within a few kiloparsecs from the Sun \citep{2005AJ....129.1993M}. The mode RM measurement in the pulsar catalogue was taken at the Parkes Observatory at a frequency of $\approx$1.4\,GHz, using 128\,MHz bandwidth \citep[e.g.][]{2006ApJ...642..868H,2008MNRAS.386.1881N,2018ApJS..234...11H}. Pulsar distances are estimated using their DMs and a model of the Galactic thermal electron density, e.g., TC93 \citep{1993ApJ...411..674T}; NE2001 \citep{2002astro.ph..7156C}; YMW16 \citep{2017ApJ...835...29Y}. However, these distance estimates can be quite uncertain, and only $\approx$200 pulsars have independent distance measurements \citep[e.g.][]{2017ApJ...835...29Y,2018arXiv180809046D}.  Therefore, it is possible that what are interpreted as field reversals along spiral arms may be caused by other intermediate-scale structures, e.g., superbubbles \citep{2015ASSL..407..483H}. RMs measured towards extragalactic sources provide complementary information about the integrated LoS through (at least) the Galaxy, which partially smooths out smaller scale structure.    %\citep[e.g.,][]{2008A&A...486..819M,2008A&A...477..573S,2011ApJ...728...97V,2012A&A...540A.122F}.

%Low frequency valuable for measuring RMs, coupled with RM-synthesis
Work towards better understanding astrophysical magnetic fields, including the GMF, is ongoing. The recent construction of low-frequency ($<$300\,MHz), next-generation aperture array telescopes (and their associated supercomputing facilities) has rejuvenated this field. These include the Low-Frequency Array \citep[LOFAR;][]{2013A&A...556A...2V} and the Long-Wavelength Array \citep[LWA;][]{2012JAI.....150004T} in the Northern Hemisphere, and the Murchison Widefield Array \citep[MWA;][]{2013PASA...30....7T} in the Southern Hemisphere.  Polarisation observations from low-frequency telescopes such as these are providing precise measurements of ISM parameters, due to the wavelength-squared dependencies from the effects of dispersion and Faraday rotation \citep[e.g.][]{0004-637X-808-2-156,2016A&A...585A.128K,2016ApJ...830...38L,2017A&A...597A..98V}. These facilities are the pathfinders or the precursor to the low-frequency aperture array component of the Square Kilometre Array (SKA-Low), which will operate at 50--350\,MHz \citep{2017arXiv171101910K}. The SKA's capabilities for pulsar discoveries, timing, and astrometry will revolutionise radio astronomy and will be invaluable for studying the GMF in 3-D on large and small scales \citep[e.g.][]{2015aska.confE..41H}. 

%The large collecting area, low frequency, large fractional bandwidth and multi-beaming capabilities
In this paper, we present RM measurements towards 137 pulsars using low-frequency ($<$200\,MHz) polarisation observations from the `LOFAR census of non-recycled pulsars' \citep[][]{2016A&A...591A.134B} and the `LOFAR census of millisecond pulsars'  \citep{2016A&A...585A.128K}. This also includes RM measurements towards PSR B0329+54, using several LOFAR (timing/monitoring) observations to further investigate the accuracy of the current ionospheric RM correction method. This is the largest low-frequency RM catalogue to date and is also the first to report a large number of RM dispersion measurements for pulsars. These measurements are complementary to similar low-frequency RM catalogues towards extragalactic sources that trace LoSs through (at least) the entire Galaxy \citep[e.g.][]{2018A&A...617A.136N,2018A&A...613A..58V,2018arXiv180909327R}, as well as polarisation observations of diffuse Galactic synchrotron emission that traces the GMF in the plane of the sky \citep[e.g.][]{2015A&A...583A.137J,2017A&A...597A..98V}. We aim to increase the number of pulsars with RM data and provide higher-precision RM measurements for those already in the literature, to study the large-scale structure of the GMF in the Galactic halo. These data also provide a baseline measurement towards monitoring fluctuations in RM over time to investigate the small-scale magneto-ionic structure in the ISM in the future. 

The LOFAR observations and data reduction are described in Section \ref{sec:obs}. In Section \ref{sec:res} we present a summary of our catalogue of low-frequency DMs and RMs. In Section \ref{sec:disc} we discuss our results, including an estimate of the scale height of the Galactic halo magnetic field, and address the limitations of the current data and methods. We provide a summary and our conclusions in Section \ref{sec:conc}. Appendix A presents the table of LOFAR RM and RM dispersion measurements, and Appendix B presents further details of the RM-synthesis analysis and the Faraday dispersion functions (FDFs or Faraday spectra) obtained for each pulsar with a significant RM detection.

%\footnote{Note: Throughout the paper, numbers in parentheses are the uncertainties corresponding to the least significant figure in the value quoted.}.

%%%%%%%%%%%%%%%%% OBSERVATIONS %%%%%%%%%%%%%%%%%%

%Normally the next section describes the techniques the authors used.
%It is frequently split into subsections, such as Section~\ref{sec:maths} below.
%Simple mathematics can be inserted into the flow of the text e.g. $2\times3=6$
%or $v=220$\,km\,s$^{-1}$, but more complicated expressions should be entered
%as a numbered equation:
%\begin{equation}
%    x=\frac{-b\pm\sqrt{b^2-4ac}}{2a}.
%	\label{eq:quadratic}
%\end{equation}
%Refer back to them as e.g. equation~(\ref{eq:quadratic}).
% Figures are referred to as e.g. Figure~\ref{fig:example_figure}, and tables as e.g. Table~\ref{tab:example_table}.

\section{Observations and data reduction}\label{sec:obs}

All of the observations were performed using LOFAR, which can observe the northern sky between 10--90 and 110--240\,MHz \citep[for a description of LOFAR see][]{2013A&A...556A...2V}. The large instantaneous bandwidth available (up to 96\,MHz) and large collecting area (up to 24 core stations in the tied-array mode) provide high-quality polarisation data at low frequencies that are well-suited to measuring precise RMs \citep[e.g.][]{2015AA...576A..62N}. 
The LOFAR observations of pulsars presented in this paper were conducted between 2012 December and 2014 November, with a typical integration time of $\ge$20 min, using the observations summarised in \citet{2016A&A...591A.134B} and \citet{2016A&A...585A.128K}.  

The signals received from between 20 and 24 of the LOFAR High-Band-Antenna (HBA) core stations were coherently combined, using the single distributed clock signal and the LOFAR correlator/beam-former, to establish a tied-array beam \citep[for a description of LOFAR's pulsar observing modes see][]{2011A&A...530A..80S}. The pulsar observations were recorded at a centre frequency of 148.9\,MHz, with 78.1\,MHz of contiguous bandwidth using the eight-bit sampling mode. The data from the observations were recorded in one of two formats. For the millisecond pulsars (MSPs), the raw complex-voltage (CV) data were recorded with a sampling time of 5.12\,$\upmu$s and a channel width of 195.3\,kHz, allowing the data to be coherently dedispersed \citep[e.g.][]{2016A&A...585A.128K}. For the slower non-recycled pulsars with larger rotational periods (allowing larger sampling times and incoherent dedispersion, resulting in smaller data volume), the Stokes $IQUV$ parameters were computed and recorded with a sampling time (1--8)$\times$163.8\,$\upmu$s and each 195.3\,kHz sub-band was split into an additional (1--8)$\times$32 channels \citep[e.g.][]{2016A&A...591A.134B}. 

The recorded data were pre-processed using the LOFAR PULsar Pipeline \citep[PULP, a Python-based suite of scripts that provides basic offline pulsar processing;][]{2016A&A...585A.128K}. The dedispersion and folding were performed using each pulsar's timing ephemeris obtained from Jodrell Bank Observatory or the Green Bank Telescope where available \citep[e.g.][]{2016A&A...586A..92P,2016A&A...591A.134B}, or the pulsar catalogue. The dedispersion and folding were performed on each frequency channel using the \textsc{DSPSR}\footnote{\href{http://dspsr.sourceforge.net}{http://dspsr.sourceforge.net}} digital signal-processing software \citep{2011PASA...28....1V} and written as a PSRFITS\footnote{\href{http://www.atnf.csiro.au/research/pulsar/psrfits/}{http://www.atnf.csiro.au/research/pulsar/psrfits/}} archive \citep{2004PASA...21..302H}. The folding produced sub-integrations of between 5\,s and 1\,min.  Radio frequency interference (RFI) in the data was initially removed in affected frequency subbands and time subintegrations using automated programs, i.e., \textsc{paz} from the \textsc{PSRCHIVE}\footnote{\href{http://psrchive.sourceforge.net}{http://psrchive.sourceforge.net}} software suite \citep{2004PASA...21..302H}, and \textsc{clean.py} from the \textsc{CoastGuard}\footnote{\href{https://github.com/plazar/coast_guard}{https://github.com/plazar/coast\_guard}} software suite \citep{2016MNRAS.458..868L}. The archives were averaged in frequency to 400 channels and stored in the LOFAR Long-Term Archive\footnote{\href{https://lta.lofar.eu/}{https://lta.lofar.eu/}}. The optimal DM and period ($P$) of the pulsars were determined previously \citep{2016A&A...591A.134B,2016A&A...585A.128K}. The data were inspected to excise obvious remaining RFI using the interactive \textsc{PSRCHIVE} \textsc{pazi} program.  Table \ref{tab:obs_table} shows a summary of the observational and corresponding RM-synthesis parameters, see Section \ref{sec:rmsynth}.

The data have not yet been polarisation calibrated, to account for the parallactic angle or the tied-array instrument beam. The polarisation calibration will be implemented in future work, which will also present the average polarisation profiles of the data described here. For aperture arrays, including LOFAR, polarisation calibration is non-trivial. There are no wavelength-squared-dependent variables in LOFAR's beam response model, especially very close to the pointing centre (where all of the observed pulsars were located), and the parallactic angle is independent of frequency \citep[e.g.][]{2015AA...576A..62N}, leaving the Faraday rotation signal in the data unaffected. Using observations of nine of the pulsars in this work, we verified that the RMs determined before the current polarisation calibration pipeline are equal to those post-calibration.  Moreover, the low-frequency data from LOFAR provide a wide bandwidth, throughout which even low RM values cause the Stokes $Q,U$ parameters to vary over many sinusoidal cycles. For example, a small RM of just 1\,rad m$^{\rm{-2}}$ causes the Stokes $Q,U$ parameters to undergo over 1.5 sinusoidal cycles across the 110--190\,MHz recorded bandwidth range. Therefore, the Faraday rotation effect that causes the sinusoidal variation in the Stokes $Q,U$ signal with wavelength squared (see Section \ref{sec:rmsynth}) is distinct compared to other possible wavelength-dependant variations due to, e.g., telescope beam effects or pulse profile evolution. LOFAR's X-Y dipoles are rotated 45 degrees away from North, so that when the  Stokes parameters are recorded, the RM sign is opposite to the IAU convention \citep{LOFARObs}. Therefore, the signs of the RMs measured in this work were flipped to be consistent with the IAU convention used in the pulsar catalogue.  The LOFAR beam model corrects this effect, so this sign flip will not be necessary post-polarisation calibration.     

 Table \ref{tab:1} provides the additional observational parameters (i.e. Modified Julian Date) for each pulsar with a measured RM.

\begin{table*}
	\centering
	\caption{Summary of LOFAR HBA data used in this work and corresponding theoretical RM-synthesis parameters.}
	\label{tab:obs_table}
%	\begin{tabular}{@{}l @{}l @{}l | @{}l @{}l} % five columns, alignment for each
	\begin{tabular}{llr} % five columns, alignment for each
		\hline
		\hline
		Parameter & Symbol & Data \\
		\hline
		Centre frequency & $\nu$ &  148.9\,MHz  \\
		Bandwidth & $\Delta\nu$ & 78.1\,MHz \\
		Channel width (averaged) & $\delta\nu$ & 195\,kHz   \\
		%\hline
		Centre wavelength squared & $\lambda^{\rm 2}$  & 4.0\,m$^{\rm 2}$ \\
		Total bandwidth in wavelength squared ($\lambda_{\rm{max}}^{\rm{2}}-\lambda_{\rm{min}}^{\rm{2}}$) & $\Delta(\lambda^{\rm 2})$ & 4.9 (7.4--2.5)\,m$^{\rm 2}$   \\
		 Resolution in Faraday space (FWHM of the RMSF) & $\delta\phi$  & 0.7\,rad m$^{\rm {-2}}$  \\
		 Largest scale in Faraday space to which one is sensitive & max-scale/$\Delta\phi$ & 1.2\,rad m$^{\rm {-2}}$  \\
		 Maximum observable Faraday depth (at $\nu_{\rm{min}}$, $\nu$, $\nu_{\rm{max}}$, respectively) & $|\phi_{\rm{max}}|$ & 66,163,327\,rad m$^{\rm {-2}}$   \\

		\hline
	\end{tabular}
\end{table*}

%%%%%%%%%%%%%%%%% RM SYNTH %%%%%%%%%%%%%%%%%%

\subsection{Determining RMs using RM-synthesis}\label{sec:rmsynth}

%Here we describe the analysis of the observations summarised in Table \ref{tb:1}. 
%Brief details on RM-synthesis.\\
%Details of RM-synthesis pipeline, error calculations, and example plots.

% Observing wavelength and RM synth parameters

A polarised wave travelling through a magnetised plasma (such as the ISM) is subject to the effect of Faraday rotation: the polarisation angle, $\chi = {0.5~{\rm tan}^{-1} (U/Q)}$\,(rad), rotates as a function of the wavelength, $\lambda$\,(m), squared:
\begin{equation}
\chi(\lambda^{\rm 2}) = \chi_{\rm 0} + {\rm RM}\lambda^{\rm 2},
\end{equation}
where the RM (rad m$^{\rm{-2}}$) characterises the magnitude of the effect, and is dependent on the integrated electron density and magnetic field strength parallel to the LoS, see Equation \ref{eq:1}. $\chi_{\rm 0}$ represents the intrinsic polarisation angle of the polarised source's emission.

\begin{figure*} %%% RMSF %%%
 \includegraphics[width=\columnwidth]{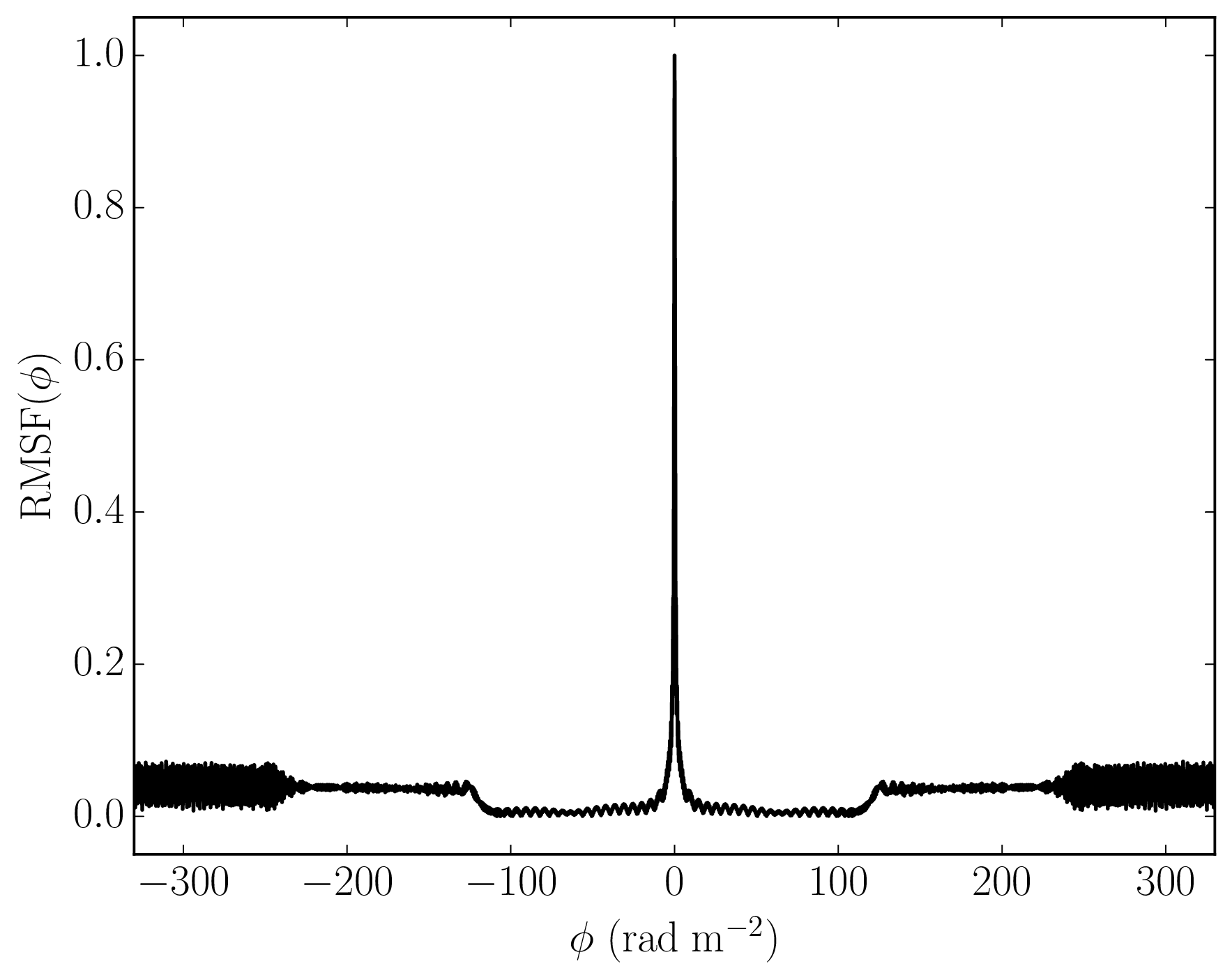}
  \includegraphics[width=\columnwidth]{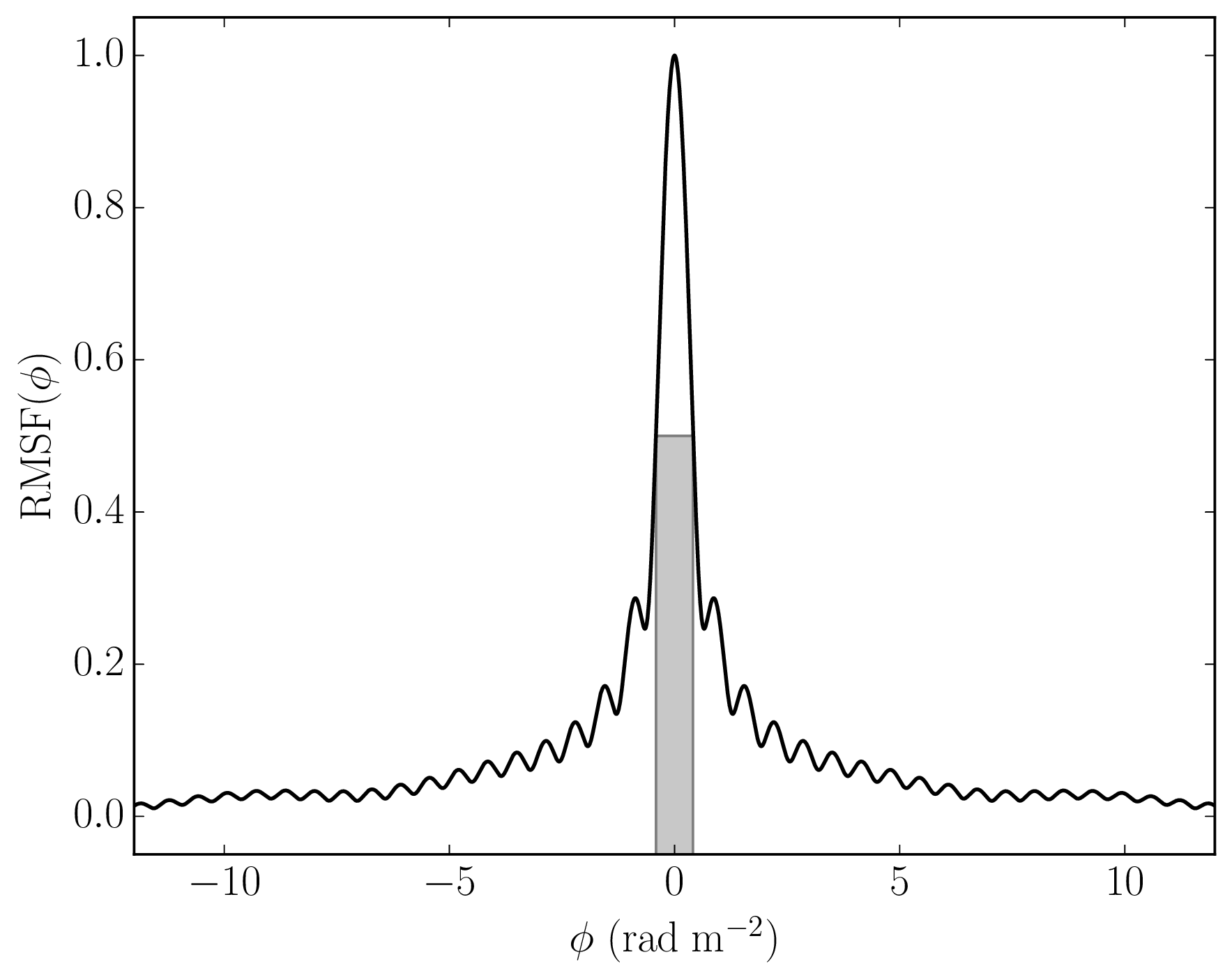}
 \caption{Theoretical, noiseless RMSF expected from RM-synthesis using the LOFAR data parameters summarised in Table \ref{tab:obs_table}, shown for both large (left) and narrow (right) ranges in Faraday space. Right: The resolution in Faraday space, $\delta\phi$, is also shown by the FWHM of the RMSF shaded in grey.    }
 \label{fig:RMSF}
\end{figure*}

It has been common practice to measure RMs by determining the gradient of the observed polarisation angle as a function of wavelength squared \citep[e.g.][]{1994MNRAS.268..497R}.  More recently, advancements in computing power have facilitated the recording of larger bandwidths from radio observations and have enabled the use of the powerful method of RM-synthesis \citep{1966MNRAS.133...67B,2005A&A...441.1217B} for investigating magnetic fields \citep[e.g.][]{2009AA...503..409H,2016ApJ...830...38L,2017A&A...597A..98V}. This method takes advantage of the Fourier-like relationship between the complex polarisation intensity vector as a function of wavelength squared, $P(\lambda^{\rm 2})=Q(\lambda^{\rm 2})+{\mathrm i}U(\lambda^{\rm 2})$, and the Faraday dispersion function (FDF; also referred to as the Faraday spectrum, $F(\phi)$) as a function of Faraday depth, $\phi$, advantageously using the entire frequency coverage of the data simultaneously for determining the Faraday depth or RM. Following \citet{2005A&A...441.1217B}, the FDF is given by:
 \begin{equation}
 \tilde{F}(\phi) \approx {K} \sum_{{i=1}}^{\rm N} \tilde{P}(\lambda^{\rm 2}_{i}) e^{{-2i\phi(\lambda^{2}_{i}-\lambda^{2}_{0})}},
 \end{equation} 
where $\lambda_{i}$ is the central wavelength of channel $i$ from the observation; $\lambda_{0}$ is the reference wavelength (ideally the weighted average of $\lambda^{\rm 2}_{i}$); $\tilde{P}(\lambda^{\rm 2}_{i}) = w_{i}{P}(\lambda^{\rm 2}_{i}) =  w_{i} \left[ Q(\lambda^{\rm{2}}_{i}) + {\rm{i}}U(\lambda^{\rm 2}_{i}) \right]$; and $K$ is the inverse sum of all weights $w_{i}$. 
We allowed the weights $w_{i}$ to be set to an equal value for all observed wavelengths.
 In the context of this work, RM and $\phi$ terms can be used interchangeably. Here, we use `$\phi$' in reference to the FDF outputs from RM-synthesis, and `RM' in reference to the measurements obtained from the FDFs (as well as from the pulsar and extragalactic catalogues). 
 See Figure \ref{fig:RMSF} for the RM spread function \citep[RMSF; analogous to an optical telescope's point spread function, formally RMTF in][]{2005A&A...441.1217B}, showing the theoretical response of the wavelength squared coverage of our LOFAR data in Faraday depth.

For each average LOFAR pulsar profile, the on-pulse Stokes parameters (for each frequency channel) were extracted using the \textsc{PSRCHIVE} \textsc{rmfit} routine. For profiles with significant signal-to-noise in linear polarisation (where S/N$_{P}>7$), the selected on-pulse region was identified as the phase bin with the largest S/N$_{P}$. 
Since the profiles were not yet corrected for Faraday rotation, many of the linear polarisation profiles were depolarised, and in these cases (S/N$_{P}\le7$) the pulse phase bin with the highest total intensity (Stokes $I$) was selected as the on-pulse phase bin.  We used a small number of on-pulse pulse profile bins to extract the Stokes parameters because this seems to minimise the instrumental polarisation fraction in the data, see below and Appendix \ref{ap:B} for further discussion.

\begin{figure*} %%% FDF for 0329 %%%
 \includegraphics[width=0.75\textwidth]{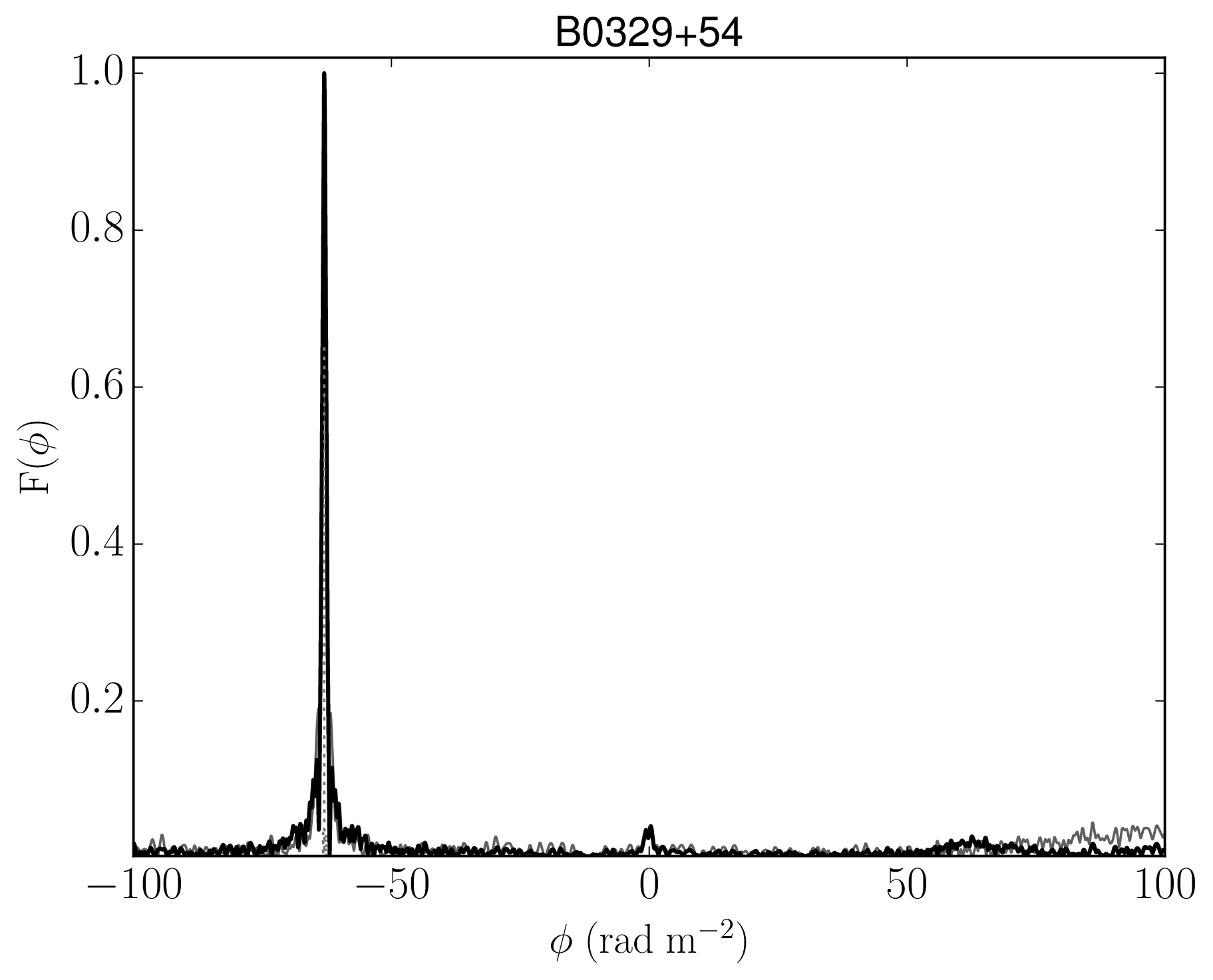}
 \caption{An example Faraday dispersion function (FDF; grey line) using a 10-minute LOFAR HBA observation towards PSR B0329+54 centred at 149\,MHz. The vertical axis is in arbitrary units (normalised) because the data are not flux or polarisation calibrated, see text for further details. The black line shows the resulting spectrum produced after ten iterations of \textsc{RM CLEAN}, stopping at the specified 8$\times$rms threshold. The dotted grey lines show the locations of the RM CLEAN components centred at RM$_{\rm{obs}} = -63.016\pm0.005$\,rad m$^{\rm {-2}}$, before correcting for the ionospheric RM. A small amount of instrumental polarisation (3 per cent) is evident near 0\,rad m$^{\rm {-2}}$. A small amount of leakage to +63\,rad m$^{\rm {-2}}$ and some RMSF structure at larger RM values can also be seen in the baseline.       }
 \label{fig:FDF_ex}
\end{figure*}

The Stokes-parameter data were used as the input to a publicly available RM-synthesis program\footnote{\href{https://github.com/gheald/RMtoolkit}{https://github.com/gheald/RMtoolkit}}, written in \textsc{python}, which computed the FDF. An example FDF obtained from a LOFAR observation of PSR B0329+54 is shown in Figure \ref{fig:FDF_ex}. The FDFs were computed in the range $-330\le\phi\le$330\,rad m$^{\rm -2}$ in steps of $\delta \phi=0.001$\,rad m$^{\rm -2}$ to oversample the FDF in Faraday space.  
The largest RM to which we have more than 50\% sensitivity to, at the centre observing frequency, is $|\phi_{\rm{max}}|=\sqrt{3}/\delta(\lambda^{\rm 2})$=163\,rad m$^{\rm -2}$.  
This quantity is dictated primarily by the observing channel width in wavelength squared $\delta(\lambda^{\rm 2})$ \citep[Equation 63 from][]{2005A&A...441.1217B}.  
The RM range of $\pm$330\,rad m$^{\rm -2}$ was computed for the FDFs because this is twice the $|\phi_{\rm{max}}|$ value. This also provided a large range in $\phi$ with which to calculate the rms noise value.
An alternative method for calculating this quantity from \citet{2015MNRAS.447L..26S} gives a somewhat similar expected value of $|\phi_{\rm{max}}|$=125\,rad m$^{\rm -2}$.  See Table \ref{tab:obs_table} for a summary of the observation and RM-synthesis parameters.

After we obtained an FDF for each pulsar, the location of the peak in Faraday space was determined and fitted using a quadratic function, providing a measurement of the RM. This was regarded as significant if the peak signal-to-noise in the FDF was greater than 4, S/N$_{{F}}>4$. This threshold was chosen because after conducting RM-synthesis with several examples of noise as the input (i.e. off-pulse data), we found that all of the peaks in the FDFs had a S/N$_{{F}}\le$3.8, although the large majority had S/N$_{{F}}<$3. 

Any possible instrumental polarisation response around 0\,rad m$^{\rm -2}$ (due to no wavelength-squared dependency) was excised from the FDFs. Instrumental polarisation can result from, for example, low levels of RFI that have not been excised in the previous data reduction steps, or from `leakage', where emission from Stokes $I$ `leaks' into the other Stokes parameters, causing the mean values of Stokes $Q,U$ to be unequal. %%check!  : $I = \frac{1}{2}(XX+YY)$ and $Q = \frac{1}{2}(XX-YY)$ 
Stokes $Q$ is generally most affected by leakage because it is formed from the same polarisation bases as Stokes $I$. This leakage can somewhat be reduced by polarisation calibrating the data, using a model for the primary beam, for example. However, the beam models and other empirical corrections used for low-frequency aperture-array telescopes such as LOFAR cannot currently completely remove this polarisation leakage \citep[e.g.][]{2017PASA...34...40L}. A major advantage in using low-frequency data to obtain RMs is that the FWHM of the RMSF is narrow, enabling identification and separation of the peaks resulting from any instrumental polarisation and the desired astronomical RM signal in the FDF. For the LOFAR data presented in this work, the theoretical FWHM of the RMSF is $\delta\phi=3.8/\Delta(\lambda^{\rm 2})\approx0.7$\,rad m$^{\rm -2}$ \citep[Equation 61 in][]{2005A&A...441.1217B,2009A&A...494..611S}. This small value for $\delta\phi$ facilitates the precise measurement of RMs, see Figures \ref{fig:RMSF} and \ref{fig:FDF_ex}. For the FDFs obtained in this work, we excised the signal within $\pm\delta\phi$ around 0\,rad m$^{\rm -2}$, i.e., $-0.7<\phi<0.7$\,rad m$^{\rm{-2}}$. This allows the measurement of absolute RMs above $\approx0.7$\,rad m$^{\rm -2}$, which applies to 99 per cent of the current pulsar catalogue. In addition, the absolute ionospheric RM is often on the order of $\approx1$\,rad m$^{\rm -2}$, see Section \ref{sec:ion}. We measured the location of the peak in the FDFs due to the instrumental polarisation near 0\,rad m$^{\rm -2}$, and a histogram of the results are shown in Figure \ref{fig:inst}. We find that the instrumental peaks in the FDFs do not deviate greatly from 0\,rad m$^{\rm {-2}}$.

The formal RM uncertainty was determined by measuring the FWHM of the signal peak in the FDF (i.e. the measured value for $\delta\phi$; FWHM$_{{F}}$), as well as the S/N$_{{F}}$. The rms noise in the FDF was calculated across the search range in RM space, excluding the peak(s) associated with the pulsar signal and the instrumental response near 0\,rad m$^{\rm -2}$.  The formal RM uncertainty was calculated as 
 \begin{equation}
\frac{\rm{FWHM_{{F}}}}{\rm{2\times S/N_{{F}}}},
 \end{equation} 
following \citet{2005A&A...441.1217B,2017MNRAS.466..378S}. 

If no significant peak in the FDF was detected, i.e., S/N$_{{F}}<4$, attempts to increase the S/N$_{{F}}$ were made by extracting the integrated Stokes parameters for an increasingly large range of the on-pulse phase bins in the average pulse profile, up to the full width at 10 per cent of the maximum of the profile.  The total number of phase bins in the pulsar profiles is mostly 1024, but range from 64 to 1024 in powers of 2, and are listed in Table \ref{tab:1}. The median number of phase bins used to extract the Stokes parameters was 3 for the `slow' pulsar census and 8 for the MSP census, these are also listed for each pulsar in Table \ref{tab:1}. The total number of bins summed were increased gradually to minimise possible effects of varying polarisation angle and/or RM across the pulse profile \citep[e.g.][]{2015AA...576A..62N}, which will be investigated in future work.  For FDFs with 4$<$S/N$_{{F}}<$8, the formal error was multiplied by 2, to reflect the larger uncertainties on the lower S/N detections. This was approximately the deviation in the RMs obtained using a subset of the pulsar data, with multiple ranges in the pulse profile bins to extract the Stokes parameters, see Appendix \ref{ap:B} for further discussion and Figure \ref{fig:bin_tests} for high and low S/N examples. FDFs with S/N$_{{F}}\ge$8 have been shown to be reliable \citep[e.g.][]{2012PASA...29..214G}.
Although RMs can also be measured as a function of pulse phase space, low-frequency observations indicate much less deviation across the pulse profile compared to higher-frequency data \citep[e.g.][]{2009MNRAS.396.1559N,2015AA...576A..62N}. 
Pulsar observations where between one and five phase bins were used to measure the RMs are good candidates for phase-resolved RM studies in future work.

%RM CLEAN down to 3$\times$rms of the noise in the FDF.
The RM-synthesis was coupled with \textsc{RM CLEAN}$^{\rm 7}$ analysis to deconvolve the FDF using the theoretical RMSF \citep{2009AA...503..409H,2018Natur.553..182M}.   \textsc{RM CLEAN} allowed us to estimate the intrinsic Faraday spectrum, to obtain the second moment of the \textsc{RM CLEAN} components (RM dispersion; $\upsigma_{\rm{RM}}$), and to investigate the extent to which the FDFs obtained are Faraday thin \citep[e.g.][]{2009AA...503..409H,2015ApJ...815...49A}. 
Faraday thickness can originate in volumes of plasmas with regular magnetic fields that both emit and Faraday rotate, or in plasmas with turbulent magnetic fields that Faraday rotate, leading to polarised flux being distributed over a range of Faraday depths \citep{1966MNRAS.133...67B}. 
A Faraday thin source satisfies $\lambda^{2}\Delta\phi\ll 1$, where $\Delta\phi$ is the extent of the source in Faraday space \citep{2005A&A...441.1217B}.  
FDFs calculated from low-frequency data provide the tightest constraint on $\Delta\phi$ because of the large $\lambda^{2}$, which is 4.9\,m$^{\rm 2}$ for the LOFAR data in this work, see Table \ref{tab:obs_table}. The largest scale in Faraday space to which one is sensitive, using these LOFAR data, is $\pi/\lambda_{min}^{\rm 2}=1.2$\,rad m$^{\rm -2}$ \citep[Equation 62 in][]{2005A&A...441.1217B}. Pulsars are not expected to be Faraday-resolved (-thick) sources because they are point sources and their magnetospheric emission is not expected to impart Faraday rotation, in particular with the usual wavelength squared dependency, due to the relativistic electron-positron pair plasma expected in their magnetospheres \citep[e.g.][]{2011MNRAS.417.1183W}. The degree of polarisation emitted often increases towards lower frequencies \citep[e.g.][]{2008MNRAS.388..261J,2015AA...576A..62N}, which is also contrary to that expected from a Faraday thick source that becomes substantially depolarised towards longer wavelengths. Despite this expectation, the RM dispersion of pulsars has not been widely explored in the literature.        

In this work, the polarisation data were normalised by a power-law model characterised by a spectral index, $\alpha$, that was fitted to the apparent Stokes I spectrum ($\mathrm{S}(\nu)\propto\nu^{-\alpha}$). 
 For each pulsar, the FDF was computed and, if a significant RM was detected, we \textsc{RM CLEAN}ed down to 3$\times$rms of the noise in the FDF. 
Any peak in the FDF near 0\,rad m$^{\rm -2}$ caused by instrumental polarisation was ignored in this analysis.  

Using the method described above, a single number for the RM (and RM dispersion) towards each pulsar was obtained for the purpose of studying the GMF, and are summarised in Table \ref{tab:1}. 
All uncertainties are labelled `$\pm$' in this paper, to avoid confusion with the RM dispersion, $\upsigma_{\rm{RM}}$.
Pulsars with published RMs in the pulsar catalogue were cross-checked with the results from this work to verify and compare the (low-frequency) RMs obtained, see Section \ref{sec:res}. 
Future work will present the calibrated polarisation profiles and phase-resolved RMs, in order to further investigate, e.g., the pulsar emission mechanism. 

%%%%%%%%%%%%%%%%% RM_ION %%%%%%%%%%%%%%%%%%

\subsection{Ionospheric Faraday rotation subtraction}\label{sec:ion}

The ionosphere, also a magneto-ionic plasma, introduces an additional RM that is both time and position dependent.   Therefore, the ionospheric RM, RM$_{\rm{ion}}$,  must be subtracted from the observed RM, RM$_{\rm{obs}}$, in order to obtain a measurement of the RM due to the ISM alone, i.e., RM$_{\rm{ISM}}$ = RM$_{\rm{obs}}$--RM$_{\rm{ion}}$.  

The RM$_{\rm{ion}}$ towards the LoS (at the corresponding ionospheric pierce point, IPP) at the time of each observation was calculated by using a previously tested and verified code, \textsc{ionFR}\footnote{https://sourceforge.net/projects/ionfarrot/} \citep[see][]{2013A&A...552A..58S}. This code models the RM$_{\rm{ion}}$ using vertical total electron content (VTEC) maps of the ionosphere (obtained using the distribution of worldwide GPS stations) plus a standard mathematical description of the Earth's main magnetic field. Several pulsar observations taken over the course of several hours have previously been used to investigate the RM$_{\rm{ion}}$ -- estimated using several publicly available VTEC maps\footnote{available at NASA's Archive of Space Geodesy Data, retrieved from \href{ftp://cddis.nasa.gov/pub/gps/products/ionex/}{ftp://cddis.nasa.gov/pub/gps/products/ionex/}.} as inputs \citep[e.g.][]{2013A&A...552A..58S,2016Sobey}.  The CODE \citep[e.g.][]{CODE2018} and IGS \citep[e.g.][]{Hernandez-Pajares2009} VTEC maps used to produce RM$_{\rm{ion}}$ estimates were found to provide a good fit to the RM$_{\rm{obs}}$. In this work, the IGS TEC maps were used because they generally have smaller uncertainties (calculated using the VTEC rms maps), which are the largest contribution to the uncertainty in RM$_{\rm{ion}}$. For this work, we also updated the \textsc{ionFR} code to use the most recent version of the International Geomagnetic Reference Field\footnote{https://www.ngdc.noaa.gov/IAGA/vmod/igrf.html} \citep[IGRF-12;][]{Thebault2015}. Comparing the \textsc{ionFR} output between using the previous and current versions of the IGRF (11 or 12, respectively), the RM$_{\rm{ion}}$ values differ by less than 2 per cent of the uncertainties (in the cases of several pulsar LoSs tested). IGRF-12 was used because it is the most recent release and provides the geomagnetic field components beyond the year 2015.

\begin{figure}
 \includegraphics[width=\columnwidth]{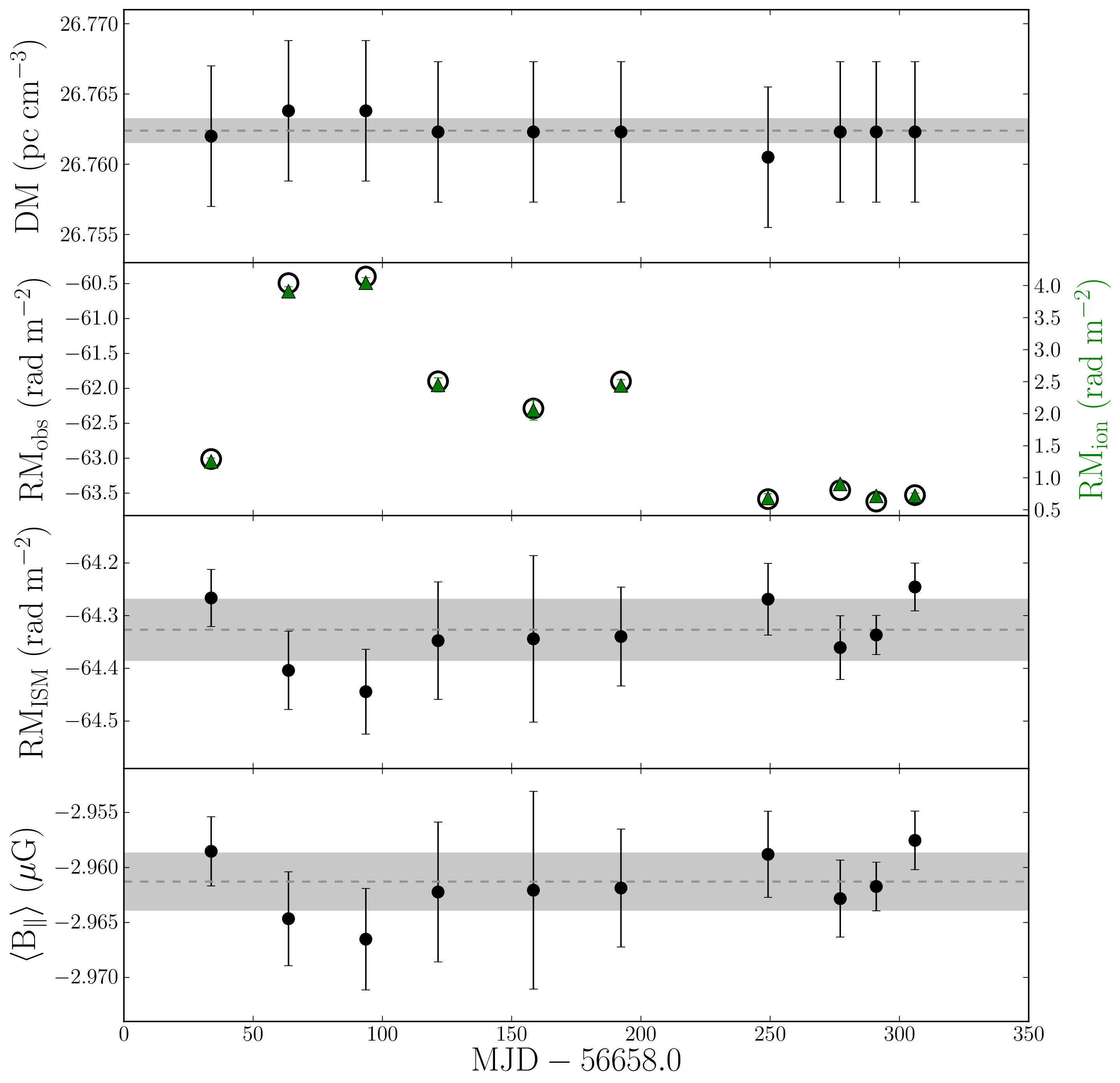}
 \caption{Measurements using the LOFAR timing observations of PSR B0329+54 centred at 149\,MHz with 78\,MHz bandwidth. From the upper to lower panels we show the following quantities plotted against Modified Julian Date from 2014 January 1 (MJD$=56658$): the measured DMs; the observed RMs (RM$_{\rm{obs}}$; open circles; left y-axis) and the estimated ionospheric RM (RM$_{\rm{ion}}$; green triangles; right y-axis) and their uncertainties; the RM$_{\rm{ISM}}$ calculated from each RM$_{\rm{obs}}-$RM$_{\rm{ion}}$; and the electron density-weighted average magnetic field parallel to the LoS, calculated using Equation \ref{eq:1} and the DM and RM$_{\rm{ISM}}$. The weighted means of the relevant measurements are shown by the grey dashed lines, and one standard deviation from the mean is shown by the grey shaded areas.    }
 \label{fig:0329_ion}
\end{figure}

In this work, we also investigated the repeatability of this method for estimating RM$_{\rm{ion}}$ over a longer timespan ($\approx$year), and the accuracy of the resulting RM$_{\rm{ISM}}$ measurements that can be expected. Ten (timing/monitoring) observations of a bright, northern pulsar B0329+54, were used to compare the RM$_{\rm{obs}}$ and RM$_{\rm{ion}}$ over the course of nine months from 2014 February 3 to 2014 November 3. These observations used the same parameters as those shown in Table \ref{tab:obs_table}. We measured the DM for each observation using the \textsc{pdmp} routine in \textsc{PSRCHIVE} and corrected the archive file for this value. We then measured the RM$_{\rm{obs}}$ using the method described in Section \ref{sec:rmsynth}. We used the IGS VTEC maps and IGRF-12 as inputs to the \textsc{ionFR} code to estimate RM$_{\rm{ion}}$ for each average observation time (and corresponding IPP) for PSR B0329+54. We do not apply the  ionospheric RM correction to the pulsar archive files on timescales shorter than one observation. This is because the data are not severely depolarised over the short integration time (just $\approx$10--20 minutes in length) and the pulsar signals are often highly polarised. Observations could be corrected on shorter timescales, if particularly active ionospheric conditions dictate that this is necessary.  However, the IGS VTEC maps we use have a time resolution of 2 hours (interpolated to every hour in the \textsc{ionFR} code) and we find that a single RM$_{\rm{ion}}$ correction per observation is sufficient for this work.   We corrected RM$_{\rm{obs}}$ using the RM$_{\rm{ion}}$ values output from \textsc{ionFR}, and also calculated the inferred electron-density-weighted $\langle B_{\parallel}\rangle$ using Equation \ref{eq:1}. The results are shown in Figure \ref{fig:0329_ion}.
 
The weighted mean of the DM measurements of PSR B0329+54, shown in Figure \ref{fig:0329_ion}, is 26.7624\,pc cm$^{\mathrm{-3}}$ with a standard deviation of 0.0008\,pc cm$^{\mathrm{-3}}$.  The DM measurements are not corrected for the ionospheric DM because the ionosphere imparts a negligible amount of DM compared to the measurement uncertainties here. Figure \ref{fig:0329_ion} shows that the variation in the RM$_{\rm{ion}}$ estimates appears to be a good fit to that for the RM$_{\rm{obs}}$ measurements.   The weighted mean of the RM$_{\rm{ISM}}$ values is --64.33\,rad m$^{\rm{-2}}$ with standard deviation 0.06\,rad m$^{\rm{-2}}$.   This RM$_{\rm{ISM}}$ value is in agreement with ($<$1.4$\times$ uncertainties), and approximately 7-times more precise than, the literature value from the pulsar catalogue: RM$_{\rm{cat}}$=--63.7$\pm$0.4\,rad m$^{\rm{-2}}$. The catalogue value was  measured using observations at several centre frequencies between 250--500\,MHz, and taken over 40 years before the LOFAR observations \citep{1972ApJ...172...43M}. \citet{1972ApJ...172...43M} corrected the data for RM$_{\rm{ion}}$ using continuous records of the total ionospheric Faraday rotation obtained towards a geostationary satellite. 
We also calculated the weighted mean of the  $\langle B_{\parallel}\rangle$ values, --2.961$\upmu$G, with standard deviation 0.003\,$\upmu$G. This is also in good agreement (within 1.3 times the uncertainties) with the value published in \citet{1972ApJ...172...43M}. 
In addition, we found that the median RM dispersion for all of the FDFs to be $\upsigma_{\rm{RM}}$=0.0005\,rad m$^{\rm{-2}}$, which is Faraday thin. The RM dispersions for the whole pulsar set are further discussed in Section \ref{sec:rmdisp}.  

The mean uncertainties on the values for RM$_{\rm{obs}}$ and RM$_{\rm{ion}}$ are 0.01 and 0.08\,rad m$^{\rm{-2}}$, respectively. This is an example of the excellent precision of the measurements that can be obtained from using the low-frequency LOFAR data. This also indicates that the RM$_{\rm{ISM}}$ uncertainty is dominated by the current method for correcting RM$_{\rm{ion}}$. The data in Figure \ref{fig:0329_ion}, along with previous work, show that we can expect reasonably high accuracy in the measurement of RM$_{\rm{ISM}}$ using the current method for correcting for RM$_{\rm{ion}}$, e.g., the standard deviation in the RM$_{\rm{ISM}}$ values for PSR B0329+54 is just 0.06\,rad m$^{\rm{-2}}$. However, it is clear that more sophisticated methods that are being developed for more accurately determining RM$_{\rm{ion}}$ will be essential towards fully realising the RM precision possible using low-frequency data \citep[e.g.][]{doi:10.1029/2018RS006559}, in particular for the LOFAR Low-Band Antenna observations.  This will also allow us to fully realise higher $\langle B_{\parallel}\rangle$ precision, since the fractional uncertainties on the RM measurements are currently a factor of $\approx$100 larger than the fractional uncertainties on the DM measurements.

The results published in Table \ref{tab:1} include the RM$_{\rm{obs}}$ measured, the RM$_{\rm{ion}}$ estimates, and the resulting RM$_{\rm{ISM}}$ and uncertainties, along with the dates and times of the observations. This is to provide the possibility of applying more advanced ionospheric Faraday rotation corrections that may become available in the future.

%The ionosphere imparts additional Faraday rotation, RM$_{\rm{ion}}$, to the total observed RM, RM$_{\rm{obs}}$, that must be subtracted in order to obtain the RM due to the ISM alone, RM$_{\rm{ISM}}$ = RM$_{\rm{obs}}$--RM$_{\rm{ion}}$. The ionospheric RM is both time and position dependent and was, therefore, calculated towards each target source line-of-sight (at the corresponding ionospheric pierce point) at the average time of each observation. We used an updated version of \textsc{ionFR}\footnote{https://sourceforge.net/projects/ionfarrot/} \citep[][]{2013A&A...552A..58S}, with input data from the latest version of the International Geomagnetic Reference Field\footnote{https://www.ngdc.noaa.gov/IAGA/vmod/igrf.html} (IGRF12; Thébault et al. 2015) and the International GNSS Service vertical total electron content maps\footnote{ftp://cddis.gsfc.nasa.gov/pub/gps/products/ionex/} for each observation date (e.g. Hernández-Pajares et al. 2009).

%%%%%%%%%%%%%%%%% RESULTS %%%%%%%%%%%%%%%%%%

\section{Results}\label{sec:res}

%Here we provide the results obtained using the LOFAR observations.
%%%%%%%%%%%%%%%%% RM %%%%%%%%%%%%%%%%%%
%\subsection{Faraday rotation measures towards pulsars using LOFAR}

Here, we summarise the results of the RMs determined towards pulsars using the LOFAR observations described. 

We determined the RMs towards 117 pulsars in the non-recycled pulsar census and 19 pulsars in the MSP census (136 in total), presented in Table \ref{tab:1}. This represents 74 and 40 per cent of all of the pulsars detected in each of the LOFAR censuses, respectively, and 86 and 60 per cent of the pulsars with total intensity S/N greater than 10, respectively. The pulsars in these censuses were not selected based on their polarisation characteristics \citep[but on position accuracy, Galactic location, see][]{2016A&A...591A.134B,2016A&A...585A.128K}. Therefore, they form a reasonably representative sample of the proportion of pulsars that can be used to measure RMs below 200\,MHz. The FDFs obtained towards each pulsar are shown in Appendix \ref{ap:B}.

The pulsars with RMs measured in this work are located between 14 and 321 degrees in Galactic longitude (although the majority are between 20--200 degrees) and --58 and 86 in Galactic latitude. The pulsar closest to the Galactic plane is PSR B1937+21, at Galactic latitude --0.29 degrees.  The DM range is between 3 and 161\,pc cm$^{\rm{-3}}$ (PSRs J1744--1134 and B2036+53, respectively). The RM$_{\rm{ISM}}$ range is between --168.7$\pm$0.1 and 154.9$\pm$0.2\,rad m$^{\rm{-2}}$ (PSRs B2210+29 and B1848+13, respectively). The smallest absolute RM measurement, RM$_{\rm{ISM}}$=0.96$\pm$0.09\,rad m$^{\rm{-2}}$, is towards PSR J0435+2749.   The range in average magnetic field strength parallel to the LoS calculated is between $-5.635\pm$0.009 and 3.524$\pm$0.004\,$\upmu$G (for PSRs J2043+2740 and B1842+14, respectively). The uncertainties on these measurements provide examples of the accuracy that can be obtained for $\langle B_{\parallel}\rangle$, towards studying the Galactic magnetic field using the methods described.  

Distance estimates for the pulsars (those without independent distance measurements) are calculated using the recently published Galactic electron density model from \citet{2017ApJ...835...29Y}, hereafter referred to as YMW16\footnote{\href{www.xao.ac.cn/ymw16/}{www.xao.ac.cn/ymw16/}}, this is further discussed in Section \ref{sec:lim}. The distance estimates for pulsars with RM measurements range from 0.2 to 25\,kpc (although the few pulsars with 25\,kpc distances are likely overestimated, as this is the limit given by the model, see Section \ref{sec:lim}). The distances of the pulsars from the Galactic plane used in the analysis of the magnetic scale height of the Galactic halo is between --3 and 4\,kpc, see Section \ref{sec:scale}. 

\subsection{RM dispersion results}\label{sec:rmdisp}

%%%RMdispersion begin

\begin{figure}
 \includegraphics[width=\columnwidth]{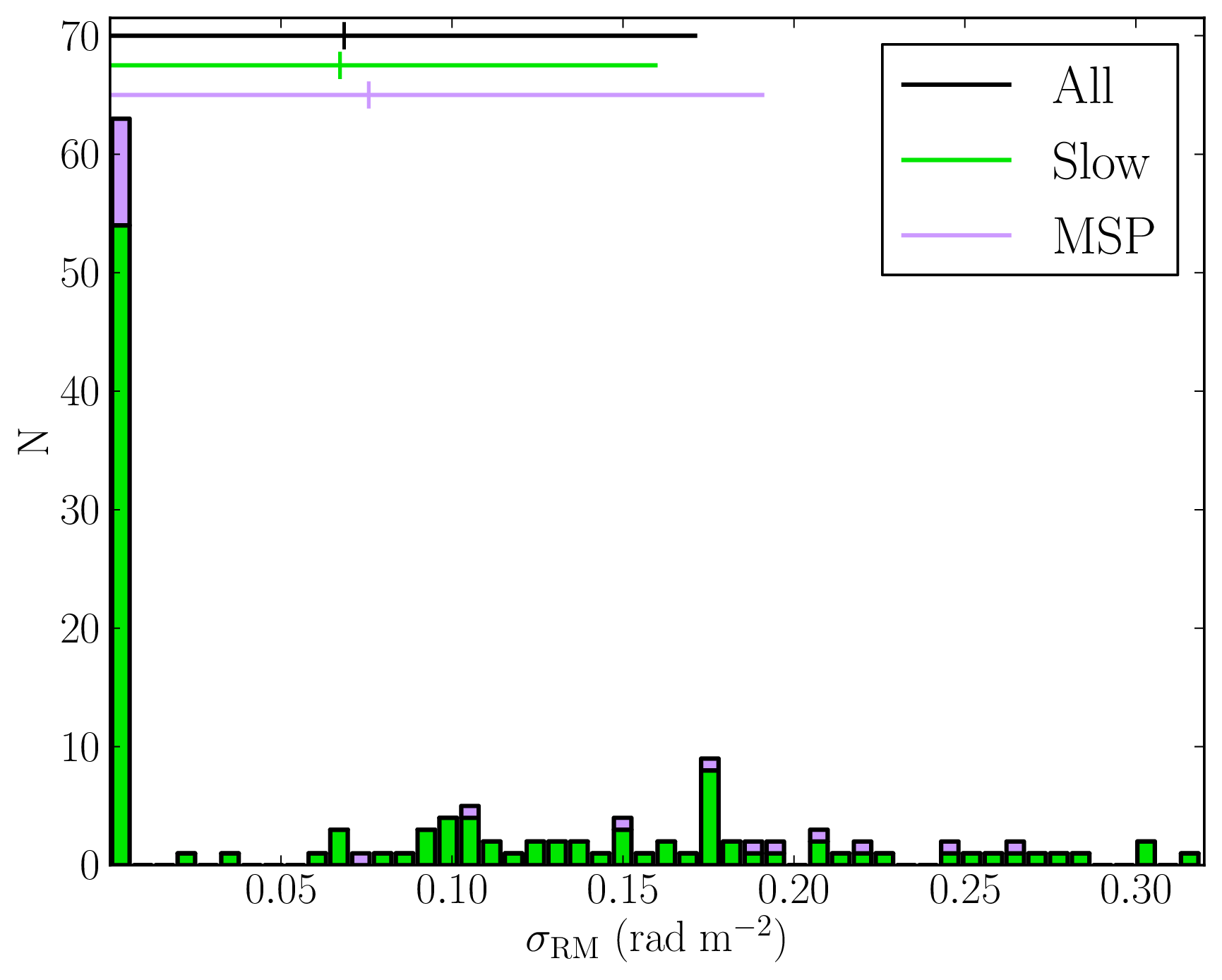}
 \caption{A stacked histogram showing the RM dispersion ($\upsigma_{\rm{RM}}$) for the pulsars in the `slow' pulsar (green) and MSP (purple) censuses. The horizonal lines with corresponding colours above mark the range between the 25th and 75th percentiles, with the median shown by the vertical line (the complete pulsar set is shown by the black line).  }
 \label{fig:disp}
\end{figure}

Figure \ref{fig:disp} summarises the RM dispersion, $\upsigma_{\rm{RM}}$, calculated by running \textsc{RM CLEAN} on the FDFs obtained towards each pulsar. The RM dispersion results for each pulsar are also included in Table \ref{tab:1}. The RM dispersion bin with the largest number of pulsars (63) represents values less than 0.0009\,rad m$^{\rm -2}$. This is $<1$ per cent of the typical RM uncertainty. Eight pulsars with S/N$_{{F}}>$7 have RM dispersions that are indistinguishable from 0\,rad m$^{\rm -2}$, i.e., all of the components from \textsc{RM CLEAN} fall within the same pixel value after at least one \textsc{RM CLEAN} iteration. These are PSRs B0331+45, B0523+11, J0611+30, B0940+16, J1612+2008, J1645+1012, J1741+2758, and B2110+27. The median RM dispersion for the RMs in this work (from both censuses) is 0.068\,rad m$^{\rm -2}$, which is less than 10 per cent of $\delta\phi$, see Table \ref{tab:obs_table}. Furthermore, all of the values obtained are less than 40 per cent of $\delta\phi$. The RM dispersions of 75 per cent of the pulsars in this work are less than 0.17\,rad m$^{\rm -2}$ and satisfy the Faraday thin criterion, $\lambda^{2}\Delta\phi\ll 1$, even at the longest observing wavelength ($\lambda^{2}=7.4$\,m$^{\rm 2}$). For the few pulsars that have larger RM dispersions, it is likely that this is largely due to \textsc{RM CLEAN}-ing too deeply in the presence of noise. All of the pulsars in this work satisfy $\lambda^{2}\Delta\phi\ll 1$ at the shortest observing wavelength ($\lambda^{2}=2.5$\,m$^{\rm 2}$). In addition, there is no evidence for emission at more than one RM value to a high degree of confidence, also as expected in FDFs obtained using pulsar data.

We investigated possible correlations between the RM dispersion and a number of parameters. We found no significant correlation between the RM dispersion and: the number of bins used to extract the Stokes parameters from the pulse profiles; the S/N in the FDFs; the DMs; the absolute value of the measured RMs; or the uncertainty on the RMs.
We also obtained published scattering measures, $\tau_{\rm{sc}}$, at 1\,GHz from the pulsar catalogue for 54 of the slow pulsars and 3 of the MSPs \citep[][and references therein]{2005AJ....129.1993M} to investigate any correlation with the RM dispersion. Again, we found no significant correlation trend between these two variables. However, two pulsars with the largest scattering times also have among the largest RM dispersions measured. These are the Crab pulsar PSR B0531+21 with $\tau_{\rm{sc}}$=1.2$\times$10$^{-4}$\,s at 1\,GHz and $\upsigma_{\rm{RM}}$=0.25\,rad m$^{\rm -2}$; and PSR B1946+35 with $\tau_{\rm{sc}}$=4.1$\times$10$^{-4}$\,s at 1\,GHz and $\upsigma_{\rm{RM}}$=0.30\,rad m$^{\rm -2}$, see Section \ref{sec:ind} for further discussion for the Crab pulsar.  Whether some of the larger RM dispersions may in fact be the result of foreground diffuse ISM structures affecting the signal will require a more comprehensive study, using the calibrated polarisation profiles as a function of observing frequency. 

The RM dispersions obtained using these LOFAR data show that the majority of the signals from the pulsars are extremely Faraday thin, as we expected. These low-frequency observations provide the most stringent information about the extent to which pulsars are Faraday thin, to date.

%%%RMdispersion end

%%%%%%%%%%%Discussion
%%%%%%%%%%%%%%%%% DISCUSSION %%%%%%%%%%%%%%%%%%

\section{Discussion}\label{sec:disc}

Here, we discuss the RM results that we obtained using the LOFAR observations and provide further analysis in combination with literature RM measurements.
In Section \ref{sec:det} we examine the nature of the data for pulsars towards which we did not detect a significant RM. 
In Section \ref{sec:comp} we compare the RMs that we measured at low-frequencies to measurements from the literature towards the same pulsars, as well as towards extragalactic sources along the same LoSs to provide further context. In Section \ref{sec:ind} we briefly comment on the results for some individual pulsars.  In Section \ref{sec:scale} we further analyse the RMs towards pulsars to give estimates of the magnetic field scale height in the Galactic halo. In Section \ref{sec:lim} we discuss the limitations of the data we currently have to study the Galactic magnetic field structure, and prospects for future improvements.

%%%discussion for not detecting RMs begin %%
\subsection{Detecting and measuring RMs}\label{sec:det}

For the pulsars where we did not detect a significant RM with S/N$_{{F}}>$4, the primary reason seems to be lower S/N in the pulse profile (resulting in lower S/N$_{\rm{P}}$ in the pulse profile). The median S/N for the slow pulsars and MSPs that we did detect an RM for are 67 and 19, respectively. Meanwhile, the slow pulsars and MSPs that we did not detect an RM for have median S/N of 10 and 8, respectively. However, the lowest total intensity pulse profile S/N for which we did detect a significant RM is just 4 (for PSR J2002+1637), as this pulsar is highly polarised. 
For 14 pulsars (10 slow and 4 MSPs), we detected a tentative RM in the FDFs, but these were below the chosen S/N threshold set, S/N$_{{F}}>$4, and so are not included in the final catalogue. Half of these pulsars do not have an RM published in the literature, and it is likely that a significant RM would be detected if a longer integration time is used to increase the S/N in the average pulse profiles. 
Furthermore, the median DM of the pulsars with significant RMs is 35\,pc cm$^{\rm{-3}}$, while the median DM of the pulsars with no significant RM detection is almost double: 61\,pc cm$^{\rm{-3}}$.  For the `slow' pulsars, this is due to larger distances to the pulsars, which also reduces the flux density: the median distance estimates for pulsars with or without a significant RM detection are 2.2\,kpc and 3.9\,kpc, respectively.  For the MSPs, this is not necessarily the case: the median distance estimates for MSPs with or without a significant RM detection are 1.2\,kpc and 1.3\,kpc, respectively.   Therefore, is likely that dispersion smearing and/or scattering in some of the pulse profiles may also have an effect on the S/N and degree of polarisation of the signals, see below.  
The S/N$_{\rm{P}}$ could be increased for pulsars that have been observed at least once (e.g. for timing/monitoring campaigns) by adding the datasets together (after correcting for ionospheric RM variations).   

Several of the pulsars show evidence of scattering tails, in some extreme cases these increase to span almost the entire pulse phase \citep[see][]{2014A&A...572A..52B,2017MNRAS.470.2659G}. Whether we can detect an RM in these data seems to vary on a case by case basis. In some cases where the pulsar is reasonably polarised at low frequencies we detected a significant RM, although more pulse phase bins may need to be summed (e.g. for PSR B1946+35 we summed 70 phase bins, or 7 per cent of the rotational period, to gain sufficient S/N). For others, the scattering may contribute to depolarising the signal towards low frequencies, e.g., PSR B2053+36 has RM$_{\mathrm{cat}}$=--68.00\,rad m$^{\rm{-2}}$, but we were unable to detect a significant RM at our observing frequency.  Furthermore, 29 of the pulsars towards which we did not detect a significant RM also do not have an RM published in the literature, listed in Table \ref{tab:non_det}. It may be that the signal from some of these pulsars is intrinsically unpolarised, possibly due to the emission-beam--LoS geometry, orthogonal polarisation modes, or spin-down luminosity \citep[e.g.][]{1969ApL.....3..225R,1975ApJ...196...83M,2018MNRAS.474.4629J}. 

For six pulsars, the peak in the FDF was within the range set for any expected instrumental polarisation ($-0.77<\phi<0.77$\,rad m$^{\rm{-2}}$) and since we cannot (using the current data processing methods) be confident whether the majority of the polarised signal is from the pulsar or the instrument, these results were not included in our catalogue. Most of these pulsars have small RM values (e.g. for PSR B1237+25 RM$_{\mathrm{cat}}$=--0.12$\pm$0.06\,rad m$^{\rm{-2}}$) or have large fractional uncertainties (e.g. for PSR J1503+2111 RM$_{\mathrm{cat}}$=--5$\pm$10\,rad m$^{\rm{-2}}$).  Limits on the RMs for these pulsars could have been included by using the ionospheric RM calculated (mostly less than 1\,rad m$^{\rm{-2}}$). However, we can obtain more reliable measurements by re-observing these pulsars when the ionosphere is more ionised and imparts a more substantial RM$_{\rm{obs}}$, e.g. during the day.  

Six of the pulsars for which we did not detect a significant RM have absolute catalogued RMs larger than the expected maximum observable RM (i.e. $\pm$163\,rad m$^{\rm{-2}}$, where $\approx$50 per cent sensitivity is lost due to bandwidth depolarisation, at the central observing frequency, see Table \ref{tab:obs_table}).  The largest absolute RMs that we measured are approximately equal to this value: PSR J2017+2043 with RM$_{\mathrm{obs}}$=--160.93\,rad m$^{\rm{-2}}$ and PSR B2210+29 with RM$_{\mathrm{obs}}$=--165.23\,rad m$^{\rm{-2}}$. This is because the upper half of the bandwidth (149--188\,MHz) can still be used to detect a significant signal (albeit with a slightly larger FWHM$_{F}$). For these pulsars, the RM measurement was verified and the S/N$_{{F}}$ increased by downloading higher frequency resolution data, available from the LOFAR Long-Term Archive (down to 30\,kHz channel width). These data raise the maximum expected observable RM to $\pm$1000\,rad m$^{\rm{-2}}$ at the central observing frequency. The higher frequency resolution data was not used for all of the census pulsars because this is not necessary for pulsars with lower absolute RMs (this applies to the majority of the pulsars in the sample) and the data volume is over 6 times larger, increasing the download and processing times.  We also downloaded the higher frequency resolution data and obtained FDFs for the six pulsars with larger catalogued RMs, but none resulted in a significant detection using the standard or higher frequency resolution data.  

%%%discussion for not detecting RMs end %%%

\begin{table}
	\centering
	\caption{Summary of pulsars towards which we did not detect a significant RM in this work, and which also do not have an RM published in the literature.}
	\label{tab:non_det}
%	\begin{tabular}{@{}l @{}l @{}l | @{}l @{}l} % five columns, alignment for each
	\begin{tabular}{llll} % five columns, alignment for each
		\hline
		\hline
		PSR name & PSR name & PSR name & PSR name  \\
		\hline
		`Slow' census & =  15 & &   \\
		\hline
		  J0006+1834 & J1834+10  & J1908+2351  &  J2040+1657    \\
		  J0324+5239 &  J1848+0826 &  J1913+3732 &  J2048+2255  \\
		  J0329+1654 & J1900+30  & J1937+2950  &  J2243+1518   \\
		  J0711+0931 &  J1901+1306 & B2025+21  &    \\
		\hline
		MSP census & = 14 &  & \\
		\hline
		 J0337+1715 &  J1544+4937 & J1905+0400  & J2215+5135    \\
		  J1023+0038 &  J1709+2313 &  J1918-0642 &  J2302+4442   \\
		   J1038+0032  & J1738+0333  & B1957+20  &   \\
		  J1231-1411 & J1816+4510  & J2019+2425  &     \\
		\hline
	\end{tabular}
\end{table}

\subsection{Comparison of results with literature measurements}\label{sec:comp}

 %% discussion about agreement 

\begin{figure}
 \includegraphics[width=\columnwidth]{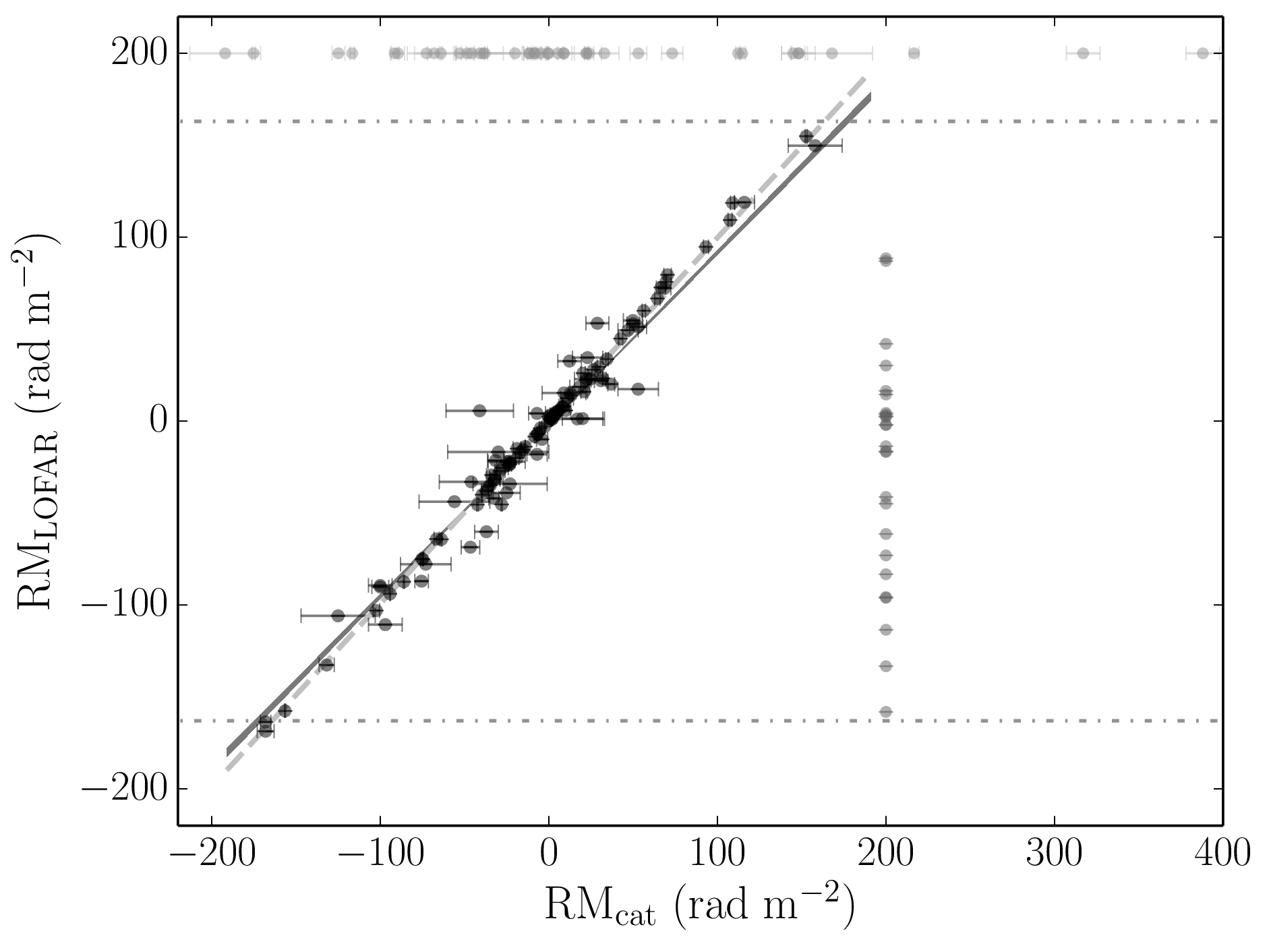}
 \caption{A comparison between the RMs obtained using LOFAR in this work, RM$_{\rm{LOFAR}}$, and the literature, RM$_{\rm{cat}}$, (black points). The expected trend RM$_{\rm{cat}}$=RM$_{\rm{LOFAR}}$ is shown by the light grey dashed line. The line of best fit with uncertainties is shown by the dark grey line. The horizontal dash-dotted lines shows the $|\phi_{\rm{max}}|$ at $|$RM$_{\rm{LOFAR}}|$=163\,rad m$^{\rm -2}$ for the centre frequency 149\,MHz, see Table \ref{tab:obs_table}. The distribution of RMs obtained in this work without previous measurements are shown by the dark grey points at a constant RM$_{\rm{cat}}$=200\,rad m$^{\rm -2}$.  The distribution of RMs not detected in this work, but with literature measurements are shown by the light grey points at RM$_{\rm{LOFAR}}$=200\,rad m$^{\rm -2}$.}
 \label{fig:RMcomp}
\end{figure}

We verified the RM results we obtained via a comparison with the RMs available in the literature (for 111 pulsars), see Figure~\ref{fig:RMcomp}. The comparison shows that there is very good agreement between the RM measurements. The line of best fit gradient is very close to the expected value of 1 (0.94$\pm$0.01; or 0.99$\pm$0.02 for an unweighted fit). Furthermore, the majority of measurements (for 102 pulsars) agree within 4 times the published uncertainties. This indicates that the RM uncertainties from the literature are likely generally underestimated. In the few cases with larger discrepancies, either the pulsar catalogue RM had no uncertainty reported (and listed as 0\,rad m$^{\rm -2}$), or both measurements have very small uncertainties and may have underestimated possible systematics due to, e.g., the ionospheric RM correction. 
This comparison of the RM measurements demonstrates that the RMs measured at low frequencies do not appear to show any disparity compared to those measured at higher frequencies. This confirms that there is no frequency-dependence in RM measurements and that the pulsar magnetosphere does not contribute to the observed RM, to the level of uncertainty to which we can currently measure, as we expected \citep[e.g.][]{2011MNRAS.417.1183W}. 
% gradient indication that higher frequency data affected by instrumental leakage?

 %% discussion about LOFAR agreement 
Furthermore, some independent RM measurements using LOFAR observations were published by \citet{2015AA...576A..62N}, 12 of which overlap with this work, see Table \ref{tab:refs}.  Comparing the LOFAR RMs from \citet[][]{2015AA...576A..62N} and this work, the measurements generally agree within the uncertainties; there is a range in agreement from 0.02--2-times the uncertainties. This indicates that we derive reasonable uncertainties on the RMs by adding (in quadrature) the formal uncertainty on the RM measurement from the FDFs and the uncertainty on the RM$_{\rm{ion}}$ corrections. 

%%%Rm_LOF vs RM_psrcat %%%

\begin{figure*}
 \includegraphics[width=1.7\columnwidth]{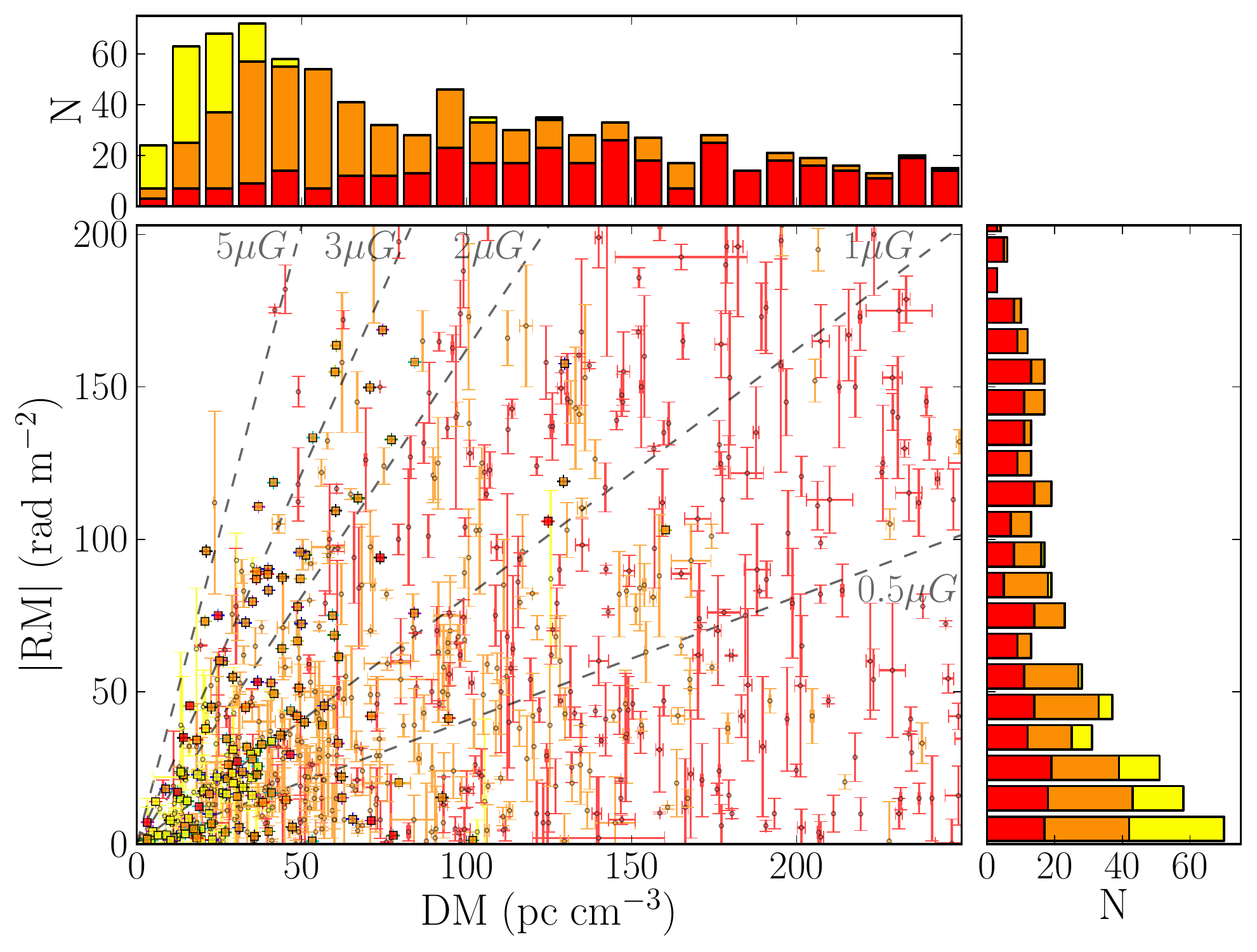}
 \caption{Modulus RM values plotted against DMs for the pulsars with measurements from LOFAR (squares) and the pulsar catalogue (points). The colours represent pulsars at different Galactic latitudes ($b$) above and below the Galactic plane: $|b|<$5 degrees (red); 5$\le|b|<$30 degrees (orange); and $|b|\ge$30 degrees (yellow). Lines of constant $| \langle B_{\parallel}\rangle |$ are shown by the grey dashed lines and corresponding labels. The stacked histograms above/right show the corresponding distribution of all DM and $|$RM$|$ measurements (with identical usage of colour). }
 \label{fig:DMRM}%Red, orange, and yellow
\end{figure*}

 %% discussion about  accuracy
The precision of the RM (and DM) measurements obtained using the low-frequency LOFAR data are illustrated in Figures~\ref{fig:RMcomp} and \ref{fig:DMRM}. For the 111 pulsars with RMs in both the literature and in this work, the mean uncertainty from the pulsar catalogue and this work are 3.4\,rad m$^{\rm -2}$ and 0.1\,rad m$^{\rm -2}$, respectively. The median fractional uncertainties on the RM measurements are 6.4 and 0.3 per cent, respectively. This indicates that the RMs measured using the low-frequency data are over 20-times more precise (on average) than literature measurements.  

 %% discussion about  uncertainties on RM
Currently, the most substantial contribution to the cumulative uncertainty on the low-frequency RM measurements is the ionospheric RM correction.  The median uncertainty on the RM$_{\rm{obs}}$, before ionospheric RM subtraction, is 0.02\,rad m$^{\rm -2}$. The median uncertainty for the RM$_{\rm{ion}}$ estimates is 0.08\,rad m$^{\rm -2}$. If future ionospheric corrections can be improved to the relative precision of the observed RM measurements, the uncertainties in the RM measurements can be improved by an order of magnitude. This will facilitate, for example, precise monitoring of RMs over time to measure and characterise small-scale magneto-ionic structures in the ISM.  

 %% discussion about  uncertainties on RM
Figure~\ref{fig:DMRM} shows the DM and modulus RM values towards the pulsars observed in this work, as well as from the pulsar catalogue, for $\rm{DM}<250$\,pc cm$^{\rm{-3}}$ and $|\rm{RM}|<204$\,rad m$^{\rm{-2}}$. Pulsars located towards low, medium, and high Galactic latitudes are identified using red, orange, and yellow colours, respectively. All of the pulsars with measured RMs from LOFAR are shown, along with the pulsars with literature measurements (64 per cent of the pulsar catalogue within the DM and RM ranges). Also shown are stacked histograms for the DMs and modulus RMs for all of the pulsars in the scatter plots. The distributions approximate log-normal and power-law distributions, respectively. Pulsars towards higher Galactic latitudes tend to have lower DM and RM values, but are still distributed across a range in Galactic magnetic field strengths ($\lessgtr\pm$4$\upmu$G), similar to the mid and low ranges ($\lessgtr\pm$6$\upmu$G).

%%%DM vs RM %%%

\begin{figure*}
  \includegraphics[width=\textwidth]{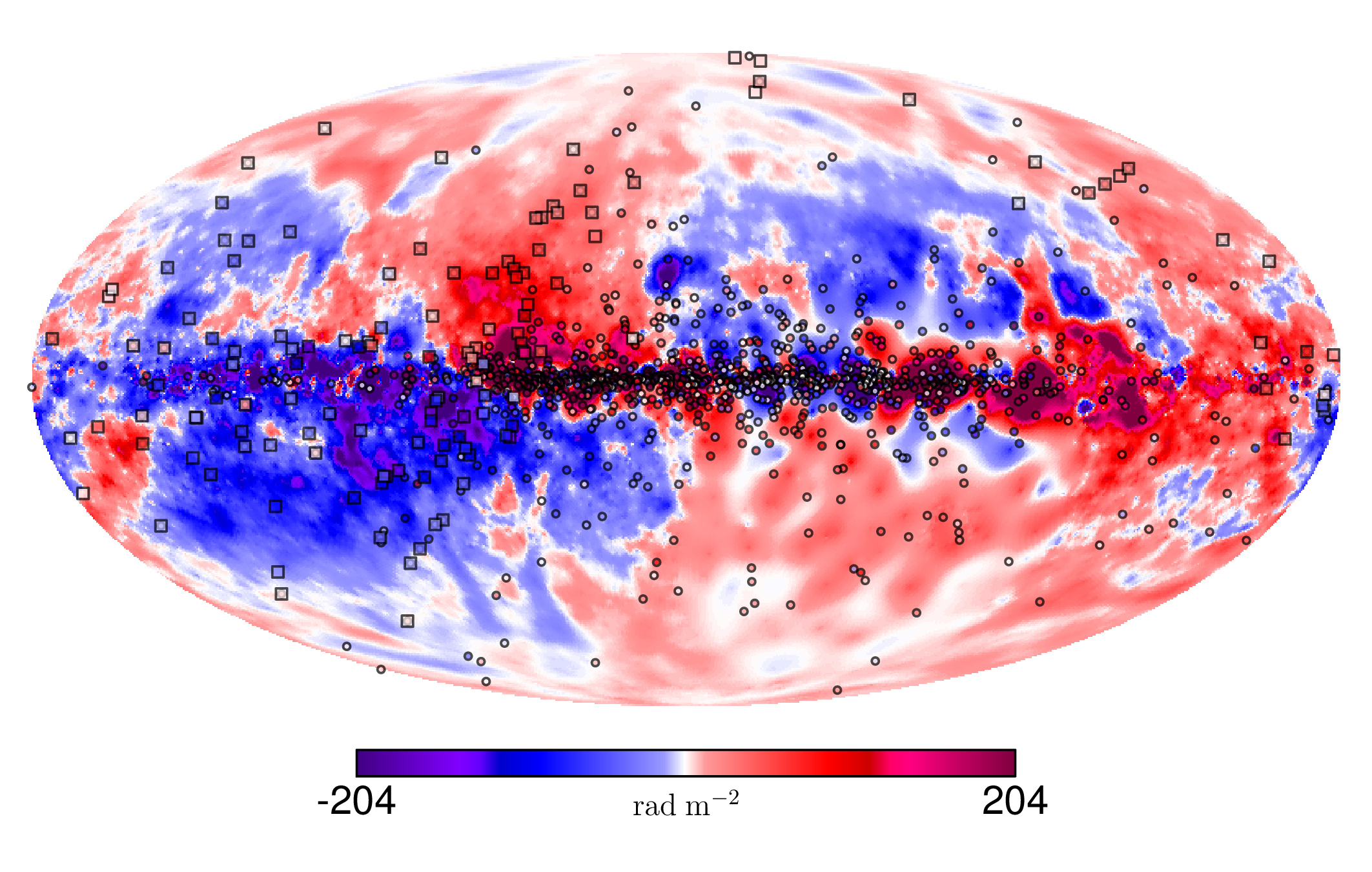}
  
  \vspace{-0.9cm}
  
    \includegraphics[width=\textwidth]{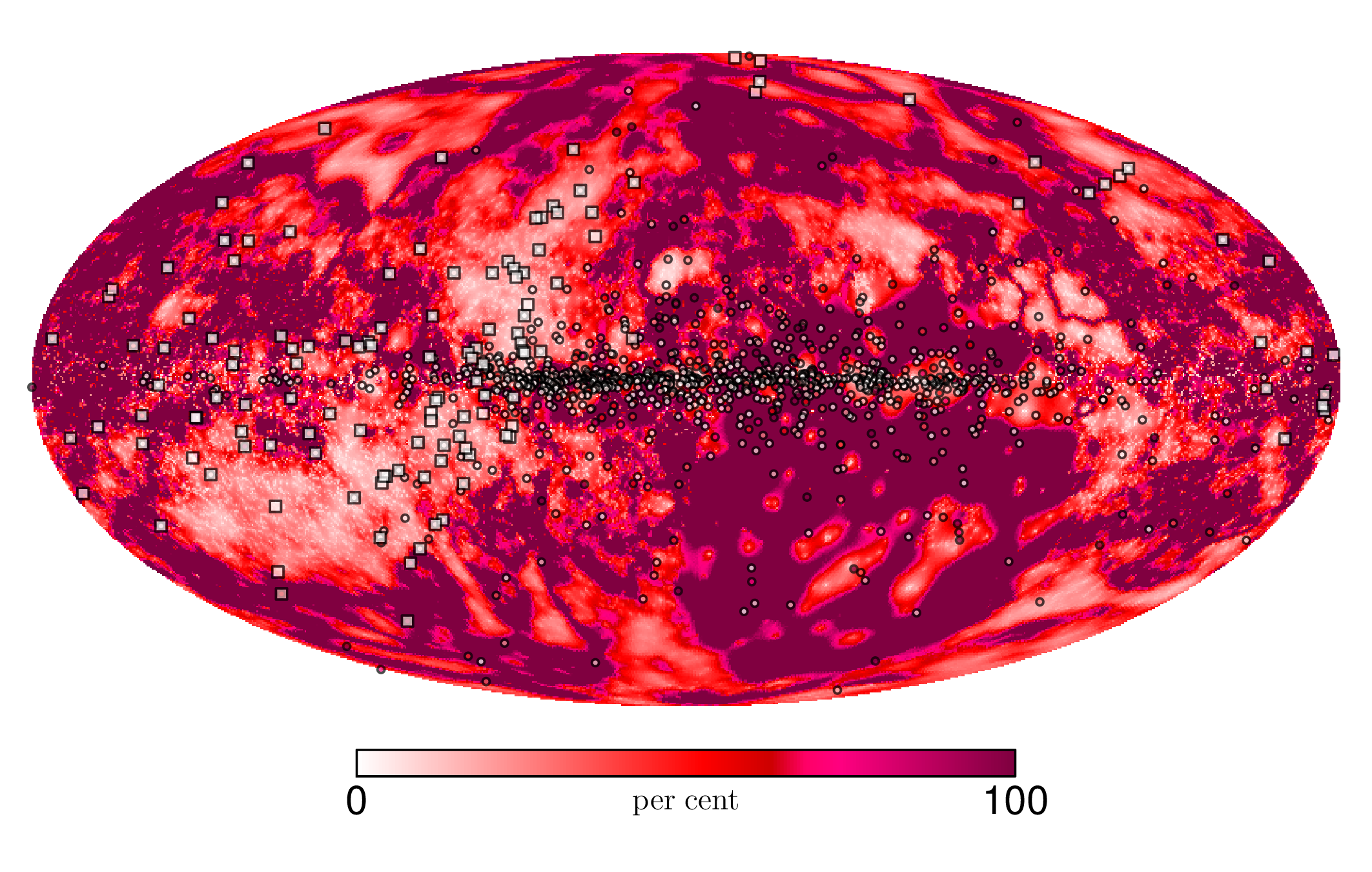}
 \caption{Upper: The RM sky shown in Galactic coordinates using Mollweide projection with $\ell,b=0$ degrees at the centre. The background shows the all-sky Galactic RM signal reconstructed using extragalactic sources. The points show the 1133 RMs from the pulsar catalogue. The squares show the 137 RMs from this work using LOFAR. All data are plotted using the same colour scale, shown by the colour bar in rad m$^{\rm -2}$, saturated at a cut-off of $\pm$204\,rad m$^{\rm -2}$ to emphasise the RM range of the LOFAR data.  Positive RMs (reds--pinks) show where the net GMF direction is towards the Earth, and negative RMs (blues--purples) show where the net GMF direction is away from the Earth. Lower: Percentage uncertainties corresponding to the measurements in the upper plot, with corresponding markers. All uncertainties are shown using the same colour scale, shown by the colour bar in per cent and truncated at 100 per cent measurement uncertainty.  }
 \label{fig:Oppermann}
\end{figure*}

 %% description of RM map and EG source RMs
Figure~\ref{fig:Oppermann} shows a summary of the current picture of the Faraday sky, along with fractional uncertainties. This includes the all-sky Galactic signal reconstructed from 41,632 RMs measured towards extragalactic sources \citep[][]{2015A&A...575A.118O}; the 1133 RMs from the current pulsar catalogue \citep[version 1.59;][]{2005AJ....129.1993M}; and the 137 RM measurements (136 from the pulsar censuses, plus PSR B0329+54) obtained in this work. The large extragalactic catalogues provide densely-packed information about the LoS through the entire Milky Way across the majority of the sky (although measurements are more sparse in the southern sky), however mostly with larger fractional measurement uncertainties. Ongoing work with low-frequency surveys promises to reduce the fractional uncertainties in the extragalactic RM catalogue \citep[e.g.][]{2018A&A...613A..58V,2018arXiv180909327R}. While future work using SKA pathfinder surveys promises to increase the number of extragalactic polarised sources further, e.g., MIGHTEE \citep[MeerKAT International GHz Tiered Extragalactic Exploration Survey;][]{2016mks..confE...6J}, POSSUM\footnote{\href{http://www.dunlap.utoronto.ca/$\sim$askap.org/possum}{http://www.dunlap.utoronto.ca/$\sim$askap.org/possum}} (Polarisation Sky Survey of the Universe's Magnetism, using ASKAP), and VLASS\footnote{\href{https://science.nrao.edu/science/surveys/vlass/vlass}{https://science.nrao.edu/science/surveys/vlass/vlass}} (VLA Sky Survey).  

 %% description of RM map and PSR source RMs - look at this para %%%%%%%%%%
Complementary to the extragalactic information, the pulsar catalogue has fewer pulsars and, therefore, LoSs with RM measurements, the vast majority of which lie within a few degrees of the Galactic disc. However, the majority of these have much lower fractional uncertainties.  Figure \ref{fig:Oppermann} also demonstrates the trend that (modulus) RMs decrease with increasing (absolute) Galactic latitude, also shown in Figure \ref{fig:DMRM}. A result of this is that the fractional uncertainties in the pulsar catalogue RMs (measured at higher observing frequencies) tend to increase with absolute Galactic latitude. The LOFAR measurements provide precise RM data towards pulsars located at a range of Galactic latitudes (above and below the plane, visible from the northern hemisphere) and show consistently low percentage uncertainties. This illustrates that the precision attainable using low-frequency observations is especially valuable for measurements where the absolute RM value is expected to be small, e.g., towards sources that are nearby or at larger absolute Galactic latitudes. 

The median pulsar catalogue RM of the pulsars for which we did not detect a significant RM in the LOFAR data is 52.8\,rad m$^{\rm -2}$, with median absolute Galactic latitude and longitude 9.6 and 54.1 degrees, respectively.  These pulsars are more likely to be in directions that tend to impart larger RMs, e.g., in the directions of the Galactic plane, the Galactic centre, the North Polar Spur and the boundary of Radio Loop \textsc{I} \citep[large structures identified in the diffuse Galactic radio emission; e.g.][]{1971A&A....14..252B,2018Galax...6...56D}. Two pulsars in the pulsar census are located towards the plane and Radio Loop \textsc{I} and have catalogued RMs greater than the maximum observable RM attainable for the LOFAR data used (PSR J1859+1526, RM=317$\pm$10\,rad m$^{\rm -2}$; PSR J1906+1854, RM=388$\pm$10\,rad m$^{\rm -2}$). For comparison, the median absolute Galactic latitude and longitude of the pulsars that have a significant LOFAR RM are 14.9 and 94.2 degrees, respectively. Therefore, for pulsars in the direction of the Galactic plane and at more considerable distances, increasingly higher frequency observations are complementary to the lower frequency data to enable us to probe the Galactic volume in its entirety. Pulsars in these areas of the Galaxy will also be increasingly difficult to detect at low frequencies due to more substantial dispersion smearing and/or scattering due to the intervening thermal electrons in the ISM. 

\begin{figure}
  \includegraphics[width=\columnwidth]{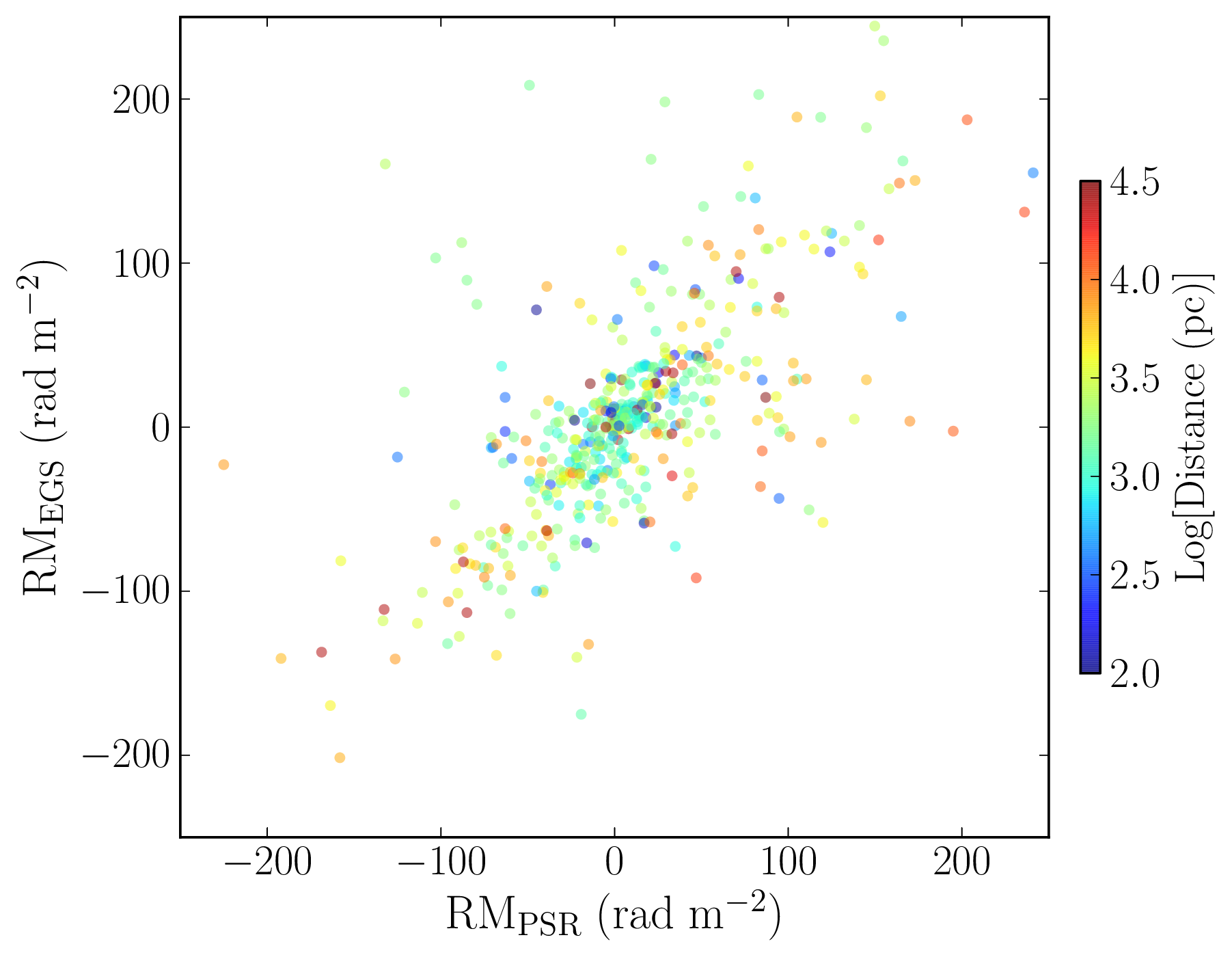}
 \caption{A comparison between the RM towards the pulsars located at $|b|>5$\,degrees, RM$_{\rm{PSR}}$, and the Galactic RM signal reconstructed using extragalactic sources, RM$_{\rm{EGS}}$, towards the corresponding LoSs.   The points are coloured according to each pulsars' distance estimate in parsecs, shown by the colour bar.    }
 \label{fig:PSR_EG_comp}
\end{figure}

 %% description of RM map and PSR source RM comparison
A reconstruction of the all-sky Galactic RM signal using extragalactic sources \citep[from][]{2015A&A...575A.118O}, shown as the background in Figure \ref{fig:Oppermann}, allows us to compare this information about the entire LoS through our Galaxy to the RMs towards pulsars (located at various distances from the Sun).  This indicates how much of the total Faraday depth along each LoS is probed by the foreground pulsars. If there are no large-scale magnetic field reversals in the Galactic halo, then we expect the RMs towards pulsars at greater distances to approach the RM values towards the extragalactic sources. %
Figure \ref{fig:PSR_EG_comp} shows the pulsar RMs (from the literature and this work; RM$_{\rm{PSR}}$) compared to the same LoS directions from the all-sky Galactic RM map (RM$_{\rm{EGS}}$) for all pulsar LoS located towards Galactic latitudes $|b|>5$\,degrees (473 points), coloured according to the pulsar's distance estimates.  Although the points seem to be somewhat scattered, 77 per cent of the RM$_{\rm{PSR}}$ and RM$_{\rm{EGS}}$ have the same sign.  Making cuts for pulsars at lower Galactic latitudes (e.g. $|b|>2$\,degrees) reduces this proportion (74 per cent), since closer to the plane there are more likely to be large-scale field reversals and small-scale structures associated with turbulence in the ISM. Making cuts for pulsars at greater Galactic latitudes (e.g. $|b|>8$\,degrees) increases this proportion (79 per cent).

%0.5, 0.3, and 0.5, 
%discussion of RM PSR/EG comparison 
We further divided the points in Figure \ref{fig:PSR_EG_comp} into three distance bins based on the pulsar distance estimates: `close', `medium', and `distant', divided by the 33rd and 66th percentiles of the distances (d$\le$1.5\,kpc, 1.5$<$d$\le$3.5\,kpc, and d$>$3.5\,kpc, respectively). The correlation coefficient (r-value) between RM$_{\rm{PSR}}$ and RM$_{\rm{EGS}}$ is positive for all distance bins. When reducing the sample to all RMs with the same signs, the correlation coefficients are: 0.76, 0.85, and 0.78, respectively.  Again, making cuts for pulsars at lower latitudes (e.g. $|b|\ge3$\,deg), the r-values decrease, except for the pulsars in the `distant' bin. Making cuts for pulsars at greater latitudes (e.g. $|b|\ge10$\,deg), the r-values increase in all bins and are 0.9 for the `medium', and `distant' bins. 
 This indicates that RM$_{\rm{PSR}}$ and RM$_{\rm{EGS}}$ tend to become more correlated for pulsars at greater distances, as expected if there are no large-scale field reversals in the Galactic halo magnetic field. However, the variables are perhaps not as well correlated as expected. This may be because at more considerable distances, the distance estimates towards pulsars become more uncertain, see Section \ref{sec:lim} for further discussion. The differences in the correlation coefficients for different cuts in $|b|$ also indicates that the RMs towards pulsars may have varying contributions from small-scale magneto-ionic foreground structures, for example, supernova remnants or H\textsc{II} regions \citep[e.g.][]{2003AA...398..993M}, which are usually more numerous closer to the Galactic plane compared to higher latitudes, also discussed in Section \ref{sec:ind}.  This highlights the requirement for collecting a large set of pulsar RMs for the purpose of studying the large-scale GMF, so that the contributions from small-scale ISM structures may be identified and down-weighted or averaged out in reconstructions of the large-scale GMF structure.

%%%%%%%%%%%%%%%%%%%%%%%%%%%%%%
\subsection{Notes on individual pulsars}\label{sec:ind}

 %% discussion about Crab RM agreement 
The RM measurements towards the Crab Pulsar B0531+21 (located within the Crab Nebula supernova remnant) show a discrepancy of 5.4-times the uncertainties: the pulsar catalogue value measured over 40 years ago is RM$_{\mathrm{cat}}$=$-42.3\pm0.5$\,rad m$^{\rm{-2}}$ \citep[][]{1972ApJ...172...43M} and the RM value measured in this work is $-45.44\pm0.08$\,rad m$^{\rm{-2}}$. It is possible that the difference in RM may be because of variations between the observing epochs. Unfortunately, there are very few published RM measurements towards the Crab pulsar for epochs besides these.  
However, there are monthly updates for other Crab ephemeris parameters\footnote{\href{http://www.jb.man.ac.uk/pulsar/crab.html}{http://www.jb.man.ac.uk/pulsar/crab.html}}, including for the DM \citep{1993MNRAS.265.1003L}. Between 1988 May to 2018 September, the minimum and maximum DMs measured were 56.734\,pc cm$^{\rm{-3}}$ and 56.921\,pc cm$^{\rm{-3}}$, respectively. The median DM measured during this time period is 56.7805\,pc cm$^{\rm{-3}}$, which is closer to the value from the LOFAR measurement on 2014 February 15 of 56.7712\,pc cm$^{\rm{-3}}$. Assuming that the magnetic field value $\langle B_{\parallel}\rangle$ stays constant, see Table \ref{tab:1}, the maximum DM measured would increase the RM by 0.12\,rad m$^{\rm{-2}}$. This does not account for the 3.1\,rad m$^{\rm{-2}}$ difference between the RM measurements, indicating that the magnetic field value may also vary along with the electron density. 

Such a variation in RM with time has also been observed for the Vela Pulsar B0833--45, located in the Vela Supernova Remnant \citep[e.g.][]{2005MNRAS.364.1397J,2017PASA...34...40L}.  In these cases, there seem to be small fractional changes in the RMs, with no change in the sign of the RM. Therefore, it is likely that we are sensing both the coherent large-scale Galactic magnetic field component, plus some contribution from a random small-scale foreground magnetic field component, e.g., within the associated supernova remnant structure.  In addition, the RM dispersion measured towards the Crab Pulsar, $\upsigma_{\rm{RM}}<0.25$\,rad m$^{\rm{-2}}$, is above the 92nd percentile of $\upsigma_{\rm{RM}}$ measurements in this work.  This may also suggest some random variations in the magnetic field along the LoS probed by different propagation paths due to scattering, but again, this can be further investigated in more detail using the polarisation profiles in future work. Repetition of observations to monitor these RM variations over long, $\sim$years, timescales could allow us to estimate the strength and variance of the random magnetic field components associated with the small-scale foreground structure, and deduce the ratio between this and the large-scale, coherent component. 

% discussion about MSP eclipsing sources  
PSR J1810+1744 is an eclipsing black widow pulsar \citep[it is in a short orbital period binary system with a low-mass companion star; e.g.,][]{2013ApJ...769..108B}. The RM towards this source has not been published in the literature. In this work, we measured the RM, $88.5\pm0.1$\,rad m$^{\rm{-2}}$, using a 20-minute observation with a total intensity S/N of 6, see Table \ref{tab:1}. Shorter integration times further reduce the S/N, making investigating changes in RM on short timescales over the binary period difficult, as has previously been studied for DM and scattering parameters \citep{2018MNRAS.476.1968P}. For the single-epoch observation used in this work, the RM dispersion measured shows that the source is Faraday thin: $<$0.0004\,rad m$^{\rm{-2}}$.

Another black widow pulsar that was observed as part of the LOFAR MSP census is PSR J2051--0827. This pulsar is also subject to repeated timing observations using LOFAR, providing data for multiple different phases of the binary period.  The polarisation profile and RM results will be presented in future work (Polzin et al., in prep.).
%Notes on individual sources end

%Show B0823 in bright and quiet modes, and the RM obtained is the same within uncertainties. Therefore, within the uncertainties, the RM measured is not dependent on emission mode. This indicates that, as expected, the RMs measured are dominated by the ISM (over long timescales) and ionospheric (over shorter timescales) magneto-ionic plasmas. 
% discussion about emission mode 
Some of the pulsars observed in this work are  known to change emission modes, e.g. PSR B0823+26 \citep{2015MNRAS.451.2493S,2018MNRAS.480.3655H}, and PSR B0943+10 \citep{2013Sci...339..436H,2014A&A...572A..52B}.  These pulsars were in the `bright' emission mode during the observations used to measure the RMs for this work. PSRs B0823+26 and B0943+10 are among the timing set that are repeatedly observed by LOFAR, and a possible correlation between RM and intrinsic emission mode (although not necessarily expected) will be explored in future work.

%%%%%%%%%%%%%%%%% SCALE HEIGHT %%%%%%%%%%%%%%%%%%

\subsection{Scale height of the Galactic halo magnetic field}\label{sec:scale}

\begin{figure}
\begin{center}
    \vspace{-1cm}

  \includegraphics[width=1.15\columnwidth]{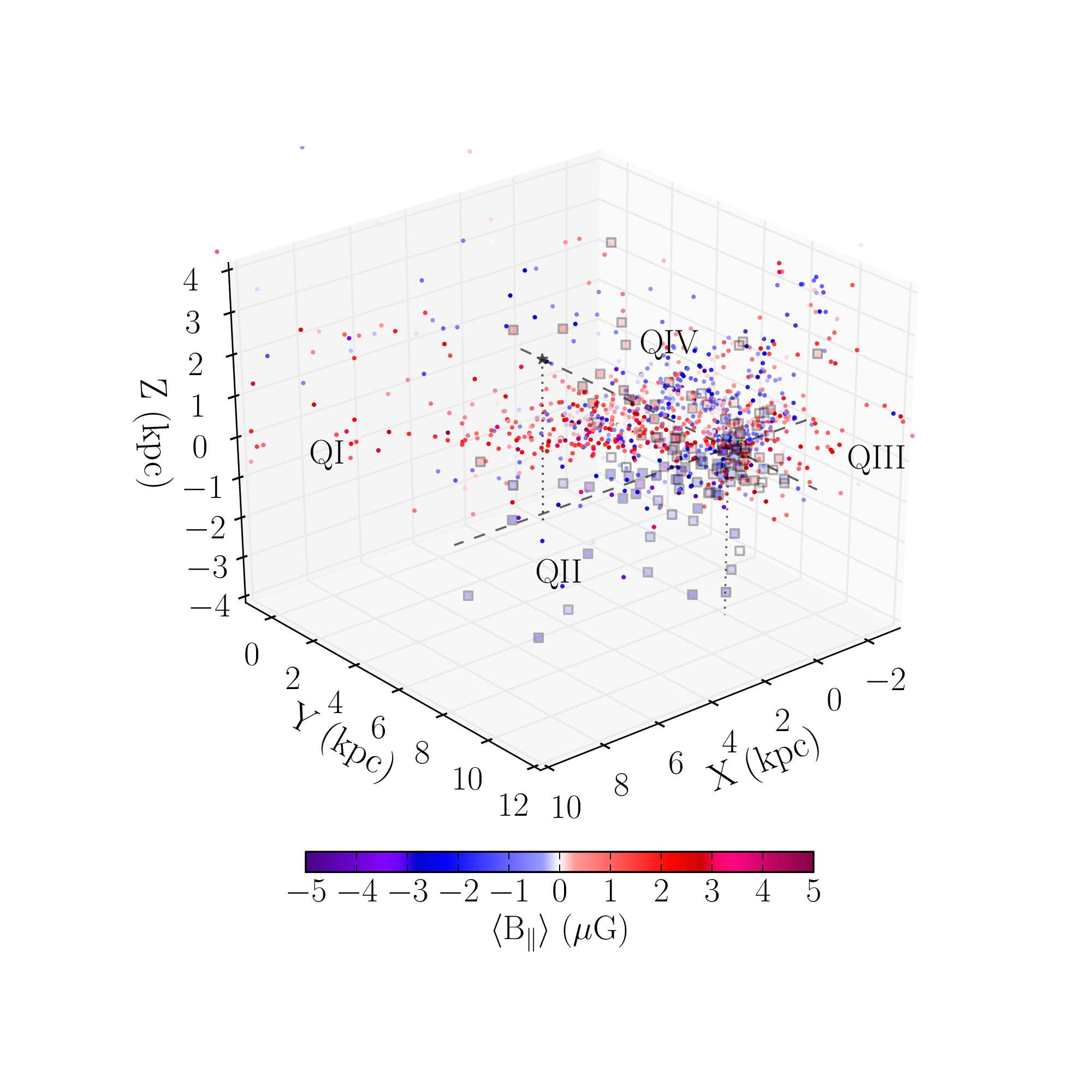} 

    \vspace{-1.1cm}
  
  \includegraphics[width=0.9\columnwidth]{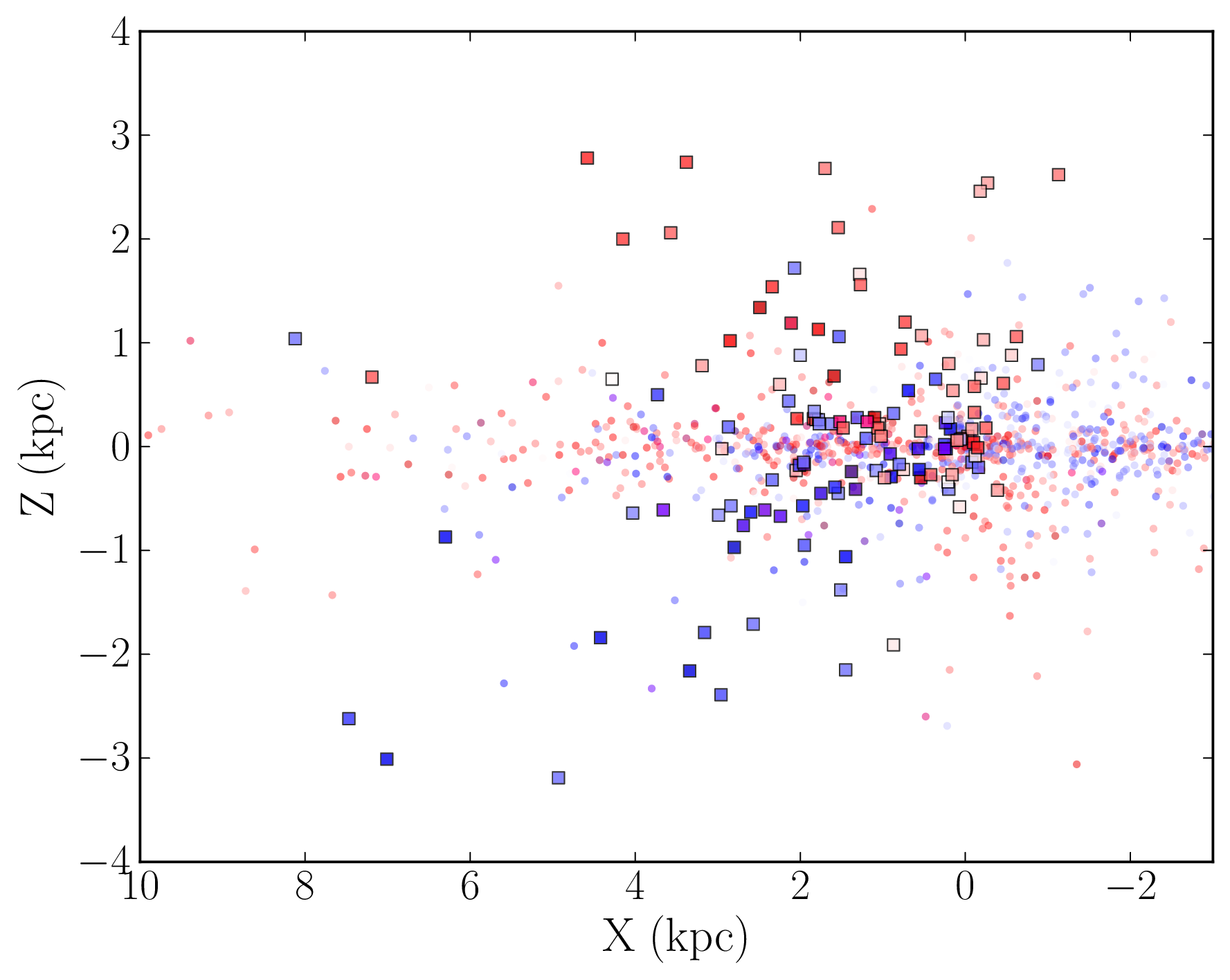}
  
     \vspace{-0.15cm}
     
  \includegraphics[width=0.9\columnwidth]{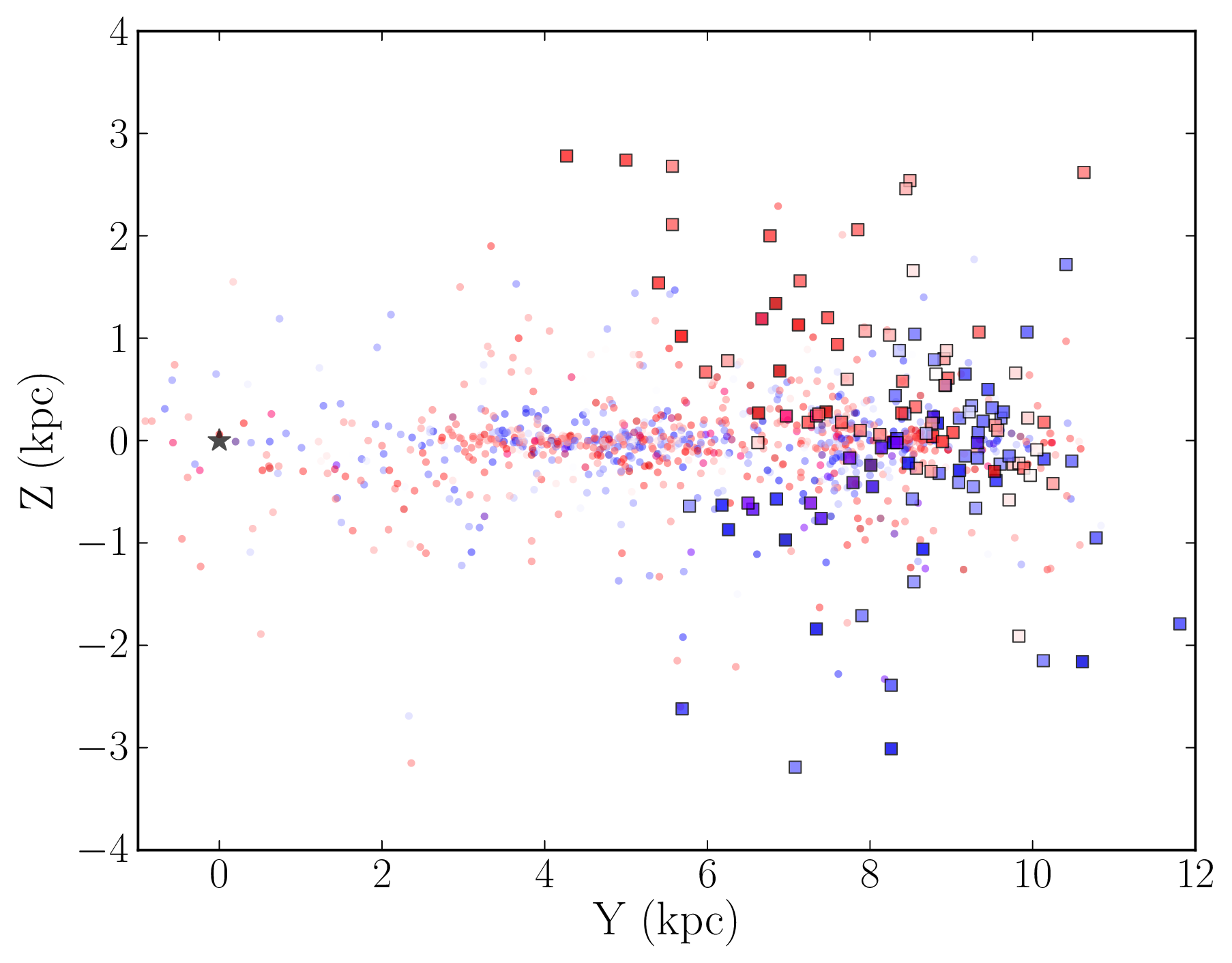}
  
       \vspace{-0.15cm}

 \caption{Upper: 3-D plot showing the location of pulsars, in Galactocentric coordinates, with colours representing $\langle B_{\parallel}\rangle$ in $\upmu$G, see colour bar. The points show the pulsars with DM and RM measurements from the literature. The squares show the pulsars with DM and RM measurements from this work using LOFAR.  The location of the Galactic Centre is shown by the black star, with the black dotted line connecting it to the x-y axis. The Galactic quadrants are labelled Q\textsc{I}--\textsc{IV}, and separated by the dashed lines. The intersection of the lines show the location of the Sun, which is also connected to the x-y plane by the grey dotted line.    Middle, Lower: The same points from the upper plot are shown in the X-Z, and Y-Z planes, respectively.  18 pulsars (10 with literature RMs and 8 with LOFAR RMs) have larger Z values ($>$8\,kpc) and are not shown here. }
 \label{fig:3D}
 \end{center}
\end{figure}

%
% discussion about 3-D data
Although there are fewer RM measurements towards pulsars compared to the extragalactic catalogue \citep[e.g.][]{2015A&A...575A.118O}, the additional DM information provided by pulsars allows us to infer the electron-density-weighted average magnetic field parallel towards each LoS. The pulsars are distributed throughout the Galaxy at a range of heights (Z) above/below the Galactic disc, obtained using the YMW16 Galactic electron density model. To demonstrate this, Figure \ref{fig:3D} shows the locations of the pulsars in the Galaxy on a 3-D plot in cartesian coordinates. The Sun is located at (X,Y,Z)=(0,8.3,0.006)\,kpc and the Galactic Centre is at (X,Y,Z)=(0,0,0)\,kpc, as in YMW16, and the Galactic quadrants are also labelled. The markers in Figure \ref{fig:3D} are coloured according to the inferred magnetic field strength $\langle B_{\parallel}\rangle$ and direction. 
The X--Z and Y--Z slices are also shown for further clarity.  Figure \ref{fig:3D} demonstrates that these data can provide a measurement of the scale height of the Galactic magnetic field in the halo.

\begin{figure} 
  \includegraphics[width=\columnwidth]{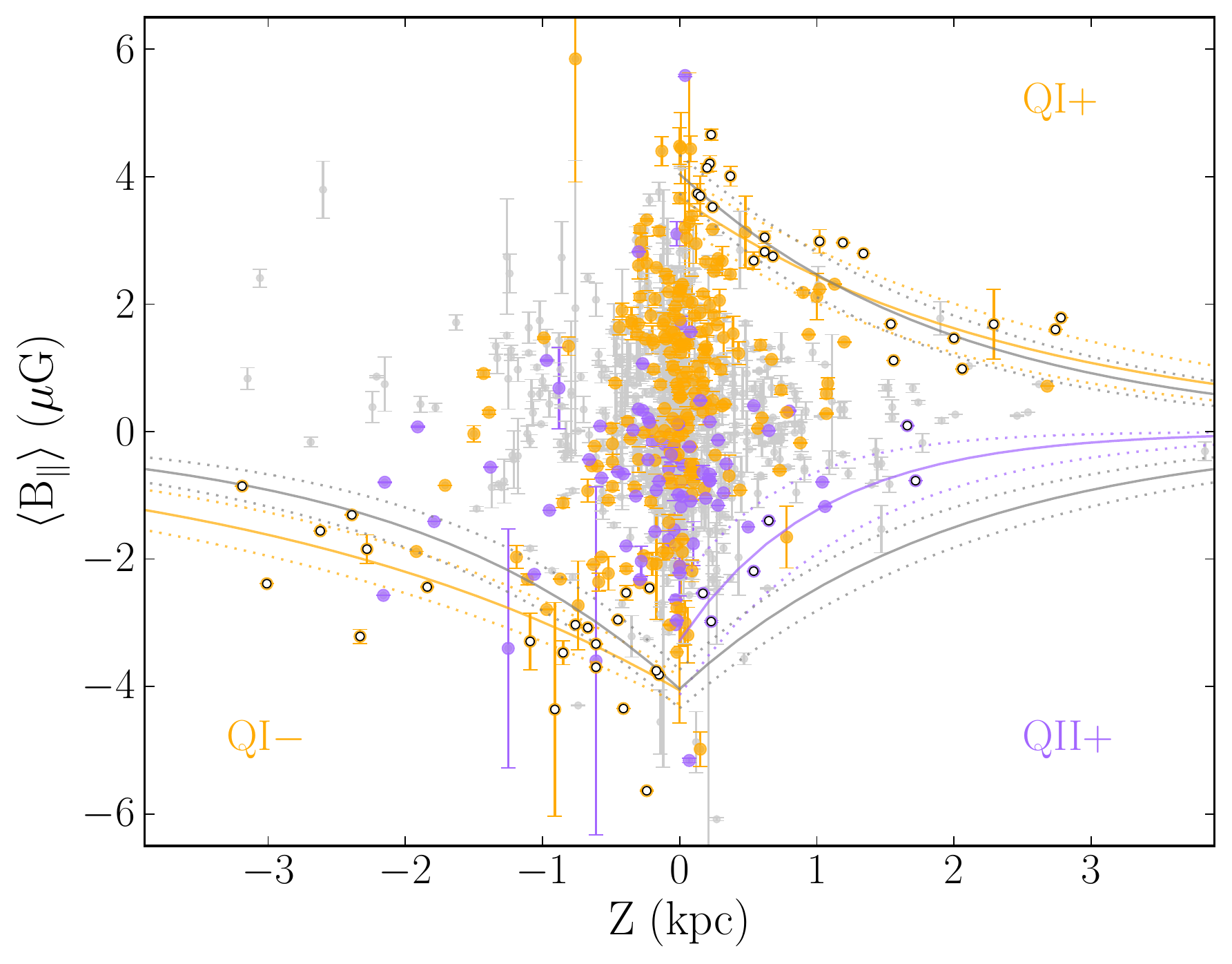}
 \caption{Using the RM and DM measurements from the pulsar catalogue and this work, $\langle B_{\parallel}\rangle$ is shown against the estimated height above/below the Galactic plane, Z, (grey points). The pulsars located in Galactic quadrant \textsc{I} in the range 30$\le\ell<$90 degrees are shown by the orange points, and pulsars located in Galactic quadrant \textsc{II} in the range 90$<\ell<$180 degrees are shown using the violet points. Three areas in the plot are identified: Galactic quadrant \textsc{I}, above the plane, with positive $\langle B_{\parallel}\rangle$ values (labelled Q\textsc{I}+ in orange); Galactic quadrant \textsc{I}, below the plane, with negative $\langle B_{\parallel}\rangle$ values (labelled Q\textsc{I}-- in orange); and Galactic quadrant \textsc{II}, above the plane, with negative $\langle B_{\parallel}\rangle$ values (labelled Q\textsc{II}+ in violet).  The `outermost' points identified (see text) in these areas are shown by the open circles. Fits of the magnetic scale height to the `outermost' points using Equation \ref{eq:scale} (solid lines) and the uncertainties (dotted lines) are shown for Q\textsc{I}+ and Q\textsc{I}-- (orange lines) and Q\textsc{II}+ (violet lines).  We also show the fit for the magnetic scale height to all of the absolute values of the outermost points (grey solid line) and the uncertainties (grey dotted lines) for comparison.  } %  
 \label{fig:ZRM}
\end{figure}

%Introduction for B vs Z plot
Figure~\ref{fig:ZRM} shows the magnetic field parallel to the LoS, $\langle B_{\parallel}\rangle$, calculated for the pulsars in this work, as well as those from the literature, as a function of their vertical distance from the Galactic plane, Z.  All pulsars located in Galactic quadrant \textsc{I} between galactic longitude 30$\le\ell\le$90 degrees are shown as orange points, and all pulsars located in Galactic quadrant \textsc{II} between galactic longitude 90$\le\ell\le$180 degrees are shown by the violet points. These ranges were chosen to investigate the directions towards the inner Galaxy (excluding regions closer towards the Galactic centre) and the outer Galaxy, respectively. These are also the ranges where the largest number of pulsars with LOFAR data are located, see Figure~\ref{fig:3D}. 

%Discussion about B vs Z plot general trends
Figure~\ref{fig:ZRM} shows that for Galactic quadrant \textsc{I}, there is a notable dichotomy between mostly positive (negative) magnetic field values for pulsars located above (below) the plane, similarly evident in Figure \ref{fig:Oppermann}. Although the large region of positive $\langle B_{\parallel}\rangle$ seems to be coincident with the North Polar Spur feature \citep[e.g.][]{2015ApJ...811...40S}, this effect may be somewhat related to the large-scale magnetic field reversal in the Galactic plane in this quadrant \citep[e.g.][]{2011ApJ...728...97V}. However, the large-scale reversal does not appear to affect the bulk sign of the RM either above or below the Galactic plane and perhaps the effect of the large-scale reversal may not be visible in the halo and confined to the plane. Conversely, in Galactic quadrant \textsc{II}, where a large-scale magnetic field reversal is not expected \citep[e.g.][]{2011ApJ...728...97V}, there is a majority of negative magnetic field values towards pulsars located both above and below the plane. Although the pulsar data are more sparse towards the Galactic anticentre, with half of the number of data points compared to quadrant \textsc{I} and 35 per cent more data points below the plane. For comparison, in quadrant \textsc{I} there are approximately equal numbers of points above and below the plane.  The median magnetic field strengths and directions in Galactic quadrants \textsc{I} and \textsc{II} using these data are 0.76 and --0.77\,$\upmu$G, respectively. These values are consistent with the strength of the regular halo magnetic field of 2\,$\upmu$G or lower from \citet{2010RAA....10.1287S}.   

%Can't you say a lot more about these magnetic fields from this plot? (Maybe you intend to anyway.)  E.g. In the first quadrant, all magnetic field values at higher distances form the plane are positive. Yet there is a reversal of large-scale B in the plane in this quadrant. Does this mean that the reversal is not visible in the halo, i.e. that it is restricted to the plane?

%Explain about B vs Z plot fits
 We use the data shown in Figure~\ref{fig:ZRM} to fit for the magnetic scale height in the Galactic halo, using the form:
\begin{equation}
\langle B_{\parallel}\rangle=\langle B_{\parallel,0}\rangle \rm{exp}(-Z/H),
\label{eq:scale}
\end{equation}
where $\langle B_{\parallel,0}\rangle$ is the largest value of the magnetic field at Z=0, and H is the magnetic scale height.  Both $\langle B_{\parallel}\rangle$ and Z can take positive and negative values depending on the areas identified in Figure \ref{fig:ZRM}. Since the RM data provide the magnetic field strength and net direction parallel to the LoS, these will provide a lower limit for the scale height of the total GMF. Therefore, we fit the `outermost' points for pulsars in: Galactic quadrant \textsc{I} above the plane where $\langle B_{\parallel}\rangle>0$ (labelled Q\textsc{I}+);  Galactic quadrant \textsc{I} below the plane where $\langle B_{\parallel}\rangle<0$ (labelled Q\textsc{I}--); and in Galactic quadrant \textsc{II} above the plane where $\langle B_{\parallel}\rangle<0$ (labelled Q\textsc{II}+), see Figure~\ref{fig:ZRM}. The `outermost' points were identified as the largest absolute values of $\langle B_{\parallel}\rangle$ in the bin ranges 0.1$<|\mathrm{Z}|\le0.5$\,kpc, 0.5$<|\mathrm{Z}|\le1.5$\,kpc, and 1.5$<|\mathrm{Z}|\le3.5$\,kpc.  For the areas Q\textsc{I}+ and Q\textsc{I}--, the number of points selected in each Z bin was chosen to be one fewer than the total number of data points in the bin with the fewest number, i.e., in the bin 1.5$<|\mathrm{Z}|\le3.5$\,kpc.   For the area Q\textsc{II}+ with fewer data points, the number of points in the 1.5$<|\mathrm{Z}|\le3.5$\,kpc bin is 2, and so 2 points were also selected from the 0.1$<|\mathrm{Z}|\le0.5$\,kpc and 0.5$<|\mathrm{Z}|\le1.5$\,kpc bins.
 Pulsars in Galactic quadrant \textsc{II} below the plane where $\langle B_{\parallel}\rangle<0$ were not fit as there are insufficient data. 
The `outermost' data points fit in each area are shown as open circles in Figure \ref{fig:ZRM}.
 Table \ref{tab:scale} summarises the parameters obtained from the magnetic scale height fit using Equation \ref{eq:scale}. We also fit the scale height for the absolute values of all of the points fit in the individual areas, shown by the grey lines in Figure \ref{fig:ZRM} and included in Table \ref{tab:scale}.

\begin{table}
	\centering
	\caption{Summary of the parameters obtained from fitting the magnetic scale height function in Equation \ref{eq:scale}. Uncertainties ($\pm$) are quoted in the columns to the right of the values. }
	\label{tab:scale}
%	\begin{tabular}{@{}l @{}l @{}l | @{}l @{}l} % five columns, alignment for each
	\begin{tabular}{lllll} % five columns, alignment for each
		\hline
		\hline
		Quadrant & Magnetic scale height (H) & $\pm$ & $\langle B_{\parallel,0}\rangle$ &  $\pm$  \\
		 & kpc & kpc & $\upmu$G &  $\upmu$G  \\
		\hline
		Q\textsc{I}+  &  2.4 &  0.4 &  3.7 & 0.4 \\
		Q\textsc{I}--  &  3.3 &  0.6 &  4.0 & 0.2  \\
		Q\textsc{II}+  &  1.0 &  0.3 &  3.3 & 0.8 \\
		All  &  2.0 &  0.3 &  4.0 & 0.3 \\
		\hline
	\end{tabular}
\end{table}

% discuss table and figure results
Table \ref{tab:scale} shows that the general magnetic scale height found using the data for all areas identified in Figure \ref{fig:ZRM} is 2.0$\pm$0.3\,kpc.  The mean of the magnetic scale heights for the identified areas is 2.2$\pm$0.5\,kpc. All of the scale heights presented in Table \ref{tab:scale} agree within the uncertainties, except for the Q\textsc{II}+ area. This indicates that the scale height in Galactic quadrant \textsc{I} towards the inner Galaxy is larger than the scale height in Galactic quadrant \textsc{II} towards the outer Galaxy.  Only six data points were fit for the Q\textsc{II}+ area, and so a larger sample would be necessary to provide a more confident outcome for the direction of the outer Galaxy.   
All of the $\langle B_{\parallel,0}\rangle$ values in Table \ref{tab:scale} agree within the uncertainties, and the value for all areas is 4\,$\upmu$G.
We extrapolated the electron-density-weighted average magnetic field to a greater distance from the galactic plane ($\pm$6\,kpc) using the results summarised in Table \ref{tab:scale}. 
At this height, the $\langle B_{\parallel,0}\rangle$ in areas Q\textsc{I}+, Q\textsc{I}--, Q\textsc{II}+, and `All'  are 0.3$\pm$0.1\,$\upmu$G, $-$0.6$\pm$0.2\,$\upmu$G, $-$0.01$\pm$0.02\,$\upmu$G, and 0.2$\pm$0.1\,$\upmu$G, respectively. 

%Explain  B vs Z plot uncertainties
The uncertainties from the fits are shown in Figure~\ref{fig:ZRM} and Table \ref{tab:scale}, and account for the uncertainties in the measurement of $\langle B_{\parallel}\rangle$, but not in Z, which are generally not known for pulsars with distance estimates but may have larger fractional uncertainties than the uncertainties on $\langle B_{\parallel}\rangle$. 
There are 5 and 3 data points in Q\textsc{I}+ and Q\textsc{I}--, respectively, that were not included in the fits because their $|Z|$ distance estimates are much larger than the general population ($>8$\,kpc) and likely overestimated.
The small uncertainties on $\langle B_{\parallel}\rangle$ from the LOFAR data at high $|Z|$ values tend to flatten the fits (especially in Q\textsc{I}--), increasing the scale height and decreasing the $\langle B_{\parallel,0}\rangle$ output. Therefore, higher numbers of pulsar data points with smaller uncertainties, especially at high $|Z|$ values, are desirable for increasing the accuracy of determining the magnetic scale height(s) in future.  Also valuable towards this goal are independent distance estimates for these pulsars, see Section \ref{sec:lim} for further discussion.

% discuss table and figure results in comparison with other MW components
The magnetic scale height summarised in Table \ref{tab:scale} is comparable, although generally somewhat larger than, the Galactic free electron vertical scale heights determined using pulsar DMs, e.g. 1.8$^{+0.1}_{-0.3}$\,kpc from \citet{2008PASA...25..184G}; 1.4$^{+0.3}_{-0.2}$\,kpc from \citet{2009ApJ...702.1472S}; 1.6$\pm0.3$\,kpc from \citet{2012MNRAS.427..664S}; or 1.67$\pm$0.05\,kpc from YMW16. This corresponds to the thick disc of the Galaxy, which has a more extensive scale height than the thin disk with scale heights between 20--70\,pc \citep[e.g.][and references therein]{2017ApJ...835...29Y}.  Since the Faraday rotation effect is caused by both the electron density and the magnetic field in the ISM, the scale heights obtained using RMs will not be independent of the electron density scale height. This may be why the scale height found here is also larger than the 0.74\,kpc exponential scale height of synchrotron emission found in, e.g., \citet{2012A&A...543A.127S}. Although, in this case, synchrotron modelling also requires ancillary parameters, such as the relativistic electron population, and is also sensitive to the magnetic field perpendicular to the LoS. This provides an incentive to use at least one magnetic field observable to derive the properties and structure of the total GMF, including the anisotropy of the Galactic halo magnetic field, see Section \ref{sec:lim}. There are complex analytical models for large-scale Galactic halo magnetic fields \citep[e.g.][]{2014A&A...561A.100F}. Which of these models best fit the RM data will be explored in future work.

% discuss table and figure results in comparison with other galaxies 
The scale heights determined for the Milky Way can also be compared to external galaxies. For example, for the edge-on galaxies NGC891 and NGC4631, \citet{1991A&A...248...23H} deduced scale heights of 0.9\,kpc and 1.3\,kpc, respectively, using the increase in the mean degree of polarisation with distance from the galactic disk. The numbers obtained here are also comparable or up to three times larger than these edge-on galaxies. The scale heights estimated here are also approximately equivalent to or greater than the radio scale heights seen in other galaxies \citep[e.g.][]{2018A&A...611A..72K}.

%\citep{2010MNRAS.401.1013J}Jaffe et al 2010
%citep{2016ApJ...830...38L} Lenc et al 2016
%Here we use the results in combination with the literature pulsar and extragalactic RM catalogue to compare and constrain models for the 3-D Galactic magnetic field proposed in the literature \citep{2008A&A...477..573S,2010RAA....10.1287S,2012ApJ...755...21M,2012ApJ...761L..11J,2001MNRAS.325..649F}. 

%%%%%%%%%%%%%%%%% LIMITATIONS AND FUTURE %%%%%%%%%%%%%%%%%%

\subsection{Limitations of current data and future prospects}\label{sec:lim}

% reversals along LoS - more pulsars with varying distances...

In future, polarisation calibration will be performed on the data described in this work, to present the low-frequency polarisation profiles. This may increase the S/N in linear polarisation for some pulsars so that an RM may be detected. In the cases where the linearly polarised S/N was less than the threshold set (i.e. S/N$_{{F}}<4$), it would be preferable to observe these sources for longer integration times to obtain more reliable detections and decrease the uncertainties. 
For pulsars that are being repeatedly observed in LOFAR timing campaigns (this includes a subset of both MSPs and `slow' pulsars) the S/N in the polarisation profiles can also be increased by concatenating observations. However, ionospheric RM corrections will have to be made before the addition in this case.  For this set of pulsars with higher S/N detections, the repeated observations can also be used to better constrain the RM (and uncertainty) after correcting for the ionospheric RM for several independent epochs, as was done as an example in this work for PSR B0329+54. The precise measurements now routinely obtained using low-frequency observations ushers us into an era where ISM parameters such as DMs, RMs, and scattering \citep[e.g.][]{2017MNRAS.470.2659G,2018MNRAS.476.2704M} can be monitored over time, also allowing us to probe small-scale turbulent structures in the ISM and other foregrounds \citep[e.g.][]{0004-637X-831-2-208,2018IAUS..337..279T}.
%avoid Ricean bias in the RM measurement.  (increases the noise but not entirely correct in this context)

The maximum absolute RM measured in this work is towards PSR B2210+29 (RM$_{\rm{obs}}$=--165.23\,rad m$^{\rm -2}$; RM$_{\rm{ISM}}$=--168.66\,rad m$^{\rm -2}$, after the ionospheric correction).
Due to the channel widths used for the data set, the largest Faraday depth to which one loses 50 per cent sensitivity, at the centre frequency, is 163\,rad m$^{\rm -2}$ (not accounting for the loss of frequency information due, e.g., to excised RFI). It is possible to use smaller channel widths for LOFAR observations, expanding this largest Faraday depth beyond 1000\,rad m$^{\rm -2}$.  These data were not used in this work, as the data volume becomes more substantial and the data reduction more cumbersome. However, for areas where larger RMs are expected (e.g. towards the Galactic plane, at larger distances, and towards the boundary of Radio Loop \textsc{I}), also usually accompanied by larger DMs, the finer channel resolution can be utilised to detect the pulsars and measure larger absolute RMs. Probing the Galaxy in the Galactic plane and at more considerable distances, DMs and RMs measured using higher frequency observations will be complementary to the measurements from low-frequency observations. This is because pulsars with large DMs and RMs may become scattered or depolarised in lower-frequency observations.  In this work, we demonstrated that the low-frequency LOFAR observations provide excellent fractional uncertainties on RMs (and DMs) towards pulsars with relatively low absolute RMs, e.g., that are located towards the Galactic halo and are relatively close to the Sun.

% Ongoing and future work with low-frequency and broadband instruments promises to increase the number of pulsars with RM data, and ongoing and future pulsar surveys continue to discover pulsars \citep[e.g.][]{2014MNRAS.439.1865N,2018MNRAS.480.3457M}, which can be followed up with polarisation observations. 
More pulsars are being discovered in ongoing pulsar surveys at various centre frequencies, including at low radio frequencies using LOFAR \citep[e.g.][]{2014A&A...570A..60C,2018MNRAS.480.3457M,2018arXiv180900965T}.  The pulsars discovered can be followed up with polarisation observations to obtain RM measurements, to increase the numbers of LoSs with which we can probe the structure of the Galaxy. The low-frequency discoveries are valuable because they are located relatively nearby or towards the Galactic halo \citep[e.g.][]{2010A&A...509A...7V}. This provides local GMF estimates, which can be compared to the data from more distant pulsars or extragalactic sources for longer path lengths through the Galaxy. Moreover, pulsars at larger distances and in the Galactic plane are preferentially discovered at higher frequencies using other large radio telescopes \citep[e.g.][]{2014ApJ...791...67S,2014MNRAS.439.1865N,2017ApJ...834..137L}, and are useful for mapping the GMF towards the more distant areas of our Galaxy. More known pulsars that are distributed throughout the Galaxy at various distances from the Sun provide more data with which to reconstruct the GMF structure (including any field reversals along the LoSs) with greater confidence.
In the future, the observational capabilities of the SKA to discover and time pulsars will allow us to approximately triple the number of known pulsars in the Galaxy \citep[e.g.][]{2015aska.confE..40K,2017PASA...34...70X}, as well as measure their parallax and proper motions using VLBI \citep[e.g.][]{2015aska.confE.143P}, enabling 3-D tomography of the electron density and GMF \citep[e.g.][]{2015aska.confE..41H}.

%Electron density models limitations. Independent distance measurements desirable. how many in this work? %% compare distances for large d in Yao et al to TC93 and NE2001.

In order to reconstruct an accurate model of the GMF (and electron density) using RMs and DMs towards pulsars, it is becoming increasingly necessary to determine independent distance measurements towards larger numbers of pulsars. For the set of pulsars with RMs measured in this work, 17 slow pulsars and 13 MSPs (15 and 68 per cent of the set, respectively) have one (or more) independent distance measurements \citep[see][and references therein]{2017ApJ...835...29Y}: via annual parallax using VLBI (14 slow and 5 MSPs, respectively), or timing (0 and 6, respectively); association with a nebula (1 and 0, respectively); kinematic distances from H\textsc{I} absorption measurements (2 and 0, respectively); or an optical counterpart association (0 and 2, respectively) \citep{2018ApJ...864...26J}.  Although pulsar distances can be estimated using a Galactic electron density model, and these are becoming more accurate with more independent distance measurements that can be used to calibrate the models, there are still large numbers of pulsars for which distances are over 40 per cent uncertain \citep[][]{2017ApJ...835...29Y}. We used the recently published YMW16 electron density model for this work because previous models had a maximum $|\mathrm{Z}|$ limit of $\le$2\,kpc. The YMW16 model allows larger heights above the Galactic plane, e.g. shown in Figure \ref{fig:ZRM}, facilitating the magnetic scale height fit. 
 There are efforts towards increasing the number of pulsars with independent distance measurements, e.g., \citet{2018arXiv180809046D} provide an annual parallax for a further 21 pulsars with RMs in this work.  For these pulsars, the median discrepancy between the DM distance estimates and the VLBI annual parallax distance measurements is 0.67\,kpc  (or 54 per cent fractional difference). However, in a couple of extreme cases, the DM distance estimates are overestimated by $\approx$20\,kpc for PSRs B2303+30 and B2210+29 towards Galactic coordinates $\ell\approx90$, $b\approx-25$.   It is essential to continue to obtain independent distance measurements for more pulsars so that 3-D tomography of the Galactic (magnetic field) structure using the precise ISM parameters measured using the SKA and its precursors can be fully realised \citep[e.g.][]{2015aska.confE..41H}.

%Using more observables
Studies using RMs alone cannot distinguish between ordered random and isotropic random field components, which are often grouped into the `random' field component \citep[e.g.][]{2010MNRAS.401.1013J}. Moreover, Equation \ref{eq:1} assumes that the magnetic field and thermal electron density are uncorrelated. However, if their fluctuations are (anti)correlated, this will yield an (under)overestimate of $\langle B_{\parallel}\rangle$, and can result in error by a factor of 2--3 in a statistically homogeneous magneto-ionic medium \citep{2003A&A...411...99B}.
 Therefore, combining results from the other magnetic field tracers is desirable, to obtain as complete a picture of the 3-D GMF structure as possible.  In the near future it may also be possible to include observations of the directions of arrival of Ultra-High Energy Cosmic Rays, but the sources and composition of these still need to be constrained \citep[e.g.][and references therein]{2018ApJ...853L..29A}.  The framework required to combine multiple observables, as well as theoretical models, to infer the most likely model of the structure of the Galaxy is being constructed and refined \citep[e.g. IMAGINE;][]{2018JCAP...08..049B}.  In the future, the SKA will also provide groundbreaking observations for many of the complementary observables of the GMF, particularly with respect to diffuse polarisation, extragalactic RM-grids, and Zeeman splitting \citep[][respectively]{2015aska.confE..96H,2015aska.confE..92J,2015aska.confE.110R}, promising to revolutionise our understanding of the GMF.

%2017A&A...598A.104S

%%%%%%%%%%%%%%%%% CONCLUSIONS %%%%%%%%%%%%%%%%%%

\section{Conclusions}\label{sec:conc}

%The last numbered section should briefly summarise what has been done, and describe
%the final conclusions which the authors draw from their work.

% Complementarity of low-freq and high-freq data (polarisation horizon)
% complementarity of Rm and synchrotron data (total B field map)
% distances
  \begin{enumerate}
      \item We have measured RMs towards 137 pulsars using the `LOFAR census of non-recycled pulsars' \citep{2016A&A...591A.134B}, the `LOFAR census of millisecond pulsars'  \citep{2016A&A...585A.128K}, and repeated timing/monitoring observations of PSR B0329+54. We present the largest low-frequency RM catalogue to date, with 25 pulsars that do not have RMs published in the literature.  For the remaining pulsars with previously published measurements, the low-frequency data generally agree with previous measurements, within uncertainties, and provide 20-times greater precision, on average. The RMs were corrected for ionospheric Faraday rotation using the \textsc{ionFR} code, with IGS VTEC maps and the IGRF-12 as inputs. This correction is essential for low-frequency data, where the largest contribution to the uncertainty is the current method for subtracting the ionospheric RM.
      \item The RMs were measured using the RM-synthesis method and the RM dispersion, $\upsigma_{\rm{RM}}$, was obtained using the \textsc{RMCLEAN} algorithm. The RM dispersions show that the majority of the pulsars are Faraday thin sources: $<$0.001\,rad m$^{\rm -2}$.  However, in some cases, e.g., the Crab Pulsar B0531+21, the larger RM dispersion found may be due to small-scale magneto-ionic structure associated with its supernova remnant.
      \item We used the current RM and DM measurements available from the pulsar catalogue and this work to measure the magnetic scale height of the GMF in the halo, for a range of Galactic longitudes towards Galactic quadrants \textsc{I} and \textsc{II}, where the majority of the LOFAR measurements are located. Fitting all points, we found a general scale height of  2.0$\pm$0.3\,kpc -- comparable to the scale height of Galactic free electrons published in the literature. Although distance estimates are available for all pulsars, independent distance measurements, e.g., annual parallax measurements, are important for the purpose of reconstructing the magnetic scale height in the halo and the GMF structure in general. 
      \item The RM measurements from this work present an initial precise catalogue, which will be expanded and also increased in accuracy by using the LOFAR timing data towards a set of slow pulsars and MSPs. Precise DMs and RMs from low-frequency instruments, such as these from LOFAR, are becoming routine, promising an era of monitoring for time variability, which can be used to further investigate, for example, small-scale foreground ISM structures.
      \item The results from the low-frequency pathfinder/precursor telescopes show the promise of the SKA. For example, the low-frequency SKA precursor in the southern hemisphere, the MWA, is also routinely observing pulsars \citep[e.g.][]{2017PASA...34...70X}. The DM and RM measurements the MWA can provide are complementary to LOFAR, from which an all-sky low-frequency catalogue can be assembled, allowing us to study the 3-D Galactic magnetic field structure in more detail. In the future, the SKA will revolutionise our knowledge of pulsars, magnetism, and our Galaxy through discovering, timing, and measuring the astrometric properties of a large number of pulsars in our Galaxy. Efforts towards an RM-grid of extragalactic sources, diffuse Galactic synchrotron emission maps and Zeeman splitting will also supply complementary magnetic field observables to reconstruct a more complete picture of the total GMF structure. Bayesian inference frameworks such as IMAGINE can enable us to combine these observables and theoretical models and to assess the likely (magnetic) structure of our Galaxy.
   \end{enumerate}

%Extrapolating the results from the low-frequency pathfinder/precursor telescopes shows the promise of the SKA.

%%%%%%%%%%%%%%%%% ACKNOWLEDGEMENTS %%%%%%%%%%%%%%%%%%

% The Acknowledgements section is not numbered. Here you can thank helpful
%colleagues, acknowledge funding agencies, telescopes and facilities used etc.

\section*{Acknowledgements}

This paper is based on data obtained with the International LOFAR Telescope (ILT) under project codes LC1\_003, LC1\_027. LOFAR (van Haarlem et al. 2013) is the Low Frequency Array designed and constructed by ASTRON. It has observing, data processing, and data storage facilities in several countries, that are owned by various parties (each with their own funding sources), and that are collectively operated by the ILT foundation under a joint scientific policy. The ILT resources have benefitted from the following recent major funding sources: CNRS-INSU, Observatoire de Paris and Universit\'{e} d'Orl\'{e}ans, France; BMBF, MIWF-NRW, MPG, Germany; Science Foundation Ireland (SFI), Department of Business, Enterprise and Innovation (DBEI), Ireland; NWO, The Netherlands; The Science and Technology Facilities Council, UK.
We thank Tom Hassall for his assistance in acquiring the data, and Gregory Desvignes for his assistance with ionospheric RM corrections. We thank Leszek B\l{}aszkiewicz and Dominic Schnitzeler for their useful comments. 
J.W.T.H., V.I.K., and D.M. acknowledge funding from an NWO Vidi fellowship and ERC Starting Grant ``DRAGNET'' (337062).
The research leading to these results has received funding from the European Research Council under the European Union's Seventh Framework Programme (FP/2007-2013) / ERC Grant Agreement n. 617199.
This work used the \textsc{Python} plotting library \textsc{Matplotlib} \citep{Hunter:2007}.
This research has made use of NASA's Astrophysics Data System Bibliographic Services.
We thank the referee, Kevin Stovall, for providing helpful comments.

%\bibliographystyle{mnras}

\bibliographystyle{mnras} 
\bibliography{RMs_mnras}

%%%%%%%%%%%%%%%%% APPENDICES %%%%%%%%%%%%%%%%%%%%%%%% 

\clearpage

\appendix

%\onecolumn
\section{Summary of observations and results}

Table \ref{tab:1} summarises the pulsar observations in the `slow' (non-recycled) pulsar and MSP LOFAR HBA censuses \citep[][respectively]{2016A&A...591A.134B,2016A&A...585A.128K}.
Columns 1--4 show published information for the pulsars: the pulsar name (based on B1950 or J2000 coordinates); and the RM published in the latest version of the pulsar catalogue (v 1.59), with uncertainty ($\pm$), and literature reference (all noted by an asterisk if no literature measurements are published in the pulsar catalogue). The pulsars are listed in order of RA (as used in the pulsar catalogue), with the pulsars in the `slow' pulsar census listed first and the pulsars in the MSP census listed after.
Columns 5--9 provide the details of the LOFAR observations: modified Julian date of the LOFAR observation (MJD); signal-to-noise (S/N) of the total intensity (Stokes $I$) pulsar profile; number of bins used to extract the Stokes parameters from the pulse profile, and the total number of pulse profile phase bins, separated by `/'; the LOFAR DM and uncertainty \citep[][]{2016A&A...591A.134B,2016A&A...585A.128K}.
Columns 10--18 show the results from this work: the total observed RM (RM$_{\rm{obs}}$), plus uncertainty; the ionospheric RM (RM$_{\rm{ion}}$), plus uncertainty; RM due to the ISM alone ($\mathrm{RM_{ISM} = RM_{obs} - RM_{ion}}$), plus uncertainty; resulting electron density-weighted average magnetic field parallel to the LoS ($\langle B_{\parallel}\rangle$), plus uncertainty; and RM dispersion output from \textsc{RM CLEAN} ($\upsigma_{\rm{RM}}$).

The references in Table \ref{tab:1} are listed in full in alphabetical order in Table \ref{tab:refs}, along with the number of pulsars in common with this work (MSPs are identified separately), the lowest centre observing frequency, and whether the RMs were corrected for RM$_{\rm{ion}}$.

\onecolumn

\begin{landscape}
\begin{center}
{\scriptsize
%\setlength\tabcolsep{2.6pt} 

%\begin{longtable}{@{}l@{}l@{}l@{}l@{}ll@{}l@{}l@{}l@{}ll@{}l@{}l@{}l@{}l@{}lll@{}}
%\begin{longtable}{llllllllllllllllll}
%\begin{longtable}{lrrlrrrrrrrrrrrrrr}
\begin{longtable}{rrrrrrrrrrrrrrrrrr}

\caption{Summary of RM measurements. See text for further description of columns. } \label{tab:1} \\

\hline \hline
PSR & RM$_{\rm{cat}}$ & $\pm$ & ref & MJD & S/N & Bins & DM & $\pm$ & RM$_{\rm{obs}}$ & $\pm$  & RM$_{\rm{ion}}$ & $\pm$ & RM$_{\rm{ISM}}$ & $\pm$  & $\langle B_{\parallel}\rangle$ & $\pm$ & $\upsigma_{\rm{RM}}$   \\
&  rad m$^{\rm{-2}}$ & rad m$^{\rm{-2}}$ &  &  &  &  & pc cm$^{\rm{-3}}$ & pc cm$^{\rm{-3}}$ & rad m$^{\rm{-2}}$ & rad m$^{\rm{-2}}$ & rad m$^{\rm{-2}}$ & rad m$^{\rm{-2}}$ & rad m$^{\rm{-2}}$ & rad m$^{\rm{-2}}$  & $\upmu$G &  $\upmu$G & rad m$^{\rm{-2}}$ \\
\hline
\endfirsthead

\multicolumn{6}{c}%
{{\bfseries \tablename\ \thetable{} -- continued from previous page}} \\
\hline
%Name & Jname & RM$^{\rm{psrcat}}$ & $\sigma_{\rm{RM}}^{\rm{psrcat}}$ & ObsID & MJD & T$_{\rm{int}}$ & S/N & DM & $\sigma_{\rm{DM}}$ & RM$^{\rm{obs}}$ & $\sigma_{\rm{RM}}^{\rm{obs}}$  & RM$^{\rm{ion}}$ & $\sigma_{\rm{RM}}^{\rm{ion}}$ & RM$^{\rm{ISM}}$ & $\sigma_{\rm{RM}}^{\rm{ISM}}$  & $\langle B_{\parallel}^{\rm{ISM}}\rangle$ & $\sigma_{\langle B_{\parallel}^{\rm{ISM}}\rangle}$  \\
%&  &  (rad m$^{\rm{-2}}$) &  &  &  & (min) &  & (pc cm$^{\rm{-3}}$) &  & (rad m$^{\rm{-2}}$) & & (rad m$^{\rm{-2}}$) &  & (rad m$^{\rm{-2}}$) &  & ($\upmu$G) & \\
PSR & RM$_{\rm{cat}}$ & $\pm$ & ref & MJD & S/N & Bins & DM & $\pm$ & RM$_{\rm{obs}}$ & $\pm$  & RM$_{\rm{ion}}$ & $\pm$ & RM$_{\rm{ISM}}$ & $\pm$  & $\langle B_{\parallel}\rangle$ & $\pm$ & $\upsigma_{\rm{RM}}$   \\
&  rad m$^{\rm{-2}}$ & rad m$^{\rm{-2}}$ &  &  &  &  & pc cm$^{\rm{-3}}$ & pc cm$^{\rm{-3}}$ & rad m$^{\rm{-2}}$ & rad m$^{\rm{-2}}$ & rad m$^{\rm{-2}}$ & rad m$^{\rm{-2}}$ & rad m$^{\rm{-2}}$ & rad m$^{\rm{-2}}$  & $\upmu$G &  $\upmu$G & rad m$^{\rm{-2}}$ \\
\hline
\endhead

\hline \multicolumn{2}{c}{{Continued on next page}} \\ \hline
\endfoot

\hline \hline
\endlastfoot

`Slow' census   &  = 117 &  &  &  &  &  &  &  &  &  &  &  &  &  &  &  &  \\
\hline

B0011+47 & --15.30 & 0.70 & hmvd18 & 56773.46 & 36 & 11/1024 & 30.4050 & 0.0130 & --13.06 & 0.03 & 2.50 & 0.09 & --15.56 & 0.10 & --0.631 & 0.005 & <0.0005 \\
B0037+56 & 9.00 & 13.00 & rl94 & 56753.42 & 38 & 43/1024 & 92.5146 & 0.0025 & 18.96 & 0.08 & 3.65 & 0.16 & 15.31 & 0.18 & 0.204 & 0.003 & <0.0003 \\
B0045+33 & * & * & * & 56703.61 & 98 & 6/1024 & 39.9220 & 0.0150 & --80.14 & 0.02 & 3.14 & 0.09 & --83.27 & 0.10 & --2.570 & 0.004 & <0.0005 \\
B0052+51 & --66.60 & 1.50 & hmvd18 & 56776.32 & 35 & 7/1024 & 44.0127 & 0.0024 & --61.69 & 0.03 & 2.40 & 0.06 & --64.09 & 0.07 & --1.794 & 0.002 & <0.2227 \\
B0053+47 & --23.00 & 22.00 & mwkj03 & 56703.59 & 19 & 51/1024 & 18.1354 & 0.0013 & --31.38 & 0.10 & 2.78 & 0.09 & --34.16 & 0.13 & --2.321 & 0.011 & <0.0004 \\
B0105+65 & --29.00 & 3.00 & man74 & 56784.4 & 88 & 5/1024 & 30.5482 & 0.0014 & --24.37 & 0.01 & 2.71 & 0.11 & --27.09 & 0.11 & --1.092 & 0.005 & <0.0241 \\
B0105+68 & --46.00 & 19.00 & mwkj03 & 56703.62 & 29 & 5/1024 & 61.0617 & 0.0036 & --30.52 & 0.02 & 2.51 & 0.08 & --33.04 & 0.08 & --0.667 & 0.002 & <0.1386 \\
J0137+1654 & * & * & * & 56703.64 & 18 & 17/1024 & 26.0838 & 0.0024 & --13.39 & 0.03 & 3.44 & 0.09 & --16.83 & 0.10 & --0.795 & 0.006 & <0.0003 \\
B0136+57 & --94.13 & 0.08 & nsk+15 & 56753.45 & 394 & 5/256 & 73.8114 & 0.0008 & --90.26 & 0.01 & 3.71 & 0.19 & --93.97 & 0.19 & --1.568 & 0.004 & <0.2819 \\
J0152+0948 & 20.00 & 12.00 & hmvd18 & 56784.43 & 12 & 13/1024 & 22.8810 & 0.0120 & 5.52 & 0.11 & 4.16 & 0.15 & 1.37 & 0.19 & 0.074 & 0.012 & <0.0005 \\
B0153+39 & --46.50 & 5.50 & hmvd18 & 56773.5 & 20 & 9/1024 & 59.8330 & 0.0110 & --65.86 & 0.02 & 2.72 & 0.08 & --68.58 & 0.09 & --1.412 & 0.002 & <0.1039 \\
J0212+5222 & * & * & * & 56773.52 & 23 & 3/1024 & 38.2356 & 0.0005 & --11.15 & 0.02 & 2.53 & 0.08 & --13.68 & 0.08 & --0.441 & 0.003 & <0.1114 \\
B0226+70 & --56.00 & 21.00 & mwkj03 & 56703.68 & 21 & 5/1024 & 46.6794 & 0.0016 & --41.56 & 0.02 & 2.25 & 0.06 & --43.81 & 0.06 & --1.156 & 0.002 & <0.0003 \\
B0301+19 & --8.30 & 0.30 & man74 & 56703.69 & 146 & 2/1024 & 15.6568 & 0.0004 & --5.48 & 0.02 & 2.95 & 0.06 & --8.43 & 0.06 & --0.663 & 0.006 & <0.1734 \\
B0320+39 & 56.30 & 1.00 & fdr15 & 56703.72 & 625 & 1/1024 & 26.1898 & 0.0009 & 62.32 & 0.02 & 2.33 & 0.04 & 59.99 & 0.05 & 2.822 & 0.003 & <0.3189 \\
B0331+45 & --41.00 & 20.00 & rl94 & 56747.65 & 143 & 3/512 & 47.1457 & 0.0003 & 10.57 & 0.05 & 4.98 & 0.08 & 5.60 & 0.09 & 0.146 & 0.003 & 0.0000 \\
B0402+61 & 9.00 & 3.00 & hl87 & 56747.67 & 48 & 2/1024 & 65.4053 & 0.0034 & 12.56 & 0.02 & 4.38 & 0.08 & 8.18 & 0.08 & 0.154 & 0.002 & <0.3028 \\
B0410+69 & --31.60 & 4.80 & hmvd18 & 56773.55 & 77 & 3/1024 & 27.4460 & 0.0004 & --19.09 & 0.02 & 2.33 & 0.06 & --21.42 & 0.06 & --0.962 & 0.004 & 0.0000 \\
J0417+35 & * & * & * & 56773.56 & 31 & 4/1024 & 48.5336 & 0.0012 & 45.04 & 0.06 & 3.03 & 0.07 & 42.01 & 0.10 & 1.066 & 0.003 & <0.0004 \\
J0435+2749 & 2.00 & 0.00 & bfrs18 & 56773.58 & 20 & 9/512 & 53.1819 & 0.0002 & 4.21 & 0.07 & 3.25 & 0.06 & 0.96 & 0.09 & 0.022 & 0.003 & <0.0005 \\
B0450+55 & 10.00 & 3.00 & hl87 & 56772.61 & 275 & 1/1024 & 14.5900 & 0.0002 & 8.74 & 0.01 & 2.94 & 0.11 & 5.79 & 0.11 & 0.489 & 0.011 & <0.0004 \\
B0523+11 & 37.00 & 2.00 & jhv+05 & 56703.77 & 32 & 28/512 & 79.4180 & 0.0130 & 22.18 & 0.13 & 2.07 & 0.10 & 20.11 & 0.17 & 0.312 & 0.003 & 0.0000 \\
B0525+21 & --39.60 & 0.20 & man72 & 56747.71 & 746 & 1/1024 & 50.8695 & 0.0013 & --34.55 & 0.01 & 5.47 & 0.06 & --40.02 & 0.07 & --0.969 & 0.002 & <0.2558 \\
B0531+21 & --42.30 & 0.50 & man72 & 56703.78 & 120 & 1/64 & 56.7712 & 0.0002 & --43.70 & 0.01 & 1.74 & 0.08 & --45.44 & 0.08 & --0.986 & 0.002 & <0.2542 \\
B0540+23 & 2.70 & 0.60 & jkk+07 & 56779.57 & 39 & 1/256 & 77.7026 & 0.0010 & 5.95 & 0.01 & 3.10 & 0.10 & 2.85 & 0.10 & 0.045 & 0.002 & <0.1797 \\
J0611+30 & * & * & * & 56772.64 & 78 & 2/1024 & 45.2551 & 0.0016 & 17.91 & 0.02 & 3.45 & 0.08 & 14.46 & 0.08 & 0.394 & 0.003 & 0.0000 \\
B0609+37 & 23.00 & 9.00 & wck+04 & 56772.66 & 50 & 1/512 & 27.1550 & 0.0003 & 37.73 & 0.02 & 3.22 & 0.05 & 34.51 & 0.06 & 1.566 & 0.003 & <0.0005 \\
B0626+24 & 69.50 & 0.20 & wck+04 & 56747.76 & 160 & 2/512 & 84.1762 & 0.0050 & 80.41 & 0.02 & 4.70 & 0.09 & 75.70 & 0.09 & 1.108 & 0.002 & <0.2716 \\
B0643+80 & --32.00 & 3.00 & hl87 & 56747.78 & 97 & 1/1024 & 33.3188 & 0.0007 & --28.87 & 0.01 & 2.89 & 0.06 & --31.77 & 0.07 & --1.175 & 0.003 & <0.2068 \\
B0656+14 & 23.00 & 0.30 & jkk+07 & 56747.75 & 15 & 14/1024 & 14.0762 & 0.0024 & 28.00 & 0.02 & 5.26 & 0.07 & 22.73 & 0.08 & 1.990 & 0.008 & <0.0004 \\
B0655+64 & --7.00 & 6.00 & hl87 & 56772.68 & 112 & 3/1024 & 8.7739 & 0.0003 & --15.36 & 0.02 & 2.71 & 0.06 & --18.07 & 0.06 & --2.537 & 0.011 & <0.0003 \\
B0751+32 & --7.00 & 5.00 & hl87 & 56772.72 & 104 & 11/1024 & 39.9863 & 0.0014 & 7.60 & 0.03 & 3.42 & 0.08 & 4.18 & 0.09 & 0.129 & 0.003 & <0.1472 \\
B0809+74 & --14.00 & 0.07 & nsk+15 & 56747.8 & 661 & 2/1024 & 5.7507 & 0.0005 & --11.07 & 0.01 & 2.84 & 0.06 & --13.91 & 0.06 & --2.980 & 0.016 & <0.1743 \\
B0823+26 & 5.38 & 0.06 & nsk+15 & 56747.82 & 1589 & 2/1024 & 19.4763 & 0.0002 & 8.93 & 0.01 & 3.87 & 0.10 & 5.05 & 0.10 & 0.320 & 0.008 & <0.0953 \\
B0841+80 & --23.20 & 3.10 & hmvd18 & 56747.83 & 23 & 7/1024 & 34.8121 & 0.0031 & --19.42 & 0.03 & 2.36 & 0.04 & --21.78 & 0.05 & --0.771 & 0.002 & <0.0004 \\
B0917+63 & --18.90 & 2.30 & hmvd18 & 56747.89 & 164 & 2/1024 & 13.1542 & 0.0002 & --12.81 & 0.01 & 2.12 & 0.06 & --14.93 & 0.06 & --1.399 & 0.007 & <0.1745 \\
B0940+16 & 53.00 & 12.00 & hl87 & 56687.1 & 21 & 2/1024 & 20.3402 & 0.0018 & 18.23 & 0.02 & 0.86 & 0.06 & 17.36 & 0.07 & 1.052 & 0.005 & 0.0000 \\
J0943+2253 & * & * & * & 56747.86 & 33 & 3/1024 & 27.2676 & 0.0016 & 19.76 & 0.02 & 3.33 & 0.08 & 16.43 & 0.08 & 0.742 & 0.004 & <0.0004 \\
B0943+10 & 13.30 & 0.50 & tml93 & 56779.82 & 882 & 1/1024 & 15.3185 & 0.0009 & 17.23 & 0.02 & 3.13 & 0.08 & 14.10 & 0.08 & 1.134 & 0.008 & <0.1470 \\
J0947+2740 & 32.00 & 0.00 & bfrs18 & 56747.88 & 24 & 5/1024 & 28.8860 & 0.0260 & 26.28 & 0.02 & 3.03 & 0.06 & 23.25 & 0.06 & 0.992 & 0.003 & <0.1104 \\
B1112+50 & --0.10 & 0.80 & fdr15 & 56747.93 & 500 & 1/1024 & 9.1863 & 0.0003 & 4.44 & 0.01 & 2.02 & 0.07 & 2.41 & 0.07 & 0.324 & 0.012 & <0.1194 \\
B1133+16 & 3.97 & 0.07 & nsk+15 & 56687.11 & 2220 & 1/1024 & 4.8407 & 0.0003 & 5.00 & 0.01 & 0.89 & 0.06 & 4.11 & 0.06 & 1.046 & 0.019 & <0.0005 \\
J1238+2152 & * & * & * & 56687.14 & 133 & 2/1024 & 17.9706 & 0.0031 & 5.21 & 0.01 & 0.80 & 0.05 & 4.41 & 0.06 & 0.303 & 0.005 & <0.1848 \\
J1246+2253 & * & * & * & 56747.99 & 21 & 29/1024 & 17.7978 & 0.0018 & 6.01 & 0.03 & 2.42 & 0.09 & 3.59 & 0.09 & 0.248 & 0.008 & <0.0004 \\
J1313+0931 & * & * & * & 56779.92 & 152 & 1/1024 & 12.0406 & 0.0000 & 4.49 & 0.02 & 2.17 & 0.10 & 2.32 & 0.11 & 0.237 & 0.013 & <0.1704 \\
B1322+83 & --23.20 & 1.10 & fdr15 & 56687.19 & 90 & 1/1024 & 13.3162 & 0.0008 & --23.32 & 0.02 & 0.35 & 0.05 & --23.67 & 0.05 & --2.190 & 0.005 & <0.0005 \\
B1508+55 & 1.28 & 0.06 & nsk+15 & 56687.26 & 3384 & 1/1024 & 19.6189 & 0.0013 & 2.04 & 0.01 & 0.55 & 0.06 & 1.49 & 0.06 & 0.094 & 0.005 & <0.1733 \\
B1530+27 & 1.00 & 0.30 & wck+04 & 56703.26 & 72 & 7/1024 & 14.6910 & 0.0160 & 4.26 & 0.02 & 0.89 & 0.08 & 3.37 & 0.08 & 0.283 & 0.008 & <0.1293 \\
B1541+09 & 21.00 & 2.00 & hl87 & 56780.02 & 544 & 1/1024 & 34.9758 & 0.0016 & 17.61 & 0.01 & 1.66 & 0.09 & 15.95 & 0.09 & 0.562 & 0.004 & <0.1028 \\
J1612+2008 & 22.00 & 3.00 & blr+13 & 56687.25 & 15 & 9/1024 & 19.5082 & 0.0032 & 23.01 & 0.02 & 0.75 & 0.06 & 22.26 & 0.07 & 1.406 & 0.005 & 0.0000 \\
J1627+1419 & 18.30 & 5.90 & hmvd18 & 56687.29 & 64 & 1/1024 & 32.1670 & 0.0008 & 19.85 & 0.02 & 1.24 & 0.06 & 18.61 & 0.06 & 0.713 & 0.003 & <0.0009 \\
B1633+24 & 31.00 & 4.00 & wck+04 & 56748.14 & 135 & 2/1024 & 24.2671 & 0.0044 & 23.81 & 0.03 & 1.85 & 0.06 & 21.97 & 0.07 & 1.115 & 0.004 & <0.0003 \\
J1645+1012 & * & * & * & 56687.33 & 87 & 1/1024 & 36.1713 & 0.0002 & 32.25 & 0.02 & 2.05 & 0.10 & 30.20 & 0.10 & 1.029 & 0.004 & 0.0000 \\
J1649+2533 & 29.70 & 1.00 & hmvd18 & 56748.12 & 25 & 11/1024 & 34.4622 & 0.0078 & 31.39 & 0.03 & 1.87 & 0.05 & 29.52 & 0.06 & 1.055 & 0.003 & <0.0987 \\
J1652+2651 & 34.50 & 0.70 & hmvd18 & 56748.11 & 33 & 21/1024 & 40.8024 & 0.0002 & 35.56 & 0.07 & 1.85 & 0.04 & 33.71 & 0.08 & 1.018 & 0.003 & <0.1472 \\
J1720+2150 & 50.00 & 3.70 & hmvd18 & 56748.15 & 16 & 5/1024 & 40.7190 & 0.0350 & 54.76 & 0.03 & 1.82 & 0.06 & 52.94 & 0.07 & 1.602 & 0.003 & <0.1456 \\
B1737+13 & 64.40 & 1.60 & wck+04 & 56687.3 & 161 & 2/1024 & 48.6682 & 0.0004 & 68.22 & 0.03 & 1.58 & 0.07 & 66.65 & 0.08 & 1.687 & 0.002 & <0.0005 \\
J1741+2758 & 49.90 & 5.70 & hmvd18 & 56773.14 & 93 & 5/1024 & 29.1449 & 0.0001 & 56.03 & 0.03 & 1.34 & 0.06 & 54.70 & 0.07 & 2.312 & 0.004 & 0.0000 \\
J1746+2245 & 68.70 & 3.70 & hmvd18 & 56784.14 & 25 & 3/1024 & 49.8543 & 0.0057 & 73.84 & 0.08 & 1.57 & 0.10 & 72.28 & 0.12 & 1.786 & 0.004 & <0.0003 \\
J1746+2540 & 93.20 & 1.50 & hmvd18 & 56784.17 & 13 & 13/1024 & 51.2044 & 0.0033 & 96.46 & 0.08 & 1.71 & 0.10 & 94.75 & 0.12 & 2.280 & 0.004 & <0.0003 \\
J1752+2359 & * & * & * & 56783.18 & 20 & 9/1024 & 36.1964 & 0.0006 & 88.64 & 0.06 & 1.59 & 0.09 & 87.05 & 0.11 & 2.963 & 0.005 & <0.0005 \\
B1753+52 & 26.60 & 1.10 & hmvd18 & 56785.18 & 17 & 5/1024 & 35.0096 & 0.0064 & 29.25 & 0.08 & 1.31 & 0.07 & 27.94 & 0.11 & 0.983 & 0.005 & <0.0672 \\
J1758+3030 & 70.40 & 2.20 & hmvd18 & 56784.19 & 157 & 4/1024 & 35.0674 & 0.0014 & 81.39 & 0.07 & 1.85 & 0.10 & 79.54 & 0.12 & 2.794 & 0.005 & <0.0005 \\
B1811+40 & 47.00 & 6.00 & hl87 & 56780.05 & 103 & 2/1024 & 41.5566 & 0.0002 & 50.53 & 0.02 & 1.15 & 0.06 & 49.38 & 0.06 & 1.464 & 0.002 & <0.1568 \\
J1821+1715 & 107.50 & 1.20 & hmvd18 & 56703.3 & 25 & 7/1024 & 60.2844 & 0.0025 & 111.21 & 0.08 & 1.85 & 0.10 & 109.36 & 0.13 & 2.235 & 0.003 & <0.0665 \\
J1838+1650 & 66.50 & 2.70 & hmvd18 & 56801.09 & 84 & 1/1024 & 32.9516 & 0.0010 & 74.36 & 0.02 & 1.71 & 0.12 & 72.65 & 0.12 & 2.716 & 0.005 & <0.0772 \\
B1839+56 & --5.00 & 0.70 & fdr15 & 56789.09 & 532 & 1/1024 & 26.7716 & 0.0002 & --2.58 & 0.01 & 1.27 & 0.06 & --3.85 & 0.06 & --0.177 & 0.004 & <0.0917 \\
B1839+09 & 53.00 & 5.00 & hl87 & 56783.19 & 110 & 1/1024 & 49.1579 & 0.0043 & 53.23 & 0.02 & 2.04 & 0.10 & 51.19 & 0.10 & 1.283 & 0.003 & <0.1000 \\
B1842+14 & 109.00 & 1.30 & jhv+05 & 56703.31 & 145 & 3/1024 & 41.4856 & 0.0006 & 120.84 & 0.04 & 2.19 & 0.11 & 118.65 & 0.12 & 3.524 & 0.004 & <0.1303 \\
J1849+2423 & 12.80 & 1.20 & hmvd18 & 56788.06 & 9 & 4/512 & 62.2677 & 0.0016 & 16.82 & 0.03 & 1.64 & 0.05 & 15.18 & 0.06 & 0.300 & 0.001 & <0.0005 \\
B1848+13 & 152.70 & 0.60 & jkk+07 & 56703.34 & 10 & 11/512 & 60.1396 & 0.0066 & 157.58 & 0.10 & 2.65 & 0.12 & 154.93 & 0.16 & 3.174 & 0.004 & <0.0005 \\
B1848+12 & 158.00 & 16.00 & rl94 & 56687.43 & 53 & 3/1024 & 70.6333 & 0.0017 & 152.53 & 0.03 & 2.73 & 0.12 & 149.80 & 0.12 & 2.613 & 0.003 & <0.2659 \\
B1905+39 & 5.20 & 0.30 & fdr15 & 56789.13 & 119 & 1/1024 & 30.9660 & 0.0140 & 6.88 & 0.01 & 1.47 & 0.08 & 5.41 & 0.09 & 0.215 & 0.004 & <0.1029 \\
J1912+2525 & 12.30 & 7.00 & hmvd18 & 56788.23 & 42 & 2/1024 & 37.8474 & 0.0016 & 34.69 & 0.02 & 2.04 & 0.09 & 32.65 & 0.10 & 1.063 & 0.004 & <0.0621 \\
B1918+26 & 20.30 & 1.30 & hmvd18 & 56748.21 & 95 & 2/1024 & 27.7088 & 0.0008 & 28.11 & 0.02 & 2.07 & 0.07 & 26.04 & 0.07 & 1.158 & 0.004 & <0.0005 \\
B1919+21 & --16.99 & 0.05 & nsk+15 & 56788.21 & 2054 & 1/1024 & 12.4440 & 0.0006 & --15.04 & 0.02 & 2.01 & 0.10 & --17.05 & 0.10 & --1.688 & 0.012 & <0.0005 \\
B1929+10 & --6.87 & 0.02 & jhv+05 & 56781.21 & 314 & 1/1024 & 3.1832 & 0.0002 & --5.27 & 0.01 & 1.86 & 0.14 & --7.14 & 0.14 & --2.762 & 0.065 & <0.0863 \\
B1944+17 & --28.00 & 0.40 & hl87 & 56801.14 & 29 & 9/1024 & 16.1356 & 0.0073 & --43.64 & 0.02 & 1.70 & 0.10 & --45.34 & 0.11 & --3.462 & 0.010 & <0.1774 \\
B1946+35 & 116.00 & 6.00 & hl87 & 56748.23 & 28 & 70/1024 & 129.3675 & 0.0008 & 121.07 & 0.09 & 2.11 & 0.06 & 118.96 & 0.11 & 1.133 & 0.001 & <0.3047 \\
B1953+50 & --23.84 & 0.05 & nsk+15 & 56789.06 & 155 & 1/1024 & 31.9827 & 0.0001 & --22.50 & 0.01 & 1.31 & 0.06 & --23.82 & 0.06 & --0.918 & 0.003 & <0.1041 \\
J1956+0838 & * & * & * & 56781.27 & 6 & 25/1024 & 67.0870 & 0.0240 & --111.05 & 0.11 & 2.44 & 0.12 & --113.49 & 0.16 & --2.084 & 0.004 & 0.0000 \\
J2002+1637 & * & * & * & 56773.22 & 4 & 8/256 & 94.5810 & 0.0480 & --39.29 & 0.09 & 1.98 & 0.10 & --41.27 & 0.13 & --0.538 & 0.002 & <0.0004 \\
J2007+0809 & * & * & * & 56779.29 & 10 & 47/512 & 53.3940 & 0.0370 & --130.49 & 0.07 & 2.84 & 0.13 & --133.32 & 0.15 & --3.076 & 0.005 & <0.0005 \\
J2007+0910 & --73.00 & 15.00 & hmvd18 & 56784.25 & 36 & 5/1024 & 48.7293 & 0.0007 & --74.57 & 0.09 & 3.24 & 0.09 & --77.81 & 0.13 & --1.967 & 0.004 & 0.0000 \\
J2017+2043 & --168.00 & 3.10 & hmvd18 & 56772.32 & 32 & 8/1024 & 60.4906 & 0.0099 & --160.93 & 0.03 & 2.68 & 0.09 & --163.61 & 0.10 & --3.332 & 0.002 & <0.0004 \\
B2016+28 & --34.60 & 1.40 & man72 & 56781.23 & 1255 & 1/1024 & 14.1839 & 0.0013 & --33.14 & 0.01 & 1.76 & 0.11 & --34.90 & 0.11 & --3.032 & 0.012 & <0.1741 \\
B2020+28 & --74.70 & 0.30 & man74 & 56748.24 & 252 & 1/1024 & 24.6311 & 0.0002 & --72.56 & 0.02 & 2.48 & 0.06 & --75.04 & 0.06 & --3.753 & 0.004 & <0.1595 \\
B2021+51 & --6.50 & 0.90 & man72 & 56789.31 & 126 & 1/1024 & 22.5497 & 0.0006 & --4.31 & 0.01 & 2.42 & 0.08 & --6.73 & 0.08 & --0.368 & 0.005 & <0.1258 \\
B2022+50 & 42.60 & 1.20 & hmvd18 & 56789.08 & 75 & 1/1024 & 32.9882 & 0.0004 & 46.11 & 0.02 & 1.28 & 0.06 & 44.84 & 0.06 & 1.675 & 0.003 & <0.0005 \\
J2036+2835 & * & * & * & 56801.17 & 23 & 3/1024 & 84.2174 & 0.0064 & --156.45 & 0.03 & 1.69 & 0.12 & --158.14 & 0.12 & --2.313 & 0.002 & <0.0006 \\
B2034+19 & --97.00 & 10.00 & wck+04 & 56773.27 & 91 & 1/1024 & 36.8916 & 0.0000 & --108.46 & 0.02 & 2.22 & 0.10 & --110.69 & 0.10 & --3.696 & 0.004 & <0.1957 \\
B2036+53 & --102.30 & 1.90 & hmvd18 & 56789.33 & 8 & 25/1024 & 160.1960 & 0.0120 & --100.56 & 0.10 & 2.49 & 0.08 & --103.05 & 0.13 & --0.793 & 0.001 & <0.0005 \\
J2043+2740 & * & * & * & 56784.28 & 233 & 1/256 & 21.0206 & 0.0002 & --93.08 & 0.01 & 3.06 & 0.12 & --96.14 & 0.12 & --5.635 & 0.009 & <0.1766 \\
B2044+15 & --100.00 & 5.00 & wck+04 & 56773.25 & 72 & 5/1024 & 39.8180 & 0.0005 & --88.02 & 0.06 & 2.18 & 0.09 & --90.20 & 0.11 & --2.791 & 0.004 & <0.0940 \\
B2045+56 & 0.60 & 1.90 & hmvd18 & 56789.16 & 54 & 1/512 & 101.7903 & 0.0001 & 2.79 & 0.02 & 1.54 & 0.08 & 1.25 & 0.08 & 0.015 & 0.001 & <0.0005 \\
B2053+21 & --100.00 & 7.00 & hr10 & 56773.34 & 100 & 1/1024 & 36.3496 & 0.0003 & --86.71 & 0.02 & 2.69 & 0.08 & --89.40 & 0.08 & --3.030 & 0.003 & <0.0961 \\
J2111+2106 & --75.30 & 0.80 & blr+13 & 56773.31 & 15 & 9/1024 & 59.2964 & 0.0035 & --72.47 & 0.10 & 2.51 & 0.09 & --74.99 & 0.14 & --1.558 & 0.003 & 0.0000 \\
B2110+27 & --37.00 & 7.00 & wck+04 & 56784.29 & 425 & 1/1024 & 25.1111 & 0.0002 & --56.95 & 0.02 & 3.23 & 0.13 & --60.18 & 0.13 & --2.953 & 0.008 & 0.0000 \\
B2113+14 & --25.00 & 8.00 & hl87 & 56703.48 & 21 & 45/1024 & 56.2044 & 0.0061 & --35.30 & 0.07 & 3.74 & 0.09 & --39.04 & 0.11 & --0.856 & 0.003 & <0.0003 \\
J2139+2242 & --86.00 & 0.40 & hmvd18 & 56753.38 & 64 & 1/1024 & 44.1597 & 0.0008 & --83.13 & 0.02 & 4.25 & 0.22 & --87.38 & 0.22 & --2.438 & 0.008 & <0.1608 \\
B2148+63 & --156.50 & 0.30 & fdr15 & 56786.34 & 54 & 26/512 & 129.7229 & 0.0055 & --155.62 & 0.03 & 1.98 & 0.09 & --157.60 & 0.09 & --1.497 & 0.001 & <0.2486 \\
J2155+2813 & --131.80 & 4.50 & hmvd18 & 56784.31 & 31 & 14/1024 & 77.1309 & 0.0043 & --129.30 & 0.06 & 3.36 & 0.13 & --132.67 & 0.14 & --2.119 & 0.003 & <0.1351 \\
B2154+40 & --32.60 & 3.00 & fdr15 & 56687.54 & 321 & 1/1024 & 71.1239 & 0.0022 & --39.68 & 0.01 & 2.35 & 0.12 & --42.03 & 0.12 & --0.728 & 0.003 & <0.2118 \\
J2205+1444 & --27.50 & 3.40 & hmvd18 & 56703.5 & 10 & 8/1024 & 36.7460 & 0.0620 & --21.76 & 0.08 & 3.76 & 0.08 & --25.51 & 0.11 & --0.855 & 0.005 & <0.0003 \\
B2210+29 & --168.00 & 5.00 & wck+04 & 56784.33 & 36 & 8/1024 & 74.5213 & 0.0015 & --165.23 & 0.07 & 3.44 & 0.12 & --168.66 & 0.14 & --2.788 & 0.003 & <0.2767 \\
J2215+1538 & --18.70 & 1.00 & hmvd18 & 56703.55 & 16 & 6/1024 & 29.2404 & 0.0012 & --16.21 & 0.03 & 3.81 & 0.09 & --20.02 & 0.09 & --0.843 & 0.005 & <0.0004 \\
B2217+47 & --35.93 & 0.06 & nsk+15 & 56687.56 & 3281 & 1/1024 & 43.4862 & 0.0060 & --33.60 & 0.01 & 2.12 & 0.10 & --35.73 & 0.10 & --1.012 & 0.004 & <0.0360 \\
J2222+2923 & * & * & * & 56773.4 & 12 & 8/512 & 49.4128 & 0.0011 & --92.96 & 0.07 & 2.75 & 0.08 & --95.71 & 0.11 & --2.386 & 0.003 & <0.0005 \\
B2224+65 & --22.99 & 0.07 & nsk+15 & 56784.36 & 232 & 1/1024 & 36.4436 & 0.0005 & --20.05 & 0.01 & 2.69 & 0.10 & --22.73 & 0.10 & --0.769 & 0.004 & <0.1735 \\
B2227+61 & --125.00 & 22.00 & mwkj03 & 56783.23 & 52 & 14/512 & 124.6388 & 0.0031 & --104.22 & 0.03 & 1.69 & 0.09 & --105.91 & 0.09 & --1.047 & 0.001 & <0.2237 \\
B2241+69 & --30.00 & 30.00 & mwkj03 & 56784.34 & 67 & 1/1024 & 40.8604 & 0.0007 & --14.16 & 0.02 & 2.64 & 0.09 & --16.80 & 0.09 & --0.506 & 0.003 & <0.0005 \\
J2253+1516 & --32.00 & 3.00 & hmvd18 & 56703.53 & 39 & 18/1024 & 29.2045 & 0.0017 & --27.14 & 0.07 & 3.84 & 0.09 & --30.98 & 0.11 & --1.307 & 0.006 & <0.0004 \\
B2303+30 & --75.50 & 4.00 & wck+04 & 56773.41 & 215 & 1/1024 & 49.5845 & 0.0012 & --84.27 & 0.03 & 2.77 & 0.10 & --87.04 & 0.10 & --2.163 & 0.003 & <0.0697 \\
B2303+46 & --25.00 & 11.00 & hmvd18 & 56773.43 & 37 & 3/1024 & 62.0676 & 0.0036 & --19.64 & 0.02 & 2.44 & 0.09 & --22.08 & 0.09 & --0.438 & 0.002 & <0.0958 \\
B2306+55 & --34.00 & 3.00 & hl87 & 56785.24 & 81 & 4/1024 & 46.5391 & 0.0004 & --27.68 & 0.02 & 1.67 & 0.07 & --29.35 & 0.08 & --0.777 & 0.003 & <0.1870 \\
B2310+42 & 4.40 & 0.10 & fdr15 & 56687.59 & 318 & 1/1024 & 17.2769 & 0.0003 & 7.03 & 0.01 & 2.06 & 0.09 & 4.97 & 0.09 & 0.355 & 0.008 & <0.2096 \\
B2315+21 & --37.00 & 3.00 & hl87 & 56687.52 & 257 & 1/1024 & 20.8696 & 0.0003 & --35.28 & 0.01 & 2.67 & 0.13 & --37.94 & 0.13 & --2.240 & 0.009 & <0.1231 \\

\hline
MSP census   & = 19 &  &  &  &  &  &  &  &  & &  &  &  &  &  &  \\
\hline
J0030+0451 & 17.00 & 16.00 & hmvd18 & 56304.694 & 31 & 6/1024 & 4.33261 & 0.00001 & 2.49 & 0.03 & 1.32 & 0.07 & 1.17 & 0.08 & 0.332 & 0.026 & <0.1489 \\
J0214+5222 & * & * & * & 56646.791 & 41 & 8/1024 & 22.03610 & 0.00010 & --15.93 & 0.03 & 0.50 & 0.07 & --16.44 & 0.07 & --0.919 & 0.005 & <0.0004 \\
J0218+4232 & * & * & * & 56473.304 & 19 & 12/256 & 61.23890 & 0.00030 & --59.81 & 0.03 & 1.60 & 0.07 & --61.40 & 0.08 & --1.235 & 0.002 & <0.0757 \\
J0407+1607 & * & * & * & 56790.515 & 30 & 13/1024 & 35.61091 & 0.00017 & 6.31 & 0.06 & 3.78 & 0.11 & 2.53 & 0.13 & 0.087 & 0.005 & <0.2229 \\
J0621+1002 & 28.80 & 6.80 & hmvd18 & 56289.023 & 9 & 12/128 & 36.53490 & 0.00810 & 53.91 & 0.11 & 0.71 & 0.07 & 53.20 & 0.13 & 1.794 & 0.005 & 0.0000 \\
J0636+5129 & * & * & * & 56648.046 & 10 & 6/128 & 11.10647 & 0.00021 & --1.72 & 0.08 & 0.41 & 0.06 & --2.14 & 0.10 & --0.237 & 0.013 & <0.0005 \\
J0645+5158 & * & * & * & 56322.009 & 19 & 6/1024 & 18.24846 & 0.00015 & --1.58 & 0.08 & 0.39 & 0.07 & --1.97 & 0.11 & --0.133 & 0.009 & <0.0005 \\
J1012+5307 & 2.98 & 0.06 & nsk+15 & 56289.149 & 32 & 1/1024 & 9.02436 & 0.00008 & 3.38 & 0.02 & 0.39 & 0.06 & 2.98 & 0.06 & 0.407 & 0.010 & <0.1889 \\
J1022+1001 & 1.39 & 0.05 & nsk+15 & 56296.126 & 49 & 3/1024 & 10.25327 & 0.00016 & 2.09 & 0.02 & 0.90 & 0.05 & 1.19 & 0.05 & 0.143 & 0.007 & <0.2638 \\
J1024--0719 & --2.40 & 0.20 & dhm+15 & 56280.16111 & 10 & 25/512 & 6.48445 & 0.00025 & --1.89 & 0.06 & 1.06 & 0.07 & --2.95 & 0.09 & --0.561 & 0.022 & <0.0005 \\
B1257+12 & 7.91 & 0.06 & nsk+15 & 56296.23 & 44 & 4/1024 & 10.15405 & 0.00011 & 8.55 & 0.02 & 0.77 & 0.07 & 7.78 & 0.07 & 0.944 & 0.011 & <0.2084 \\
J1640+2224 & 24.60 & 9.40 & hmvd18 & 56289.368 & 15 & 6/512 & 18.42766 & 0.00001 & 24.22 & 0.03 & 1.41 & 0.07 & 22.81 & 0.07 & 1.525 & 0.006 & <0.1926 \\
J1744--1134 & 2.20 & 0.20 & dhm+15 & 56293.44 & 15 & 23/512 & 3.13815 & 0.00008 & 4.22 & 0.07 & 2.63 & 0.08 & 1.59 & 0.11 & 0.625 & 0.052 & <0.0005 \\
J1810+1744 & * & * & * & 56293.456 & 65 & 24/512 & 39.66000 & 0.00020 & 90.15 & 0.07 & 1.62 & 0.06 & 88.53 & 0.10 & 2.750 & 0.004 & <0.0004 \\
J1923+2515 & 10.80 & 3.80 & lbr+13 & 56318.472 & 12 & 51/512 & 18.85567 & 0.00012 & 15.44 & 0.10 & 3.10 & 0.09 & 12.34 & 0.13 & 0.807 & 0.011 & 0.0000 \\
B1937+21 & 8.30 & 0.10 & dhm+15 & 56434.134 & 25 & 1/256 & 71.02373 & 0.00005 & 9.45 & 0.02 & 1.72 & 0.06 & 7.73 & 0.06 & 0.134 & 0.001 & <0.2463 \\
J2043+1711 & * & * & * & 56311.522 & 14 & 3/256 & 20.71232 & 0.00019 & --70.94 & 0.03 & 2.09 & 0.07 & --73.04 & 0.08 & --4.344 & 0.006 & <0.1028 \\
J2214+3000 & * & * & * & 56457.159 & 7 & 13/128 & 22.54310 & 0.00420 & --43.13 & 0.11 & 1.79 & 0.08 & --44.91 & 0.13 & --2.455 & 0.009 & <0.0005 \\
J2317+1439 & --4.00 & 3.30 & hmvd18 & 56304.635 & 40 & 13/256 & 21.89876 & 0.00026 & --8.40 & 0.01 & 1.53 & 0.07 & --9.93 & 0.07 & --0.559 & 0.005 & <0.1760 \\
\hline

  \end{longtable}
  }    \end{center}
  
  %long with the number of pulsars with published RMs in common with this work, and the method of correcting for ionospheric Faraday rotation
\begin{table*}
	\begin{center}
	\caption{Summary of the references for published RMs in Table \ref{tab:1}.}
	\label{tab:refs}
%	\begin{tabular}{@{}l @{}l @{}l | @{}l @{}l} % five columns, alignment for each
	\begin{tabular}{llll} % five columns, alignment for each
		\hline
		\hline
		Reference &  N pulsars in common & (lowest) Centre frequency & Ionospheric RM corrected  \\
		 &  `slow' + MSP & (MHz) &  (Yes/No) \\
		\hline
bfrs18 \citep[][]{2018MNRAS.474.2012B} & 2 & 327 & N \\
 blr+13 \citep[][]{2013ApJ...763...80B} & 2 & 820 &  N \\
 dhm+15 \citep[][]{2015MNRAS.449.3223D} & 3 MSPs & 730 & Y \\ % IRI
 fdr15 \citep[][]{2015MNRAS.453.4485F}  & 8  & 1100 & N \\ %Weisberg acknowledged
 hl87 \citep[][]{1987MNRAS.224.1073H}  &  14  & 408 &  Y  \\ % ionosondes
 hmvd18 \citep[][]{2018ApJS..234...11H}  & 32 + 4 MSPs & 774  & Y \\ % IRI
 hr10 \citep[][]{2010AJ....139..168H} & 1 & 50 & N \\
 jhv+05 \citep[][]{2005MNRAS.364.1397J}  & 3 & 1369 & Y \\
 jkk+07 \citep[][]{2007MNRAS.381.1625J}   & 3 & 690  & N \\
 lbr+13 \citep[][]{2013ApJ...763...81L}  & 1 MSP & 350/820 & N \\
 man72 \citep[][]{1972ApJ...172...43M} & 4 &  250  & Y \\
 man74 \citep[][]{1974ApJ...188..637M} & 3 & 250  &  Y  \\
 mwkj03 \citep[][]{2003AA...398..993M} & 5 & 1400  & N \\
 nsk+15 \citep{2015AA...576A..62N} & 9 + 3 MSPs & 150 & Y \\
 rl94 \citep[][]{1994MNRAS.268..497R} & 3 & 1400  & N \\
 tml93 \citep[][]{1993ApJS...88..529T} & 1 & various & various \\ %B0943+10 Hamilton+1981 unpublished
 wck+04 \citep[][]{2004ApJS..150..317W} & 10 & 430  & Y  \\
		\hline
	\end{tabular}
	\end{center}
\end{table*}

\end{landscape}

%\twocolumn

\newpage
\twocolumn

\section{FDF summary}\label{ap:B}

Here, we present some additional results from analysis of the FDFs output by the RM synthesis method.   

\begin{figure*}
 \includegraphics[width=0.49\columnwidth]{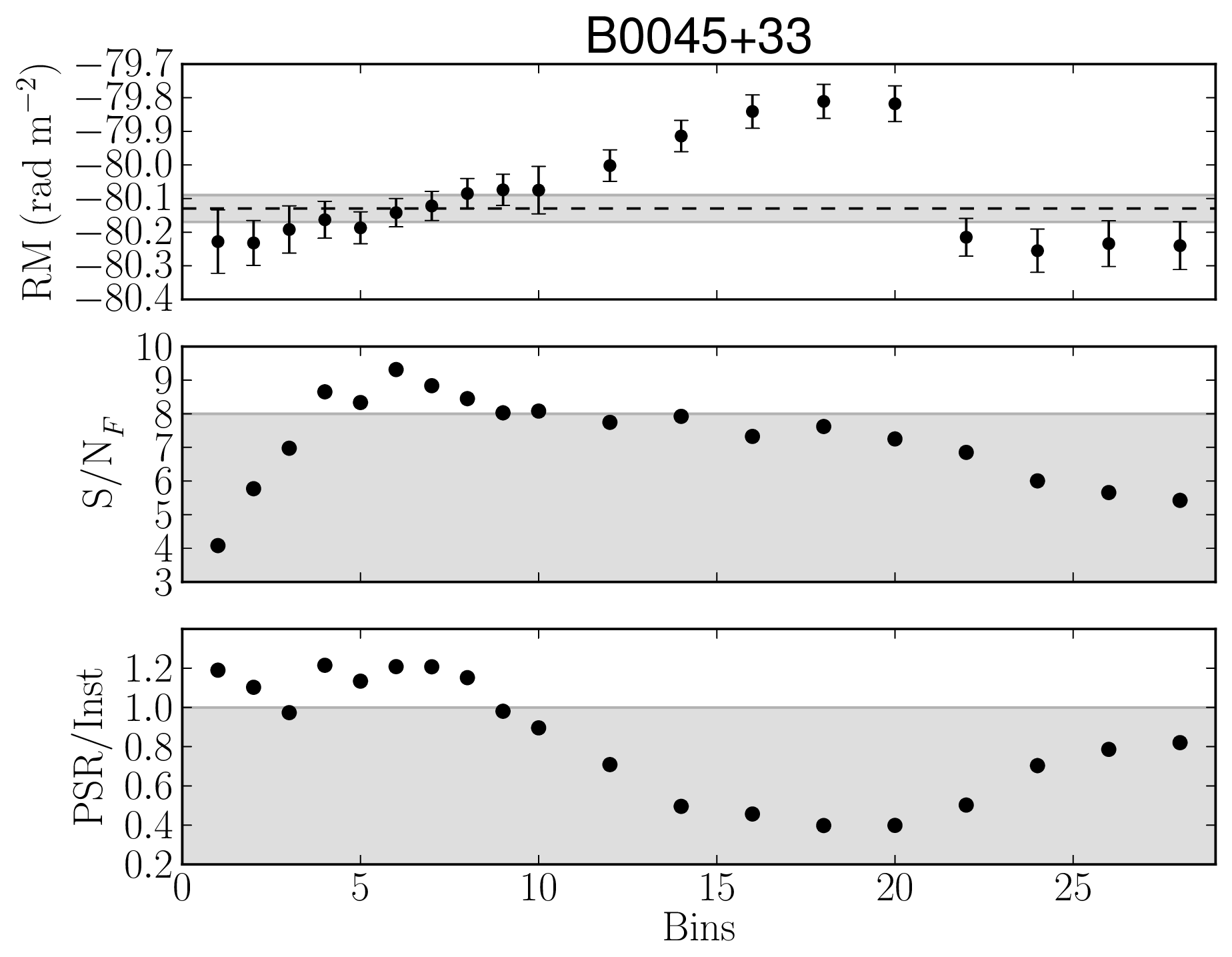}
 \includegraphics[width=0.505\columnwidth]{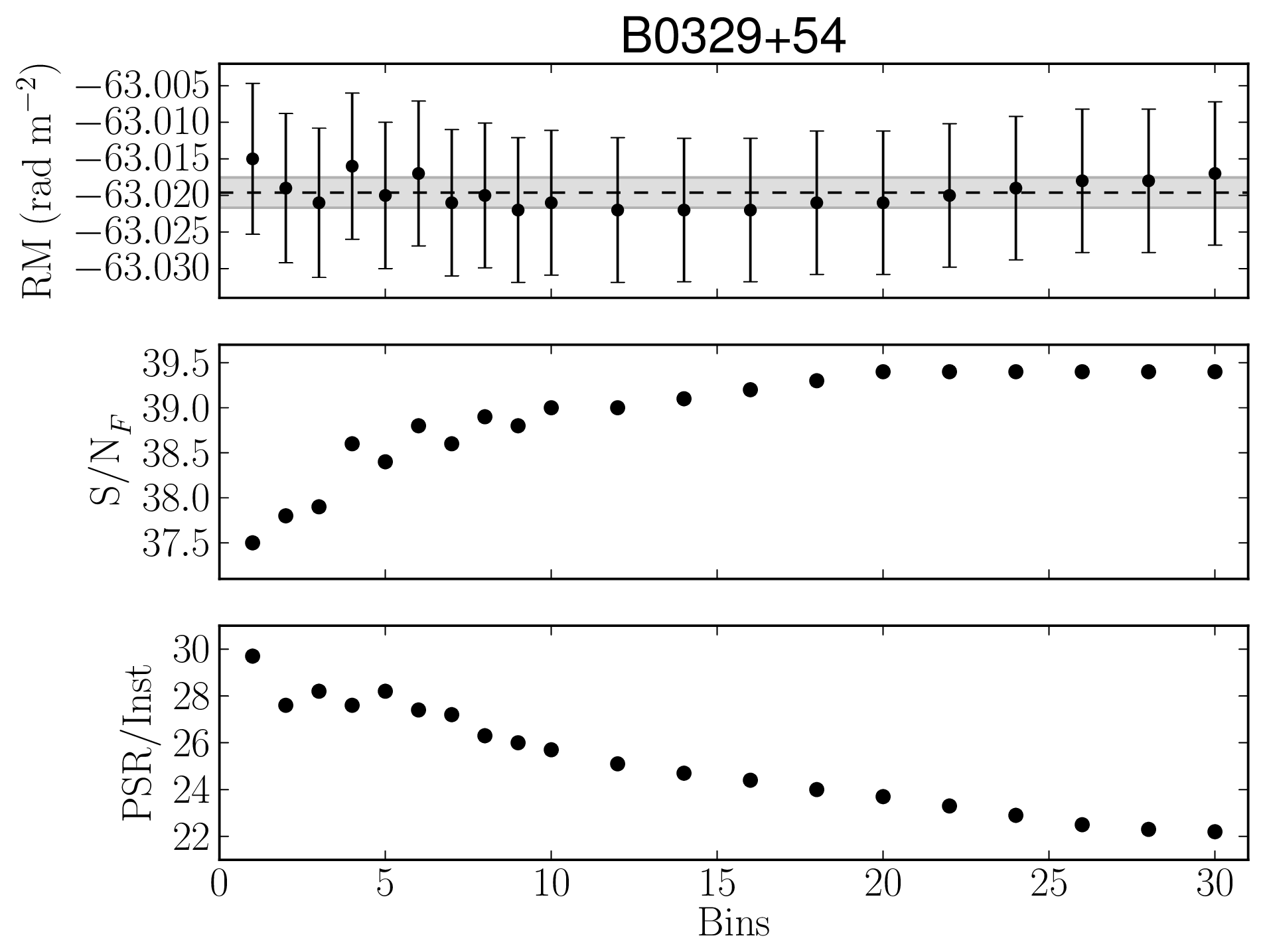}
 \caption{FDF statistics as a function of increasing numbers of pulse profile bins used to obtain the Stokes parameters from the average pulse profiles of PSR B0045+33 (left) and PSR B0329+54 (right). Top panel: measured RMs and formal uncertainties.  The weighted mean RM for all measurements with S/N$_{F}>$8 is shown by the dashed line, and the weighted mean $\pm$ the standard deviation of the measurements is shaded in grey.   Middle panel: measured S/N of the signal associated with the pulsar in the FDFs, S/N$_{F}<$8 is shown by the shaded region. Lower panel: the ratio between the intensity of the signal associated with the pulsar and the intensity of the instrumental polarisation near 0\,rad m$^{\rm{-2}}$ in the FDF (PSR/Inst). The region where PSR/Inst$<$1 shaded. }
 \label{fig:bin_tests}
\end{figure*}

Figure \ref{fig:bin_tests} shows examples of some of the statistics output from testing the numbers of pulse phase bins used to output the Stokes parameters for the RM synthesis step. 
This includes the RM measured from the FDF, the S/N in the FDF (S/N$_{{F}}$), and the ratio between the peak signal associated with the pulsar and the peak due to the instrumental polarisation  in the FDF (PSR/Inst). Figure \ref{fig:bin_tests} provides an example of two pulsars: one with with relatively low S/N (PSR B0045+33); and one with high S/N (PSR B0329+54).
 Table \ref{tab:bin_tests} summarises the main results from the tests. For PSR B0045+33, the RM with the highest S/N$_{{F}}$ was used for the final results presented in this work (6 bins). Cumulative bin numbers 4--8 would also be suitable, based on S/N$_{{F}}\ge$8 and PSR/Inst$>$1, and these measured RMs agree within the uncertainties. For PSR B0329+54, the RM value for the final catalogue was chosen based on a compromise between relatively high S/N and PSR/Inst (5 bins). However, due to the high S/N and PSR/Inst for the pulsar, we see little variation between the measured RMs, which are all in excellent agreement within the uncertainties. Table \ref{tab:bin_tests} shows that for pulsars with lower S/N, the standard deviation for the RMs measured using different ranges of pulse phase bins is larger than for the higher S/N case. In the low S/N case, the standard deviation is approximately equal to the formal errors output from the RM measurement in the FDF. Due to the larger variation identified in these cases, the uncertainties from the RM measurements for pulsars with $4<\rm{S/N}_{{F}}\le8$ were doubled. This increased the uncertainties for 37 pulsars in the sample, but nonetheless, the uncertainties remain relatively small -- the maximum value is 0.13\,rad m$^{\rm {-2}}$.
 The standard deviations for the instrumental signals in both low and high S/N cases are identical. 

\begin{figure*}
 \includegraphics[width=0.5\columnwidth]{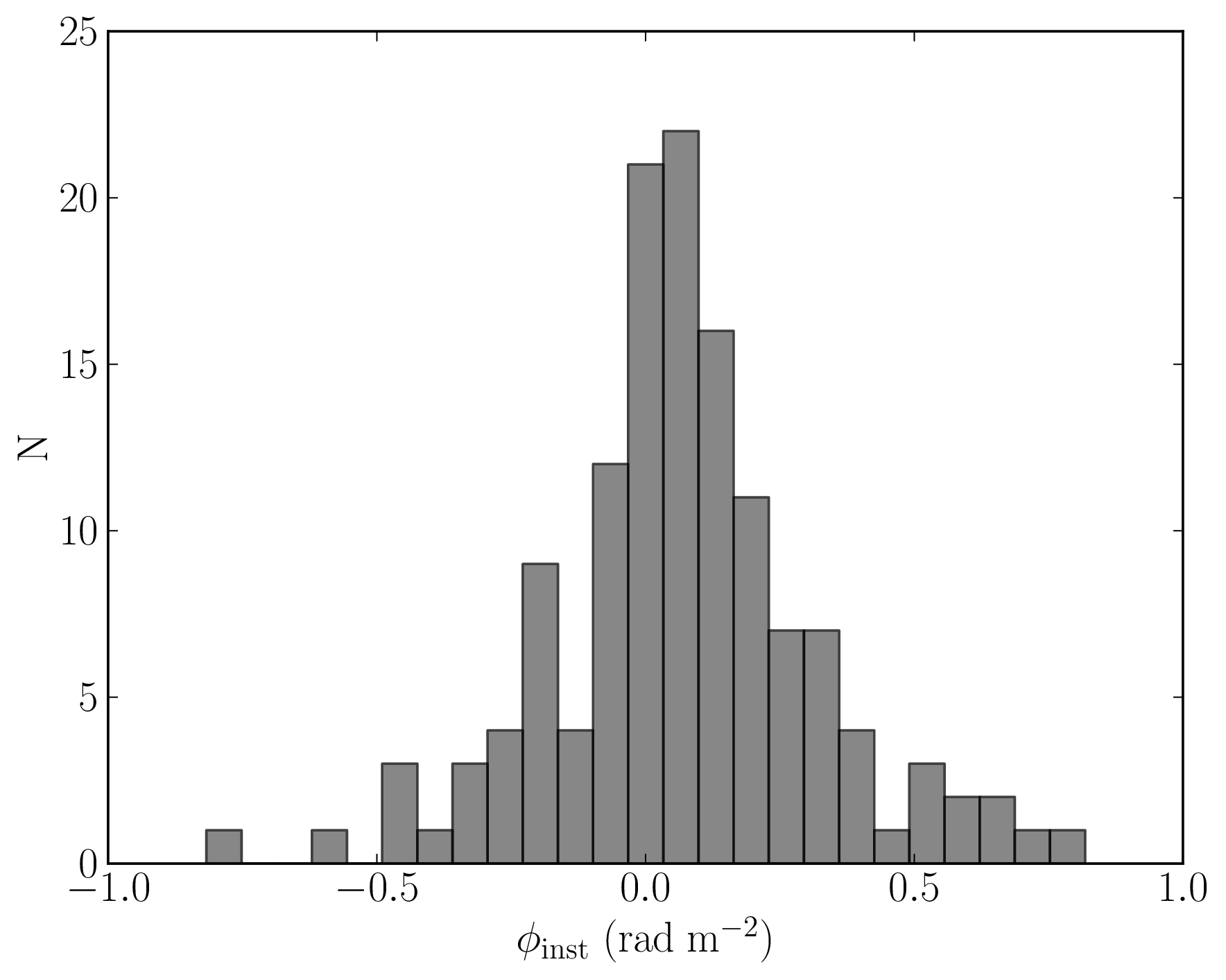}
 \caption{A histogram showing the positions of the instrumental peak measured in the FDFs ($\phi_{\rm{inst}}$) determined for all pulsars for which we obtained an RM measurement. }
 \label{fig:inst}
\end{figure*}

Figure \ref{fig:inst} shows the position of the instrumental peak ($\phi_{\rm{inst}}$) measured from the FDFs where a significant RM was detected. The median value is 0.04\,rad m$^{\rm {-2}}$, and the standard deviation is 0.25\,rad m$^{\rm {-2}}$ (32 per cent of $\delta\phi$).  The histogram and statistics show that the instrumental peak in the FDFs does not deviate greatly from 0\,rad m$^{\rm {-2}}$. This is generally expected since instrumental polarisation is not likely to vary with observing wavelength squared.
%the 25th and 75th percentiles are -0.06 and 0.18\,rad m$^{\rm {-2}}$

Figure \ref{fig:FDFs} presents the FDFs obtained for each pulsar summarised in Table \ref{tab:1}, highlighting the RM$_{\rm{obs}}$ measured. 

\onecolumn

\begin{table*}
	\centering
	\caption{Summary of RM statistics (in units of rad m$^{\rm {-2}}$) from the Stokes parameters output using a range in pulse profile bin numbers, shown in Figure \ref{fig:bin_tests}. For PSR B0045+33, the RM weighted mean and standard deviation were calculated using values with S/N$>$8. }
	\label{tab:bin_tests}
%	\begin{tabular}{@{}l @{}l @{}l | @{}l @{}l} % five columns, alignment for each
	\begin{tabular}{lrr} % four columns, alignment for each
		\hline
		\hline
		Parameter &  PSR B0045+33  &  PSR B0329+54 \\
		\hline
		Pulsar RM (RM$_{\rm{obs}}$) weighted mean &  --80.13 & --63.020  \\
		Pulsar RM standard deviation  & 0.04 & 0.002 \\
		Instrumental ($\phi_{\rm{inst}}$) mean &  --0.07 & --0.33   \\
		Instrumental standard deviation  &   0.02 &  0.02  \\
		%\hline
		
		\hline
	\end{tabular}
\end{table*}

%\begin{figure}
%	\includegraphics[width=\columnwidth]{example}
%    \caption{This is an example figure. Captions appear below each figure.
%	Give enough detail for the reader to understand what they're looking at,
%	but leave detailed discussion to the main body of the text.}
 %   \label{fig:example_figure}
%\end{figure}

%

\begin{figure*}
\centering

 \includegraphics[width=0.246\columnwidth]{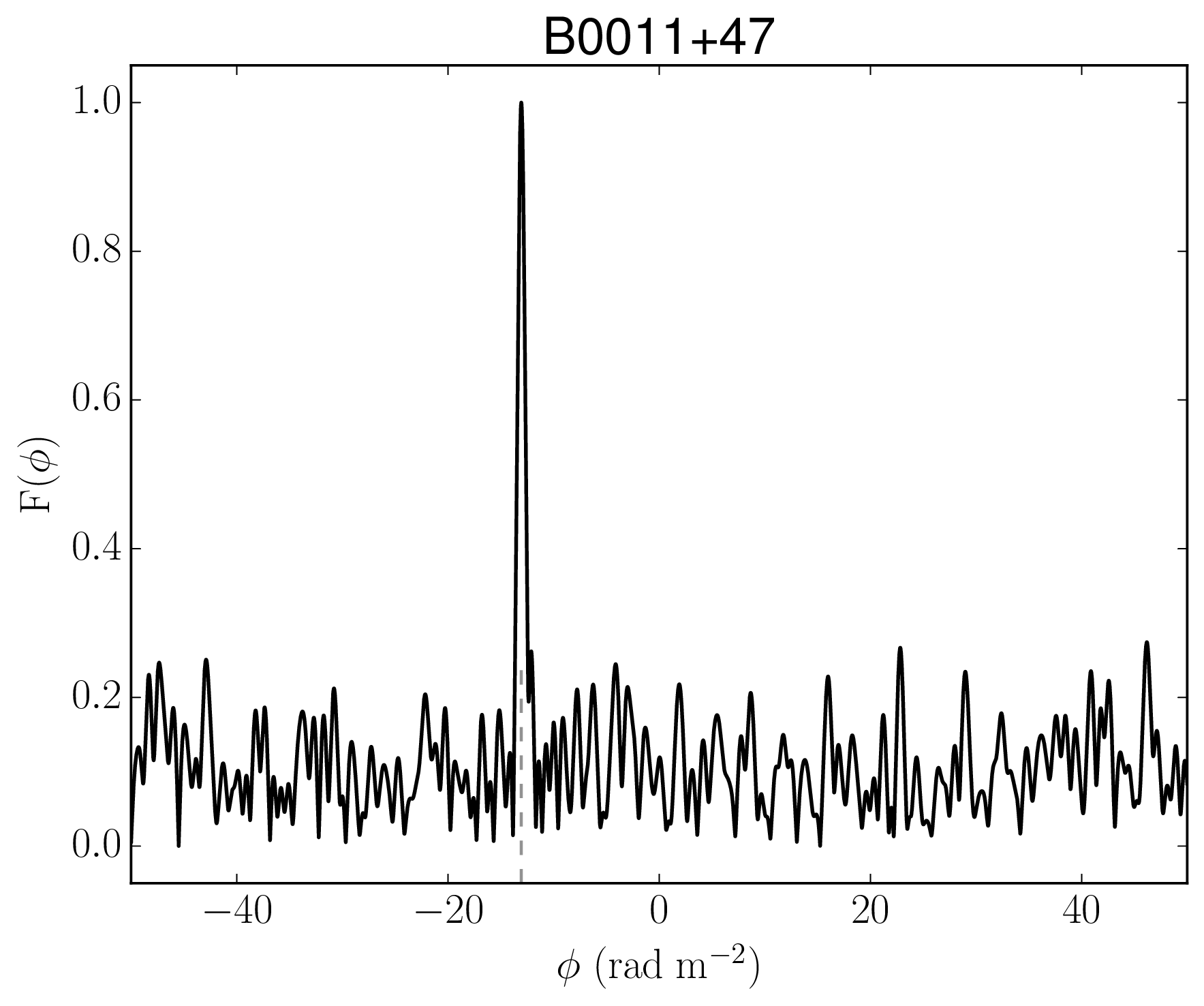}
  \includegraphics[width=0.246\columnwidth]{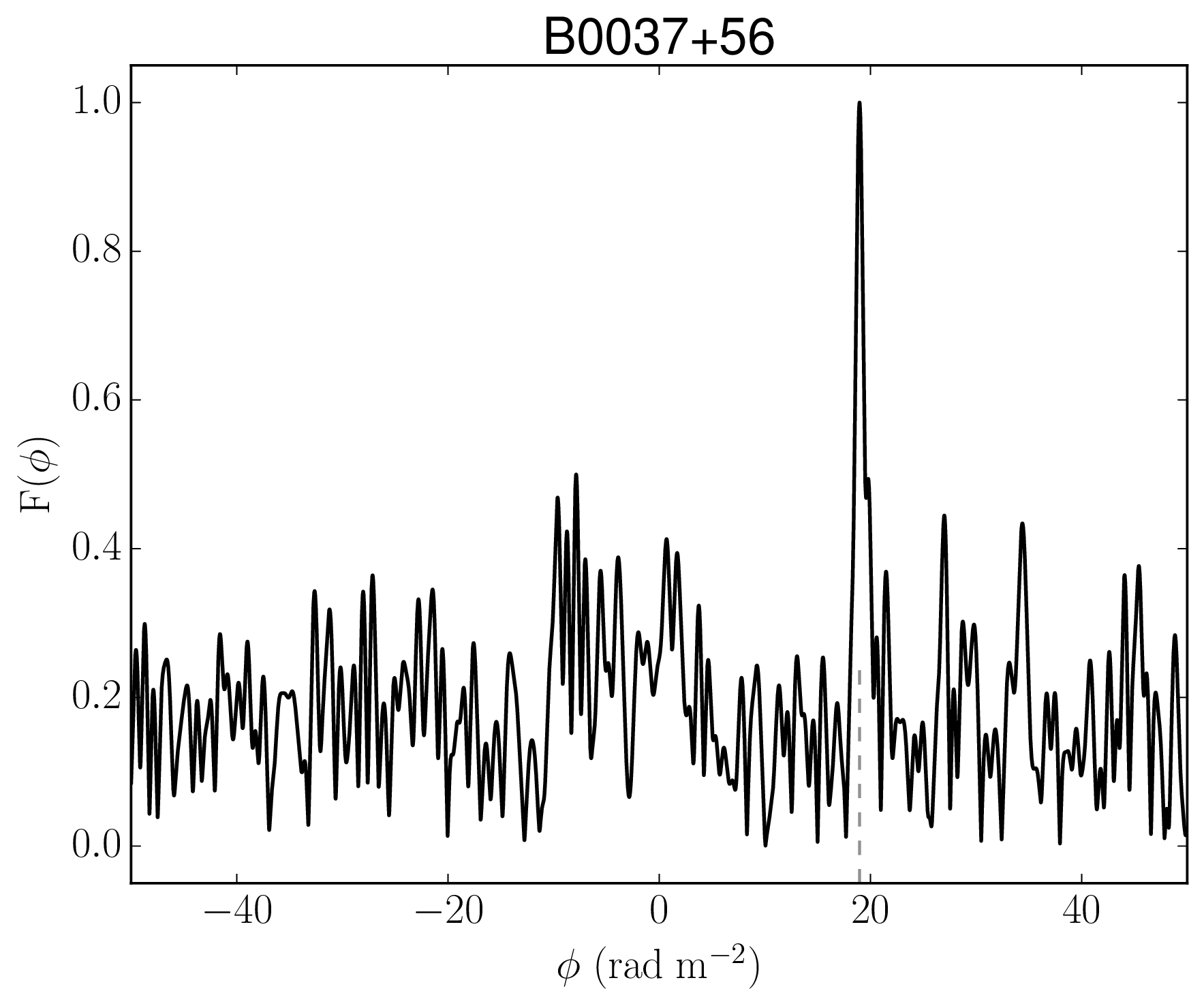}
 \includegraphics[width=0.246\columnwidth]{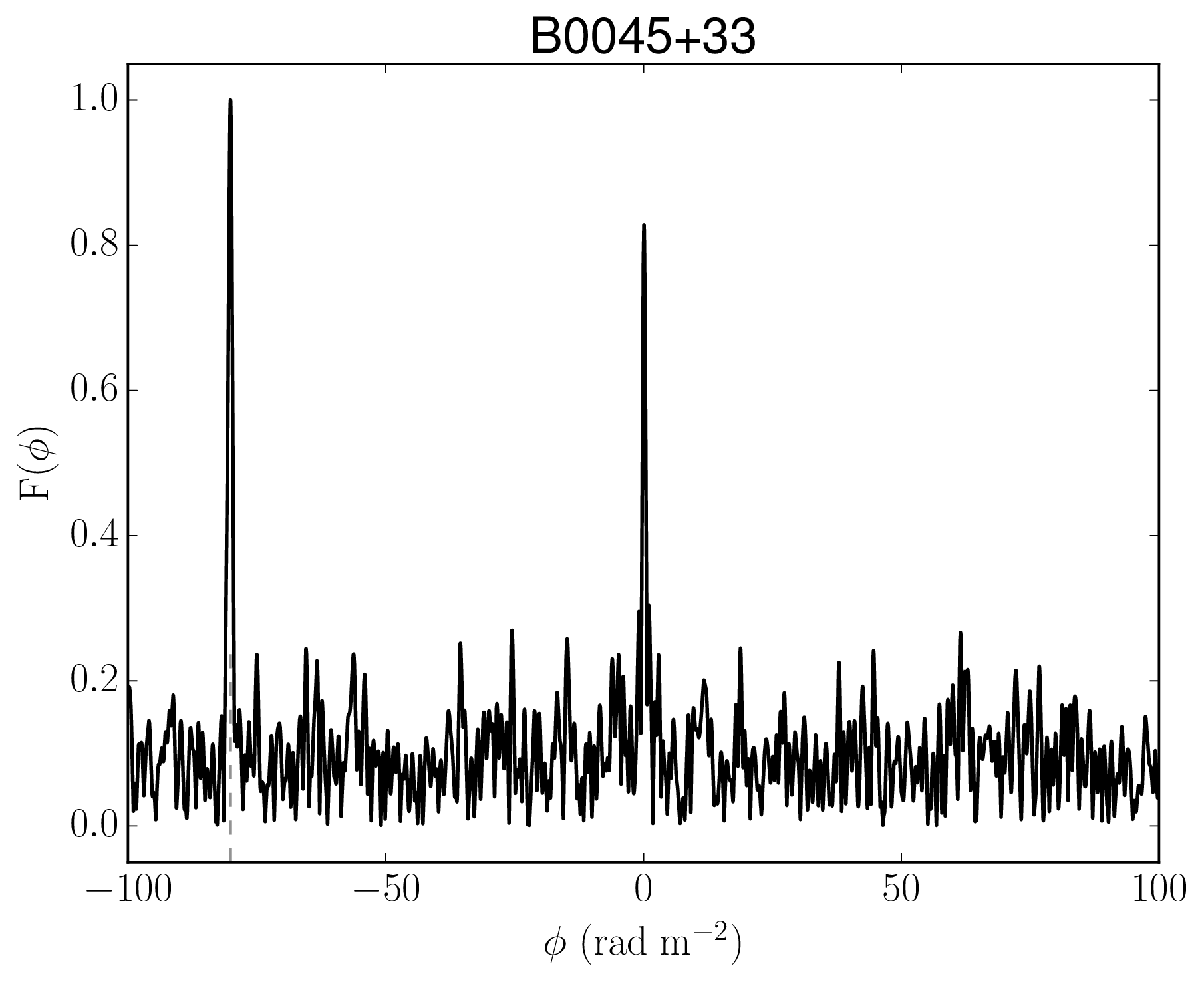}
  \includegraphics[width=0.246\columnwidth]{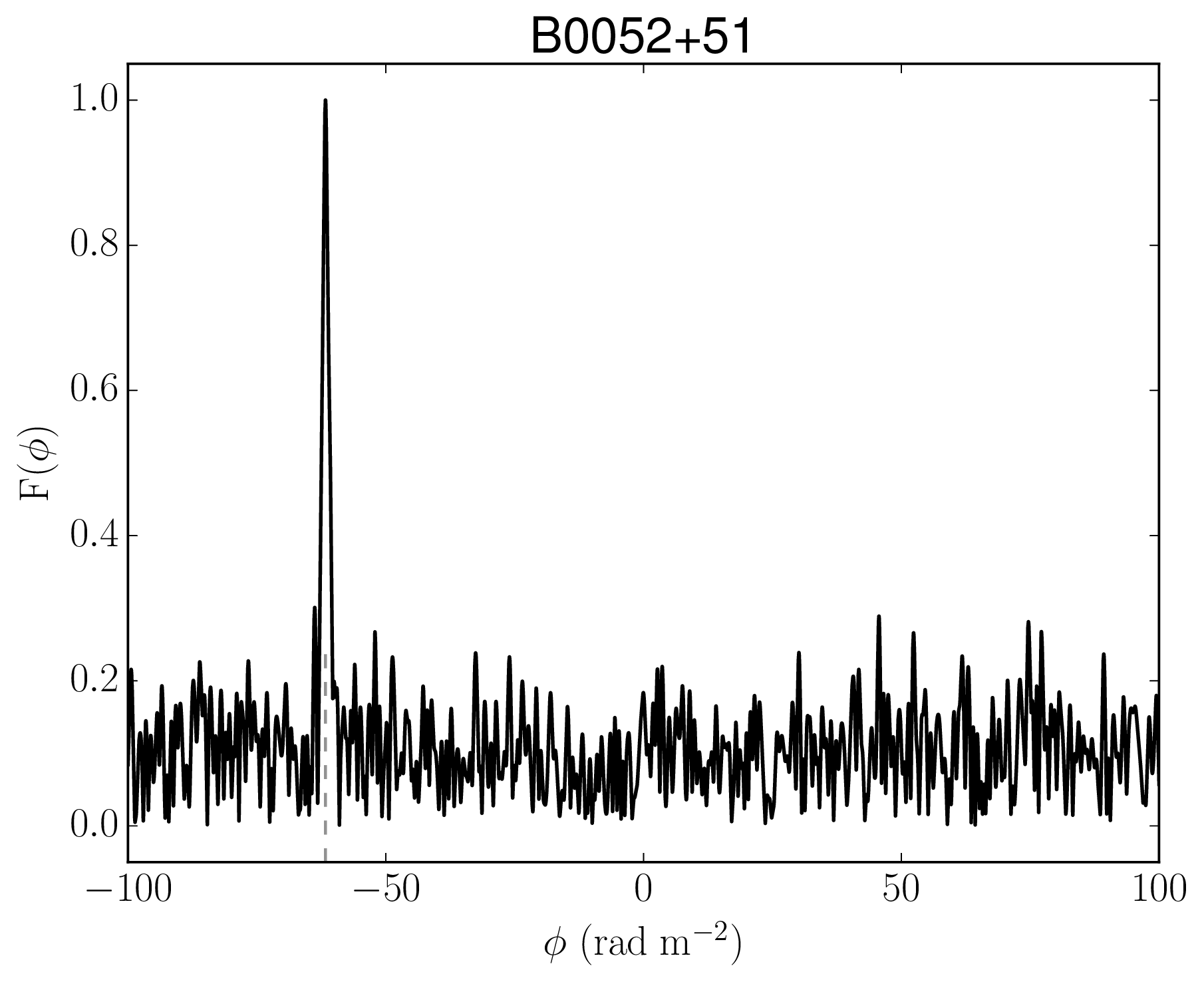}

  \includegraphics[width=0.246\columnwidth]{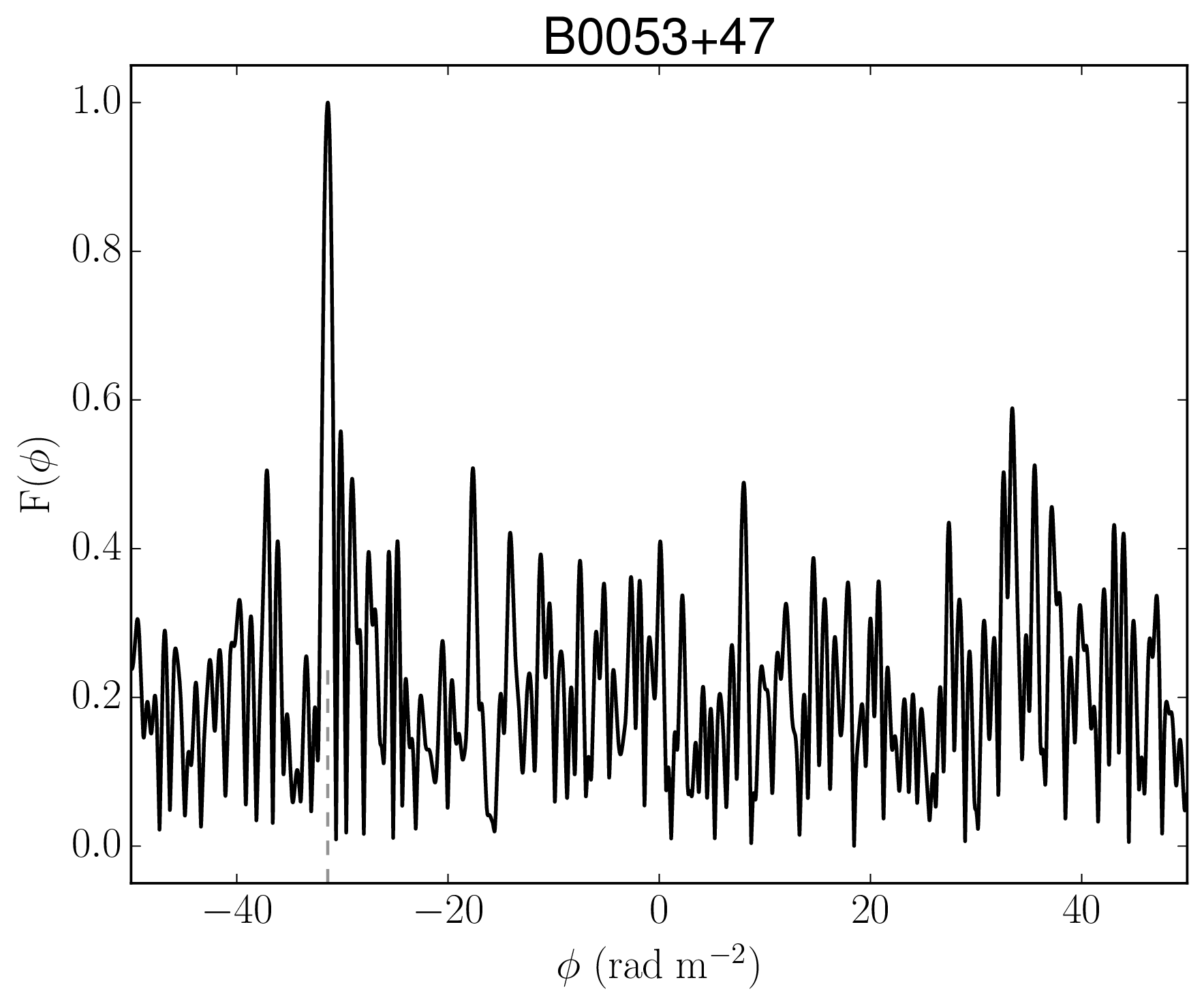}
  \includegraphics[width=0.246\columnwidth]{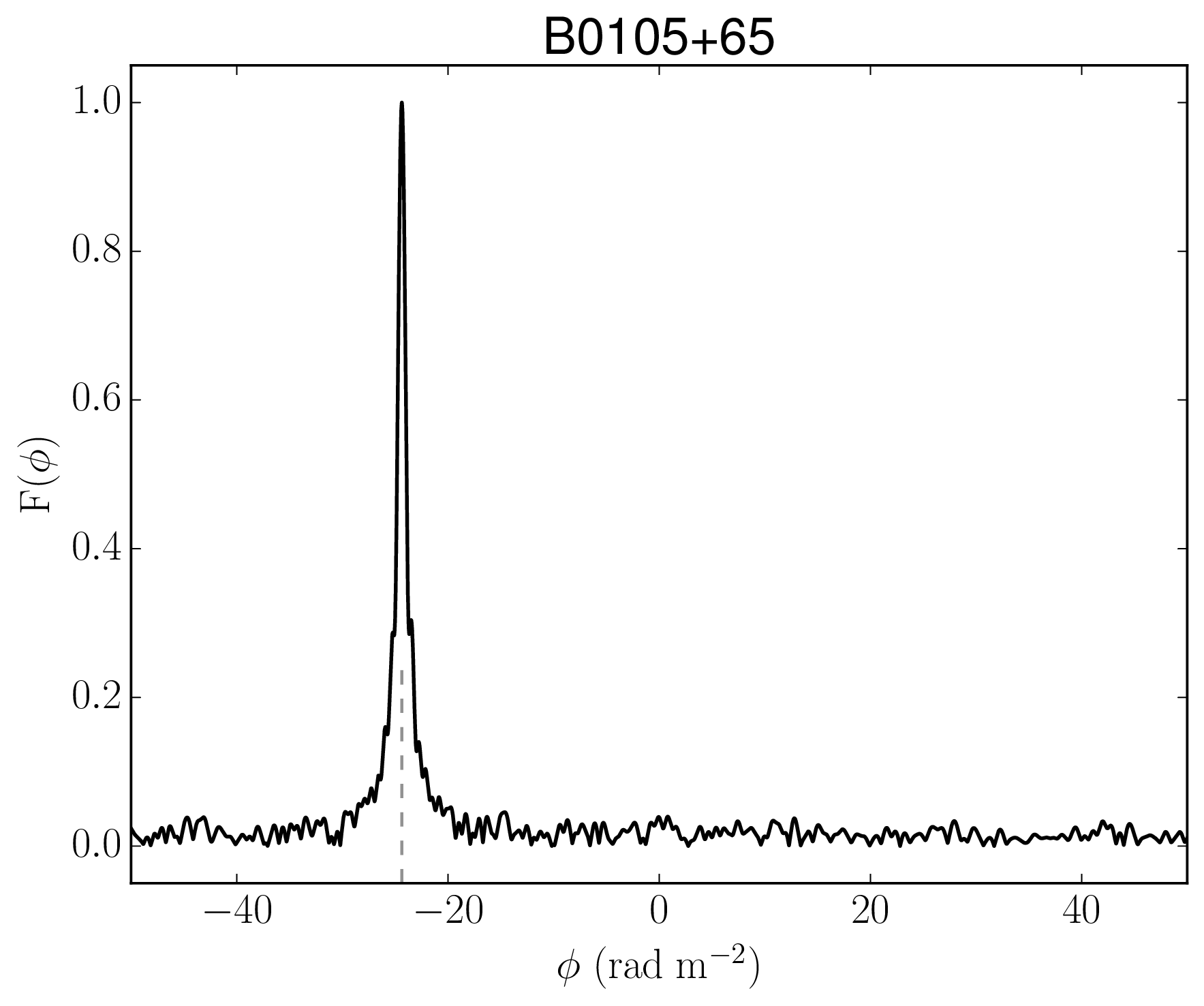}
   \includegraphics[width=0.246\columnwidth]{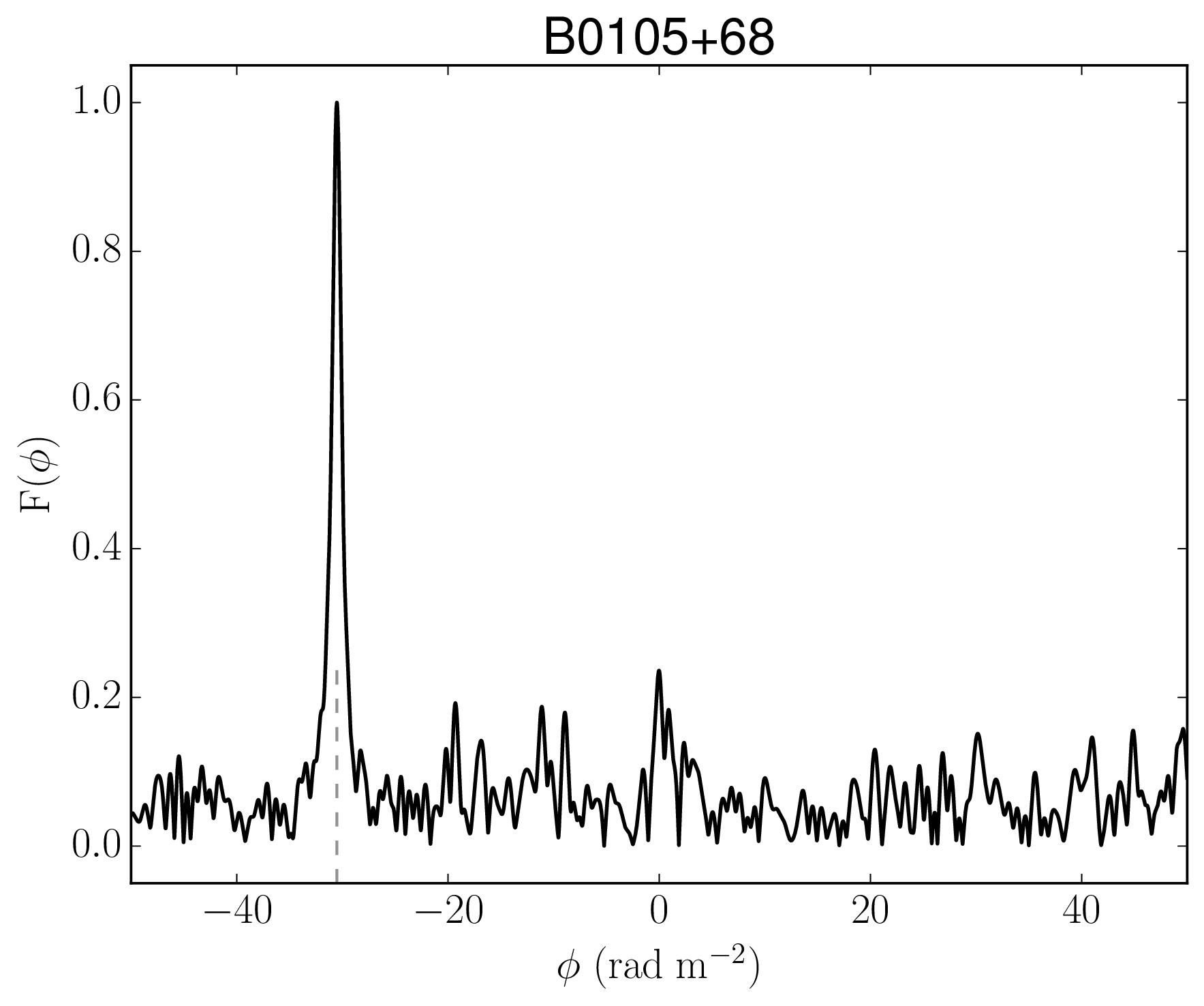}
    \includegraphics[width=0.246\columnwidth]{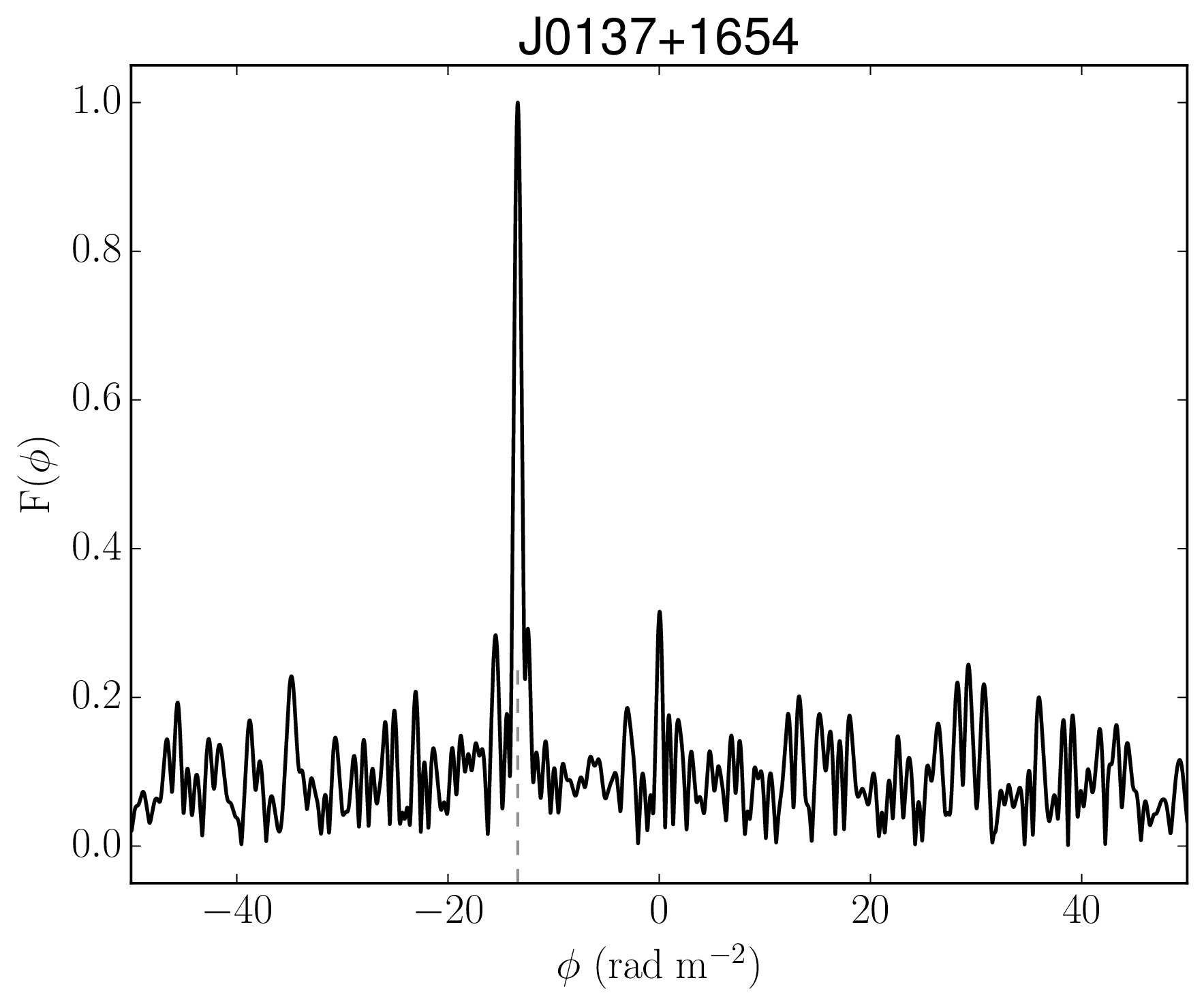}

 \includegraphics[width=0.246\columnwidth]{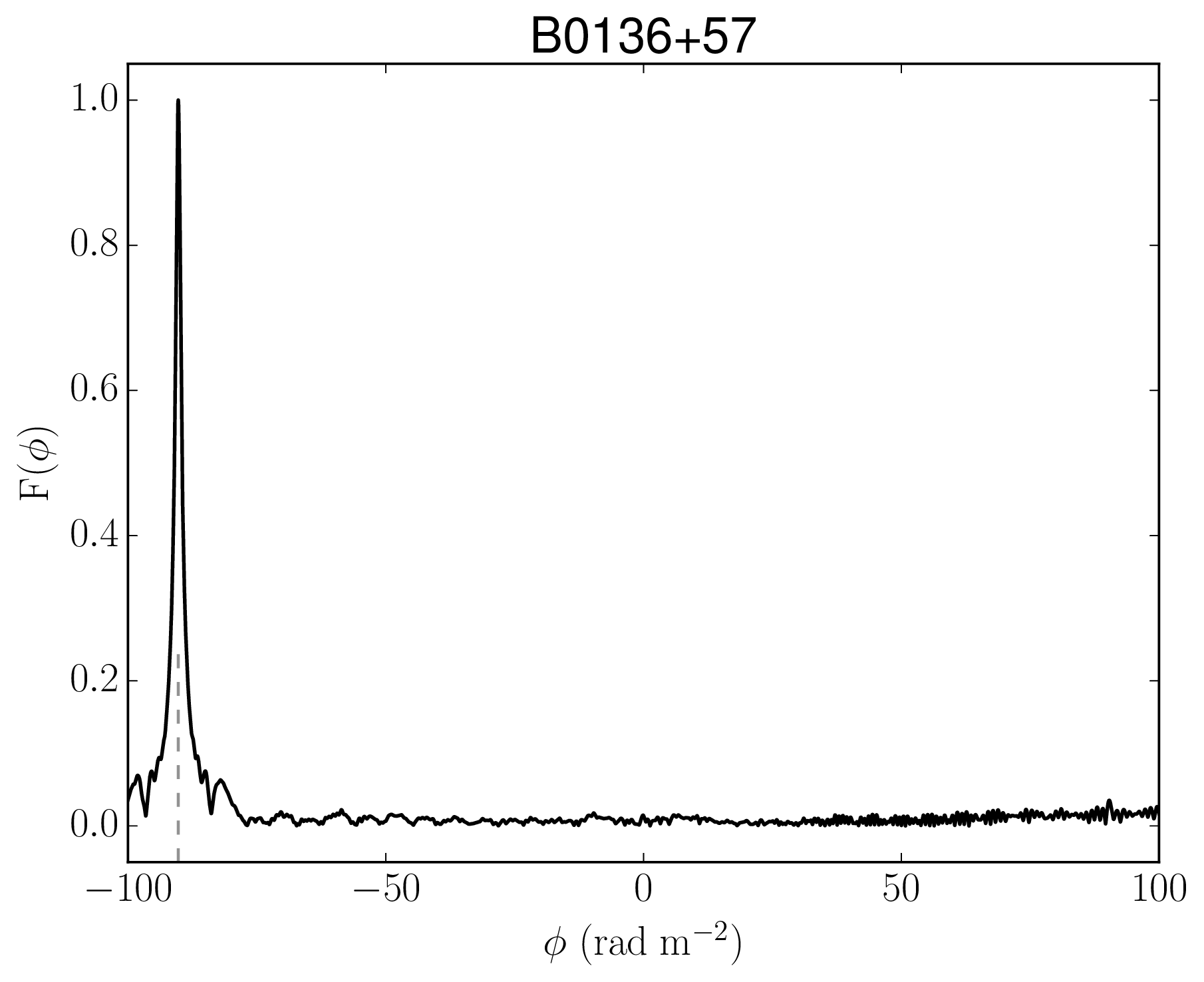}
 \includegraphics[width=0.246\columnwidth]{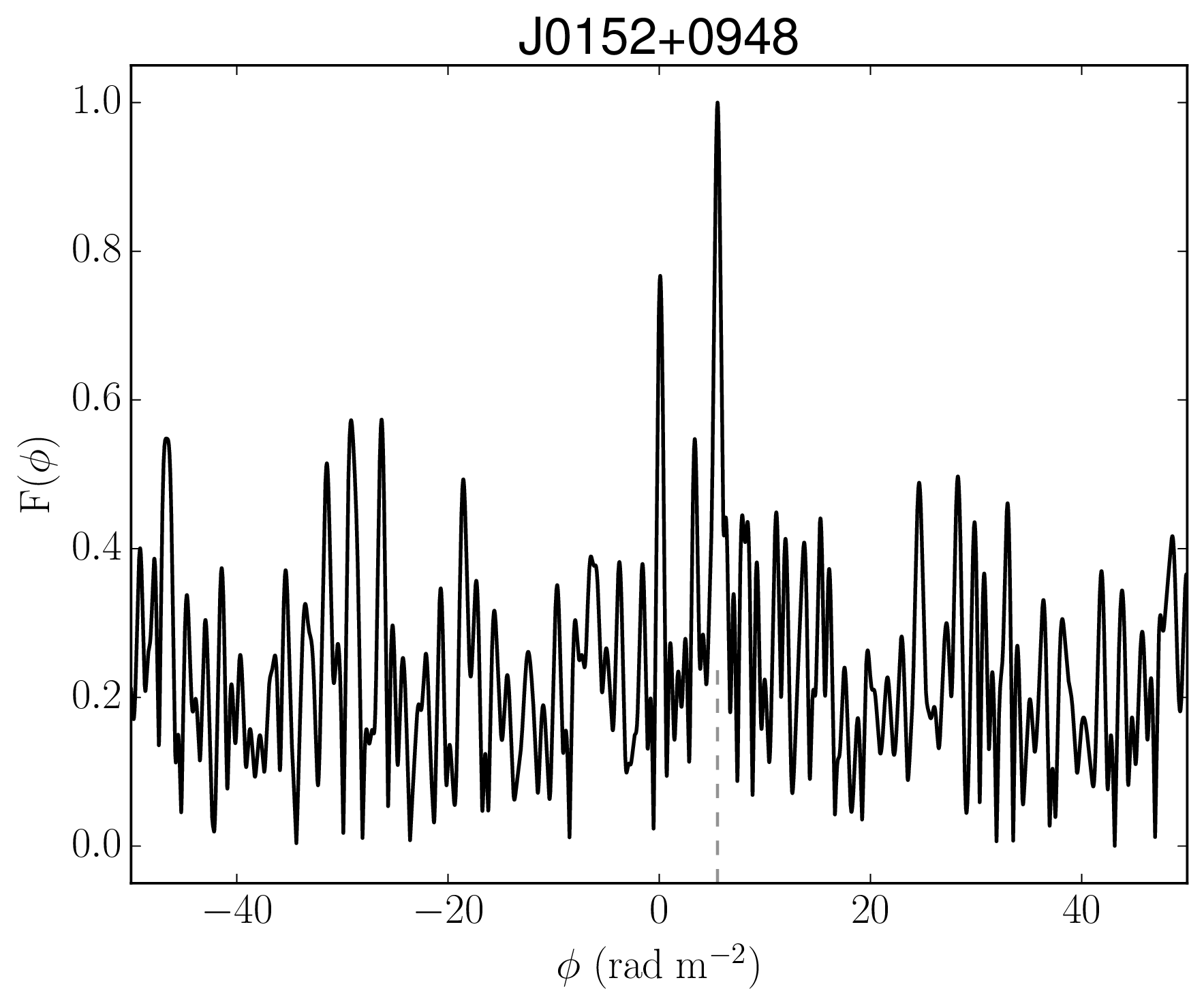}
 \includegraphics[width=0.246\columnwidth]{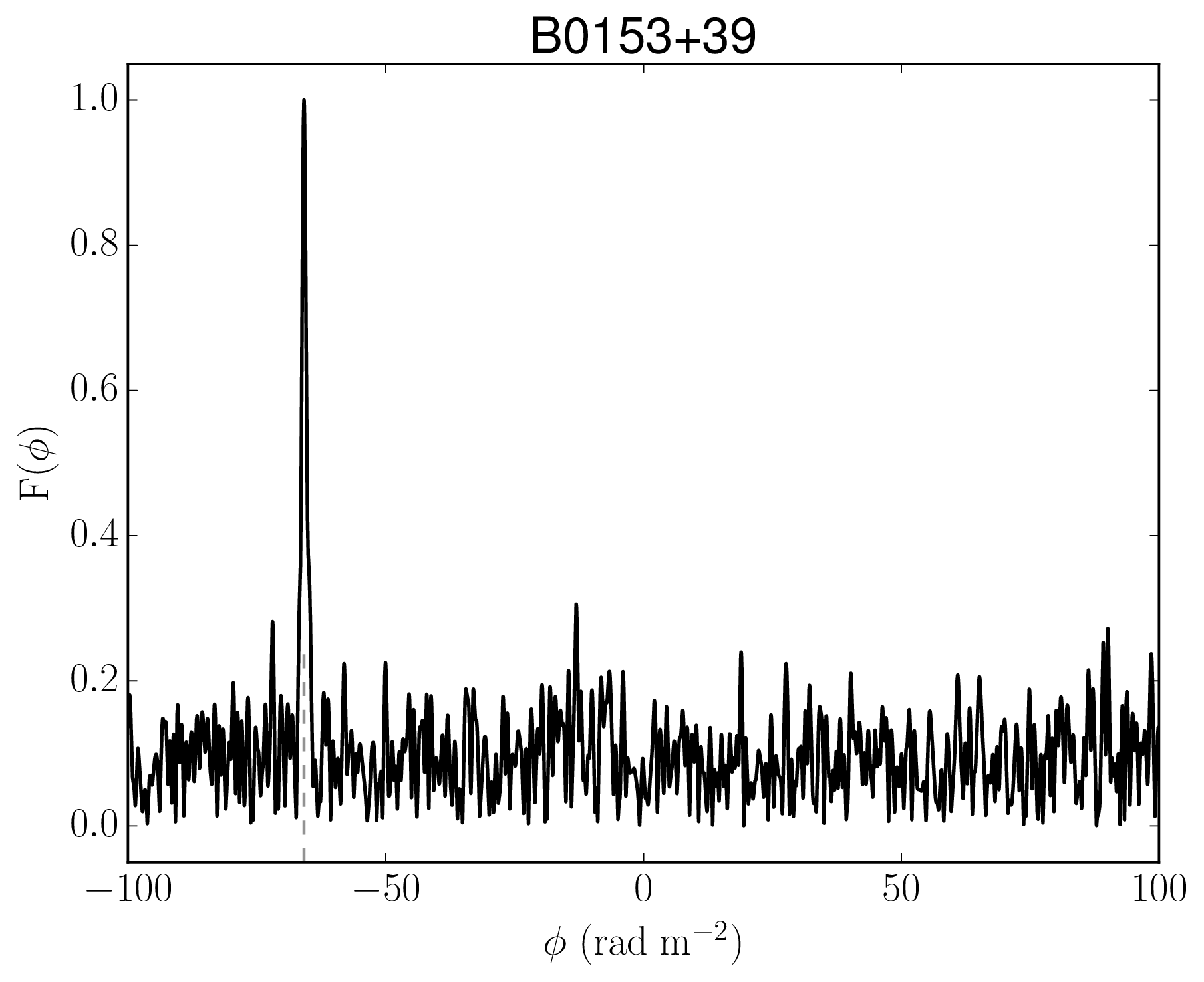}
\includegraphics[width=0.246\columnwidth]{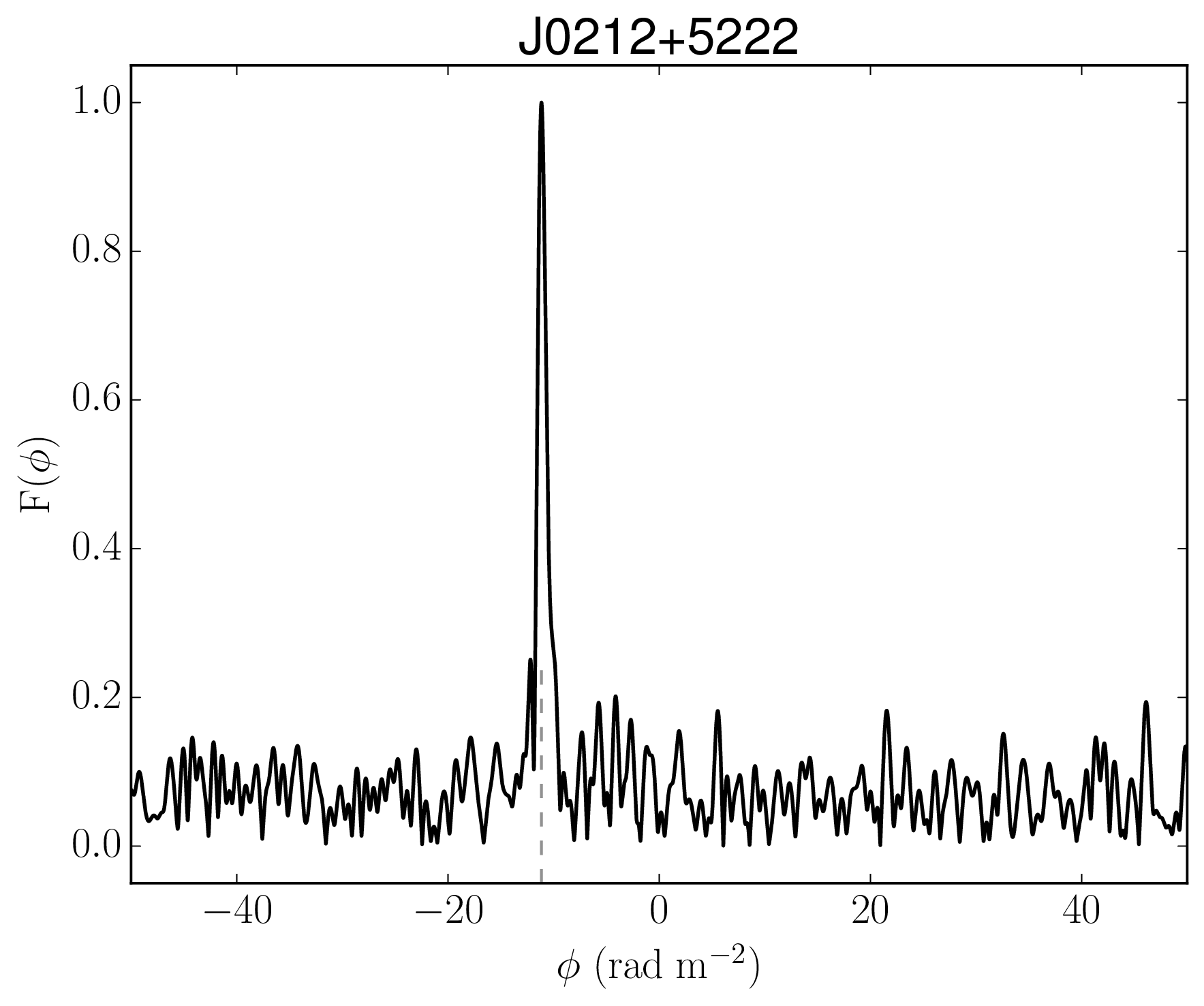}

\includegraphics[width=0.246\columnwidth]{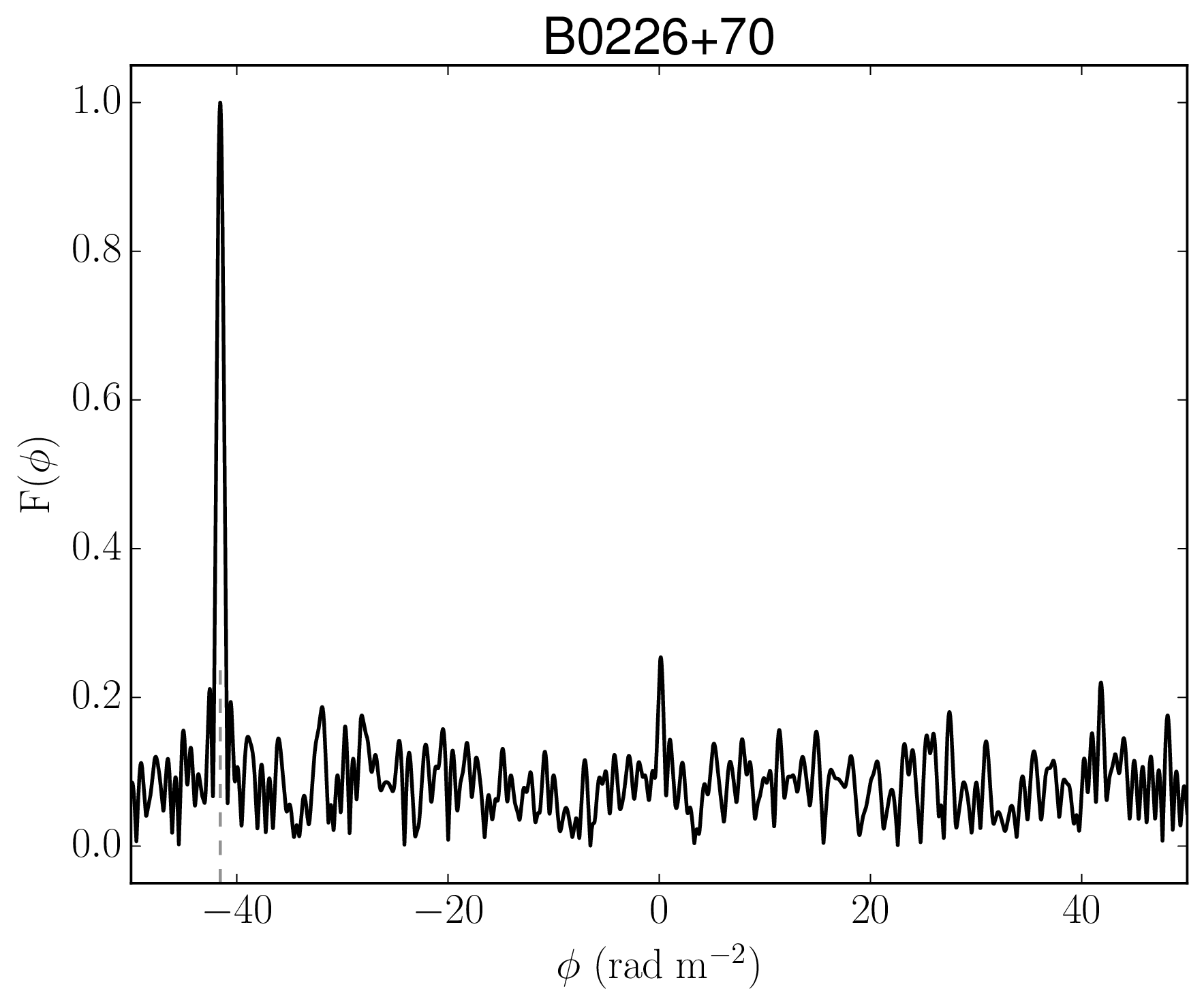}
\includegraphics[width=0.246\columnwidth]{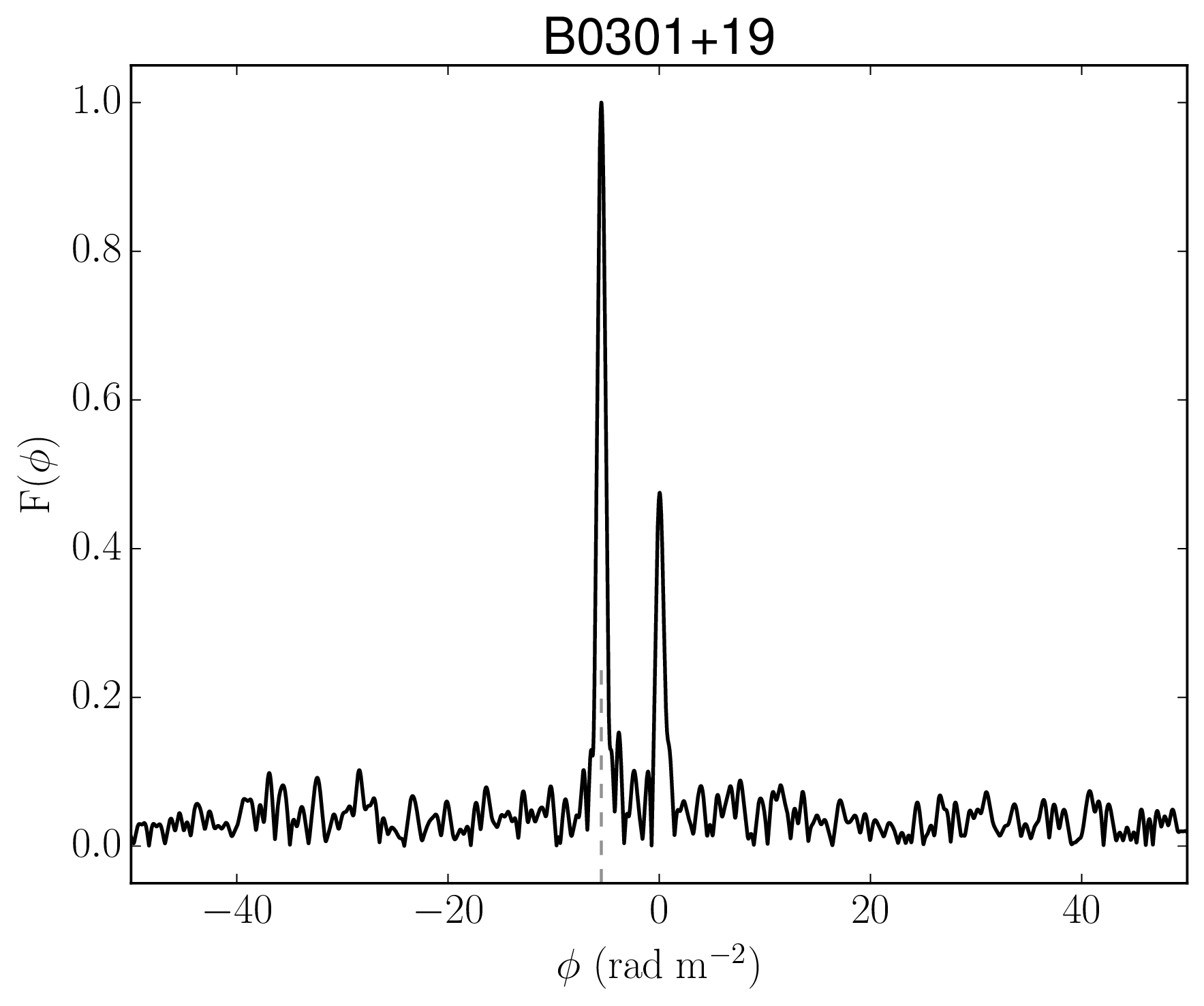}
\includegraphics[width=0.246\columnwidth]{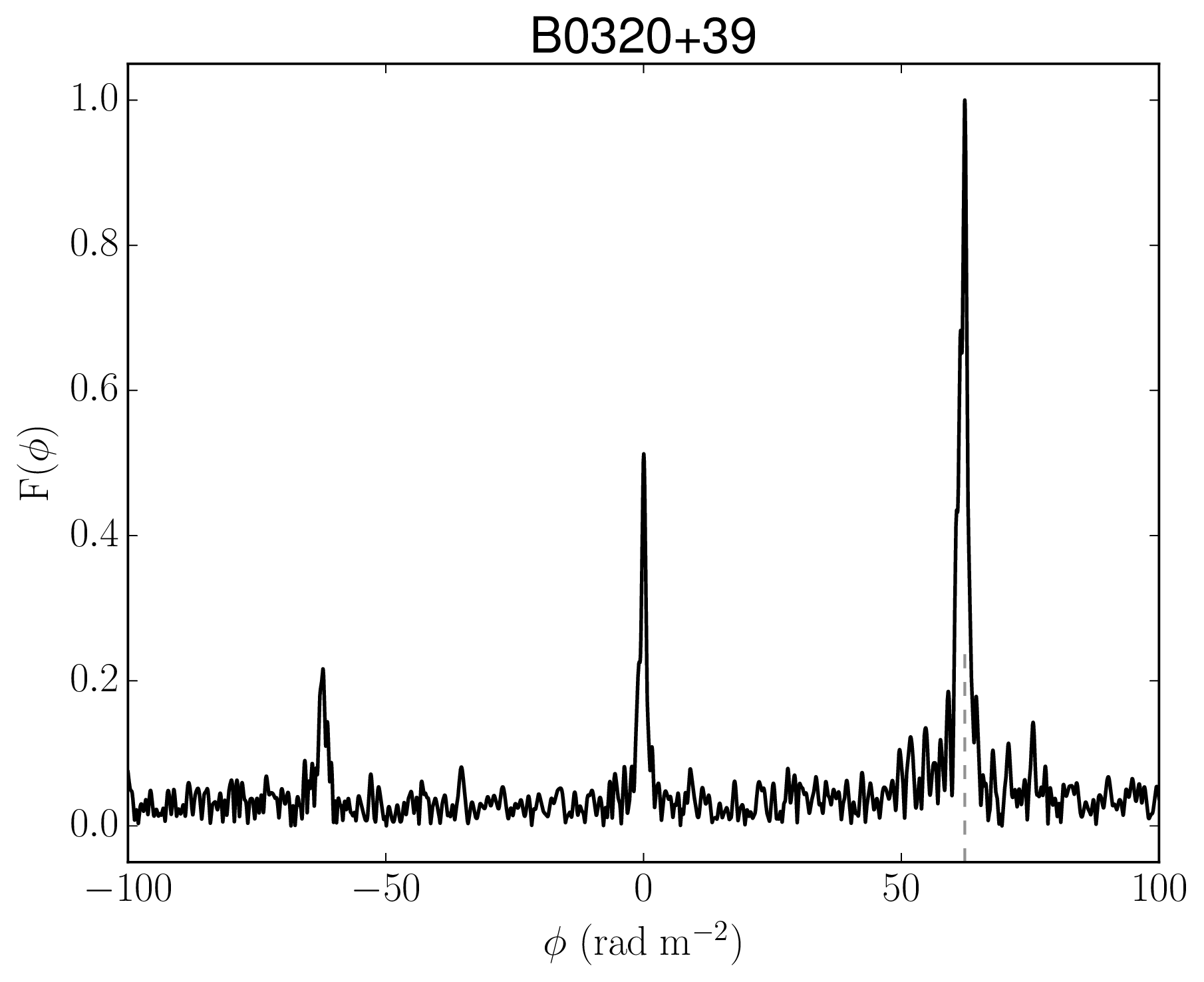}
\includegraphics[width=0.246\columnwidth]{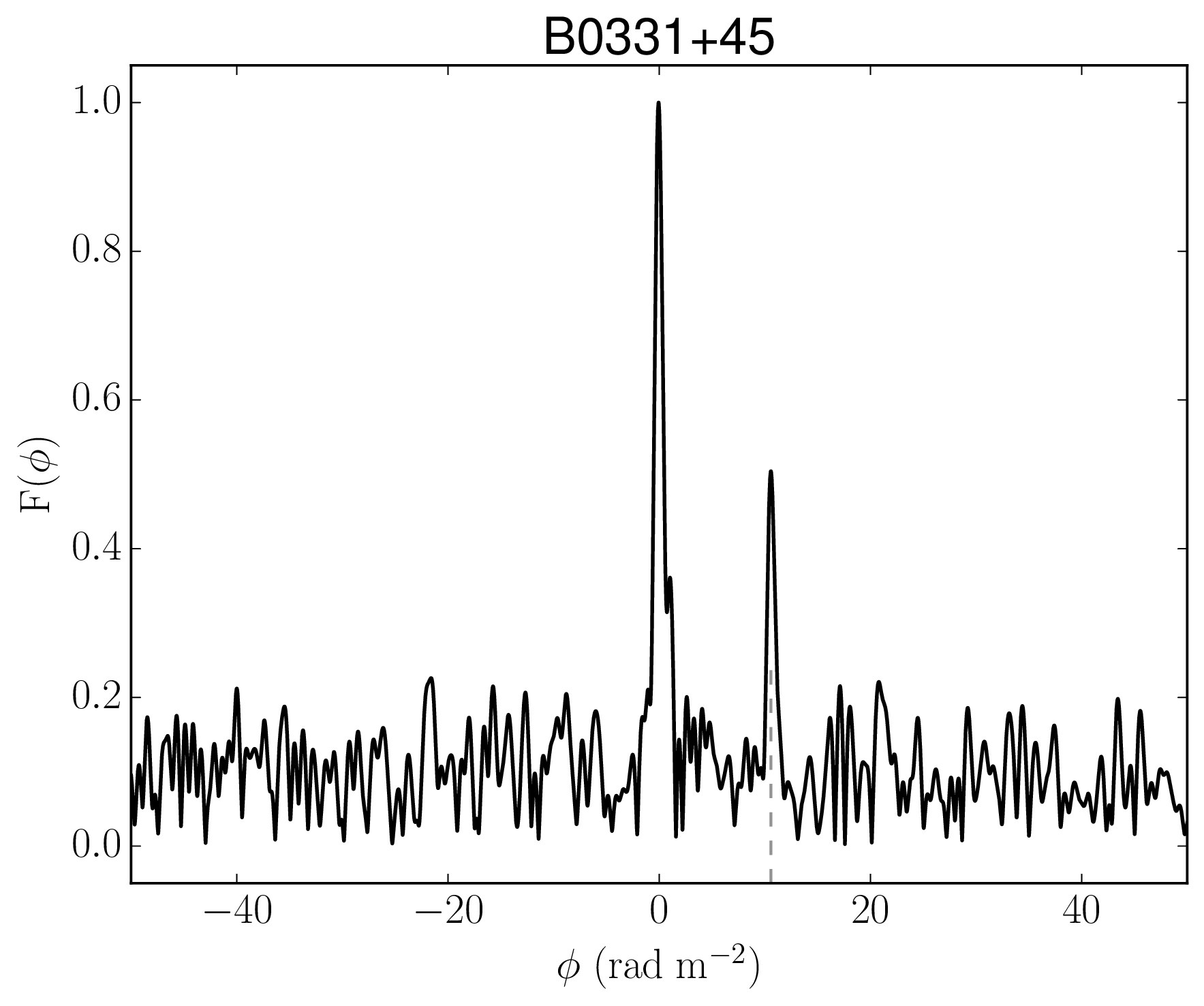}

\includegraphics[width=0.246\columnwidth]{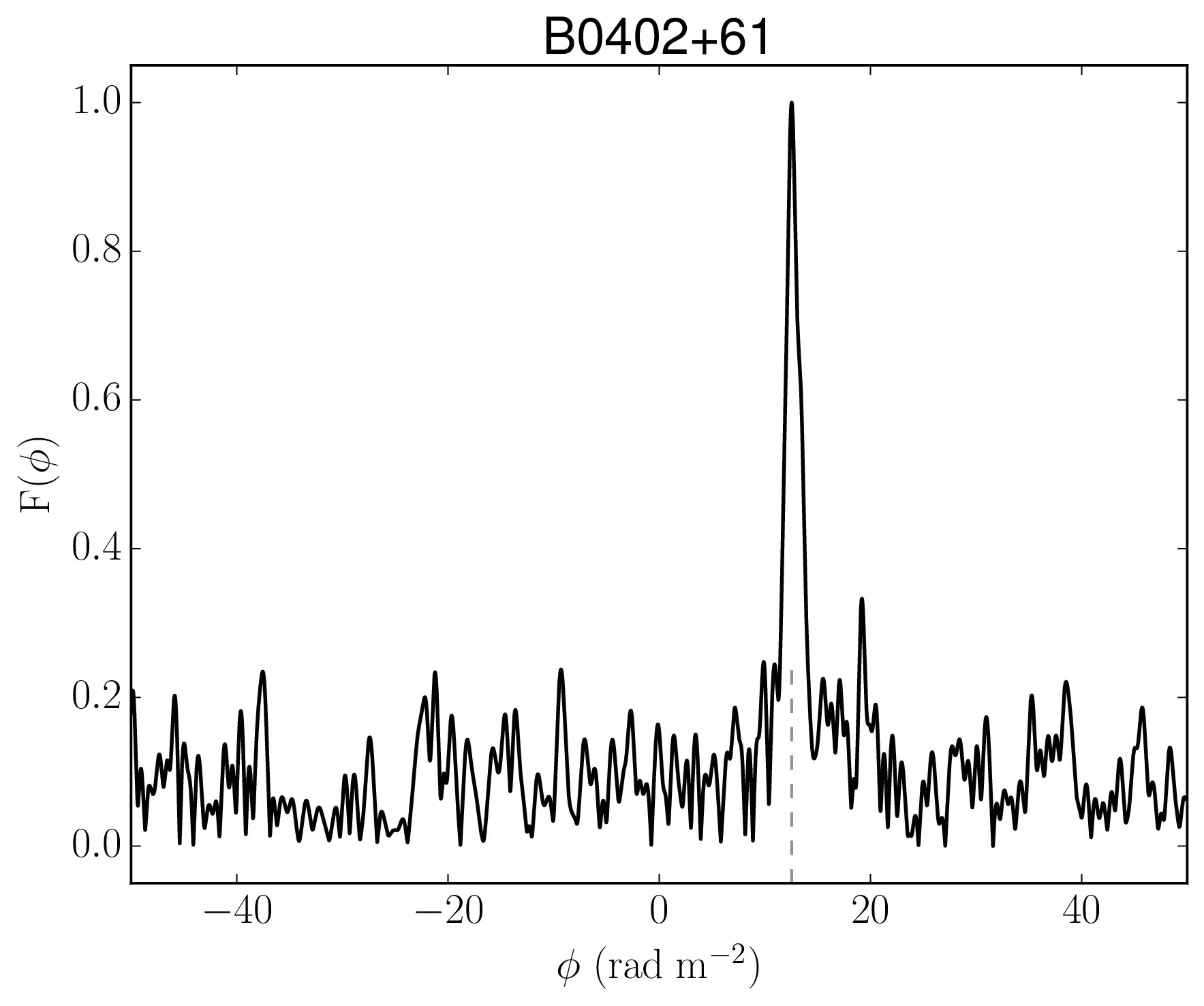}
\includegraphics[width=0.246\columnwidth]{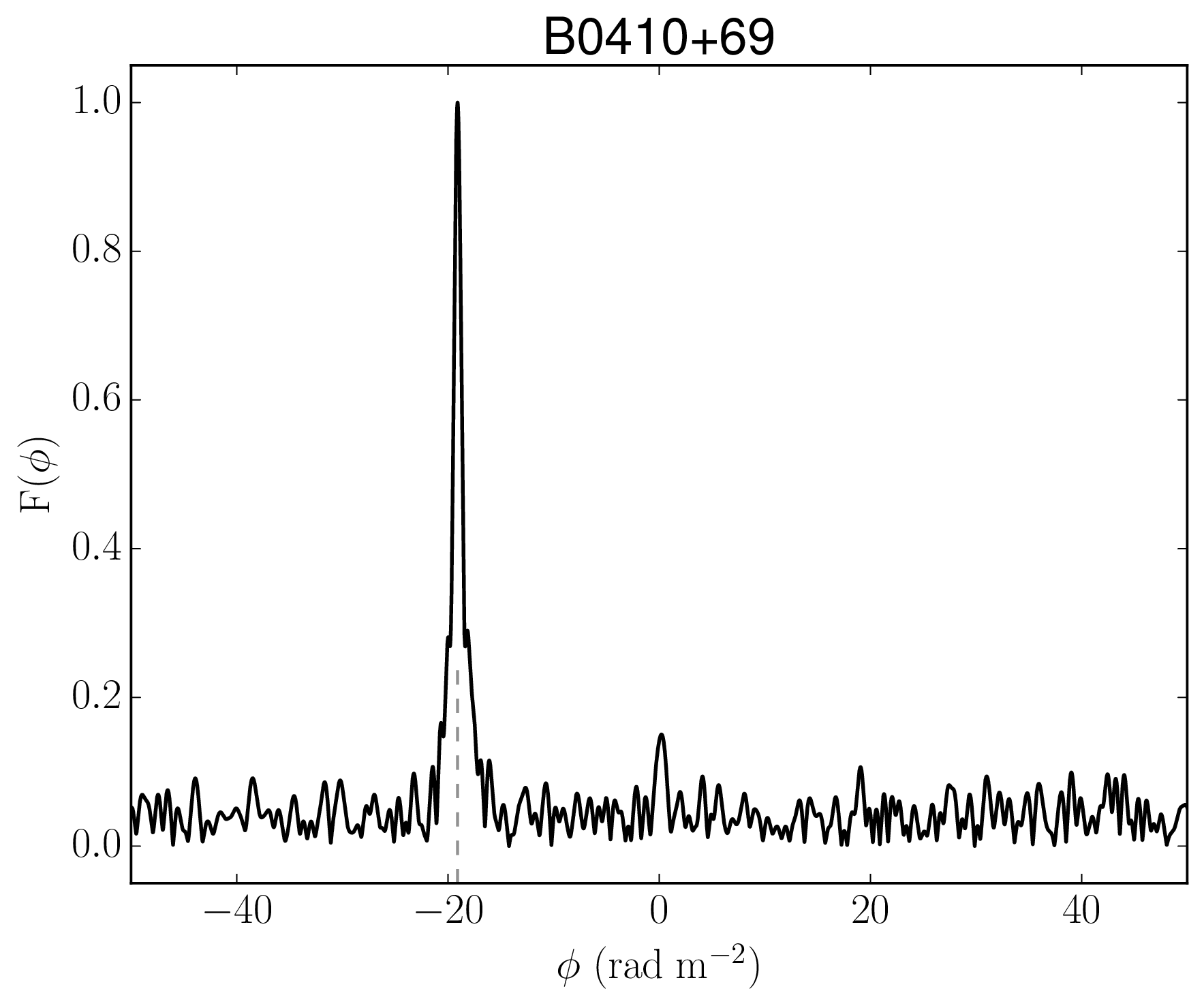}
\includegraphics[width=0.246\columnwidth]{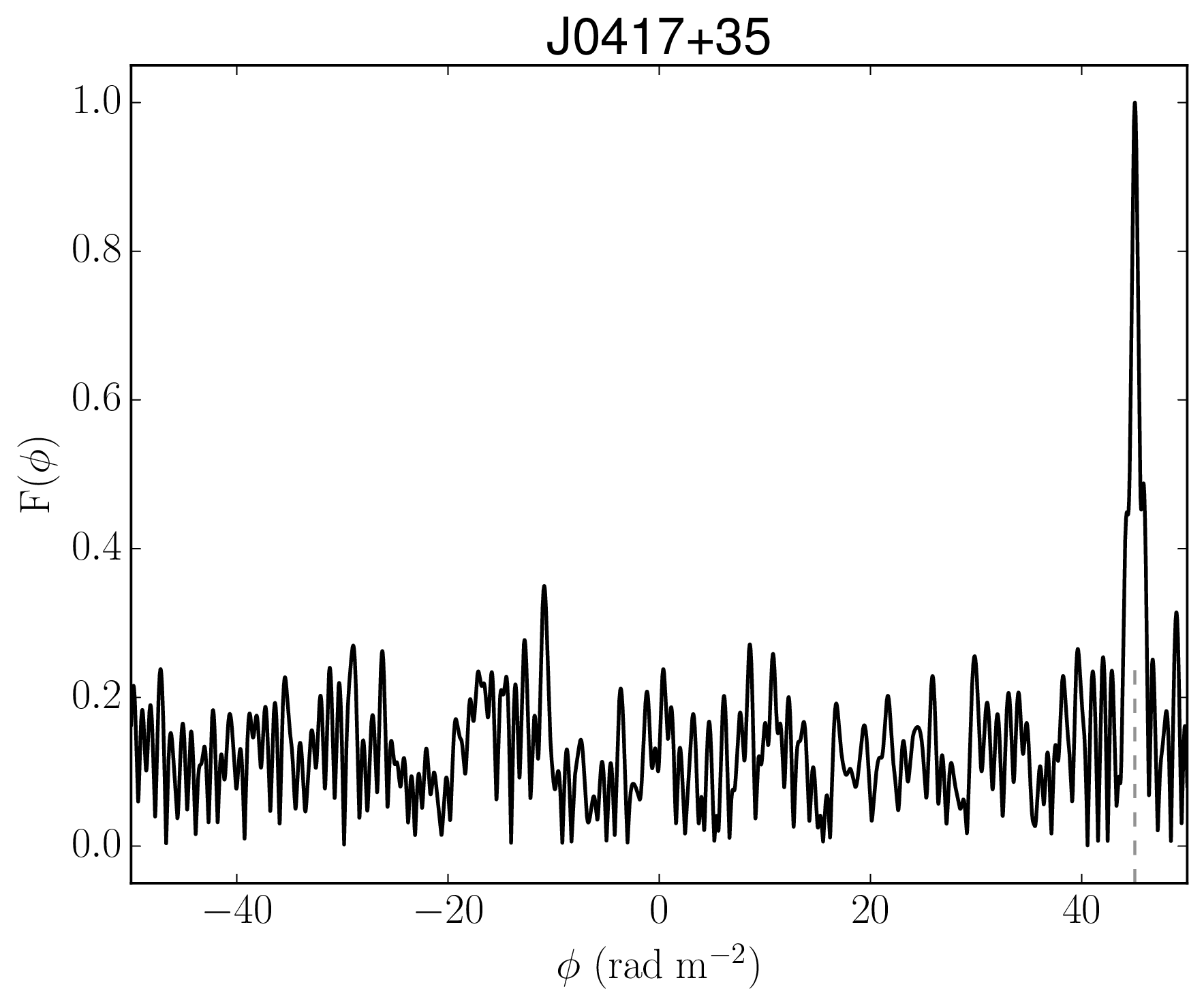}
\includegraphics[width=0.246\columnwidth]{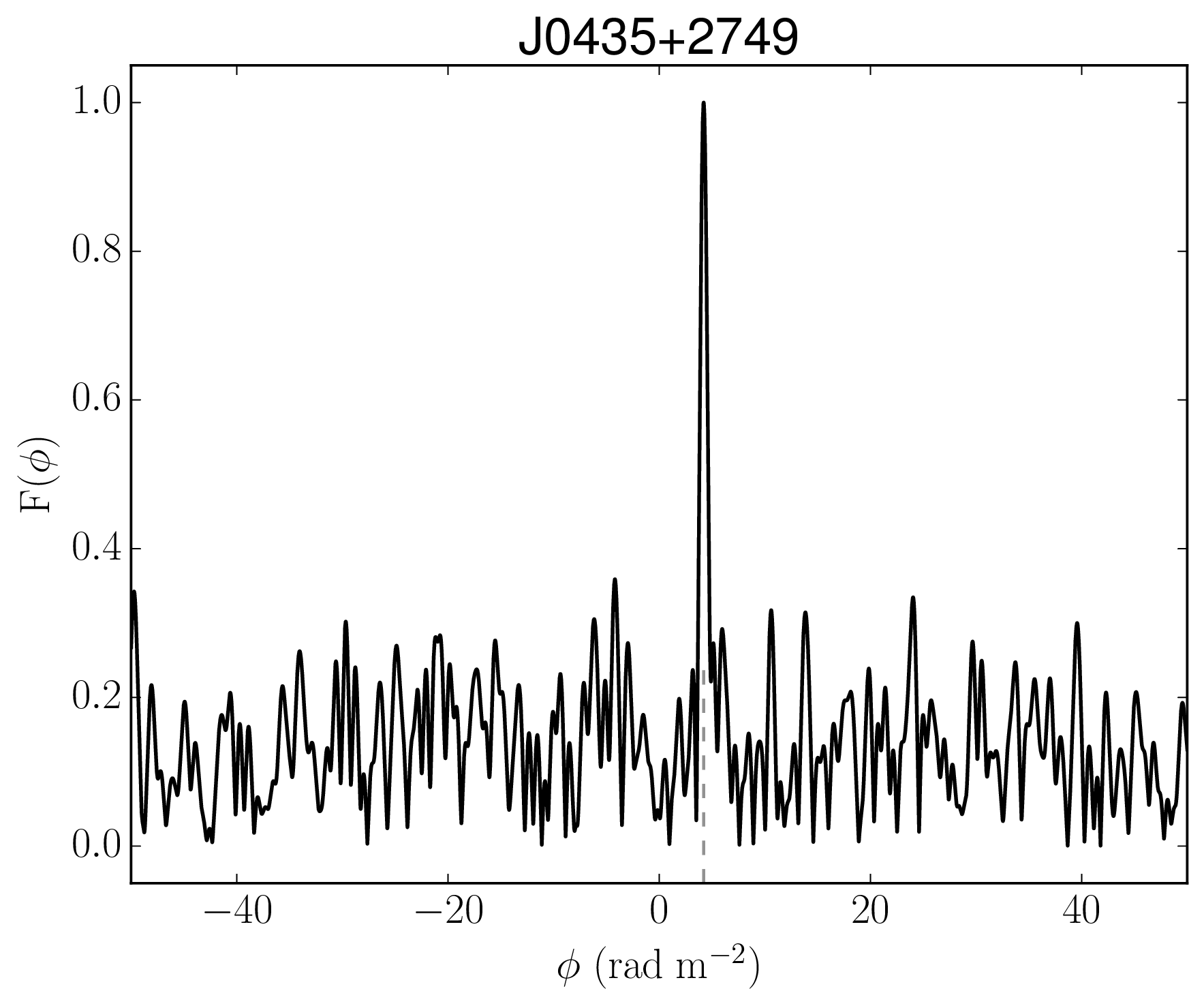}

\includegraphics[width=0.246\columnwidth]{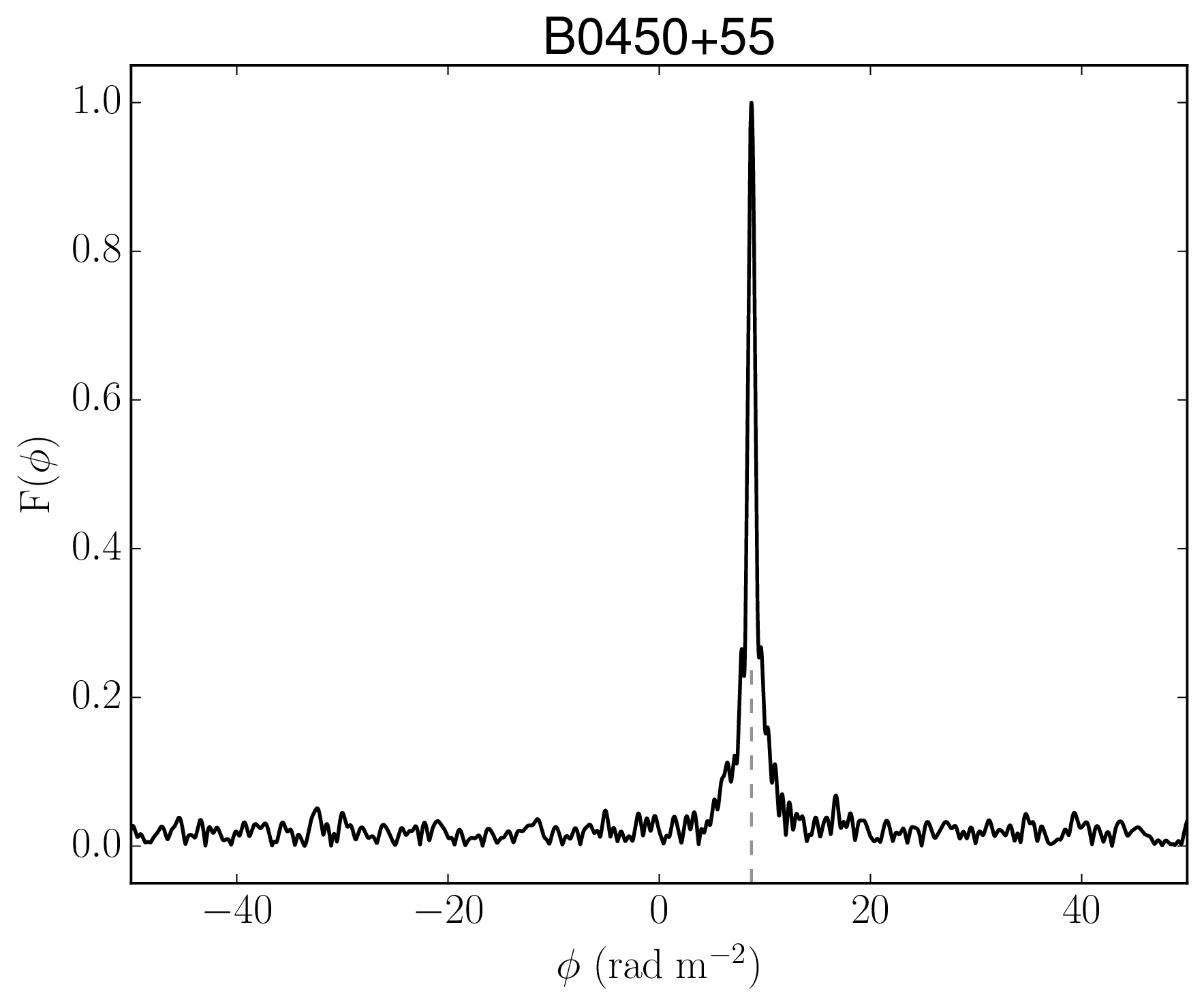}
\includegraphics[width=0.246\columnwidth]{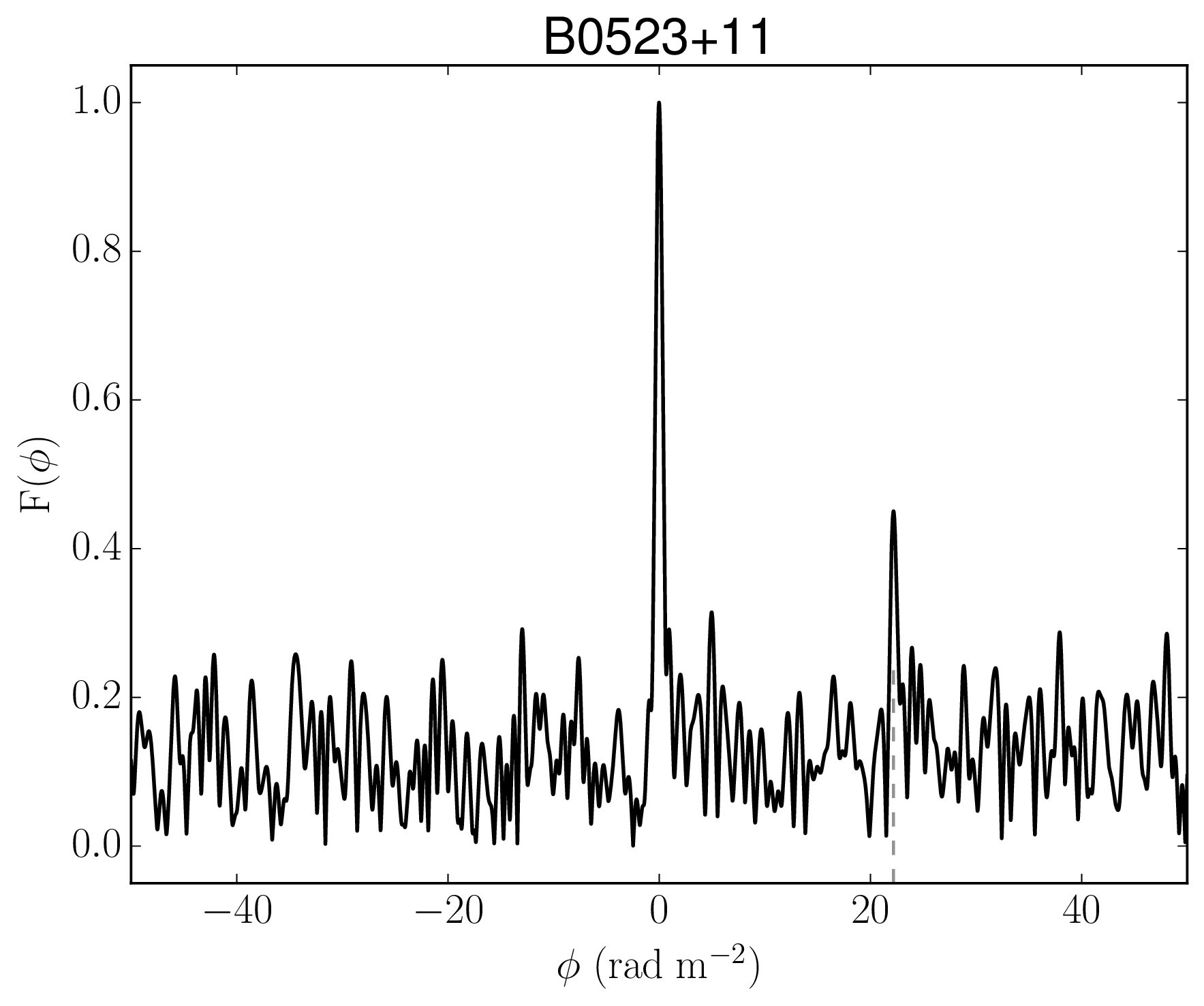}
\includegraphics[width=0.246\columnwidth]{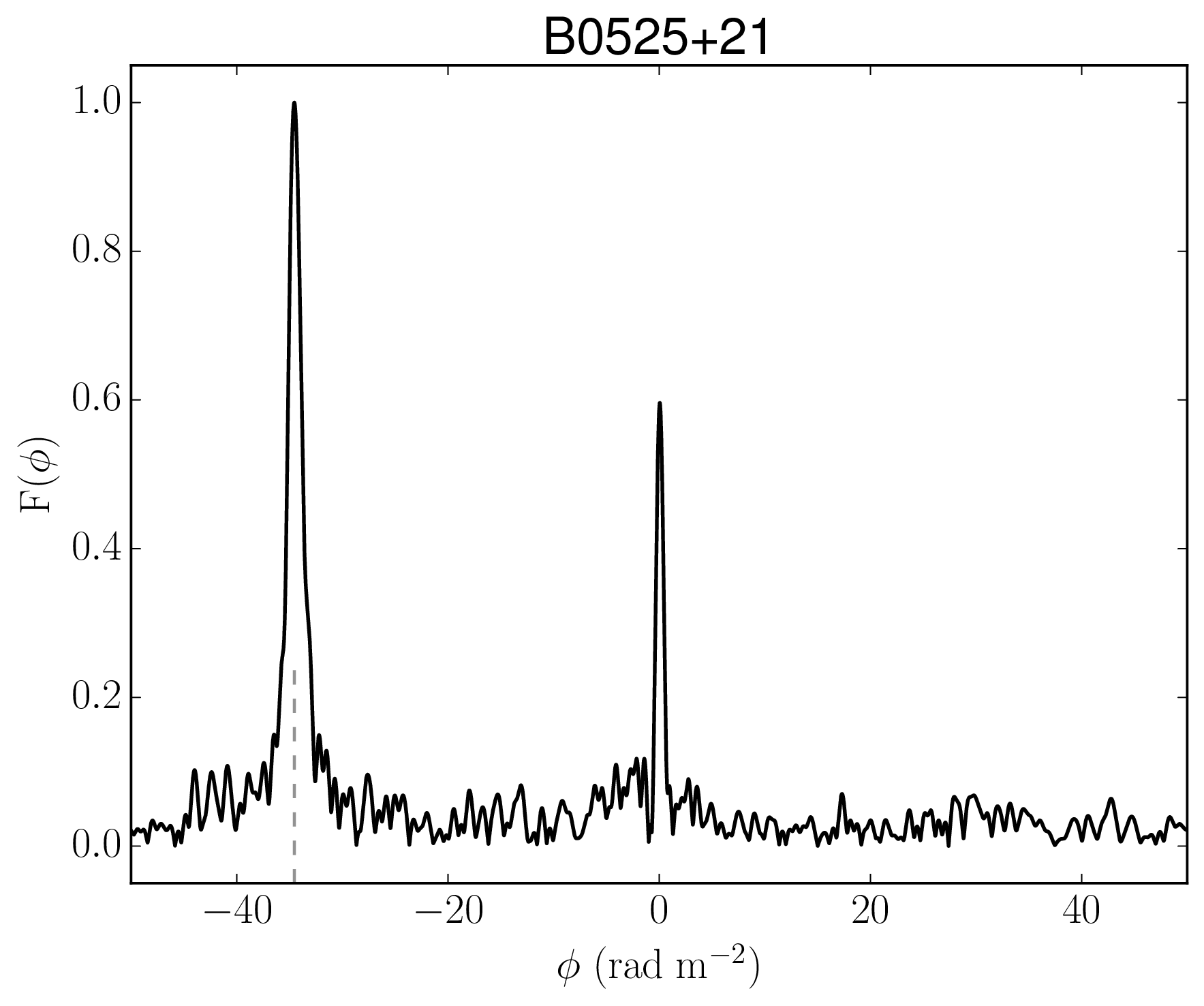}
\includegraphics[width=0.246\columnwidth]{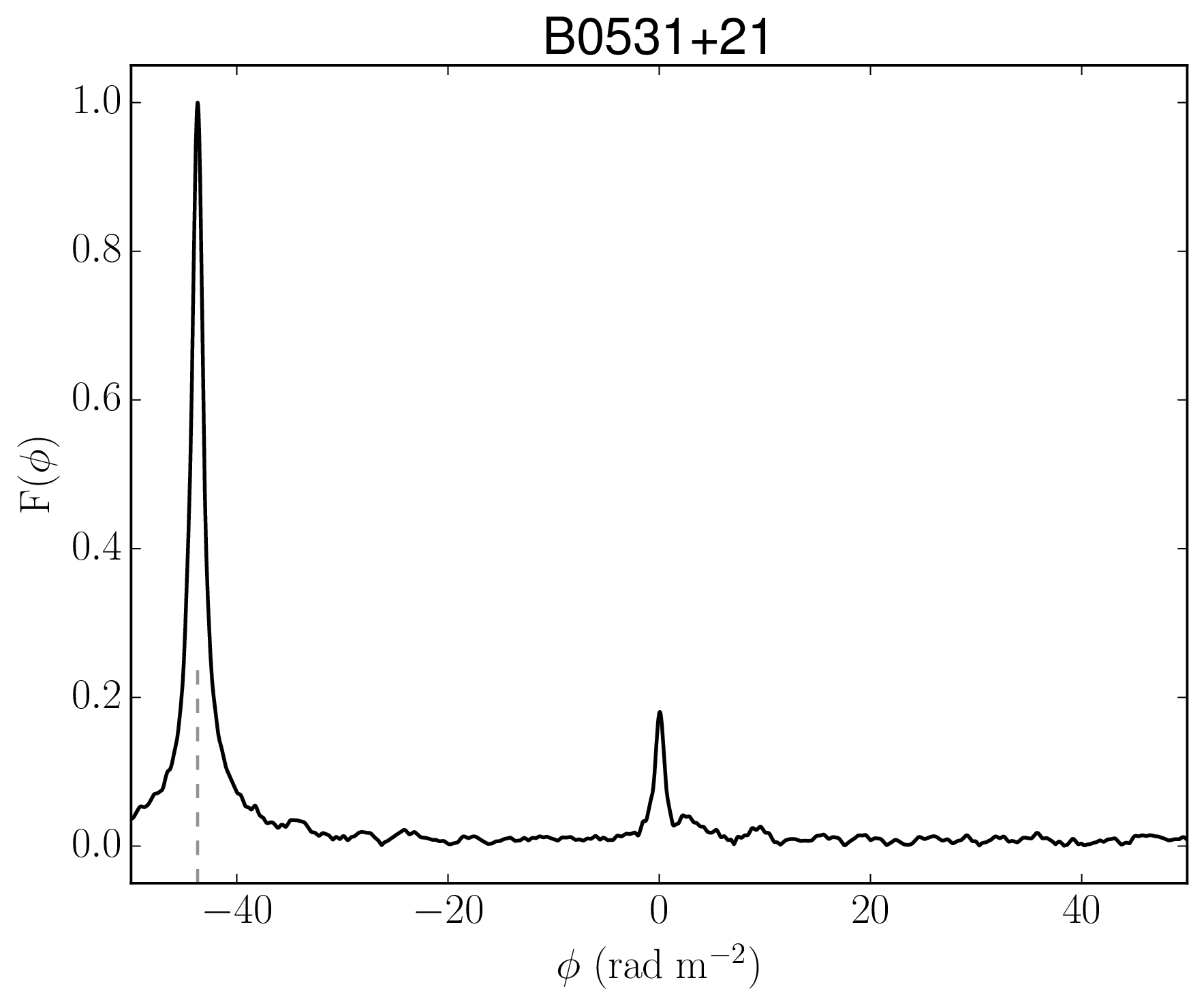}

 \caption{Faraday spectra (modulus values) for the pulsars towards which we measured an RM, ordered as in Table \ref{tab:1}. The grey dashed line indicates the position of the peak in Faraday space, RM$_{\rm{obs}}$, prior to any subtraction of the ionospheric RM.  The x-axes show the Faraday depth between $\pm$50, $\pm$100, $\pm$150, or $\pm$200\,rad m$^{\rm{-2}}$, depending on the measured RM towards each pulsar. The y-axes show the intensity of linearly-polarised emission in arbitrary units (normalised) because the data are not flux or polarisation calibrated. Peaks near 0\,rad m$^{\rm {-2}}$ are due to instrumental polarisation, and smaller symmetric peaks about 0\,rad m$^{\rm {-2}}$ are due to polarisation leakage. }

 \label{fig:FDFs}
\end{figure*}

\begin{figure*}
 \centering
 %\ContinuedFloat
%\setcounter{subfigure}{2}    

\includegraphics[width=0.246\columnwidth]{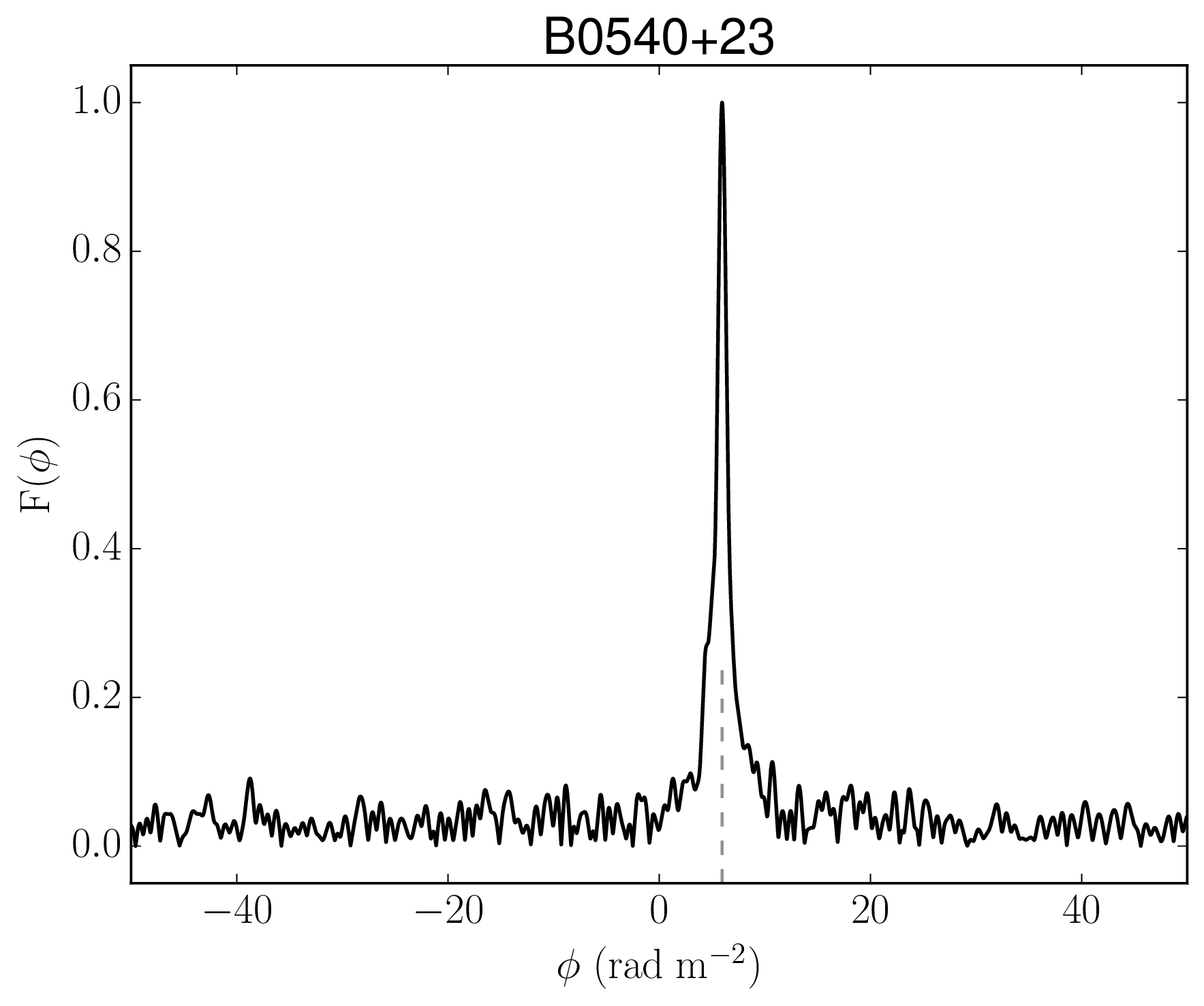}
\includegraphics[width=0.246\columnwidth]{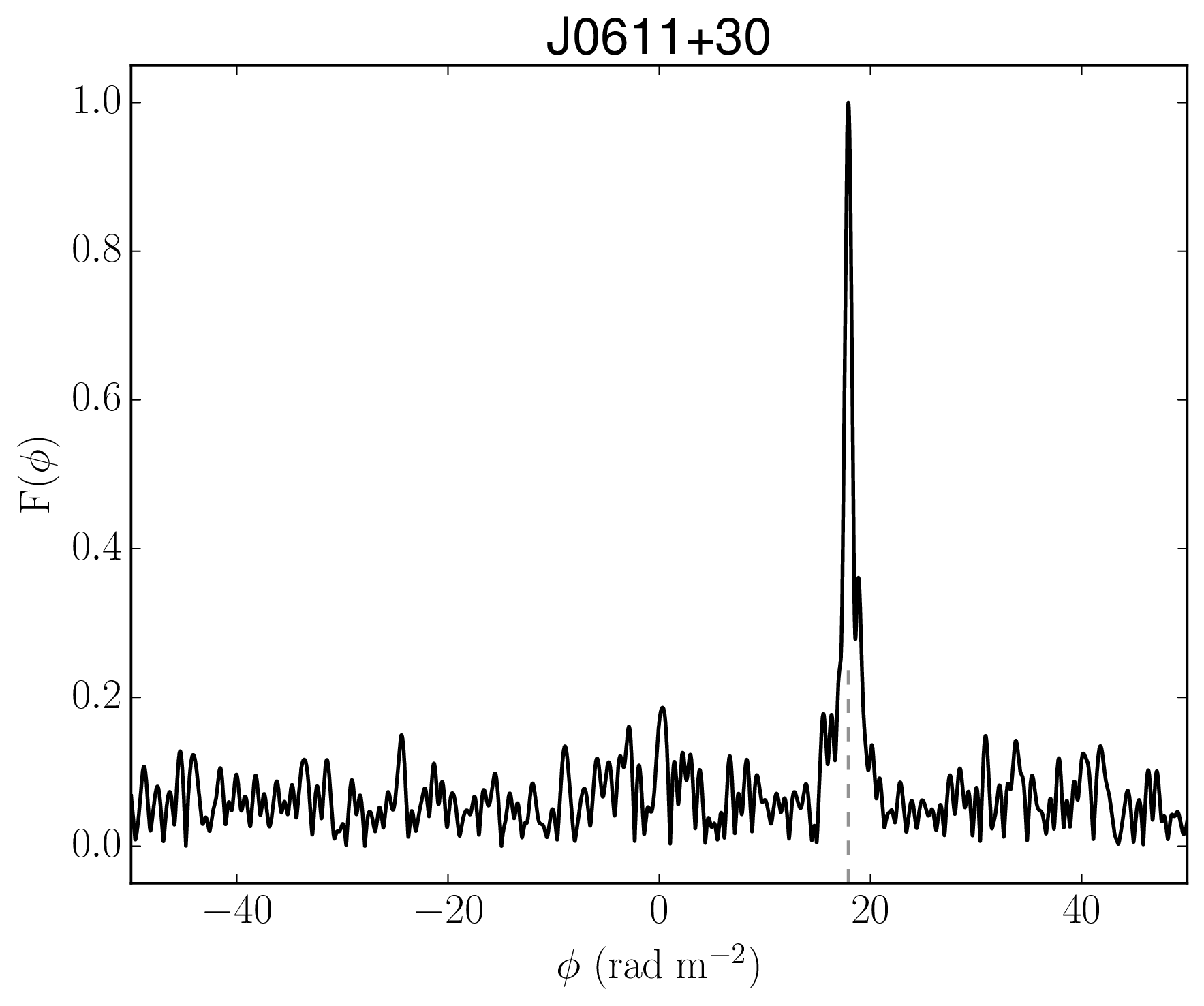}
\includegraphics[width=0.246\columnwidth]{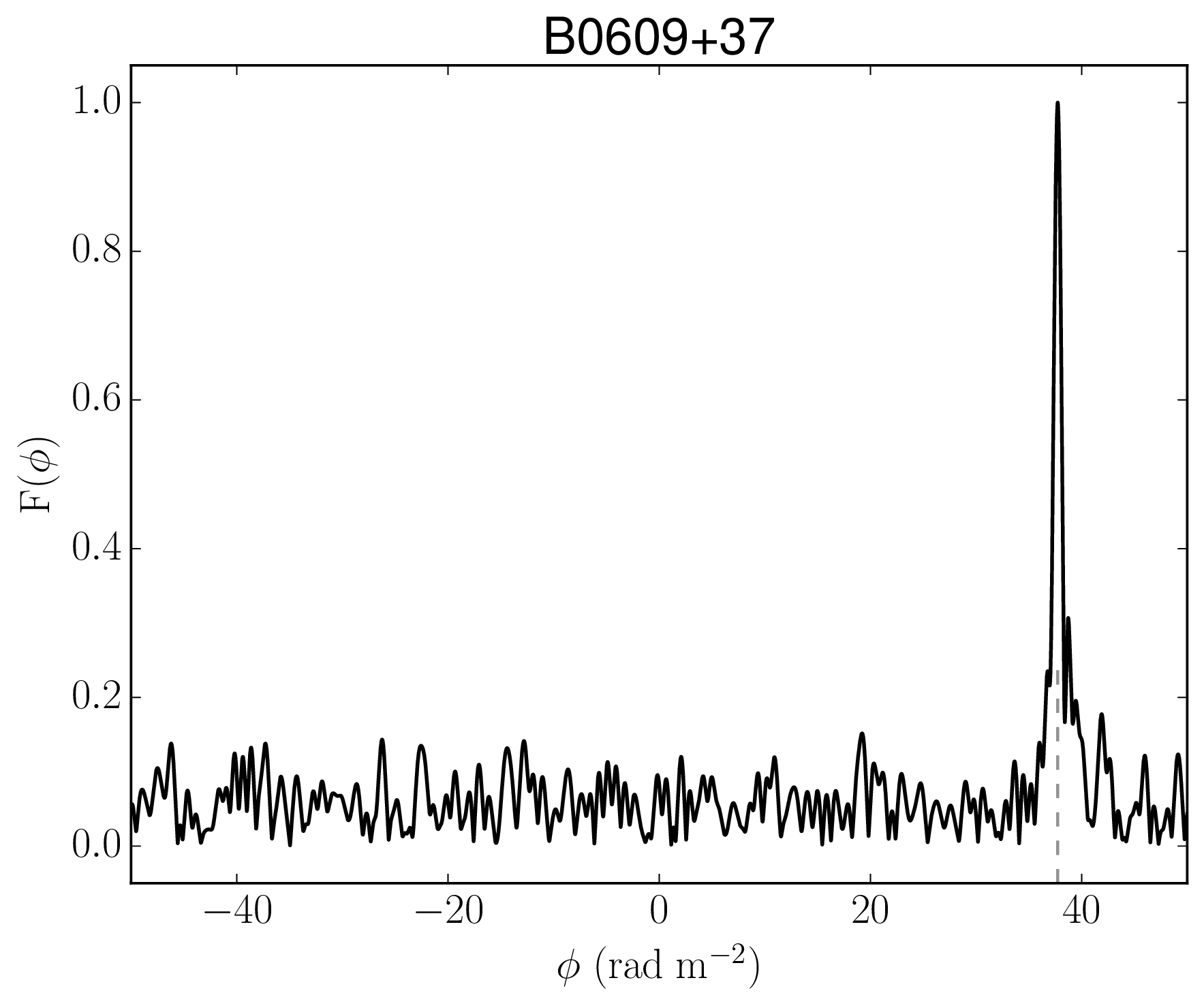}
\includegraphics[width=0.246\columnwidth]{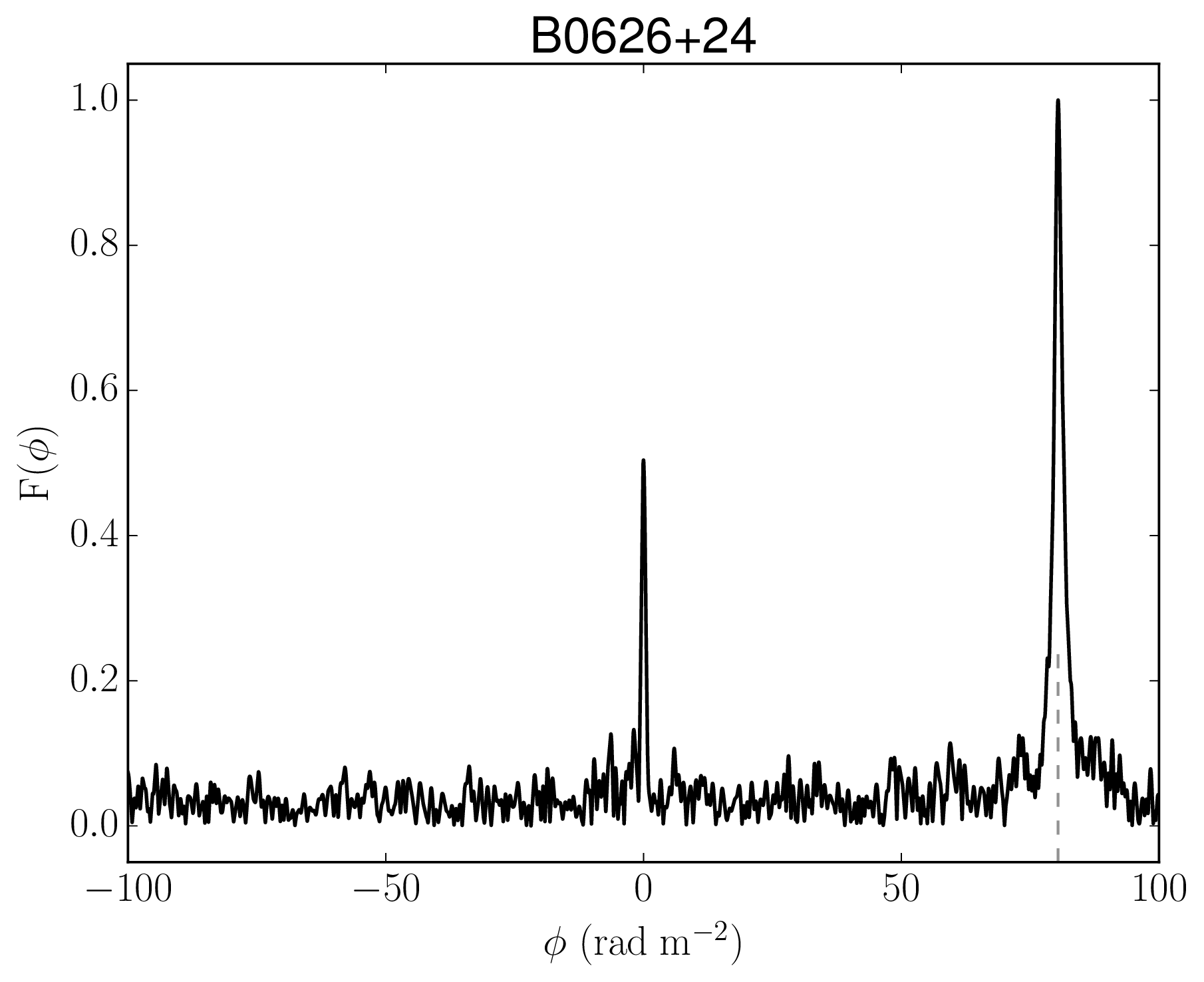}

\includegraphics[width=0.246\columnwidth]{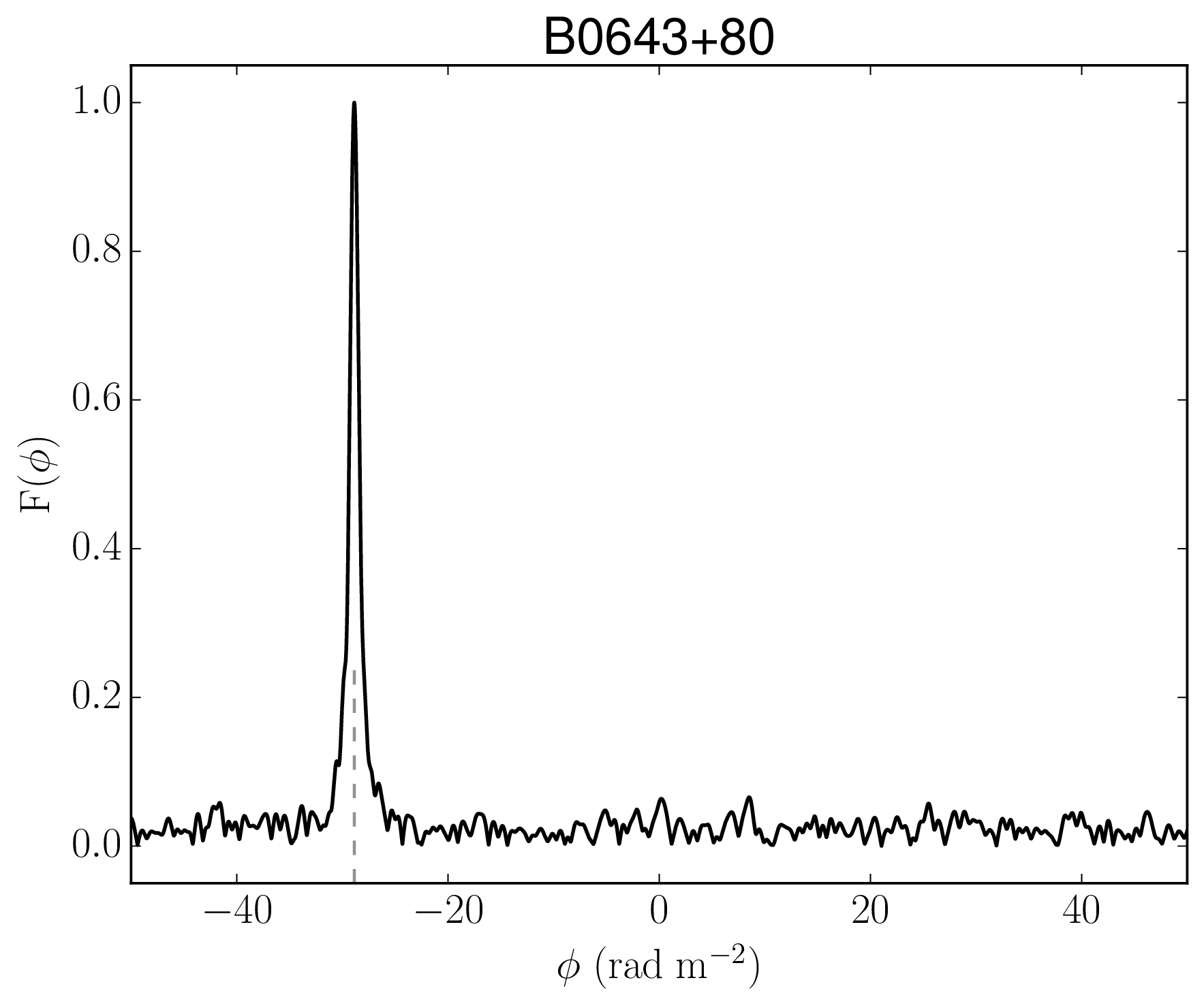}
\includegraphics[width=0.246\columnwidth]{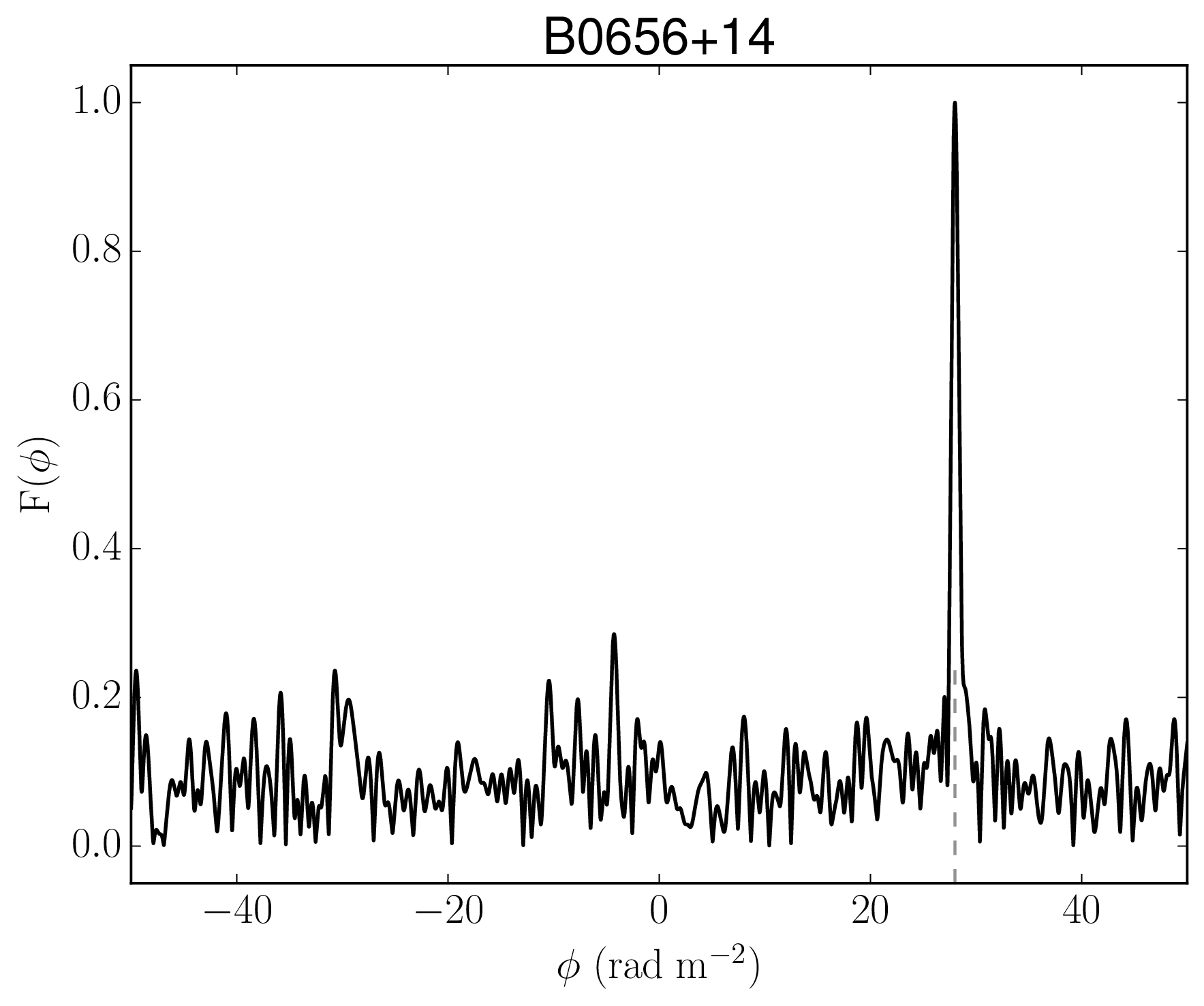}
\includegraphics[width=0.246\columnwidth]{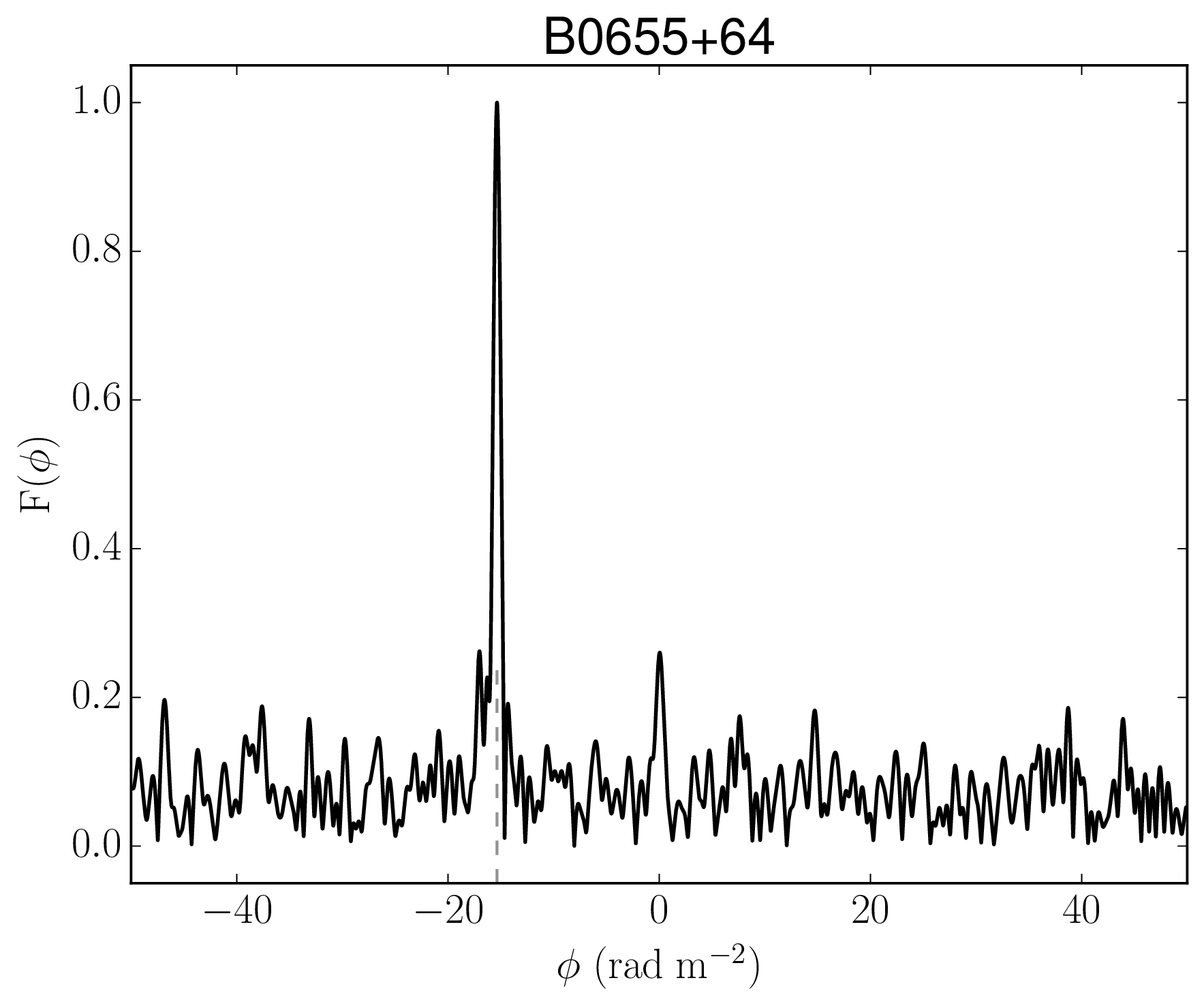}
\includegraphics[width=0.246\columnwidth]{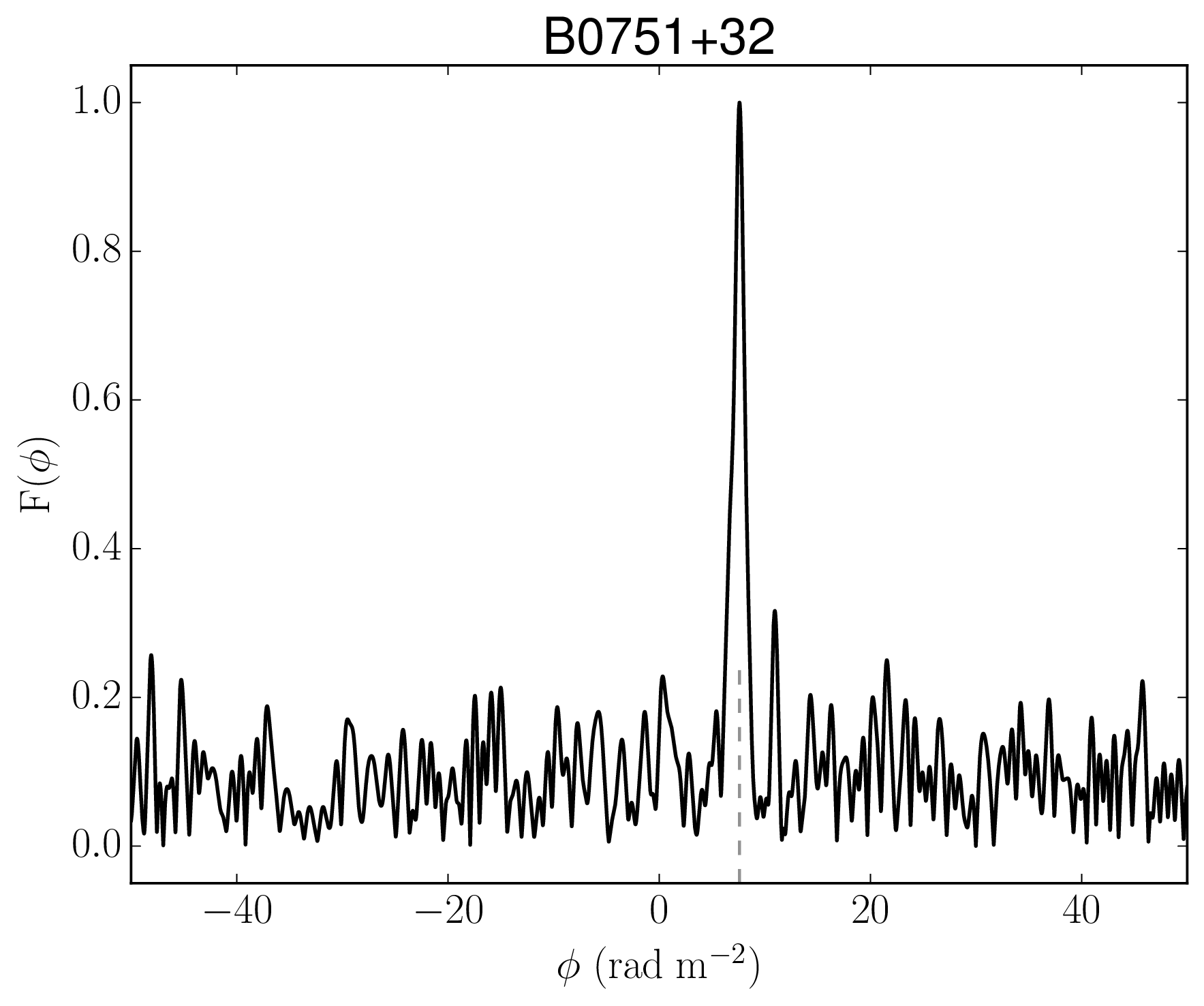}

\includegraphics[width=0.246\columnwidth]{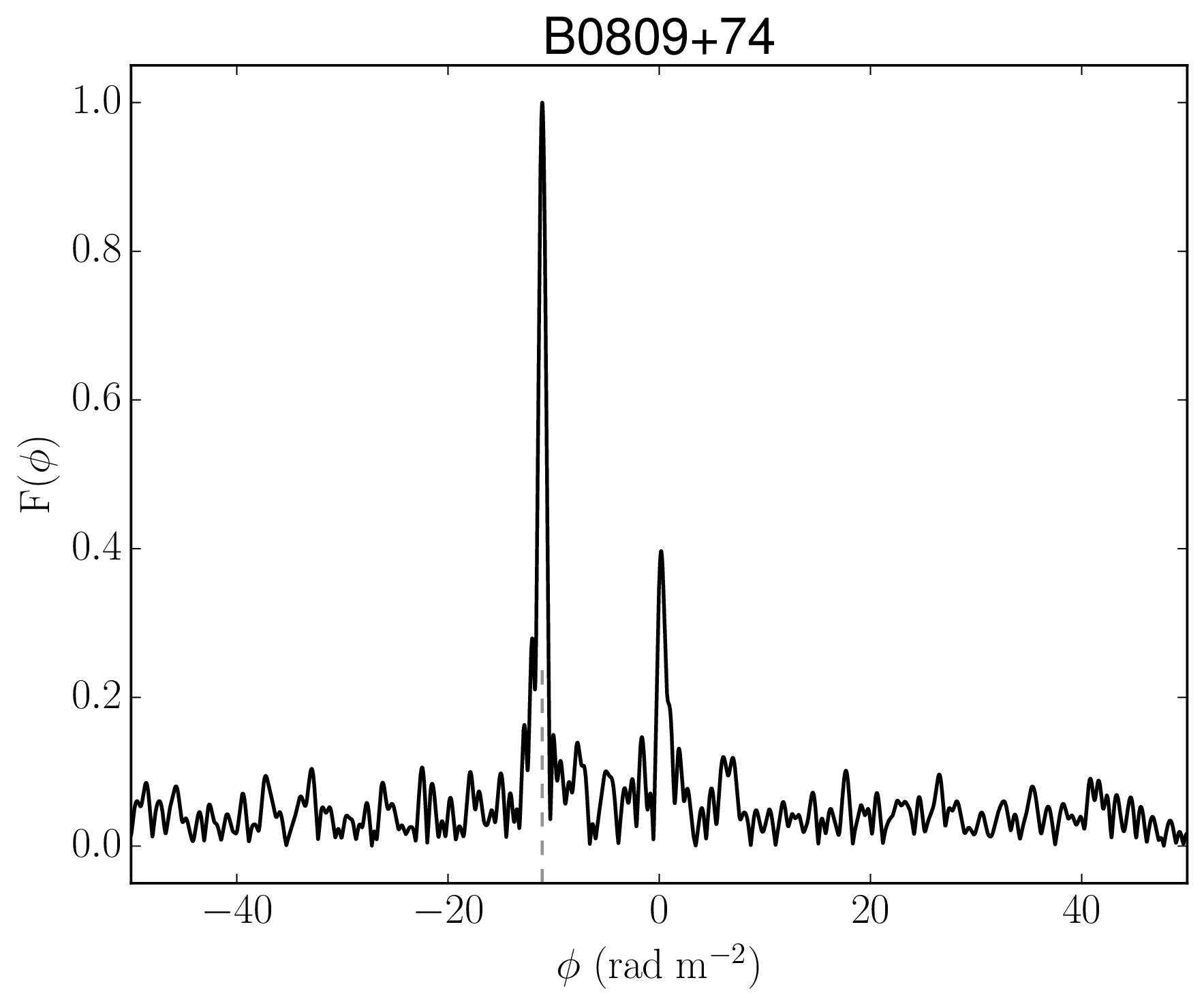}
\includegraphics[width=0.246\columnwidth]{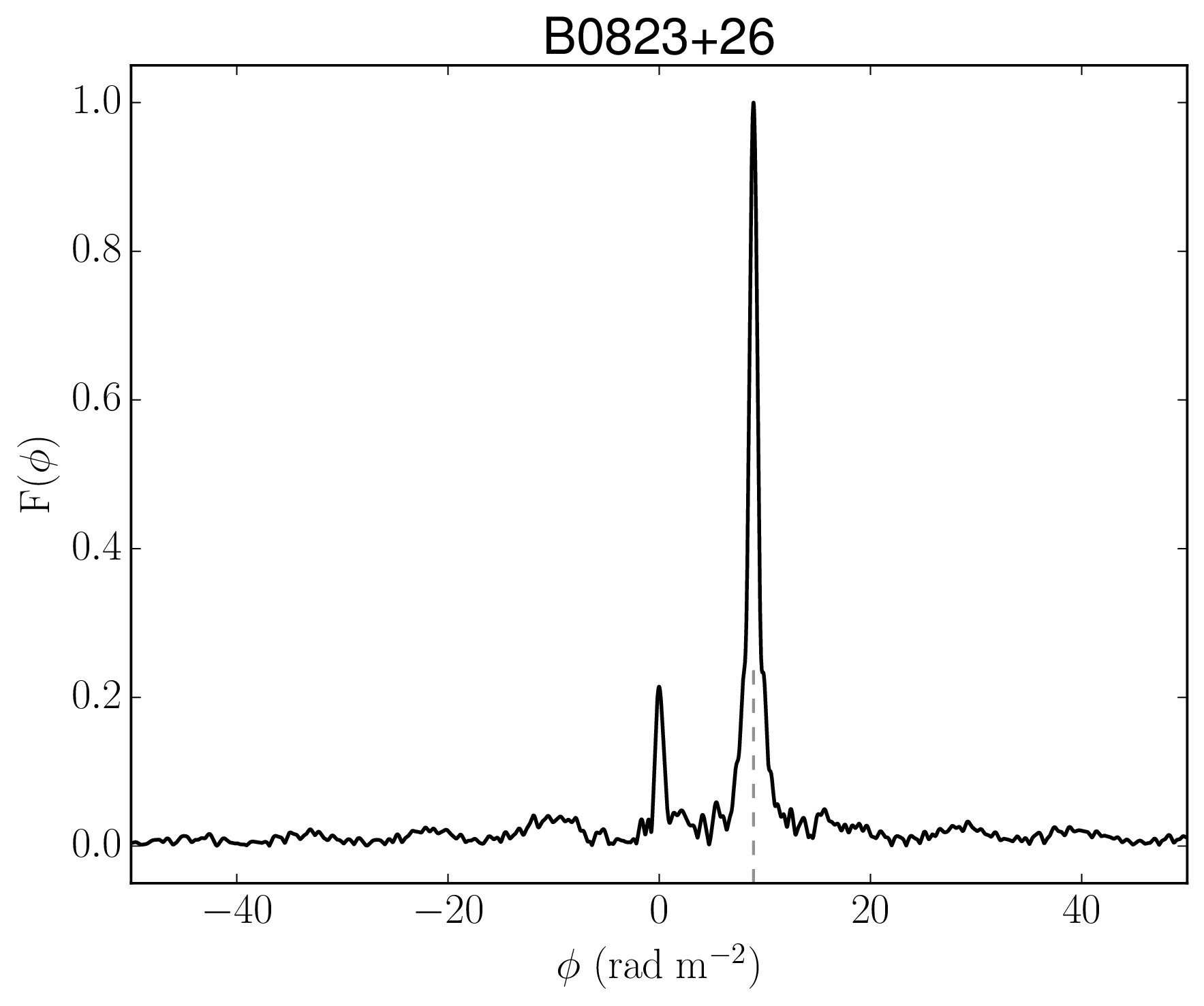}
\includegraphics[width=0.246\columnwidth]{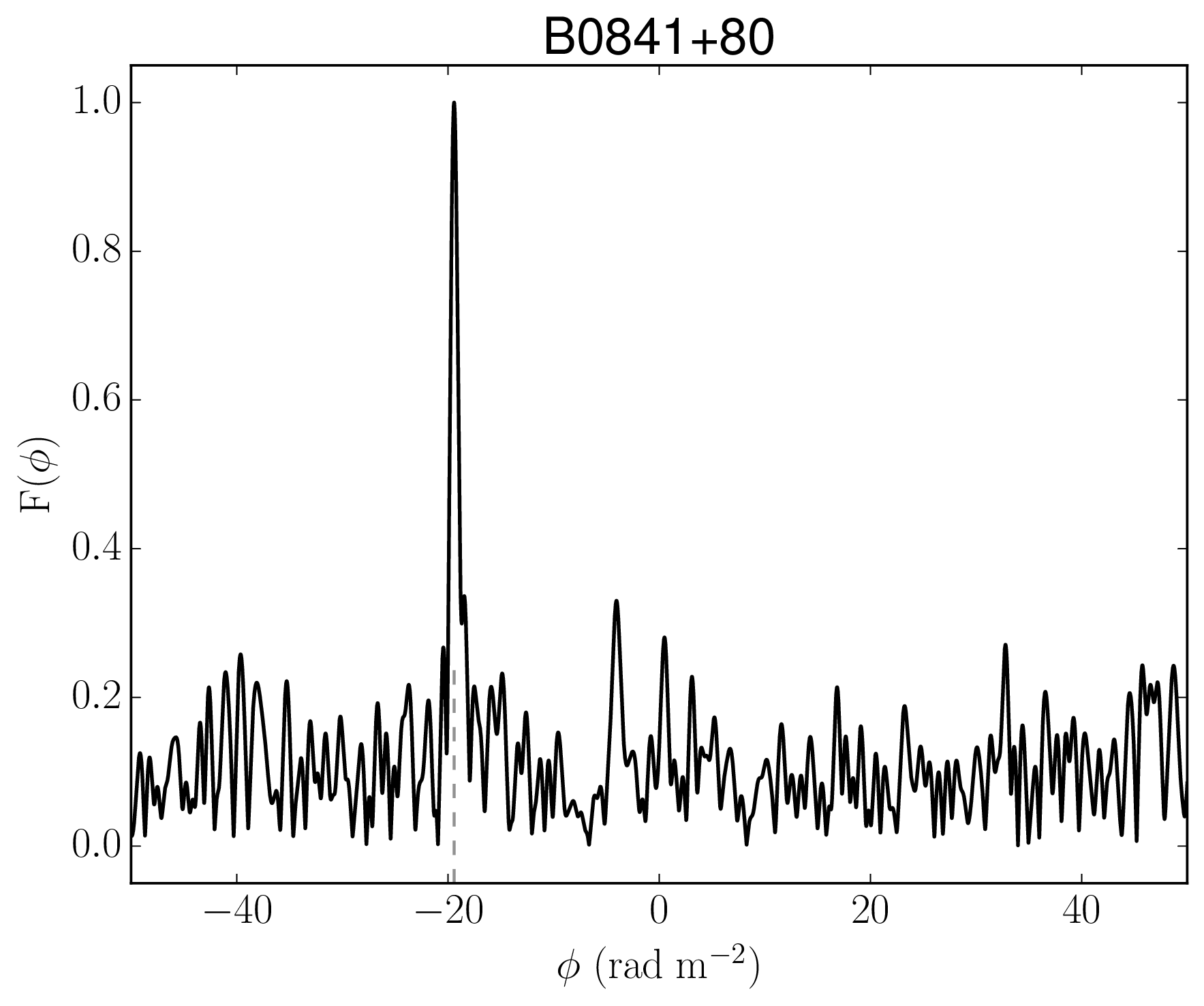}
\includegraphics[width=0.246\columnwidth]{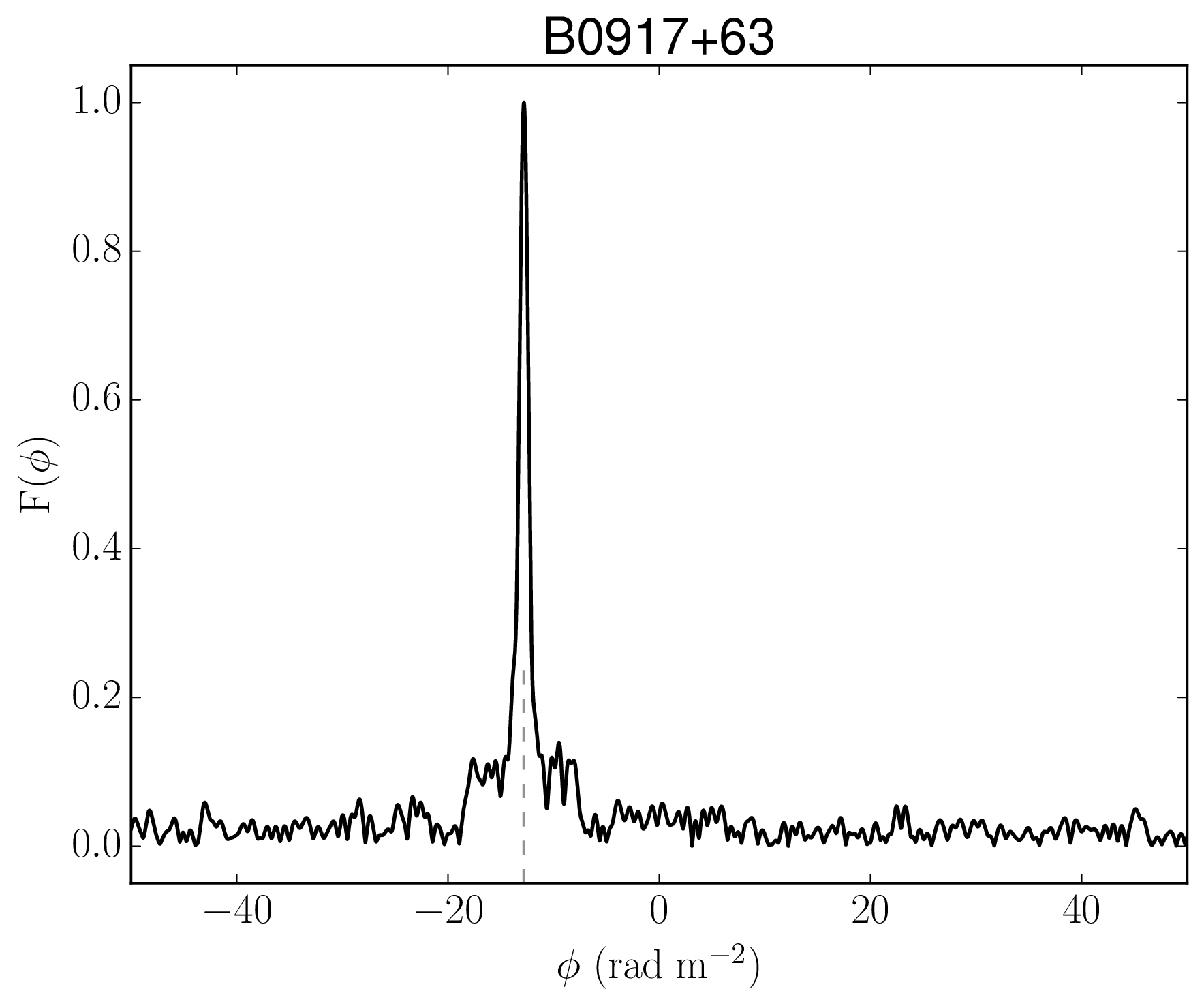}

\includegraphics[width=0.246\columnwidth]{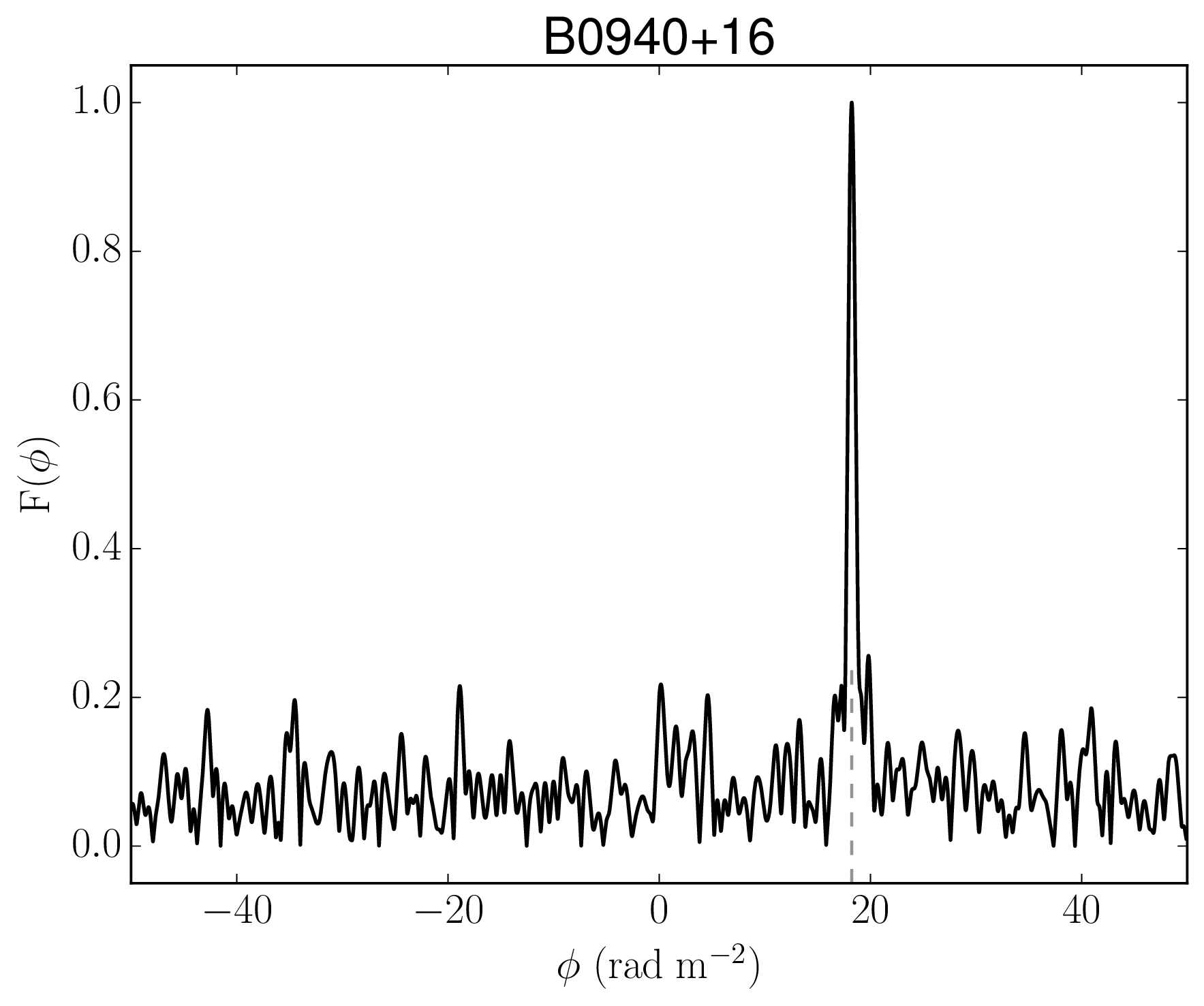}
\includegraphics[width=0.246\columnwidth]{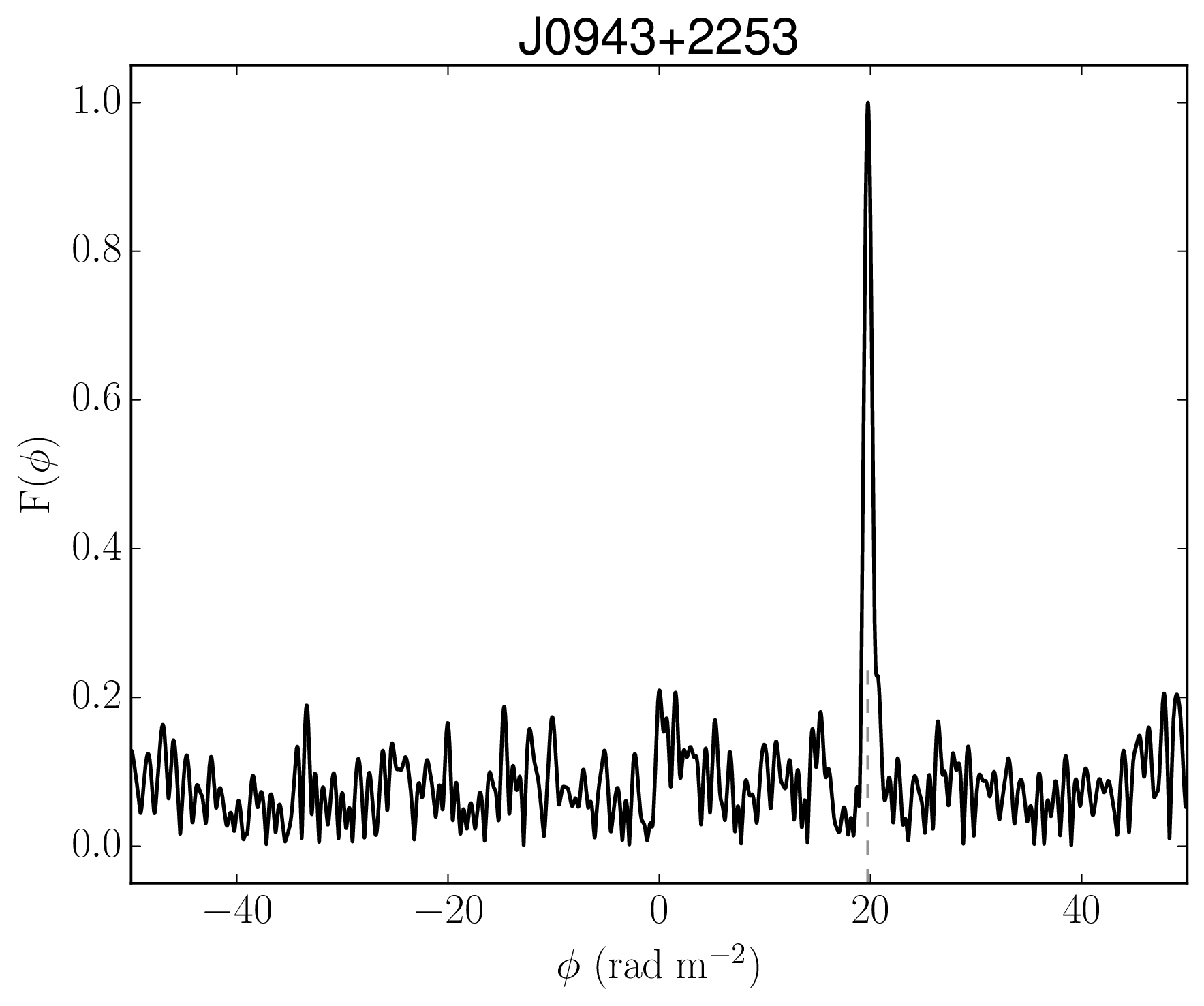}
\includegraphics[width=0.246\columnwidth]{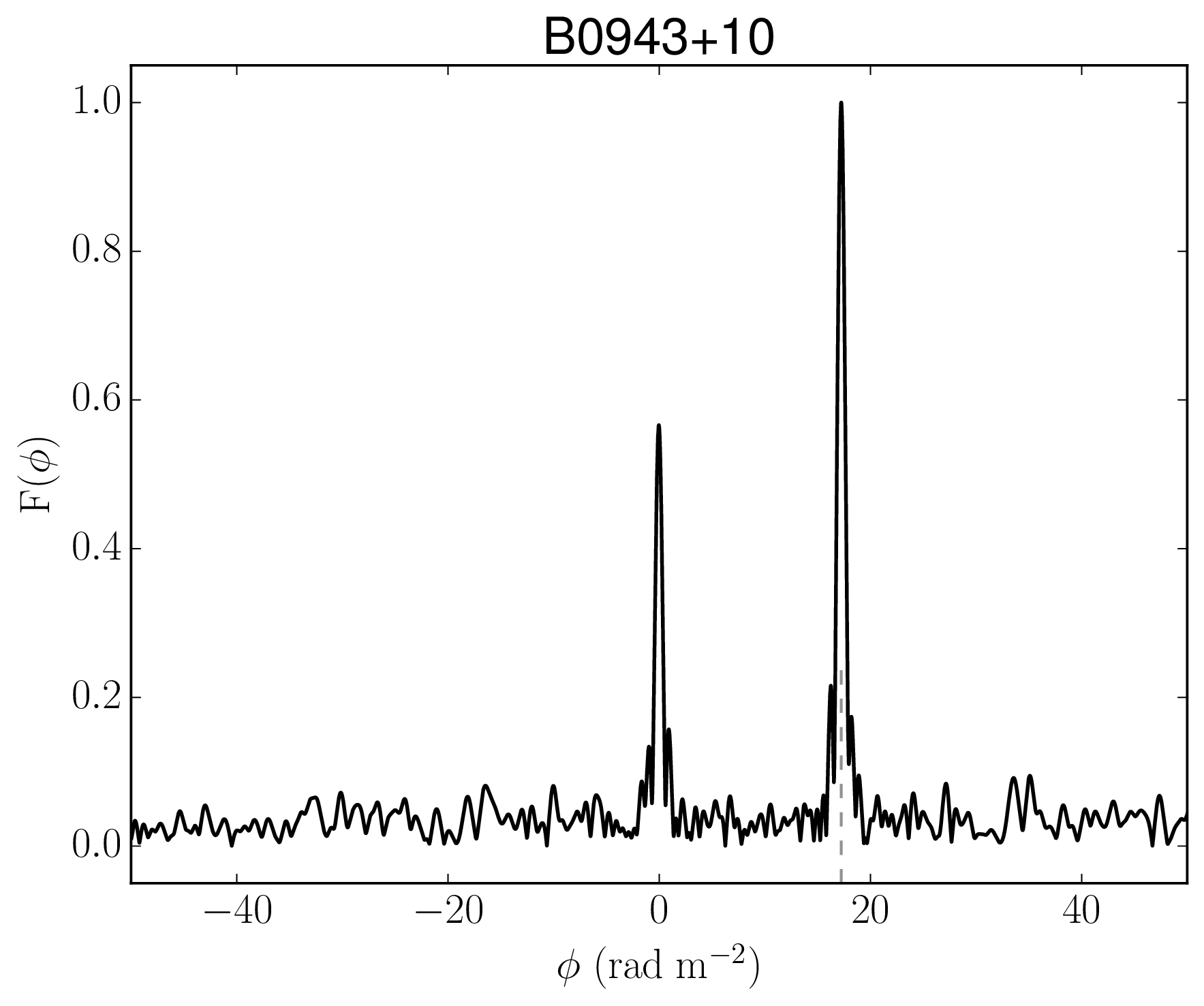}
\includegraphics[width=0.246\columnwidth]{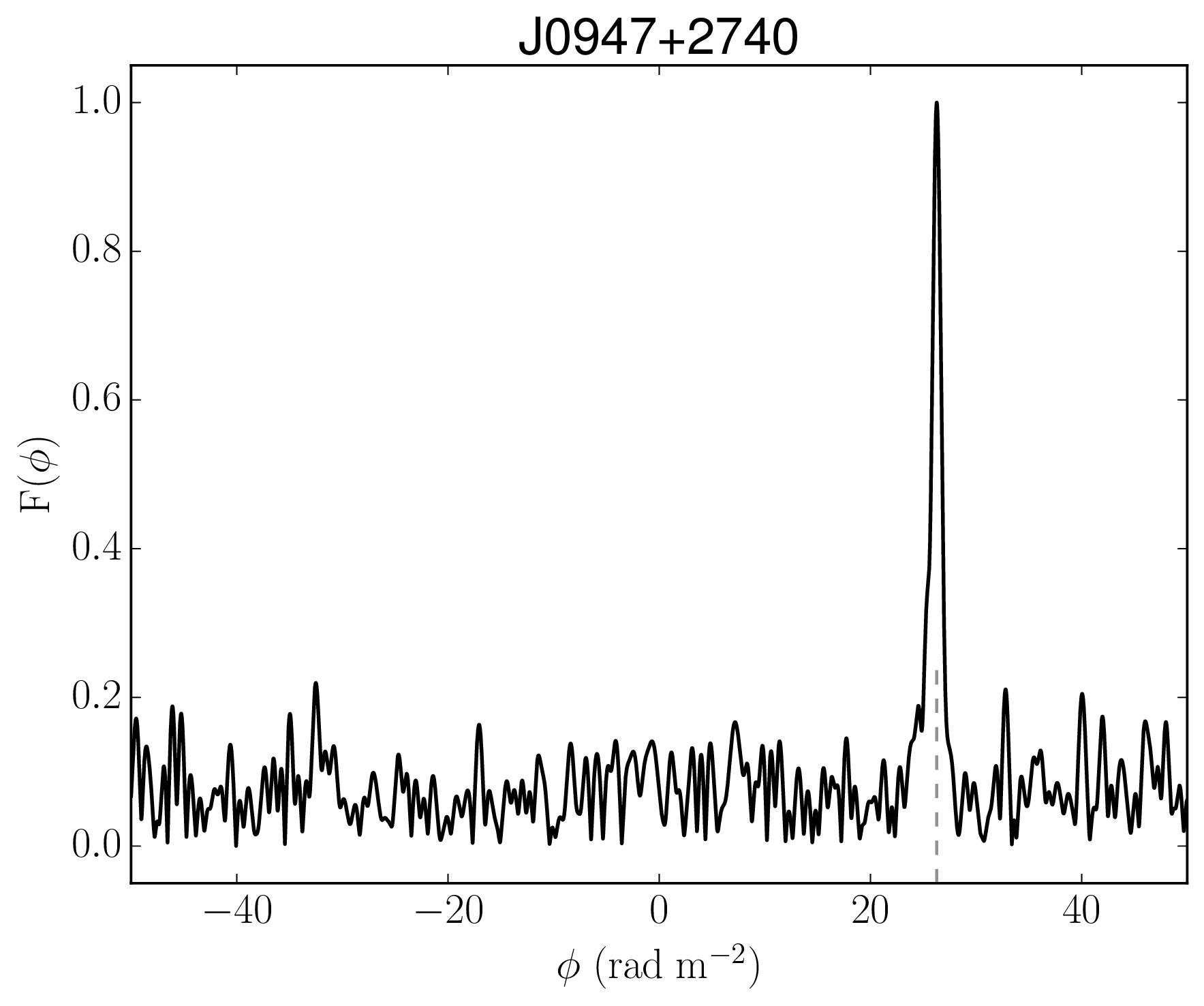}

\includegraphics[width=0.246\columnwidth]{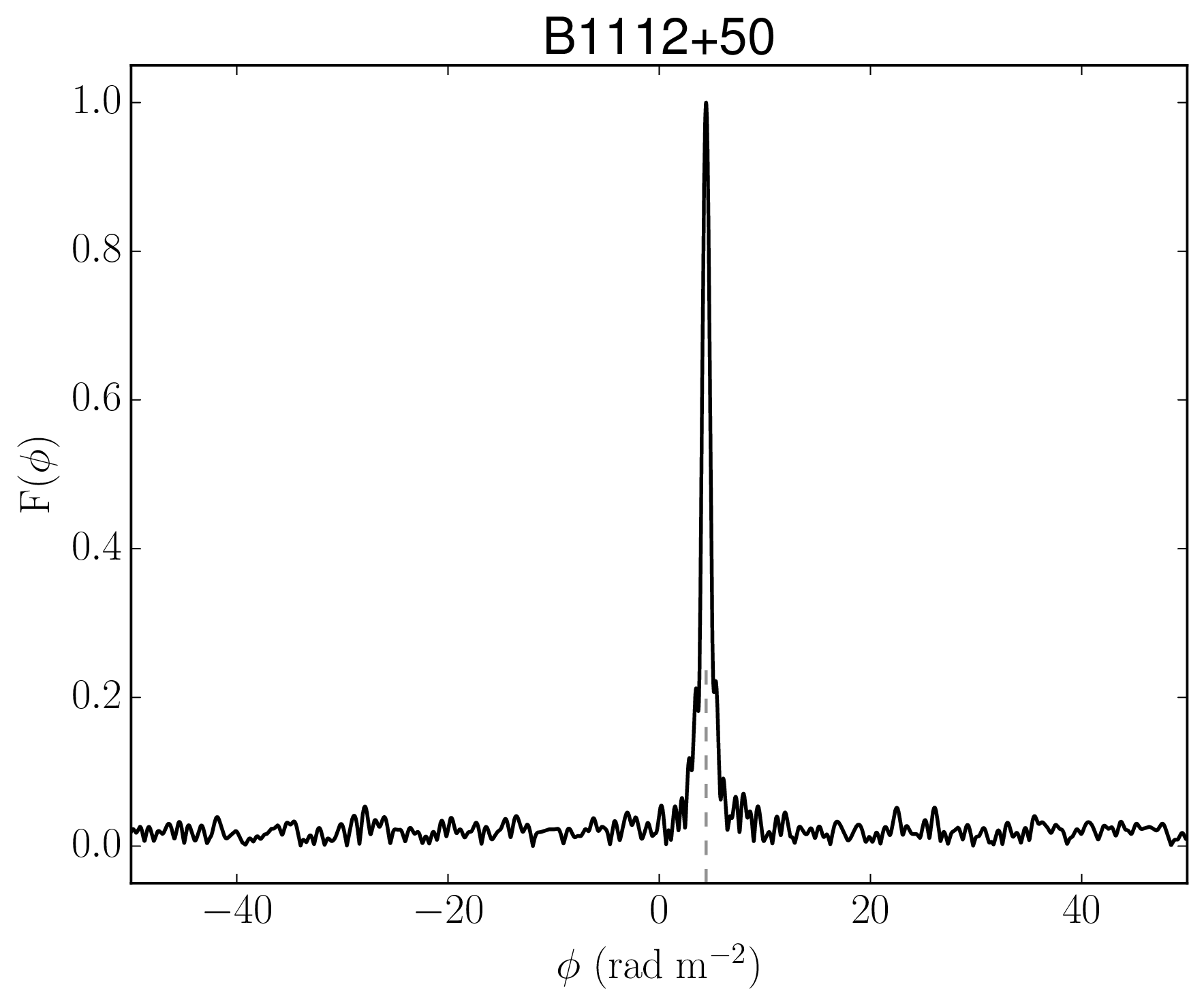}
\includegraphics[width=0.246\columnwidth]{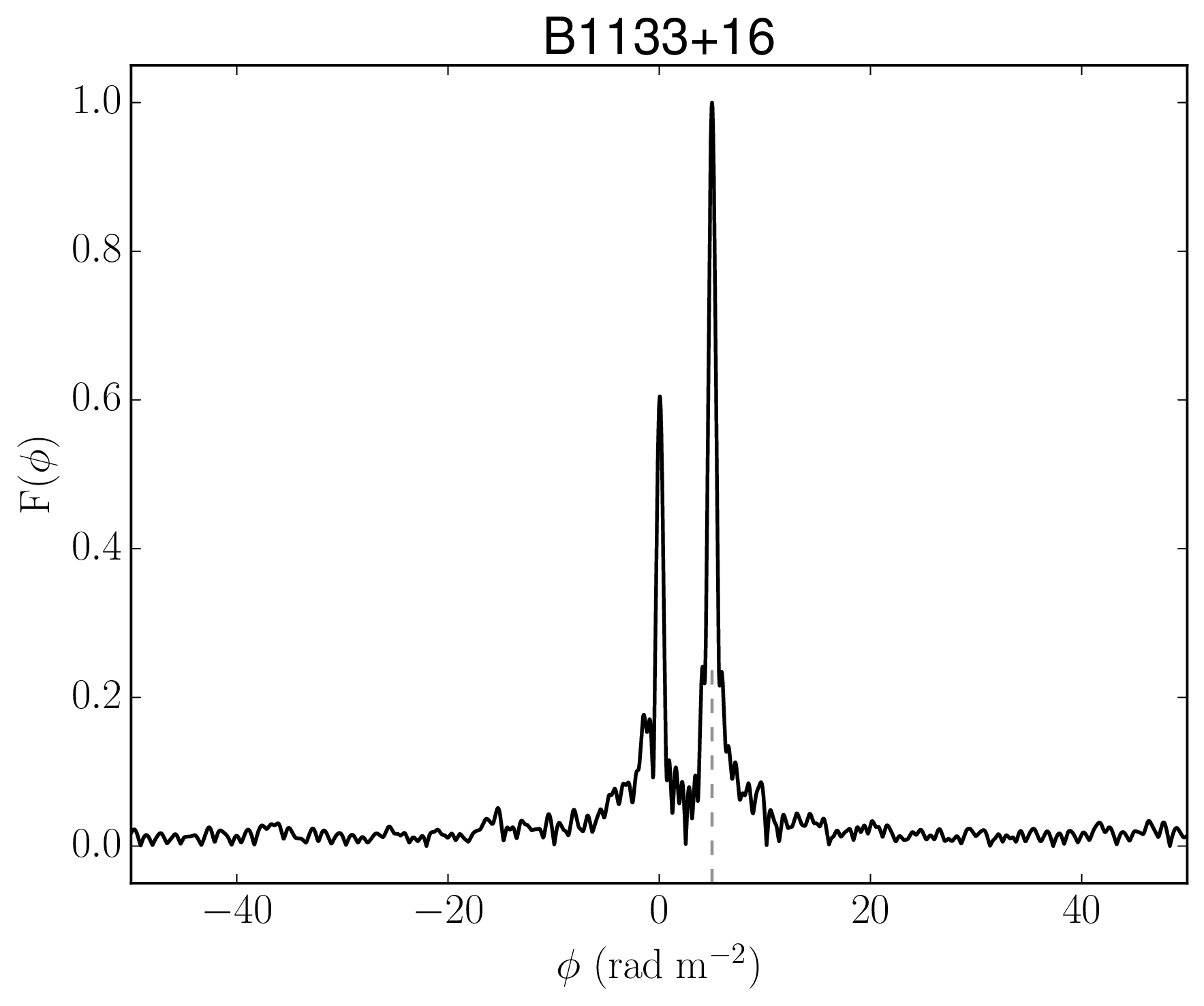}
\includegraphics[width=0.246\columnwidth]{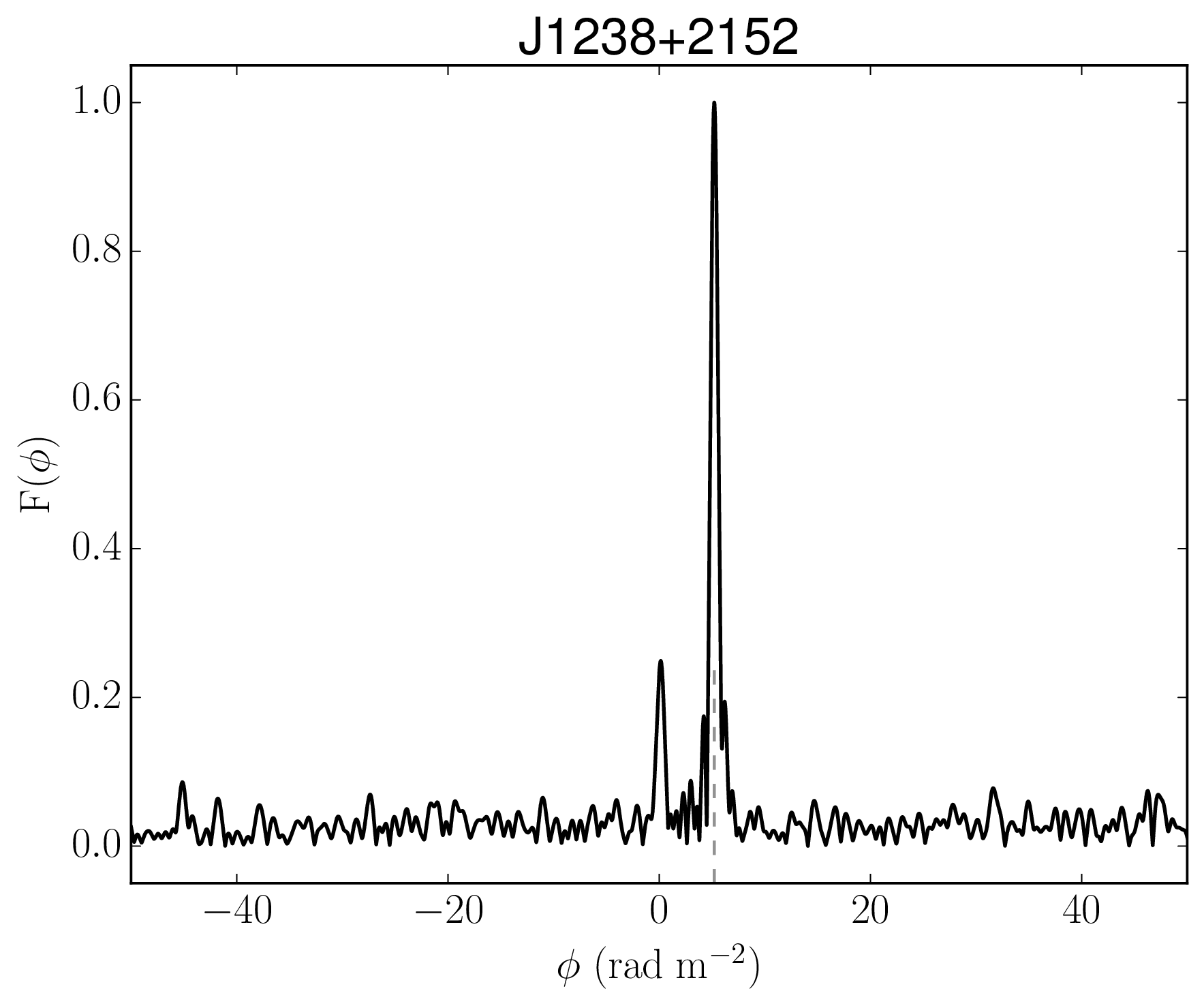}
\includegraphics[width=0.246\columnwidth]{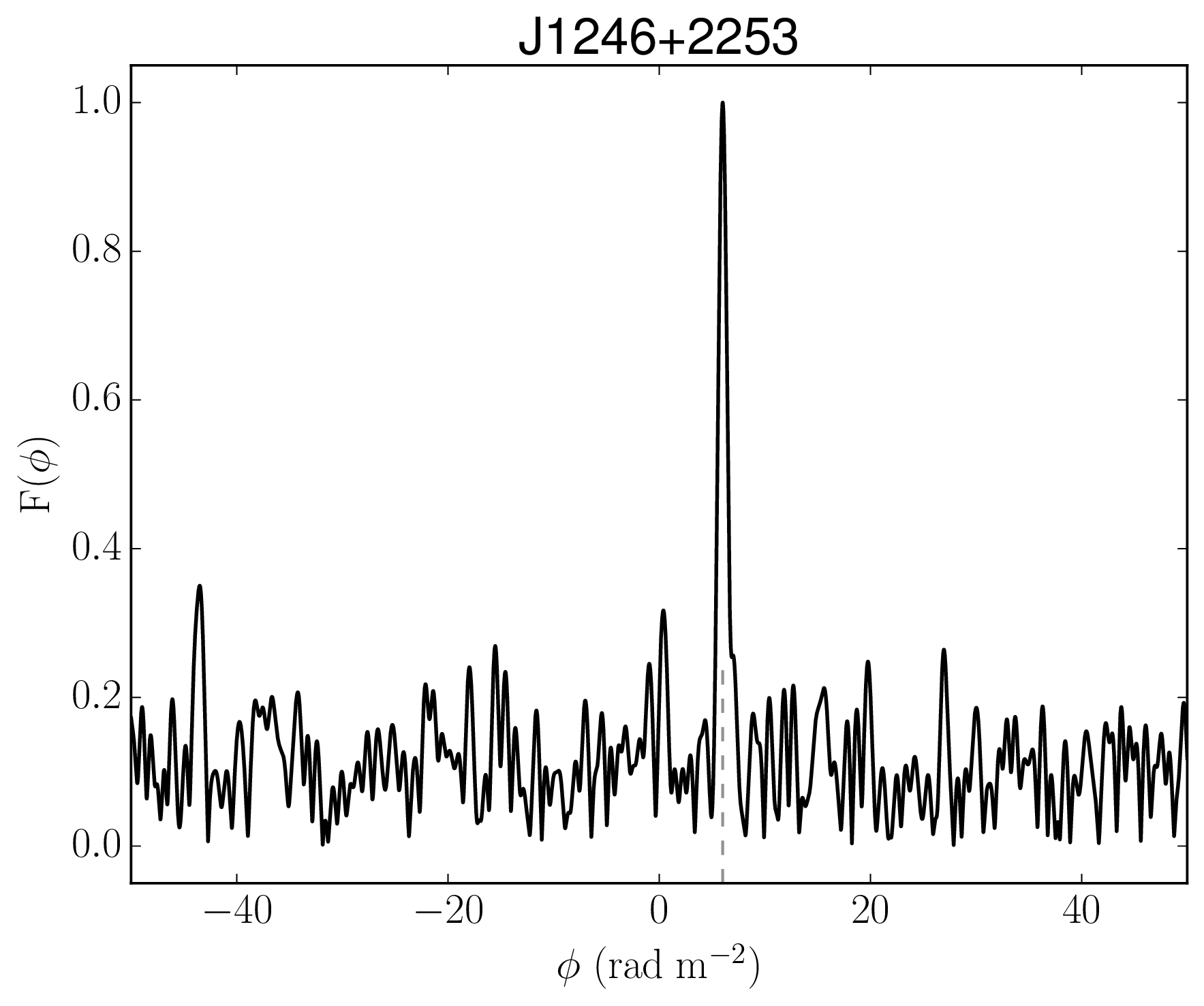}

\includegraphics[width=0.246\columnwidth]{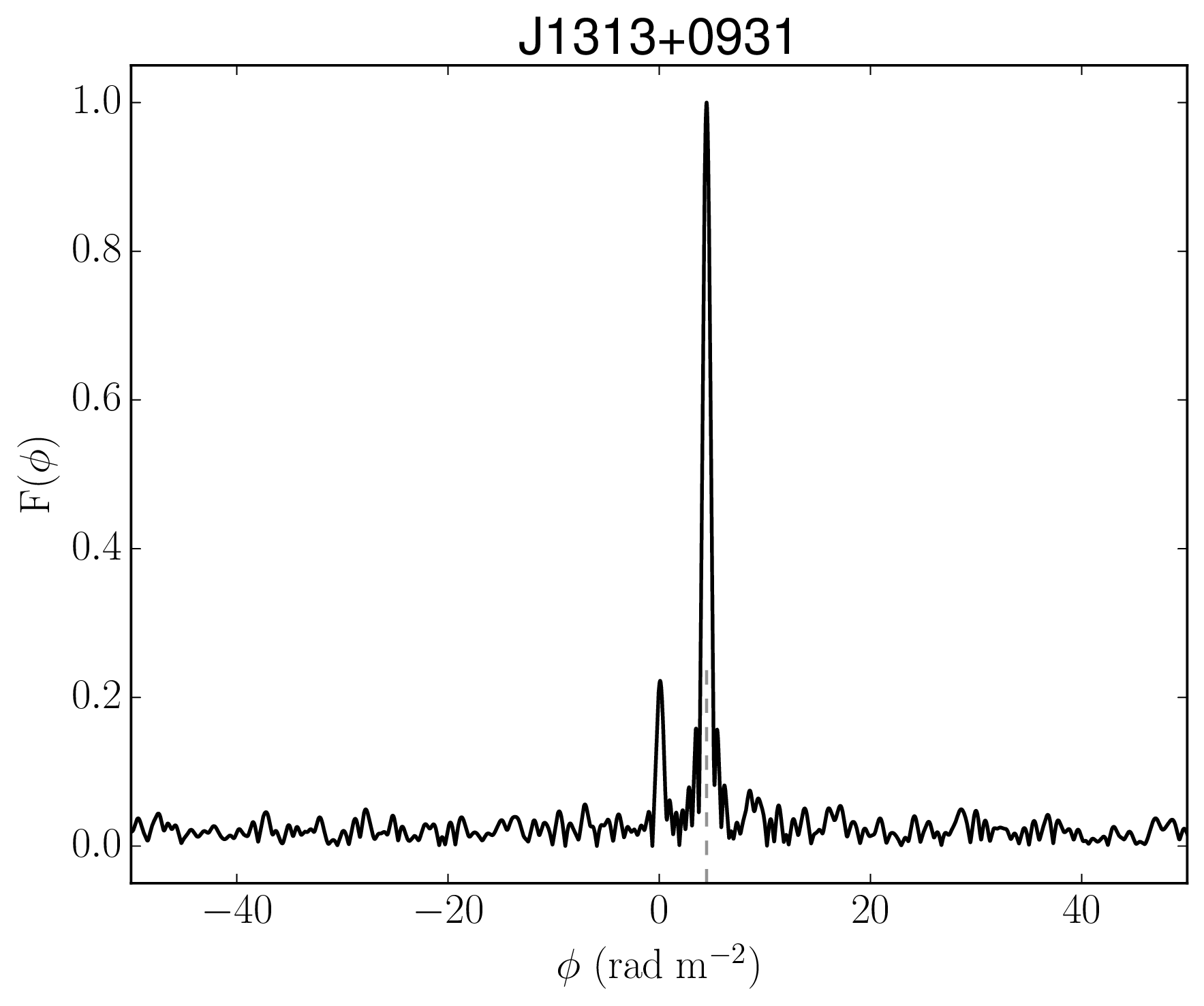}
\includegraphics[width=0.246\columnwidth]{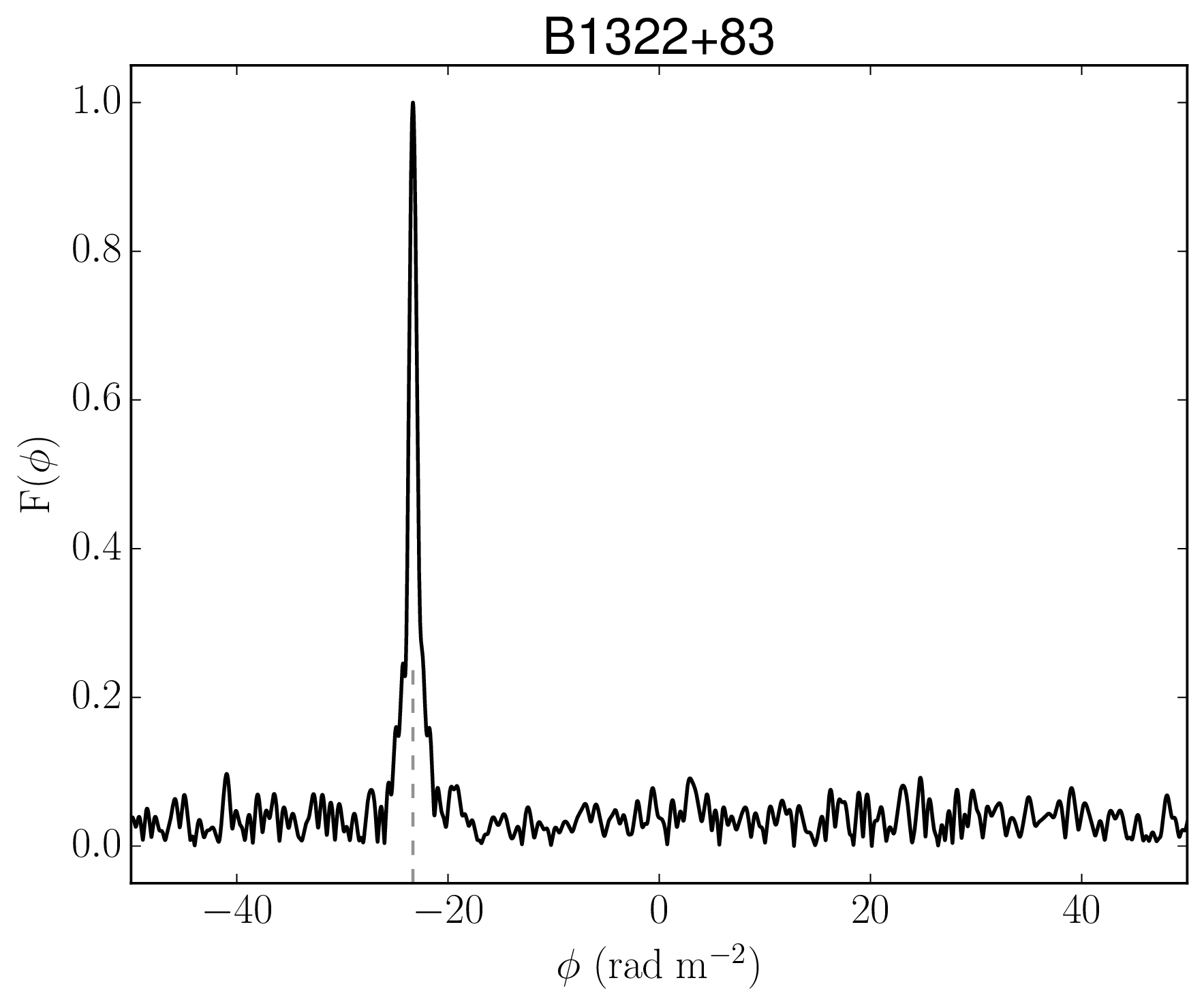}
\includegraphics[width=0.246\columnwidth]{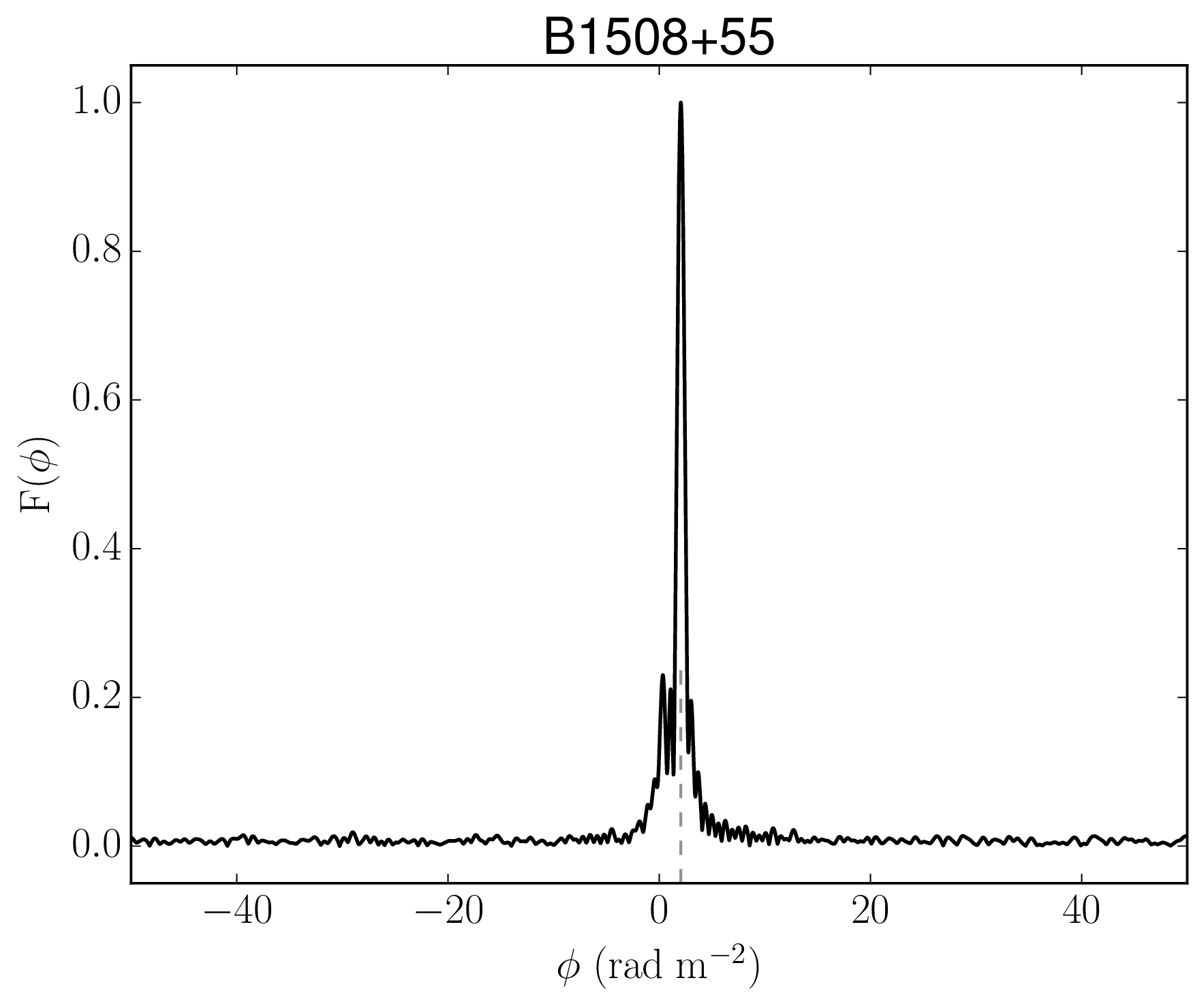}
\includegraphics[width=0.246\columnwidth]{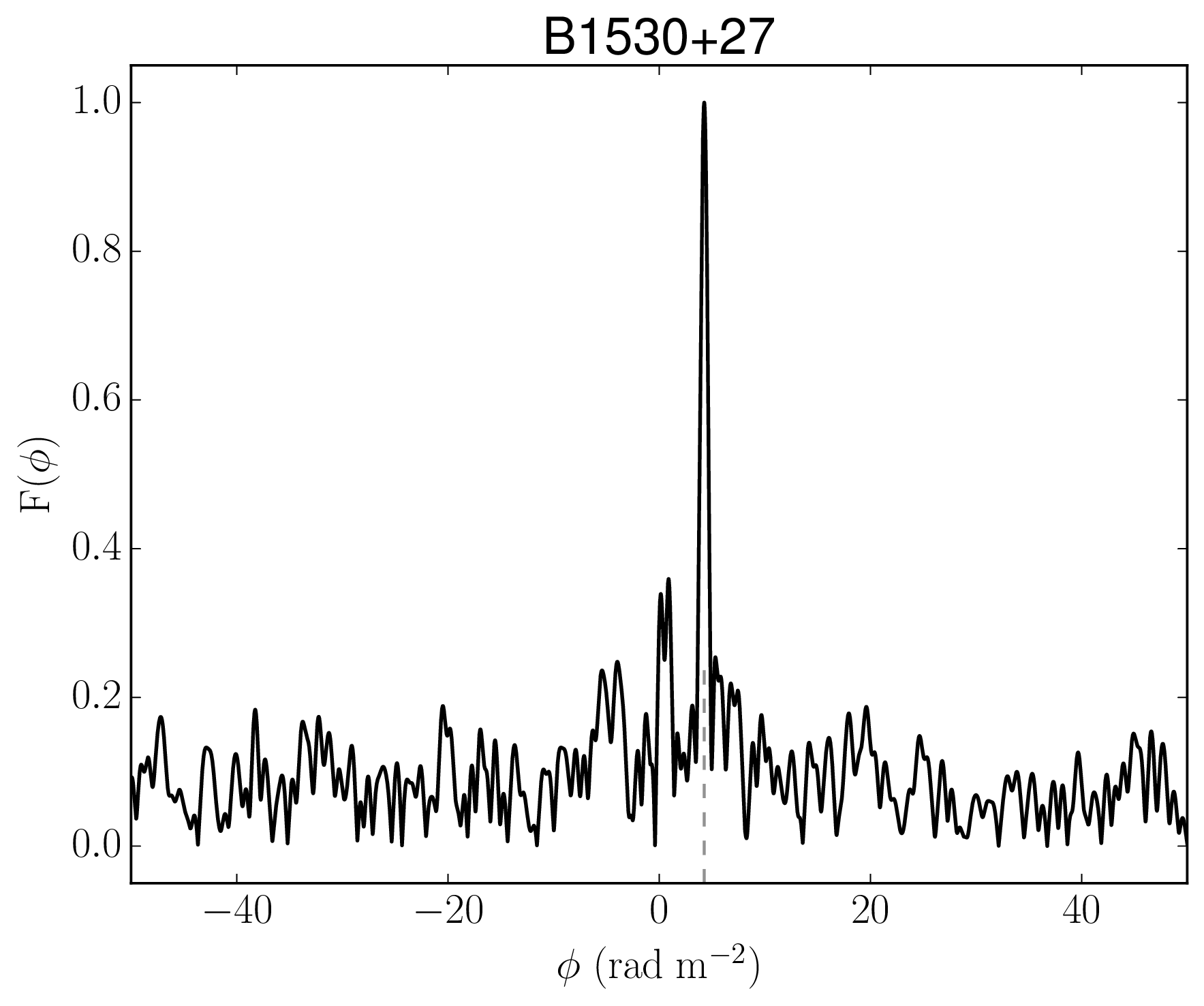}

 \caption{Continued from the previous page.}
% \label{fig:FDFs}
\end{figure*}

\begin{figure*}
 \centering
 %\ContinuedFloat
%\setcounter{subfigure}{2}    

\includegraphics[width=0.246\columnwidth]{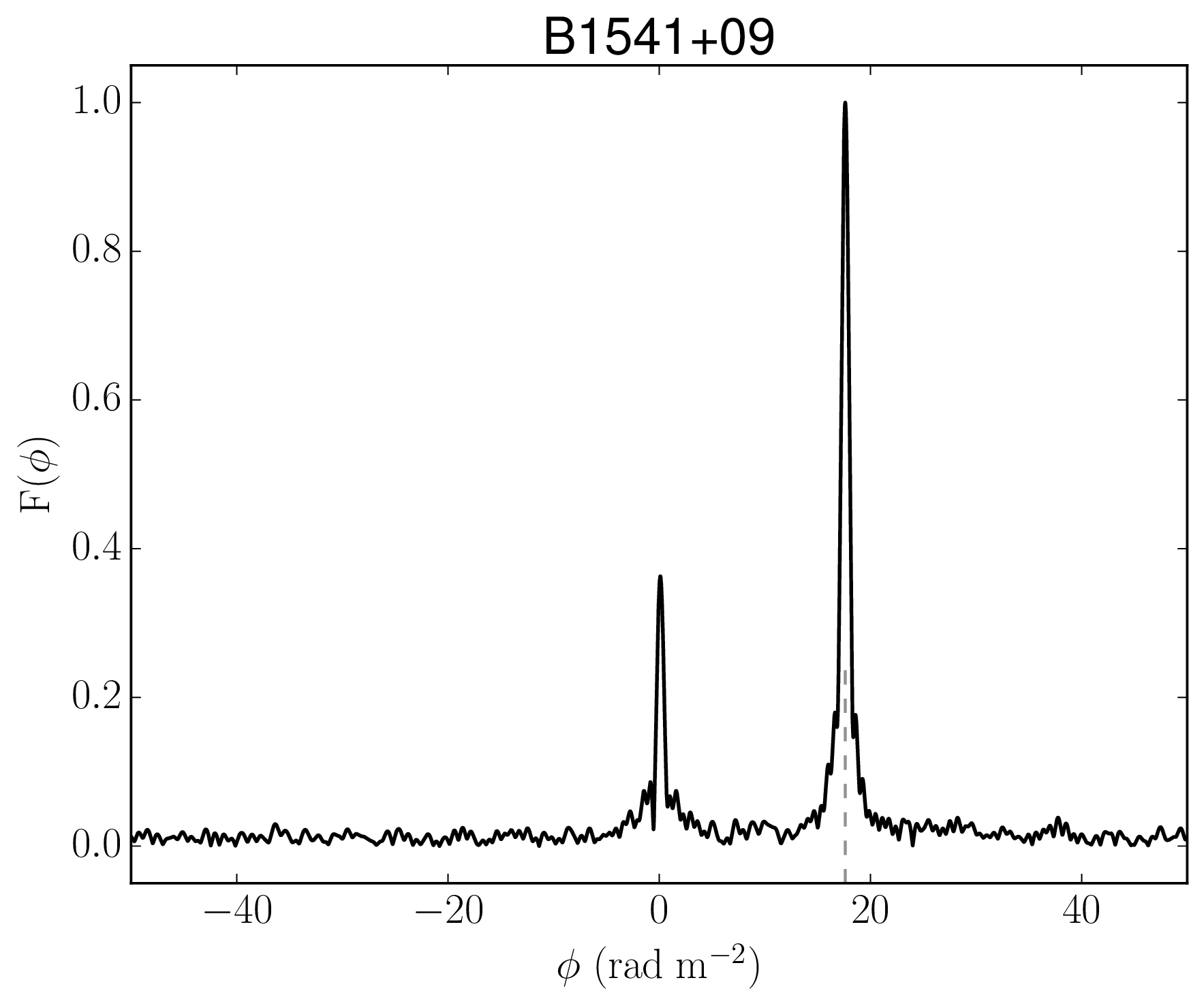}
\includegraphics[width=0.246\columnwidth]{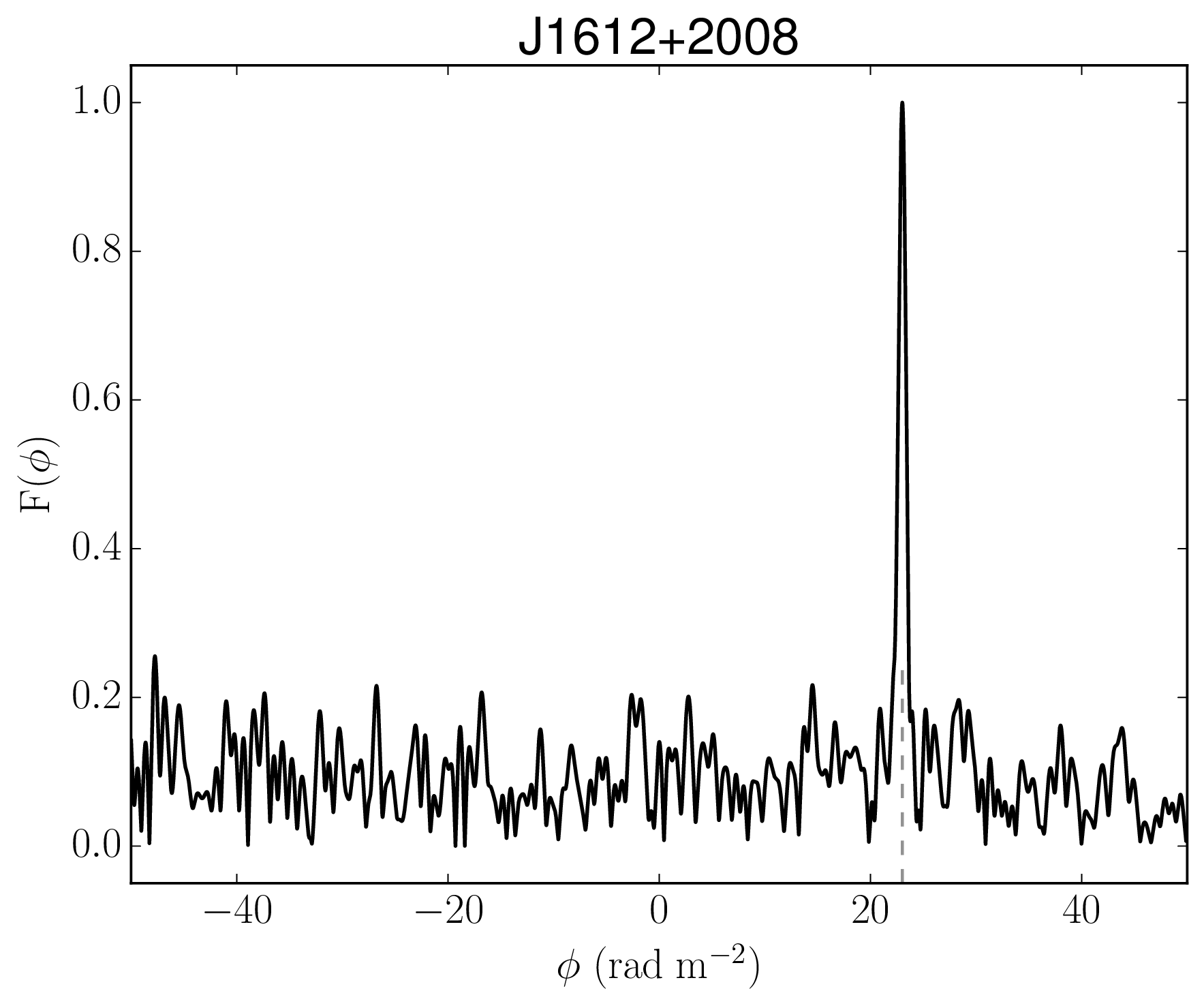}
\includegraphics[width=0.246\columnwidth]{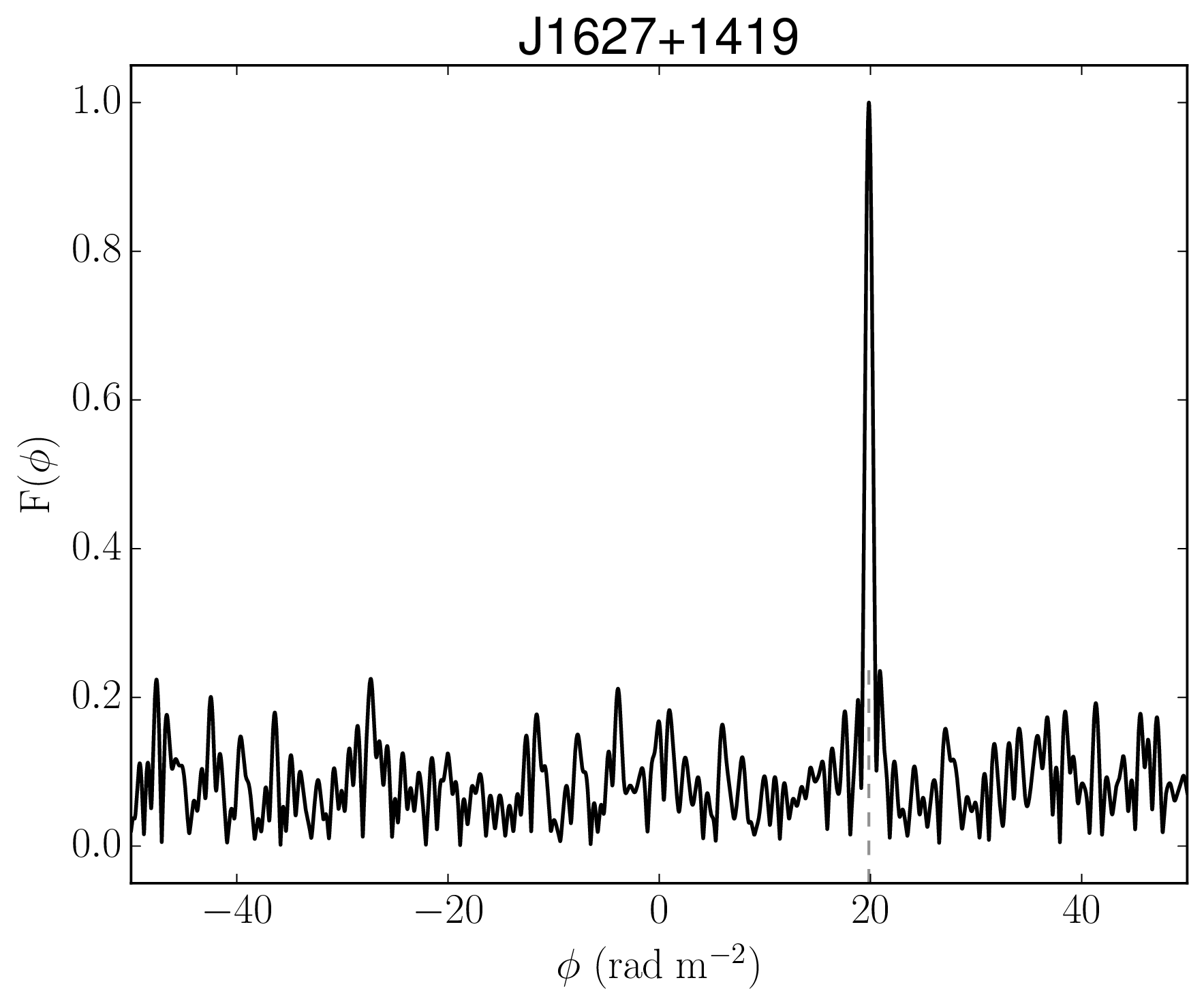}
\includegraphics[width=0.246\columnwidth]{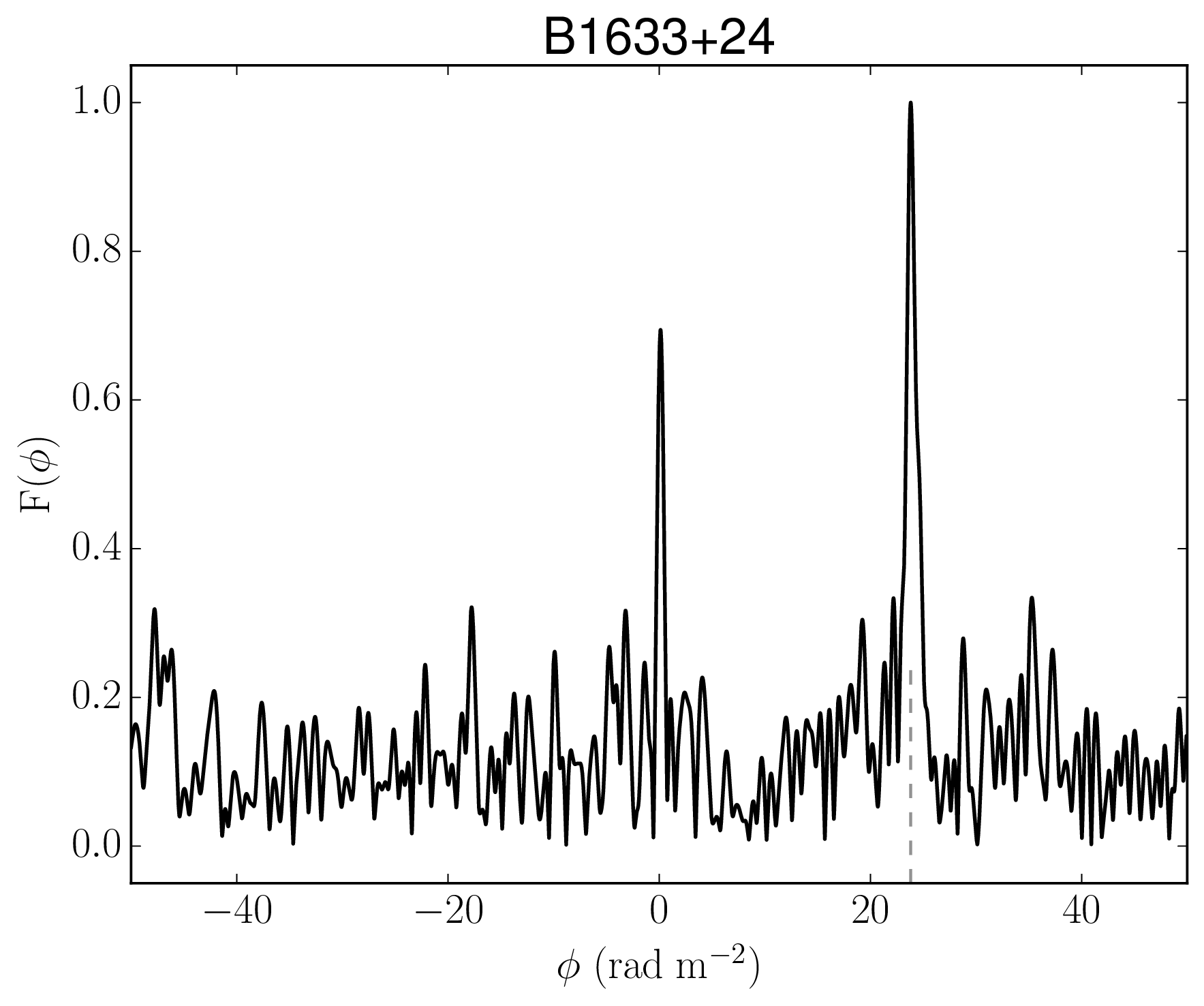}

\includegraphics[width=0.246\columnwidth]{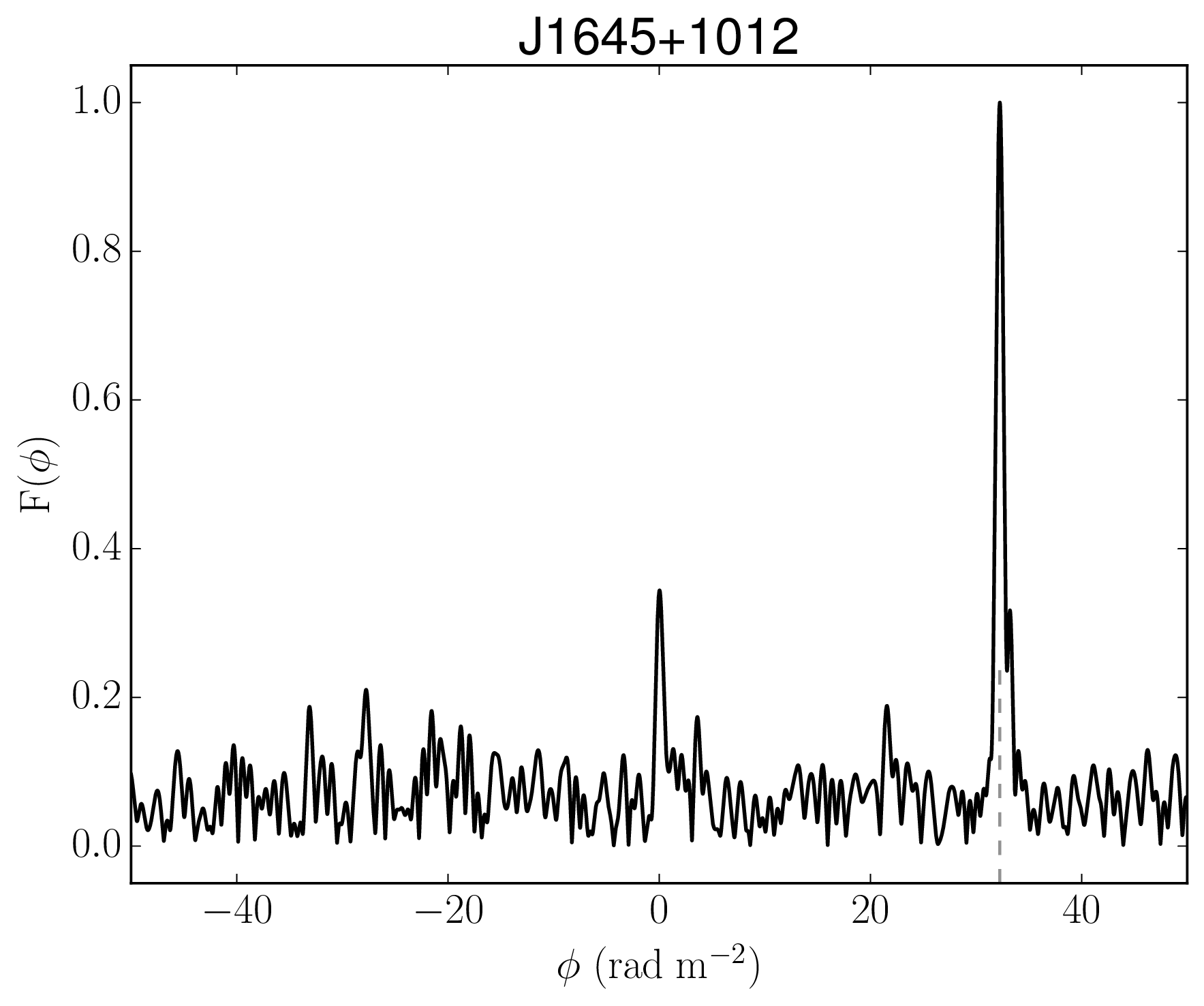}
\includegraphics[width=0.246\columnwidth]{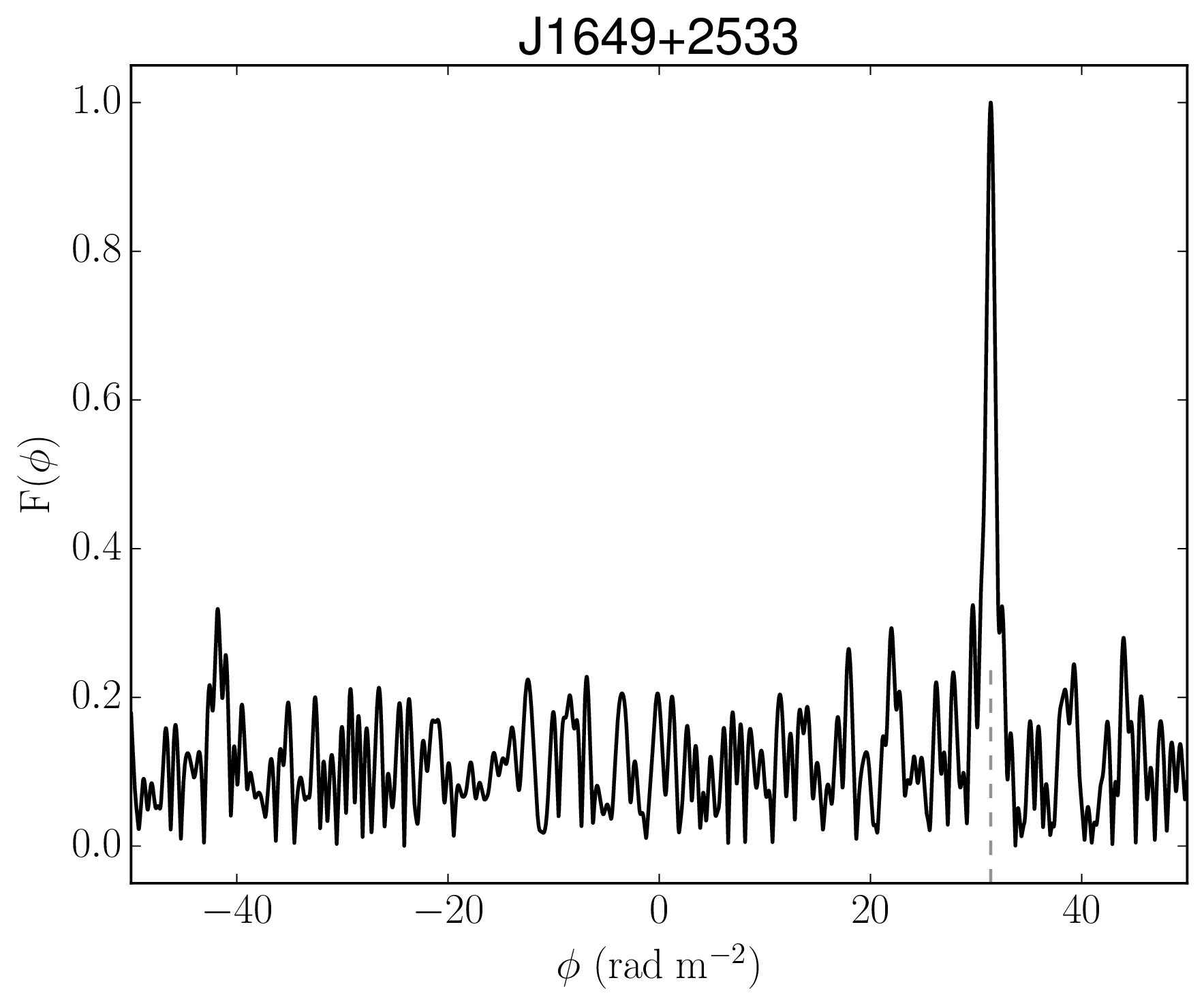}
\includegraphics[width=0.246\columnwidth]{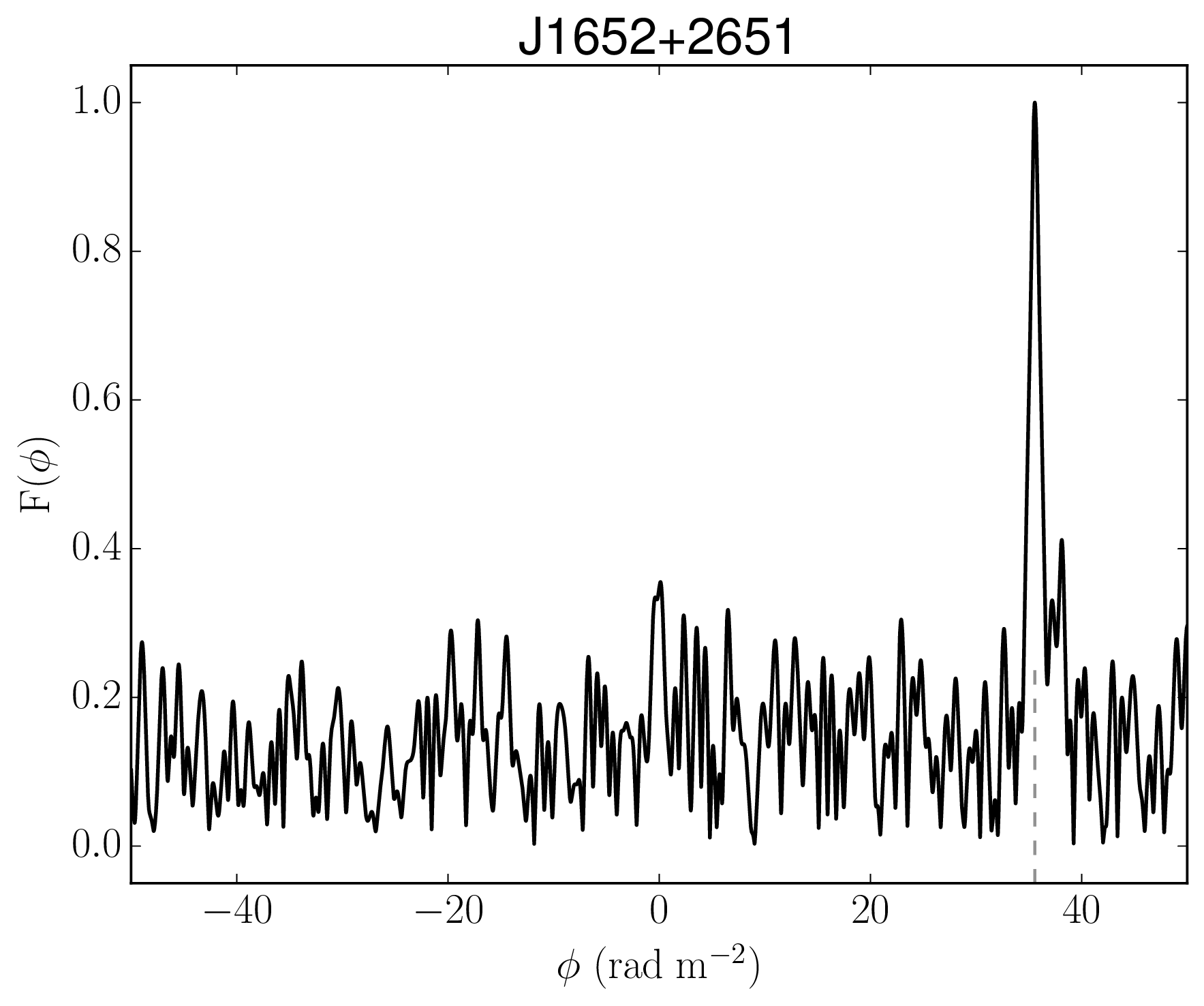}
\includegraphics[width=0.246\columnwidth]{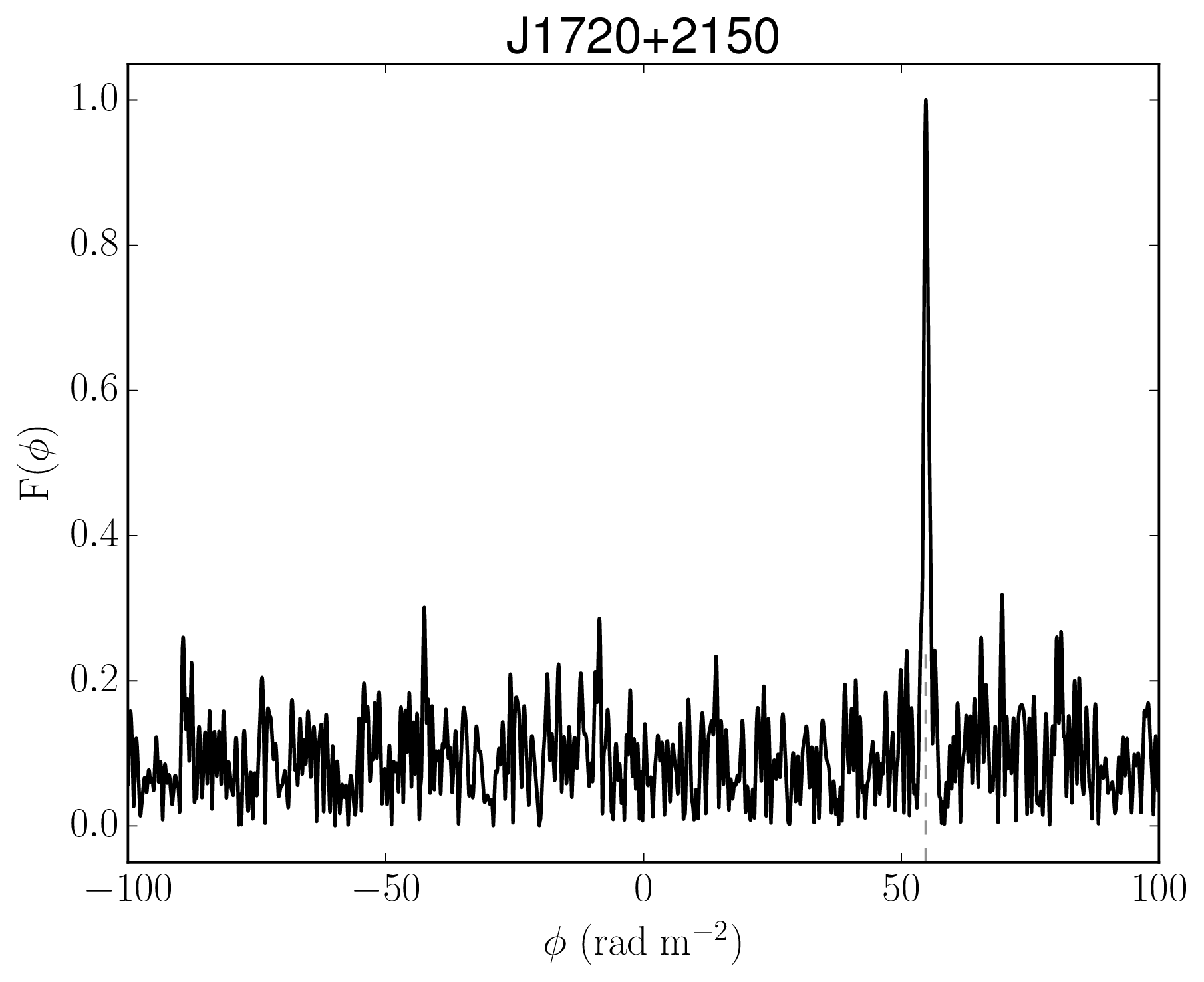}

\includegraphics[width=0.246\columnwidth]{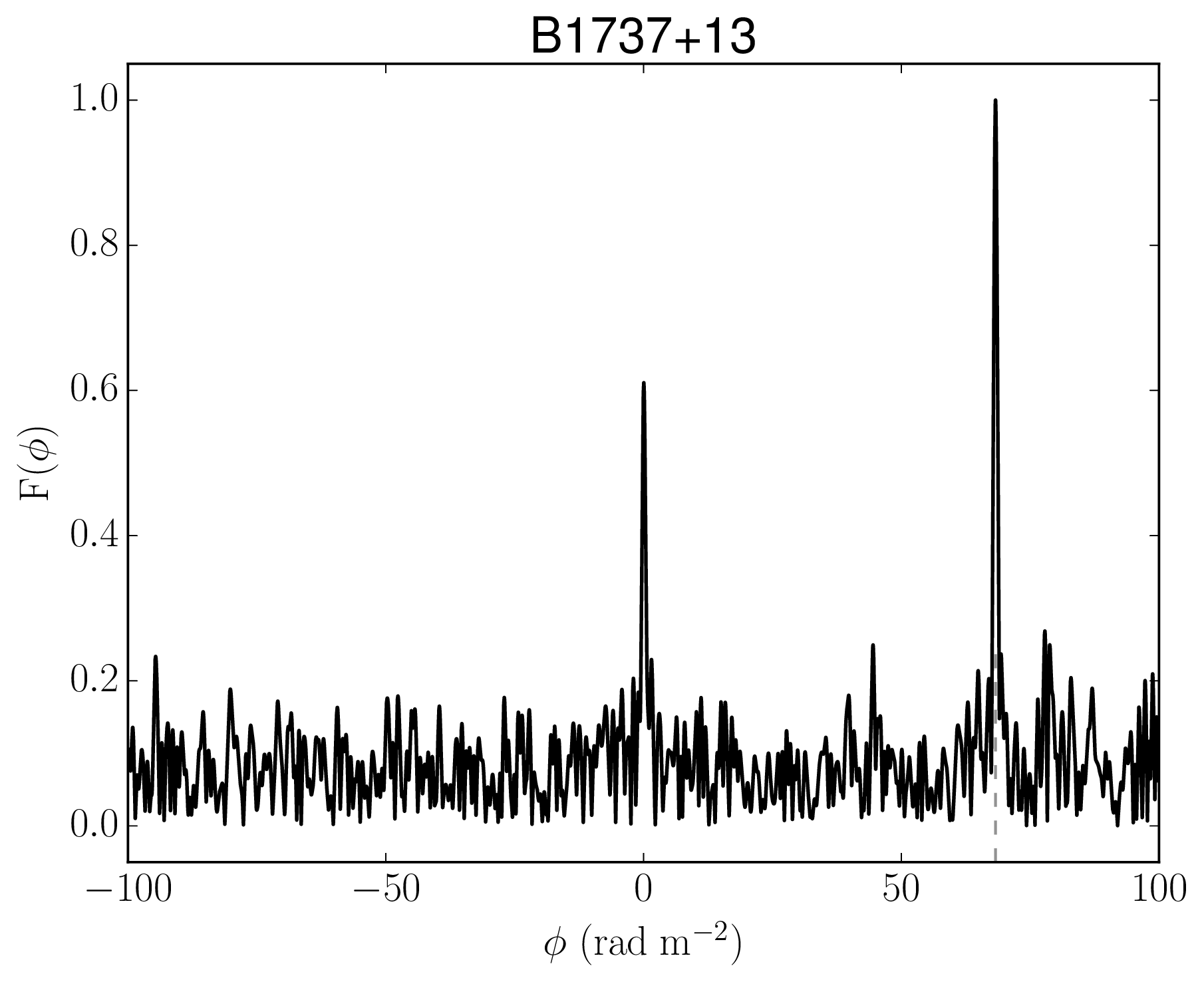}
\includegraphics[width=0.246\columnwidth]{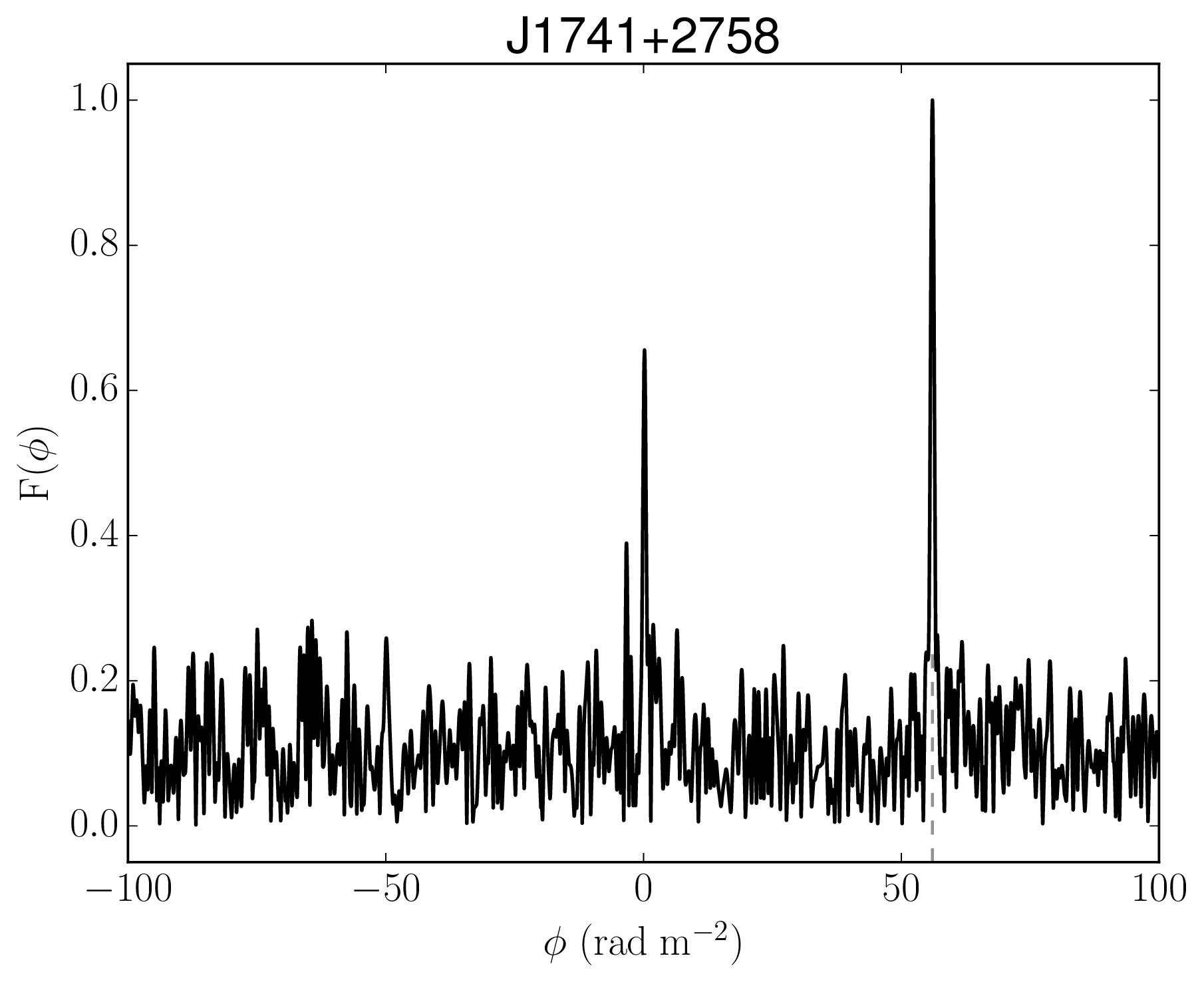}
\includegraphics[width=0.246\columnwidth]{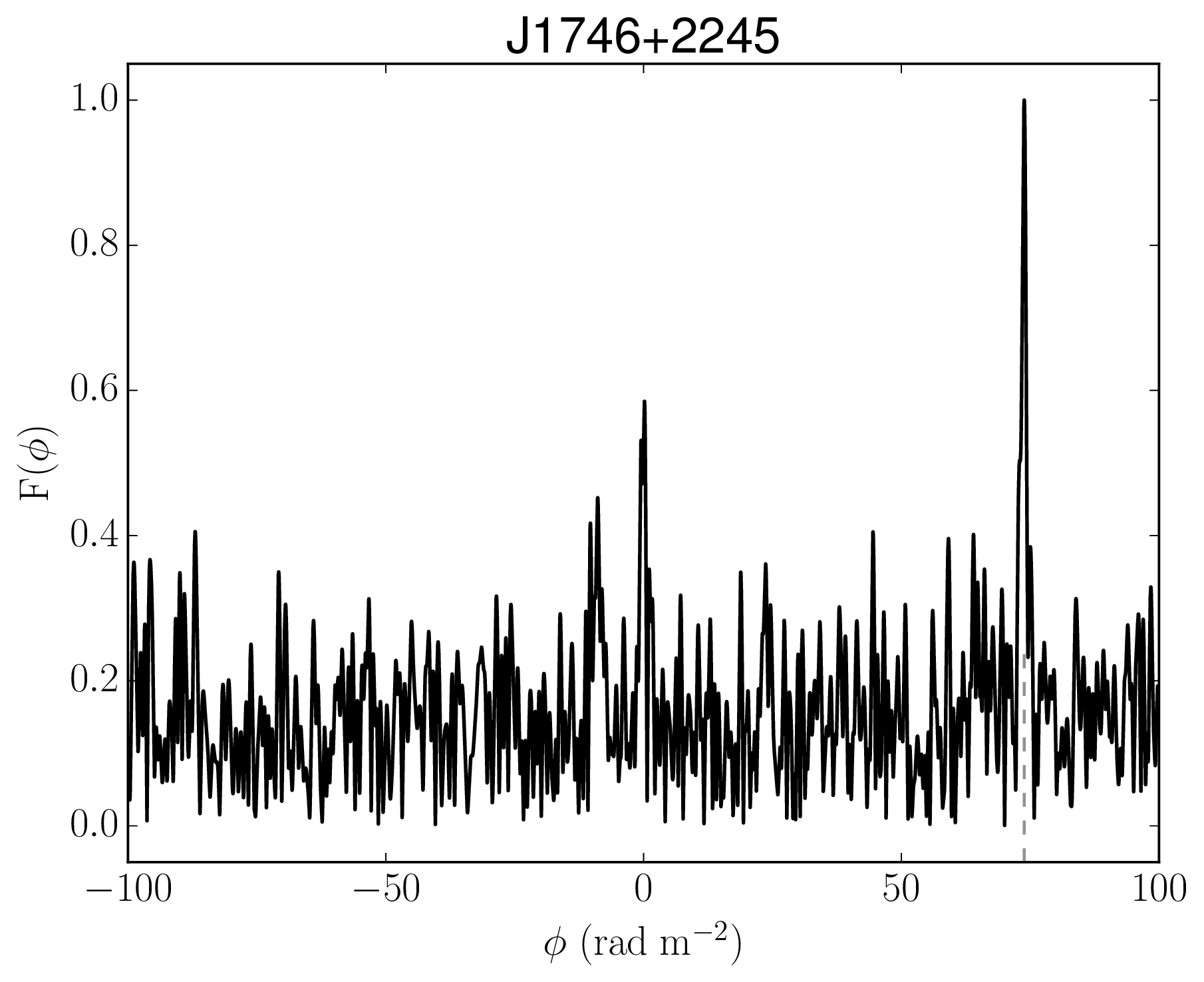}
\includegraphics[width=0.246\columnwidth]{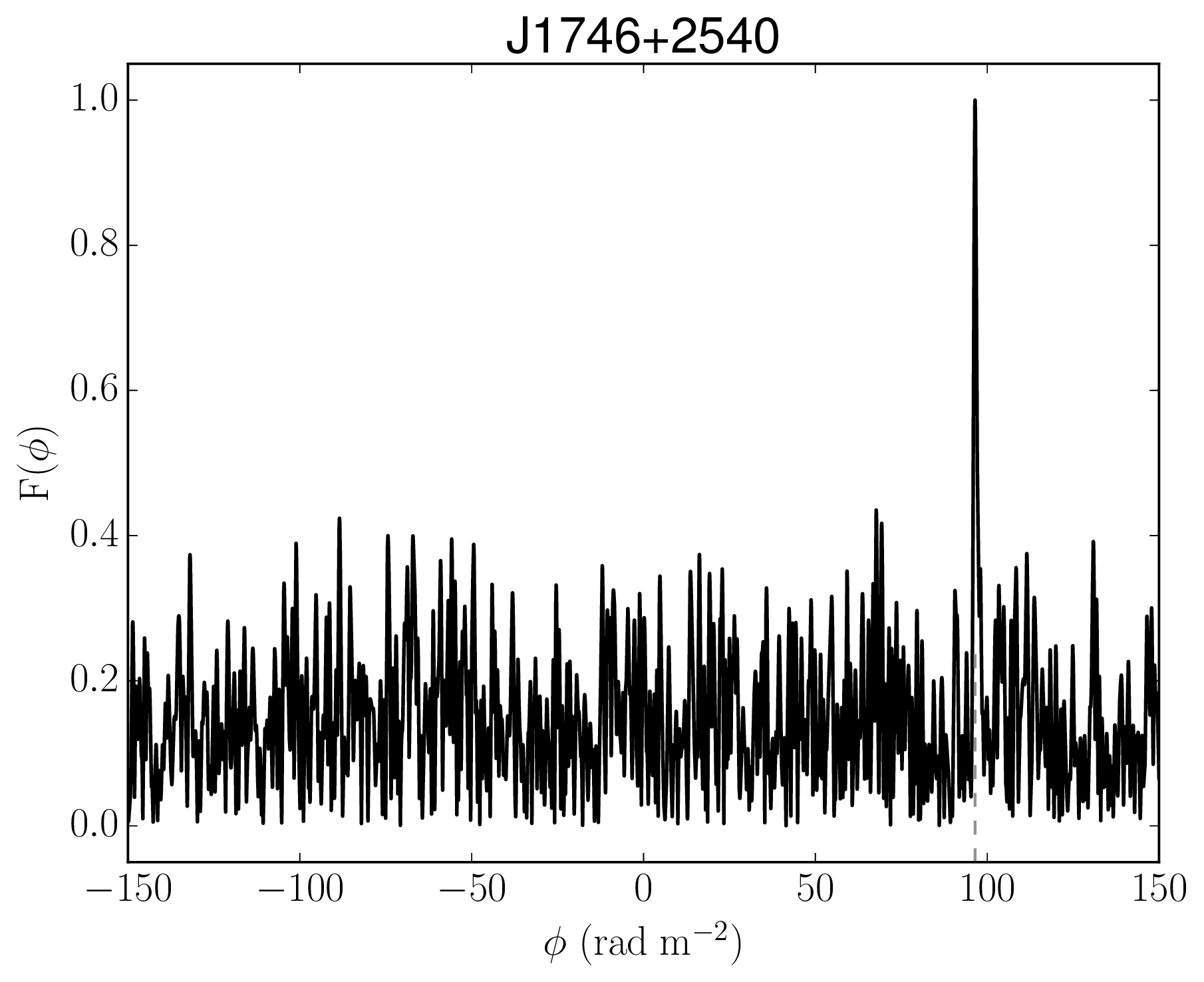}

\includegraphics[width=0.246\columnwidth]{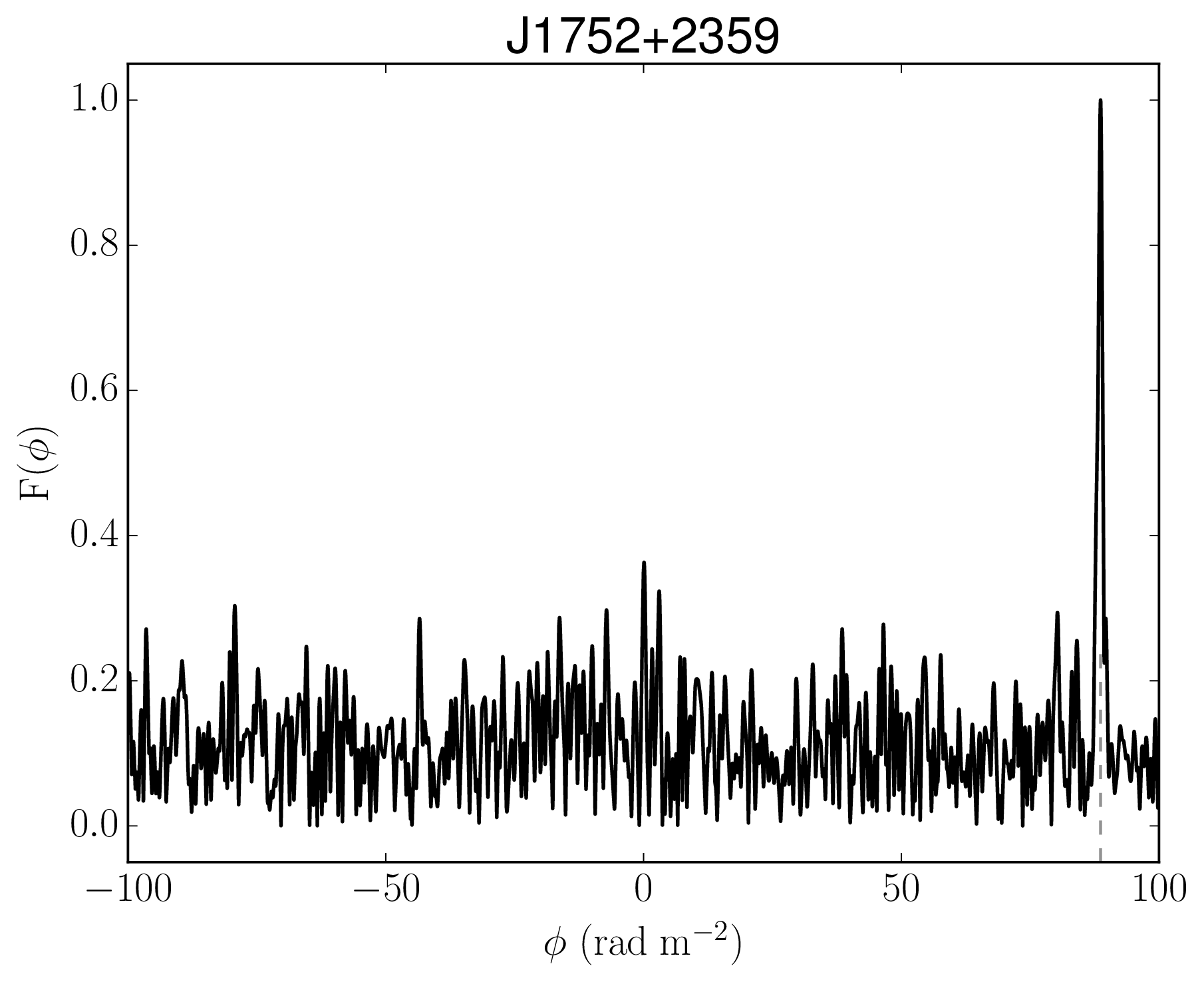}
\includegraphics[width=0.246\columnwidth]{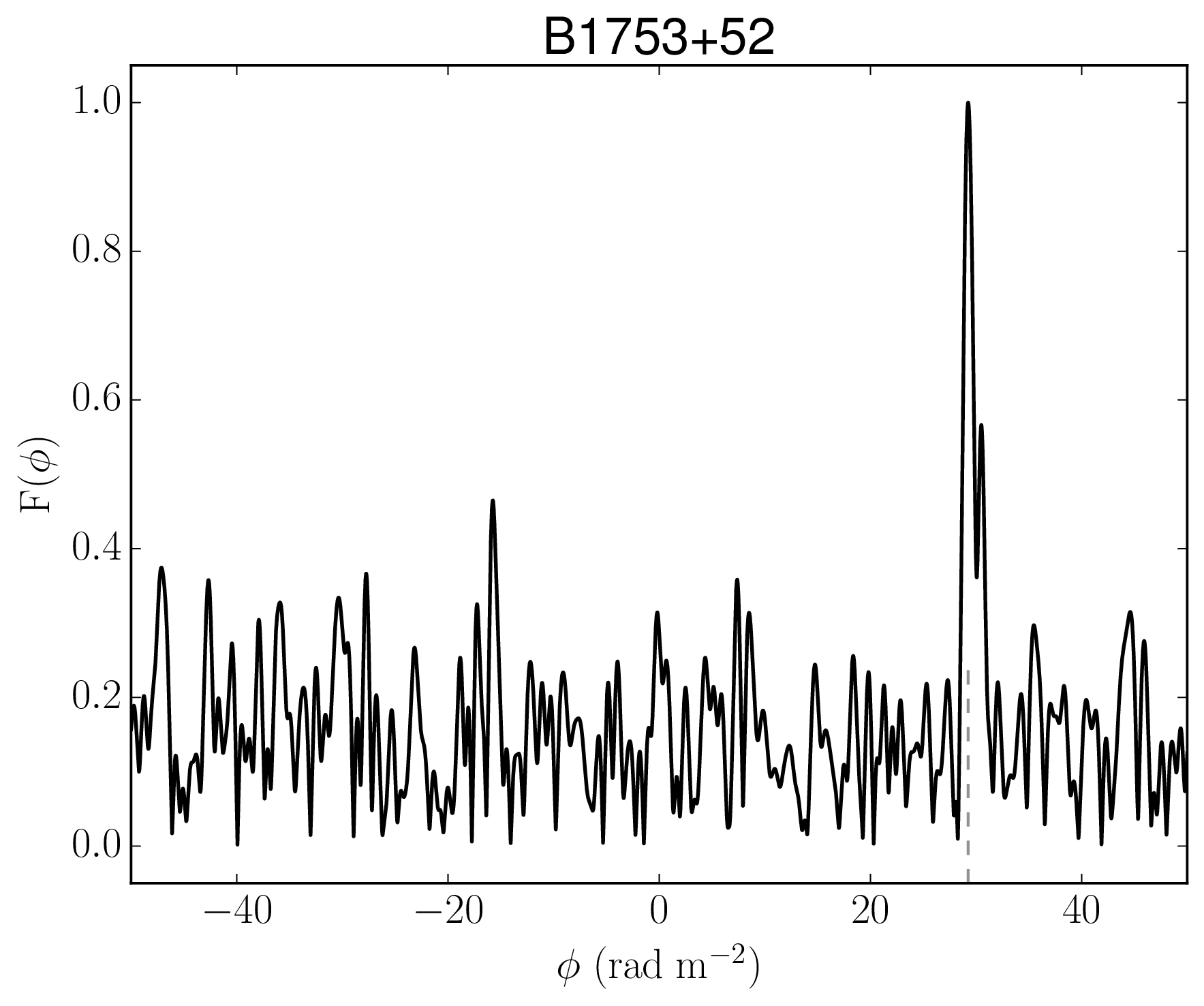}
\includegraphics[width=0.246\columnwidth]{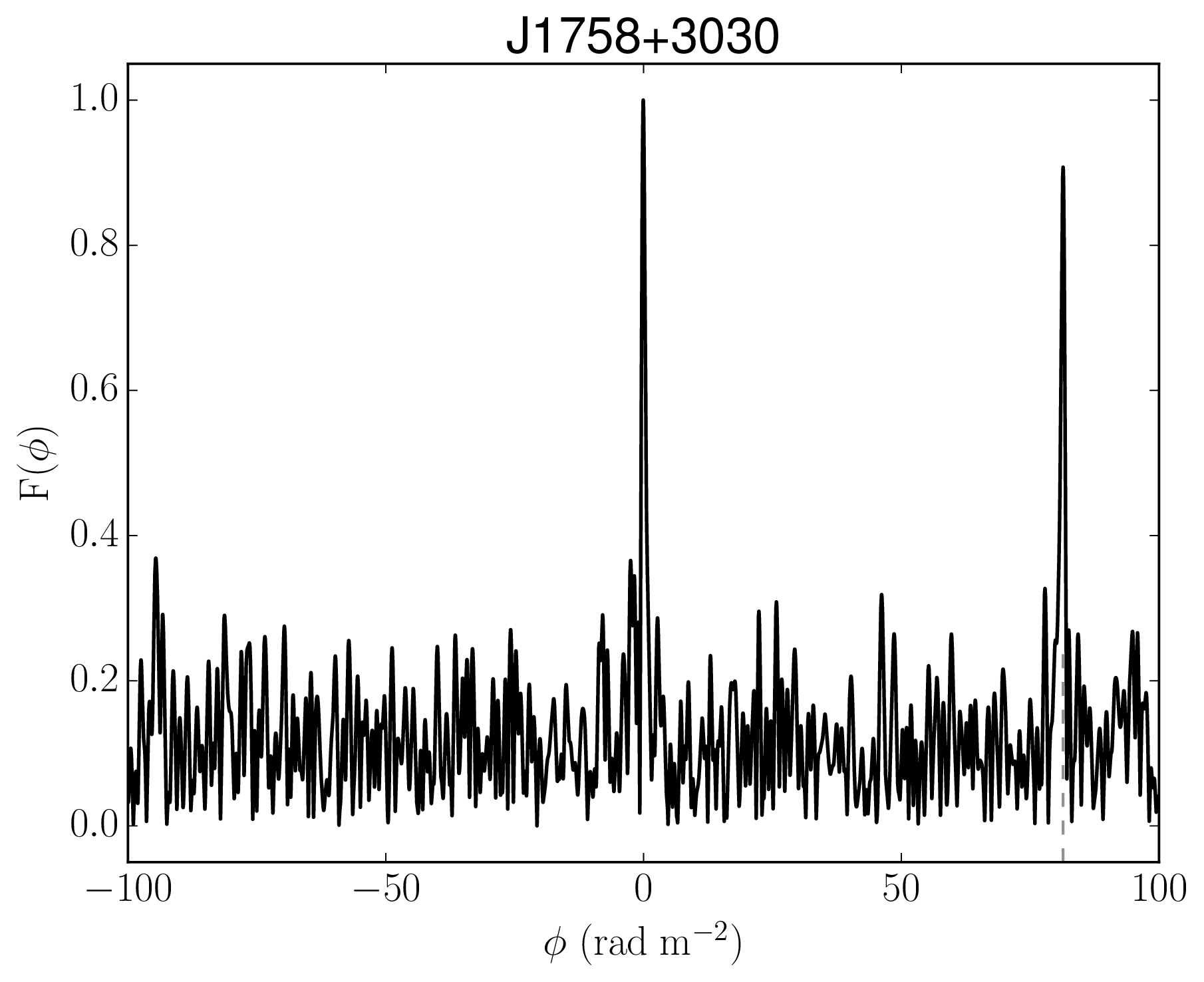}
\includegraphics[width=0.246\columnwidth]{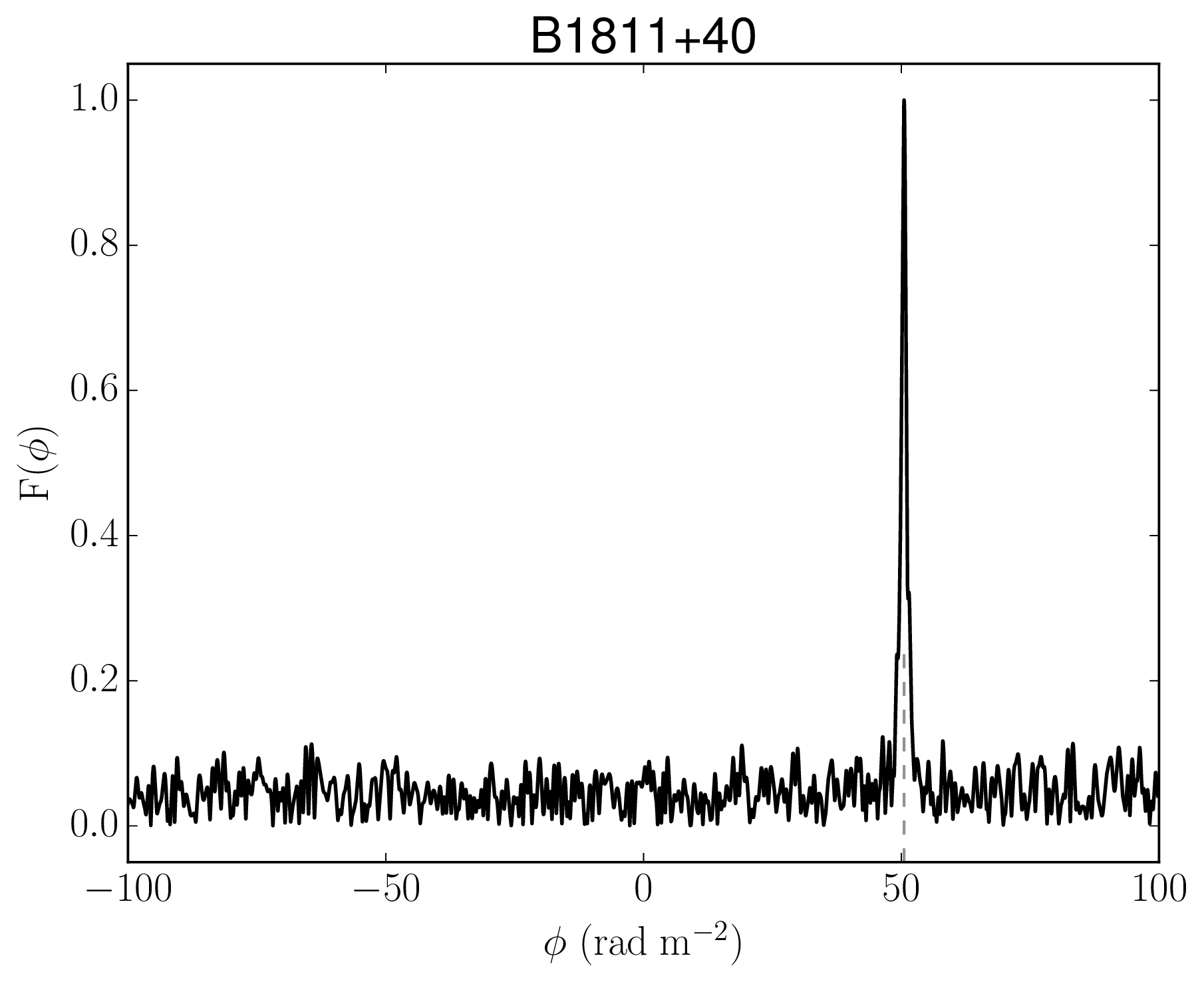}

\includegraphics[width=0.246\columnwidth]{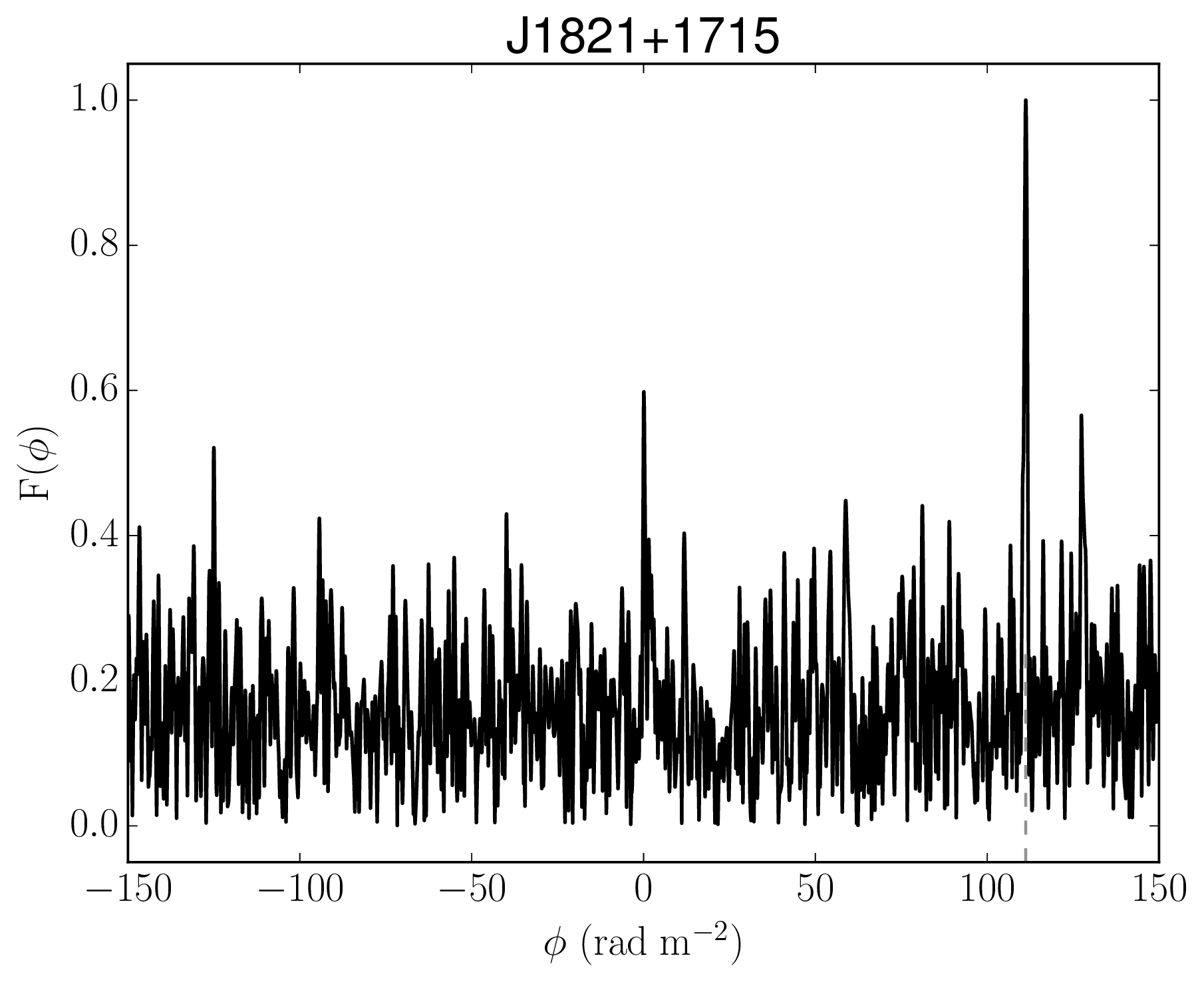}
\includegraphics[width=0.246\columnwidth]{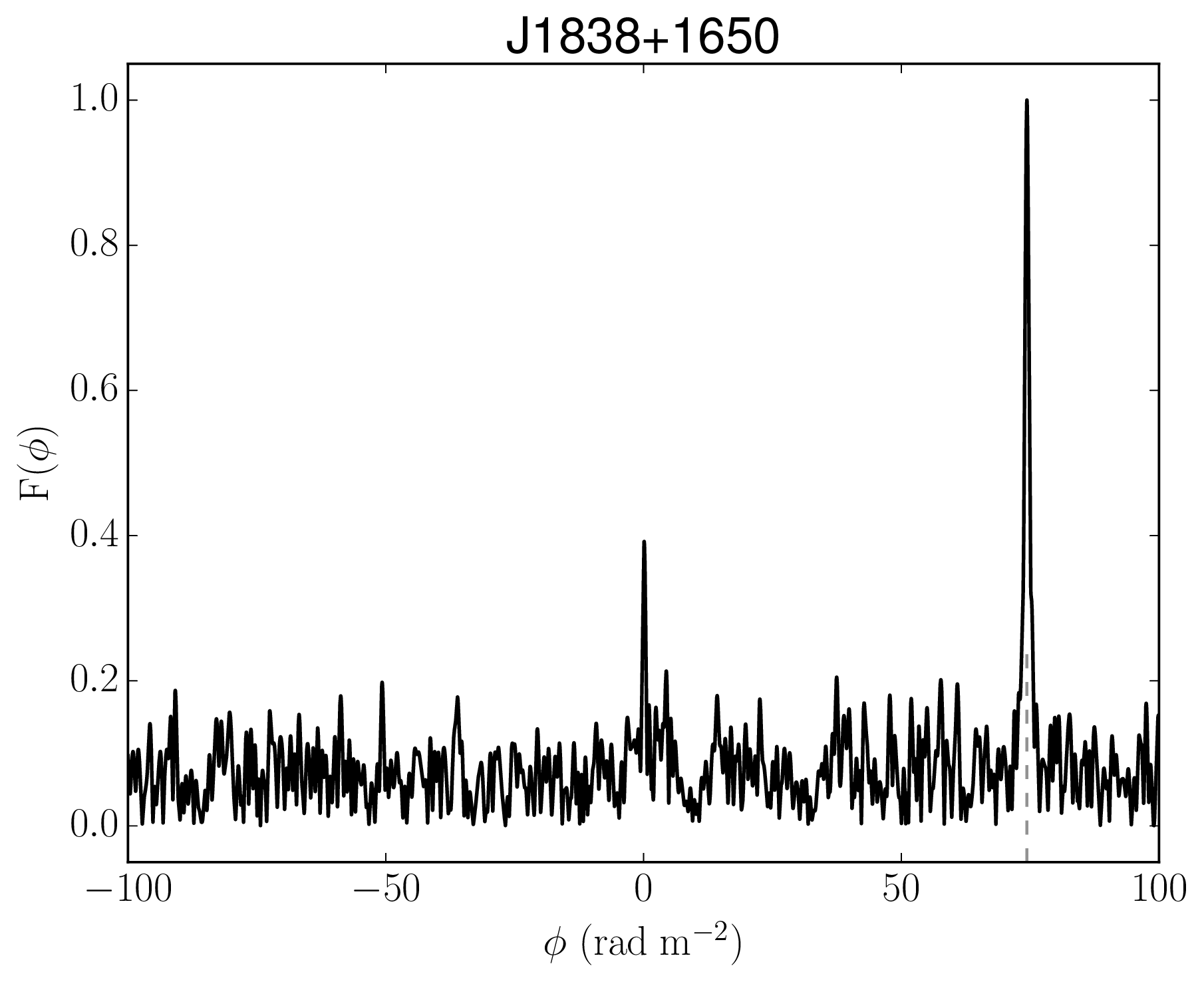}
\includegraphics[width=0.246\columnwidth]{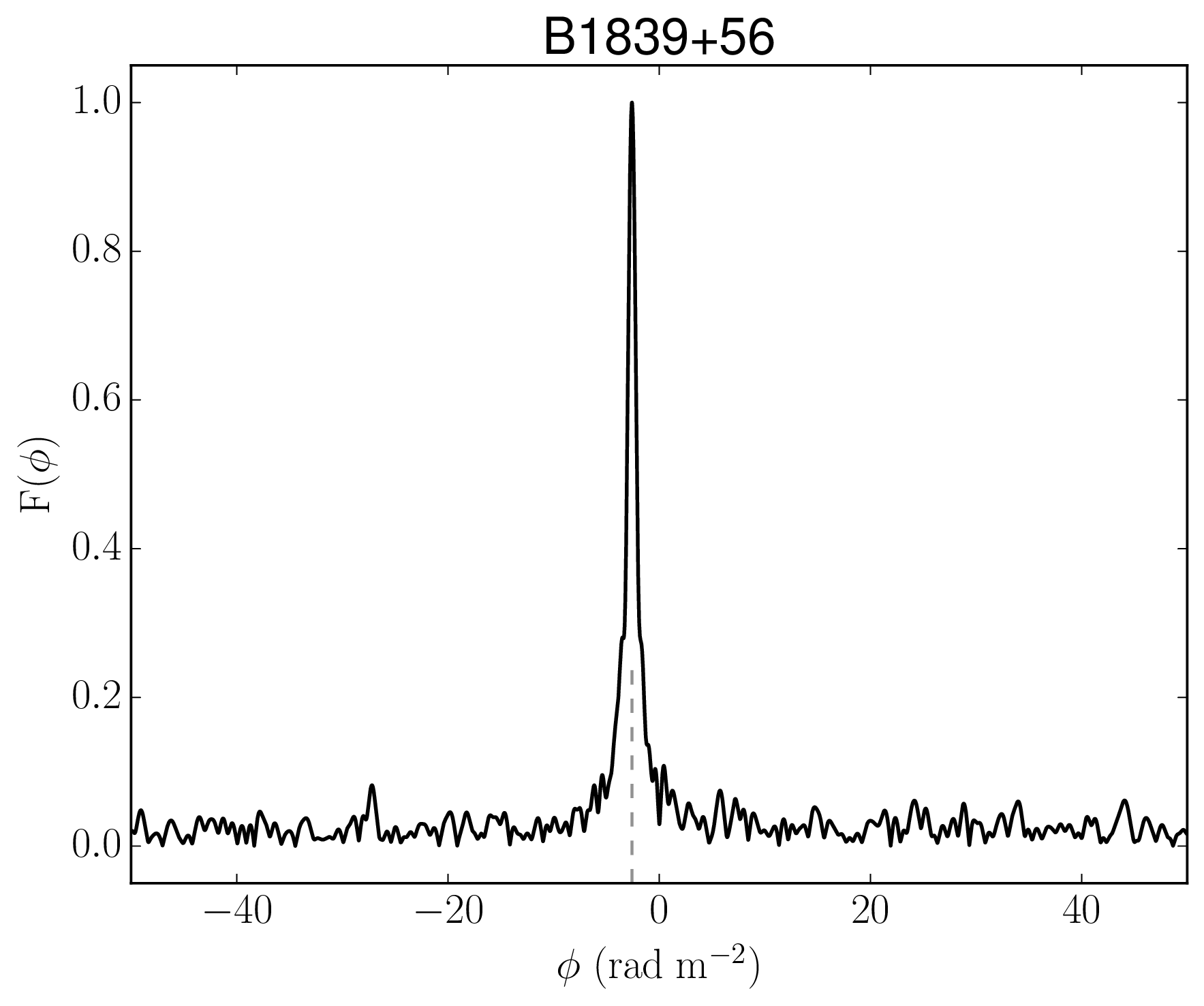}
\includegraphics[width=0.246\columnwidth]{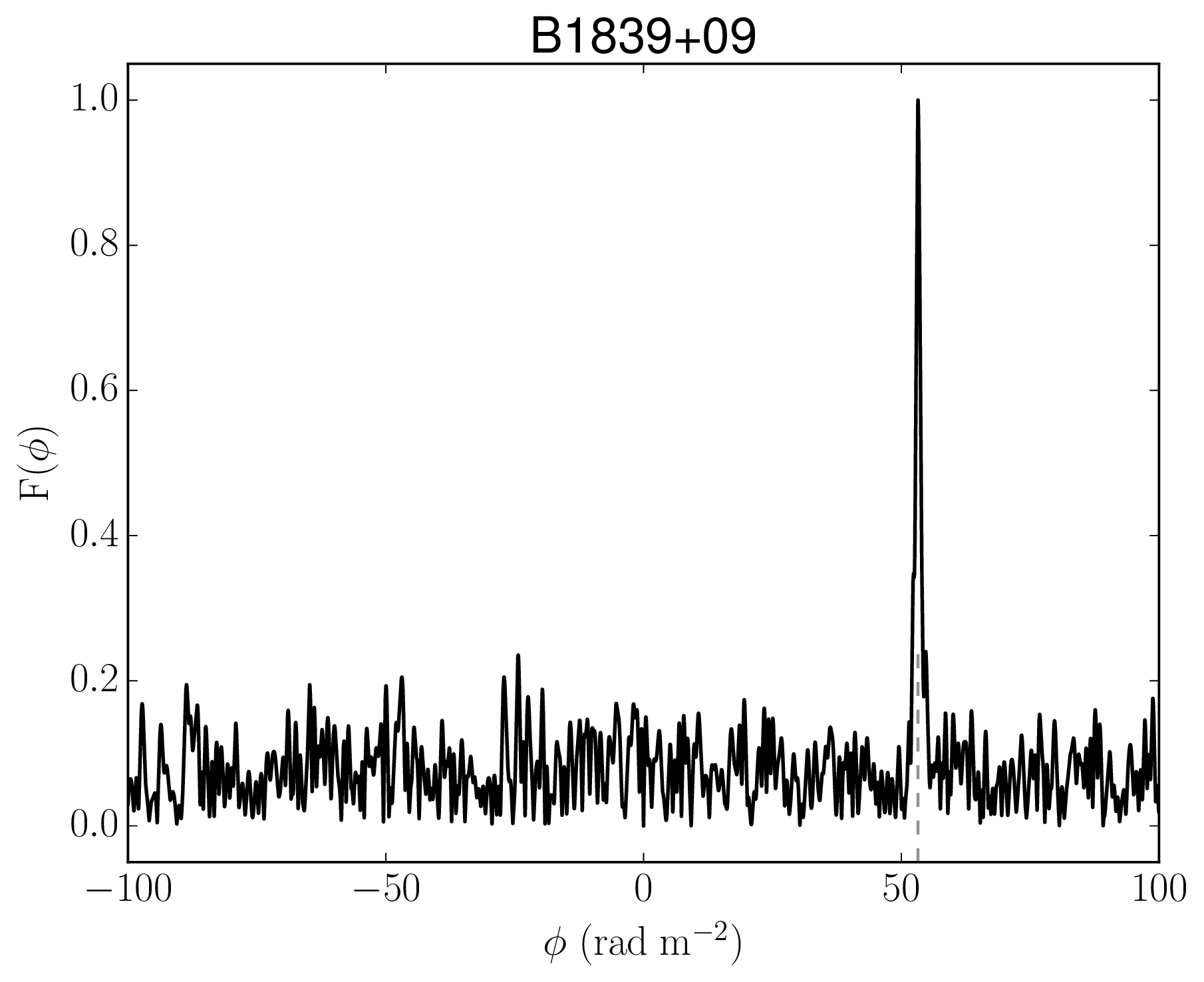}

\includegraphics[width=0.246\columnwidth]{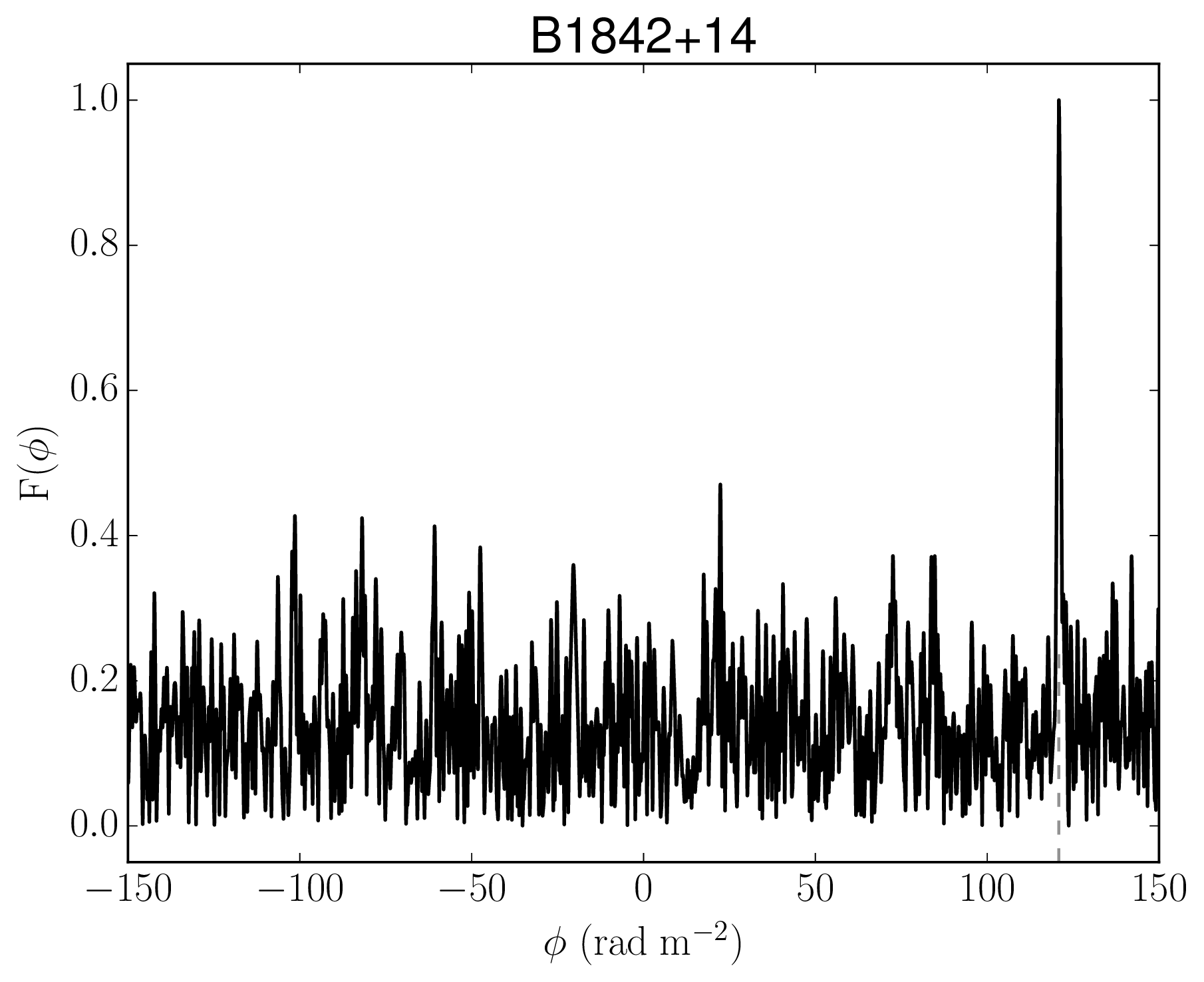}
\includegraphics[width=0.246\columnwidth]{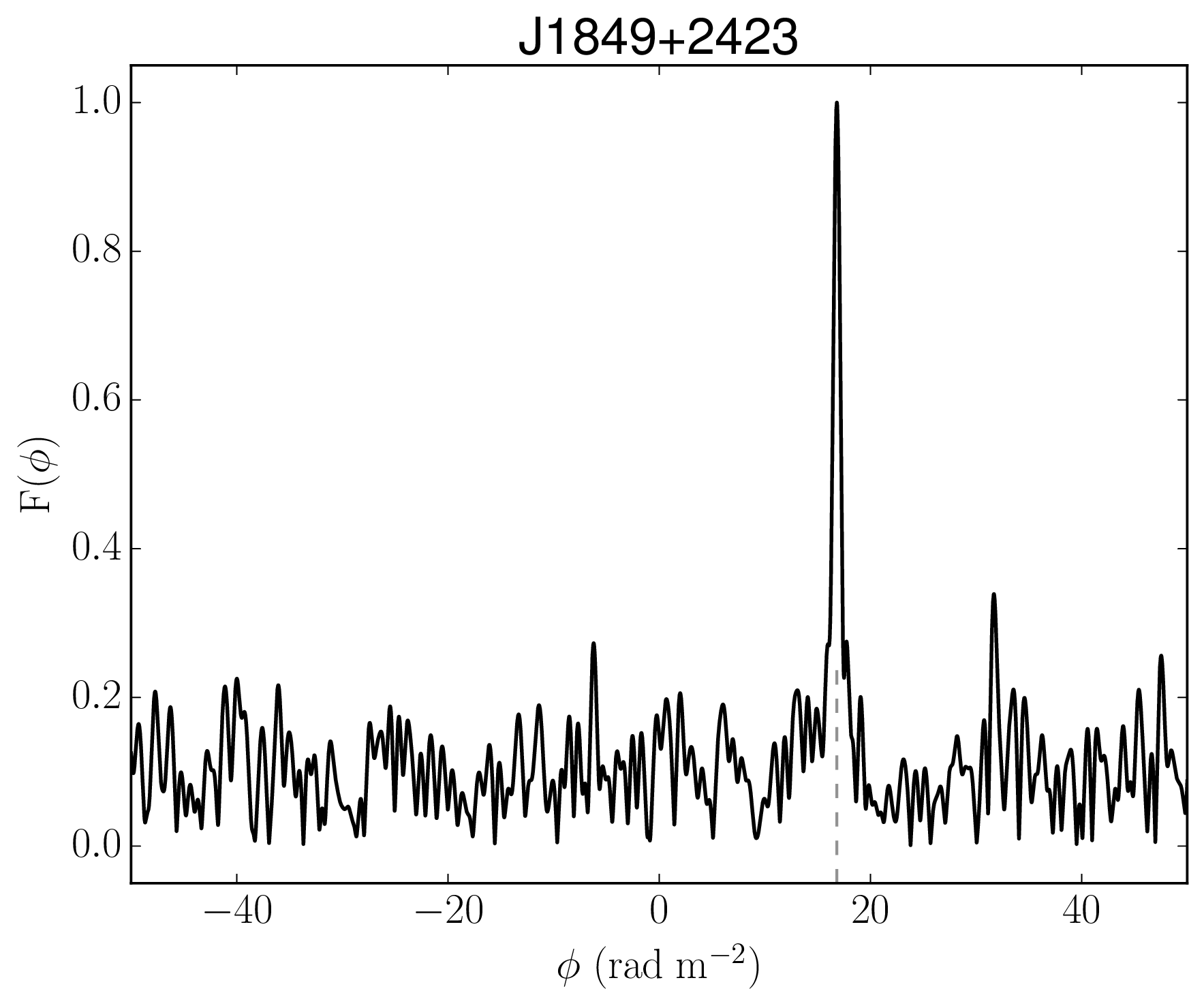}
\includegraphics[width=0.246\columnwidth]{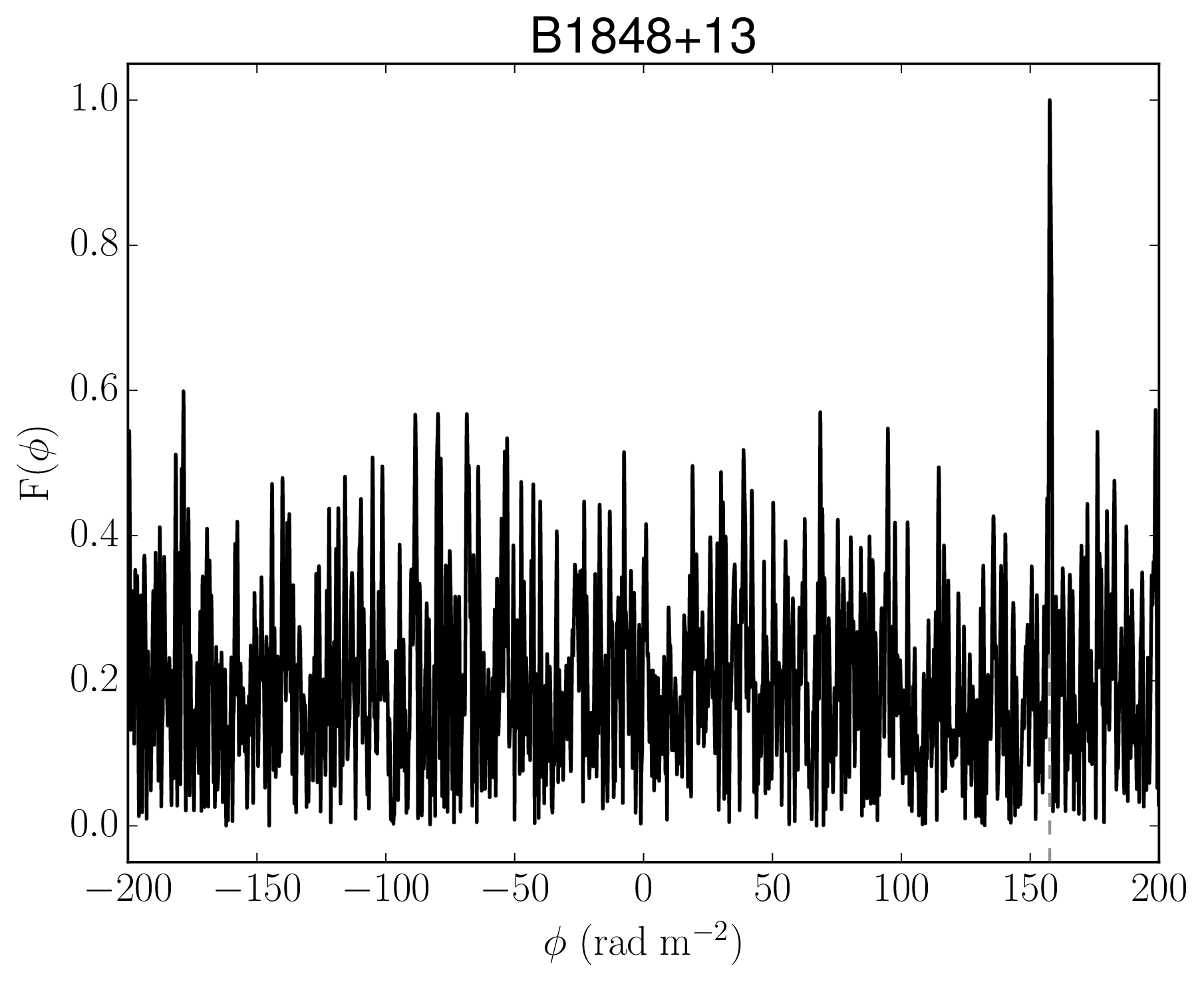}
\includegraphics[width=0.246\columnwidth]{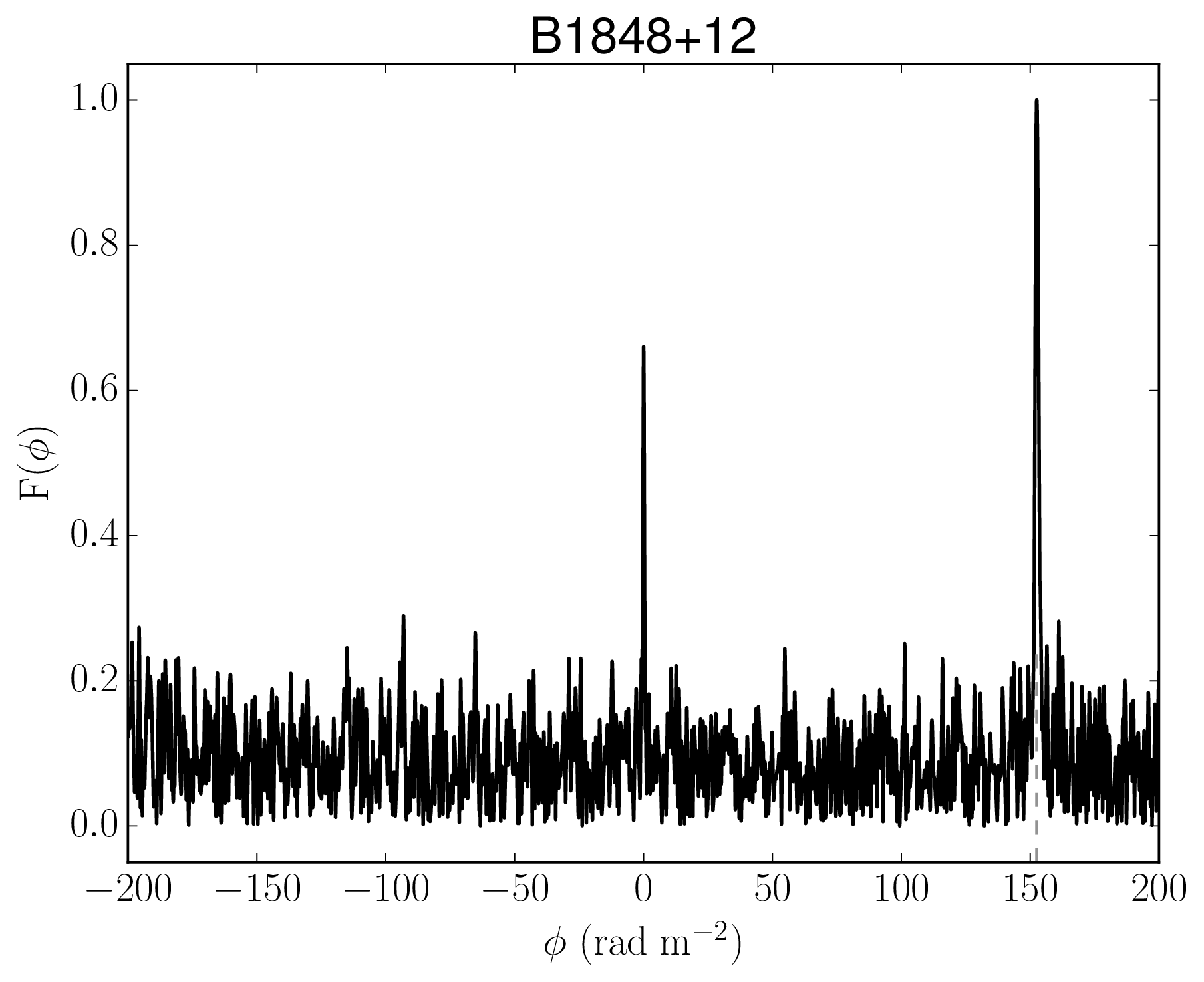}

 \caption{Continued from the previous page.}
% \label{fig:FDFs}
\end{figure*}

\begin{figure*}
 \centering
 %\ContinuedFloat
%\setcounter{subfigure}{2}    

\includegraphics[width=0.246\columnwidth]{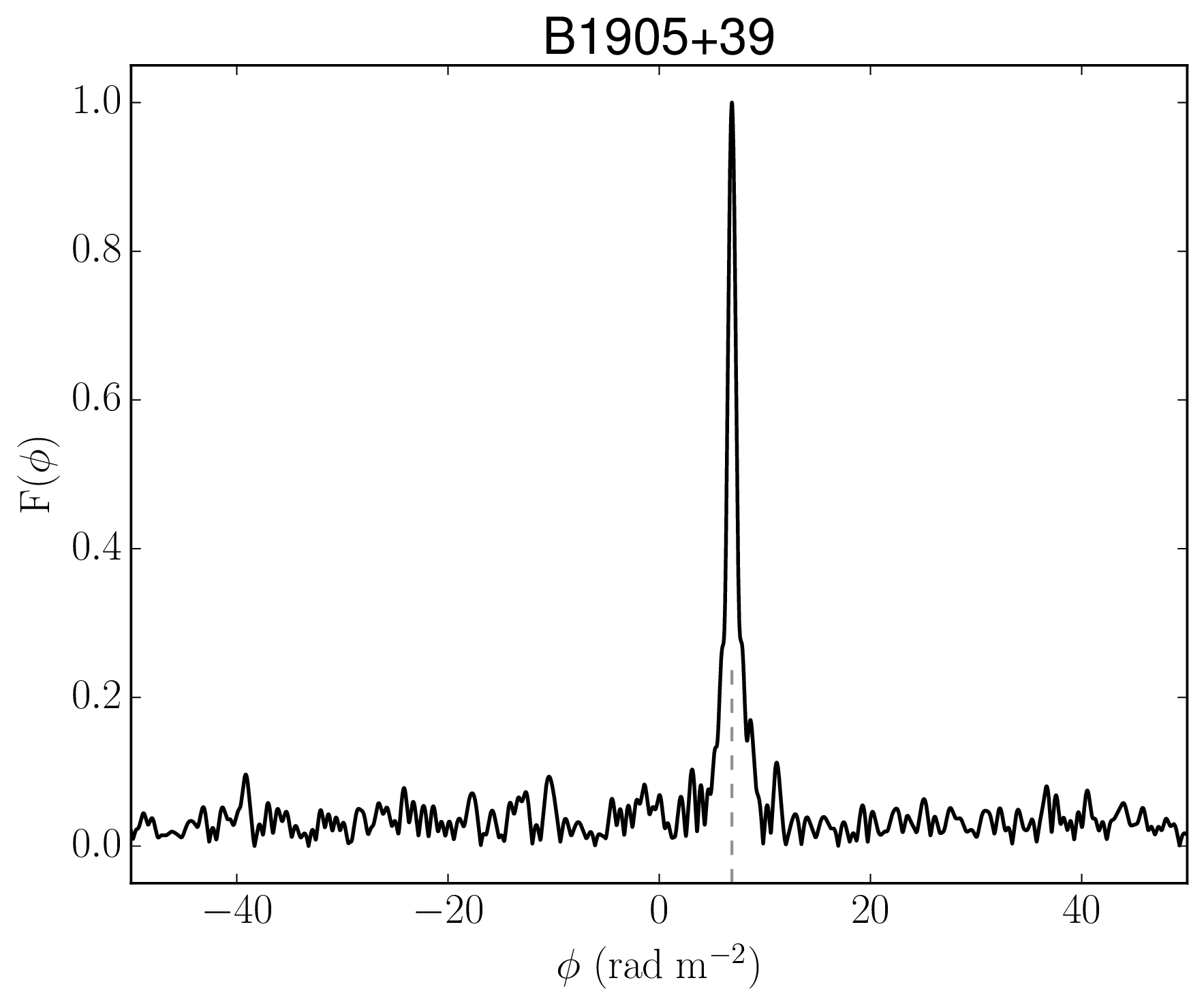}
\includegraphics[width=0.246\columnwidth]{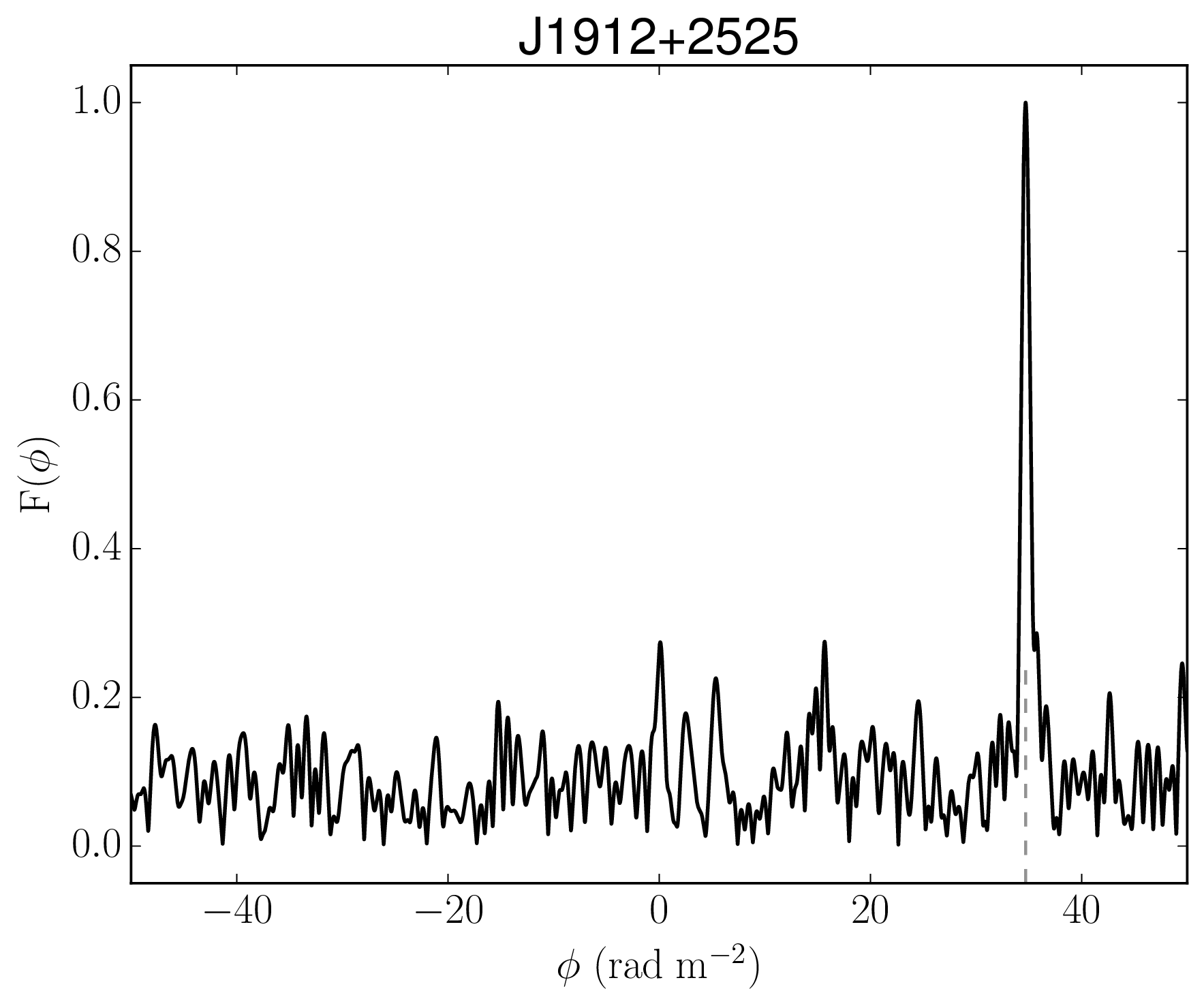}
\includegraphics[width=0.246\columnwidth]{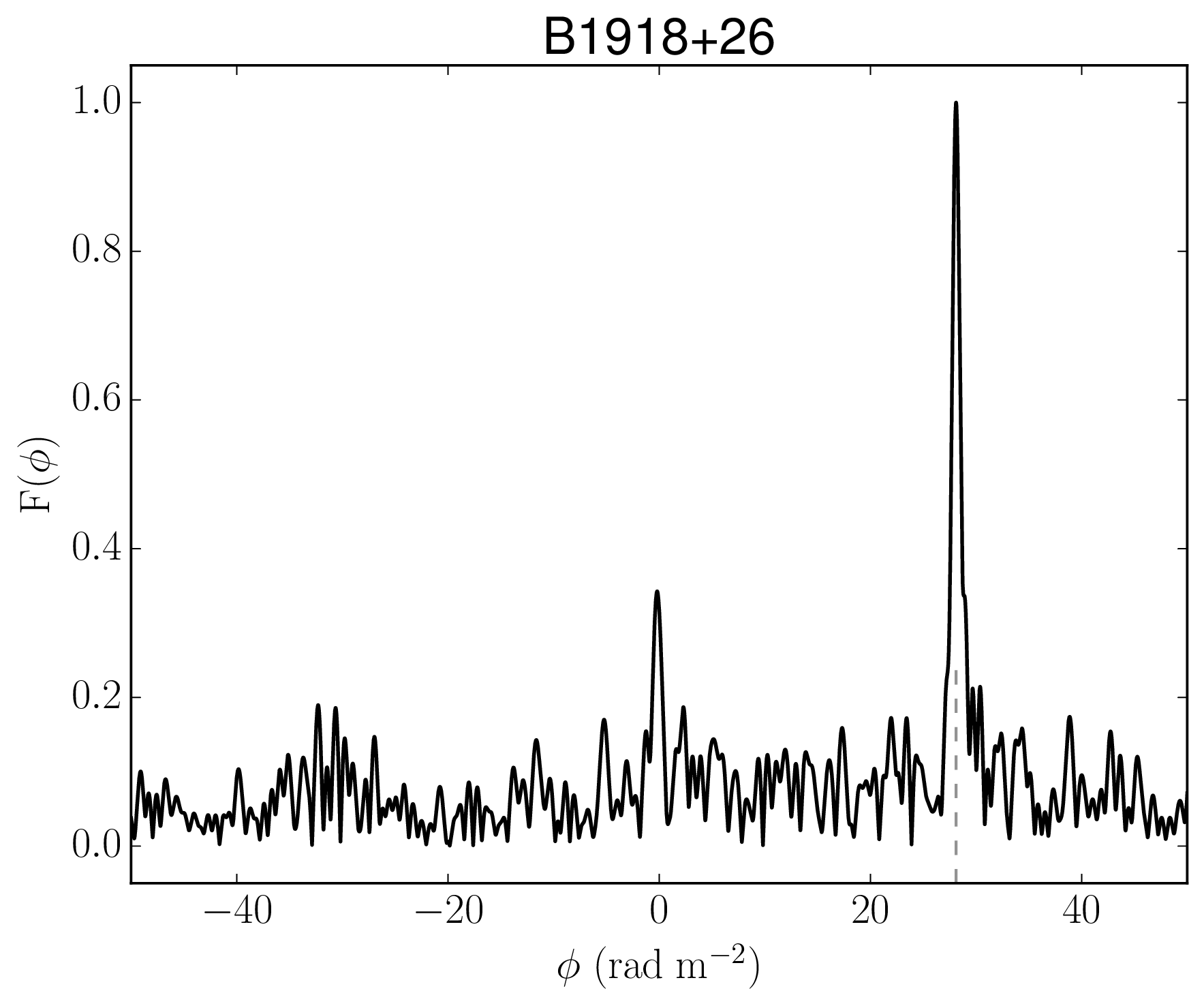}
\includegraphics[width=0.246\columnwidth]{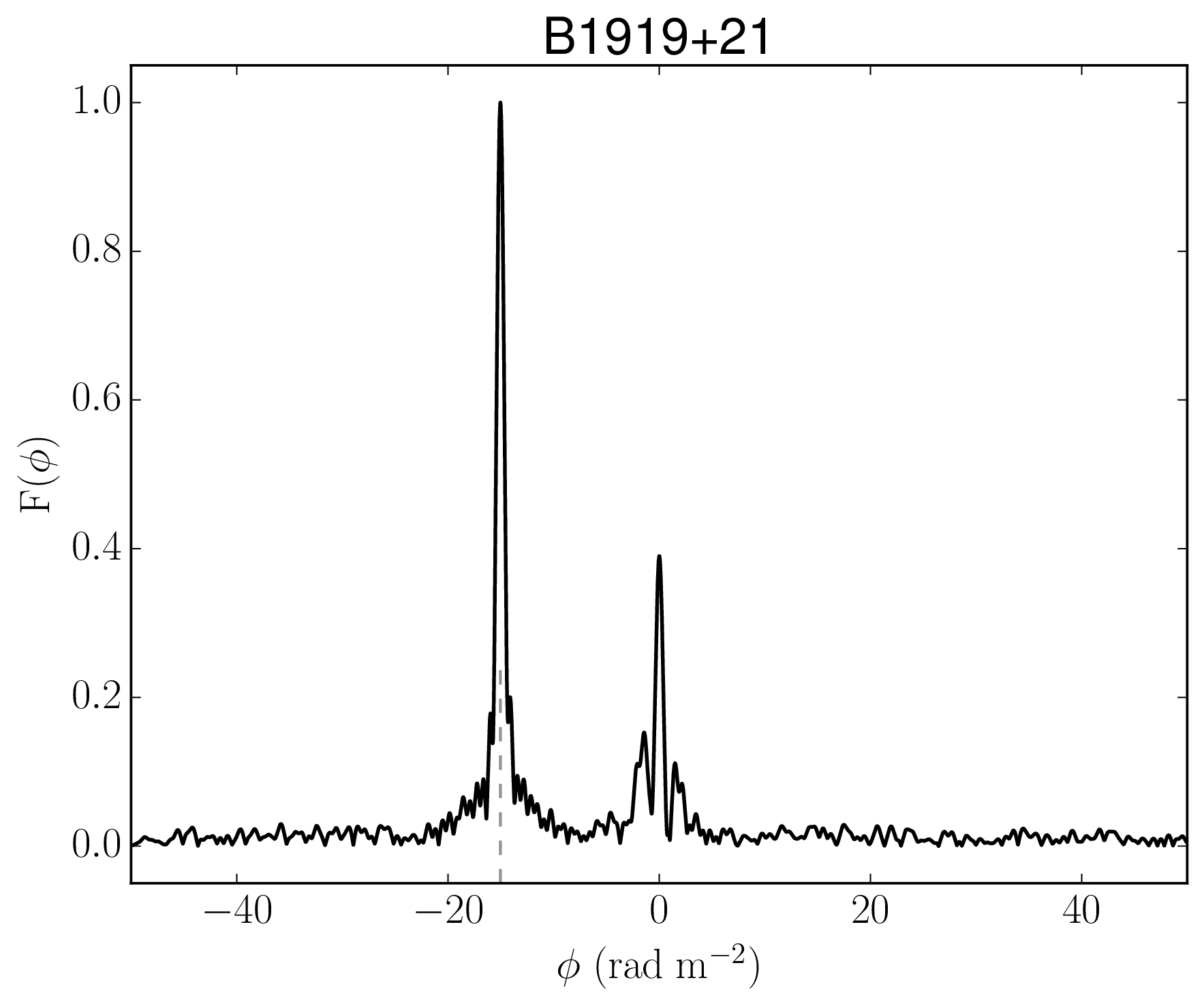}

\includegraphics[width=0.246\columnwidth]{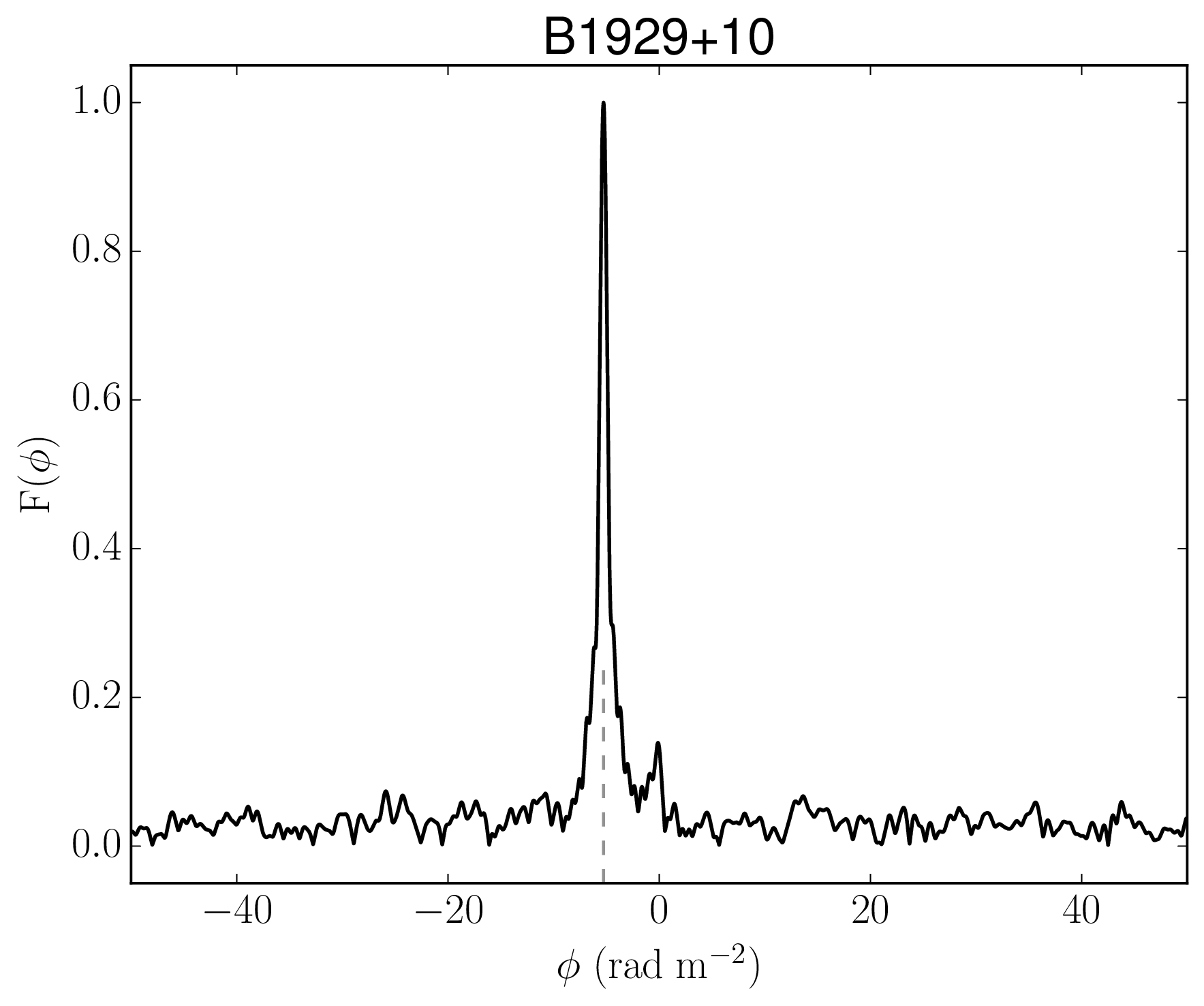}
\includegraphics[width=0.246\columnwidth]{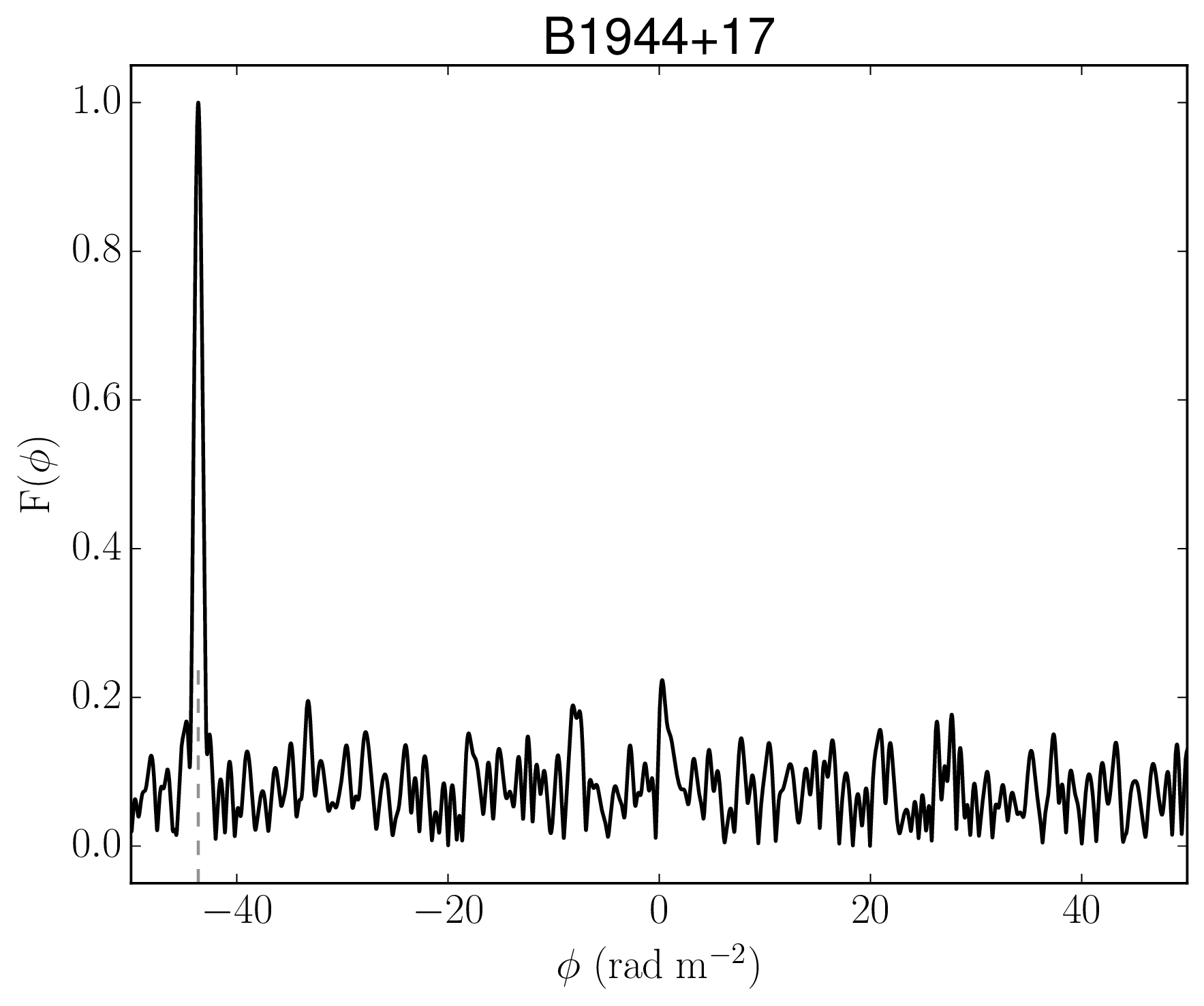}
\includegraphics[width=0.246\columnwidth]{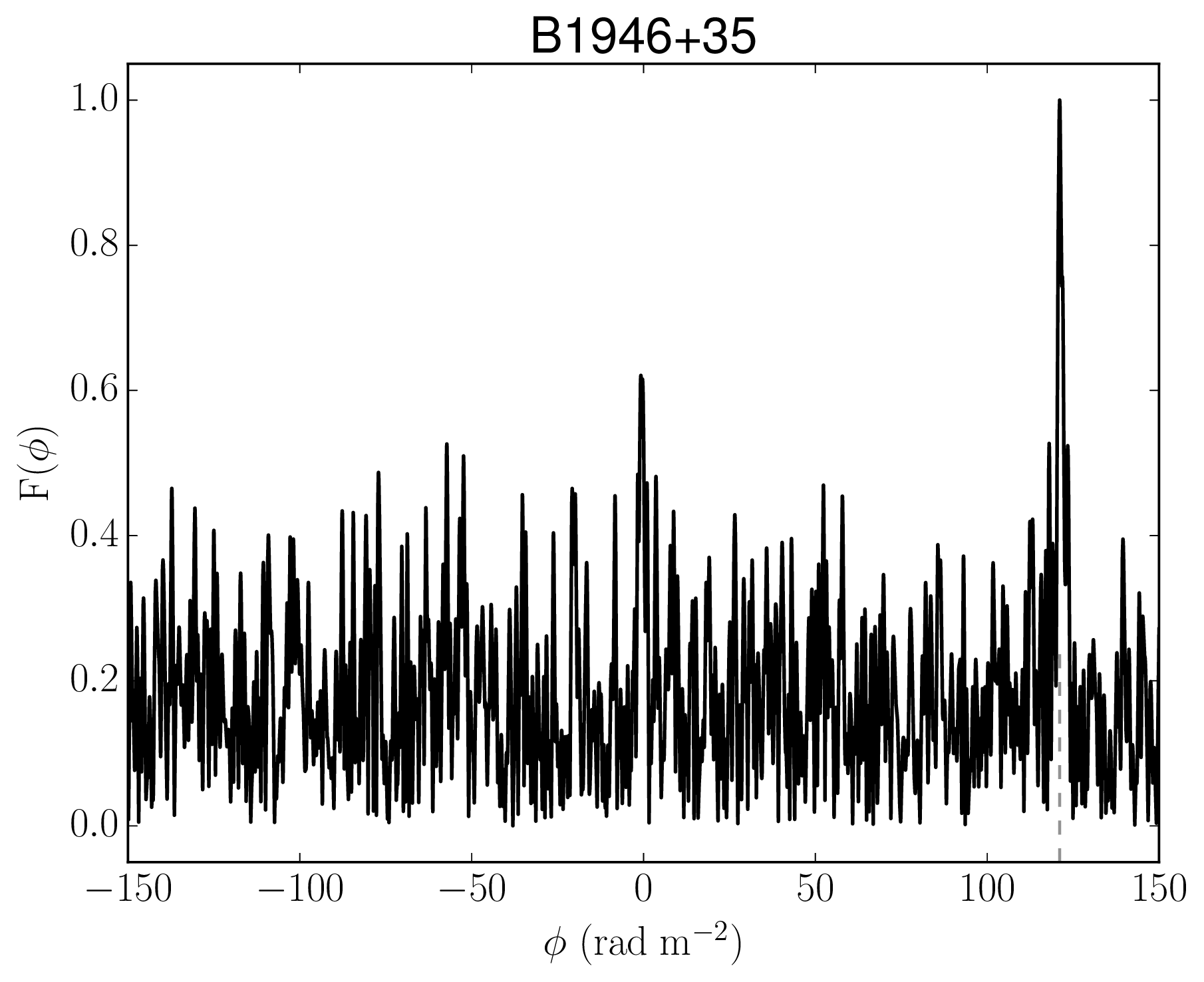}
\includegraphics[width=0.246\columnwidth]{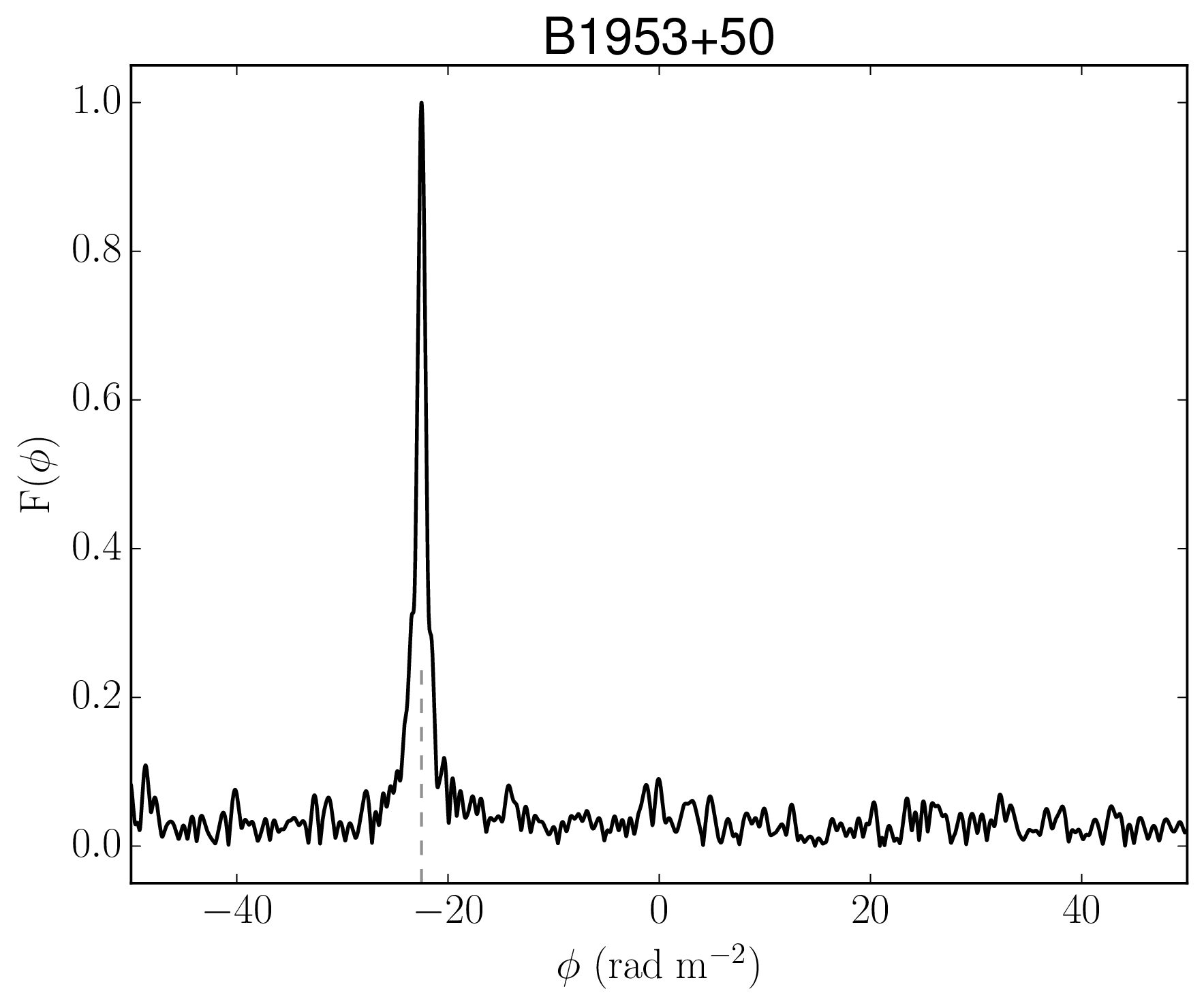}

\includegraphics[width=0.246\columnwidth]{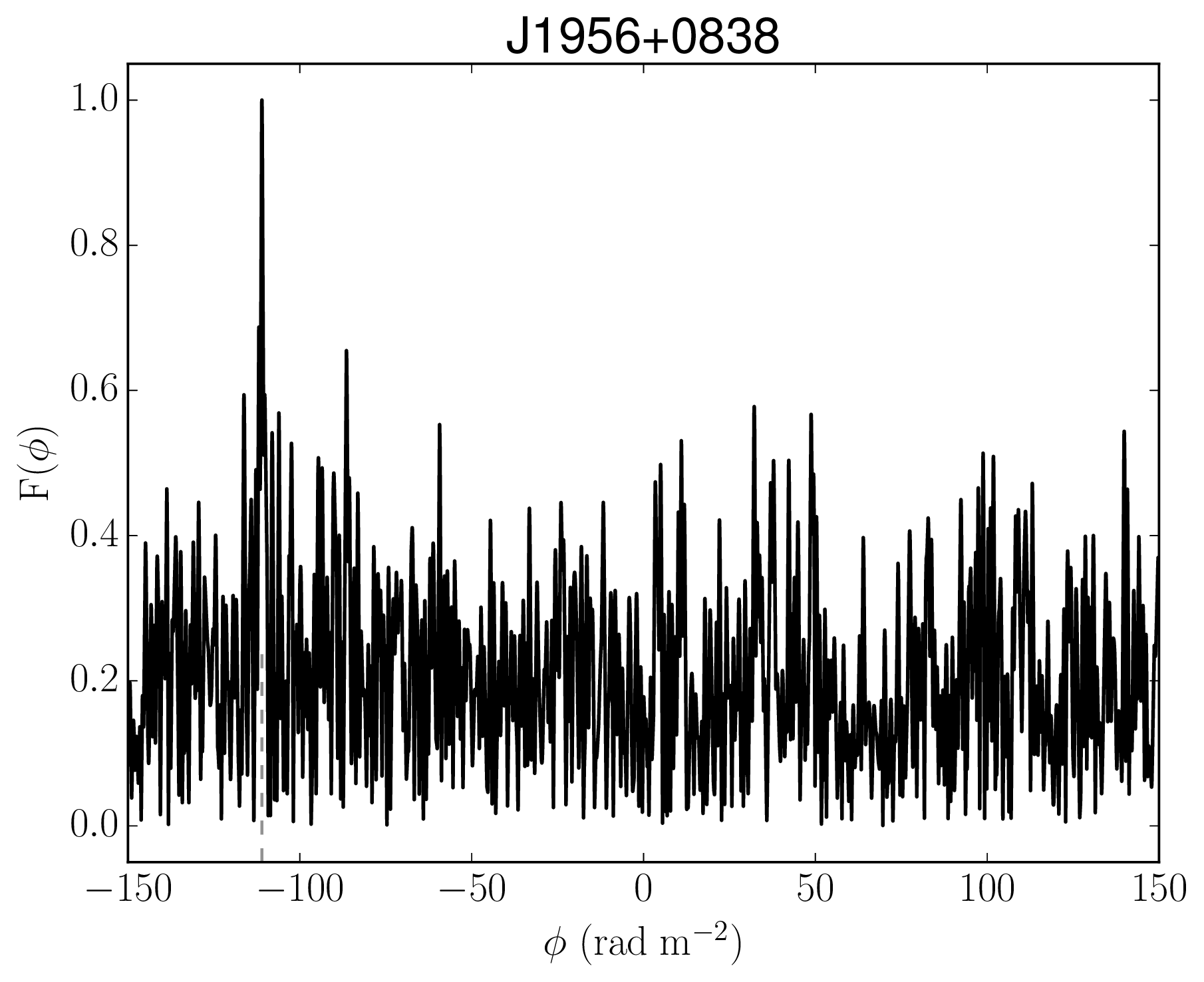}
\includegraphics[width=0.246\columnwidth]{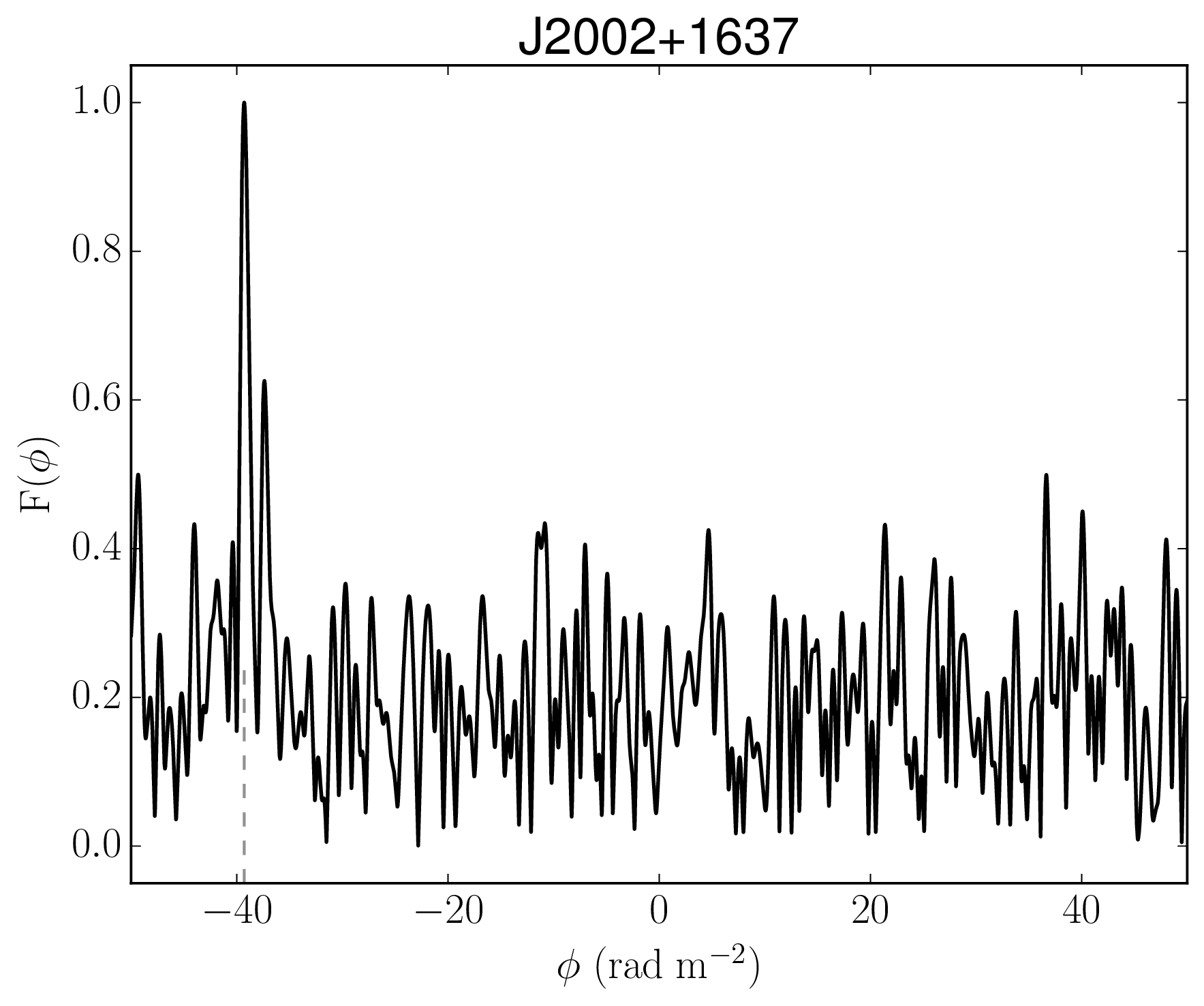}
\includegraphics[width=0.246\columnwidth]{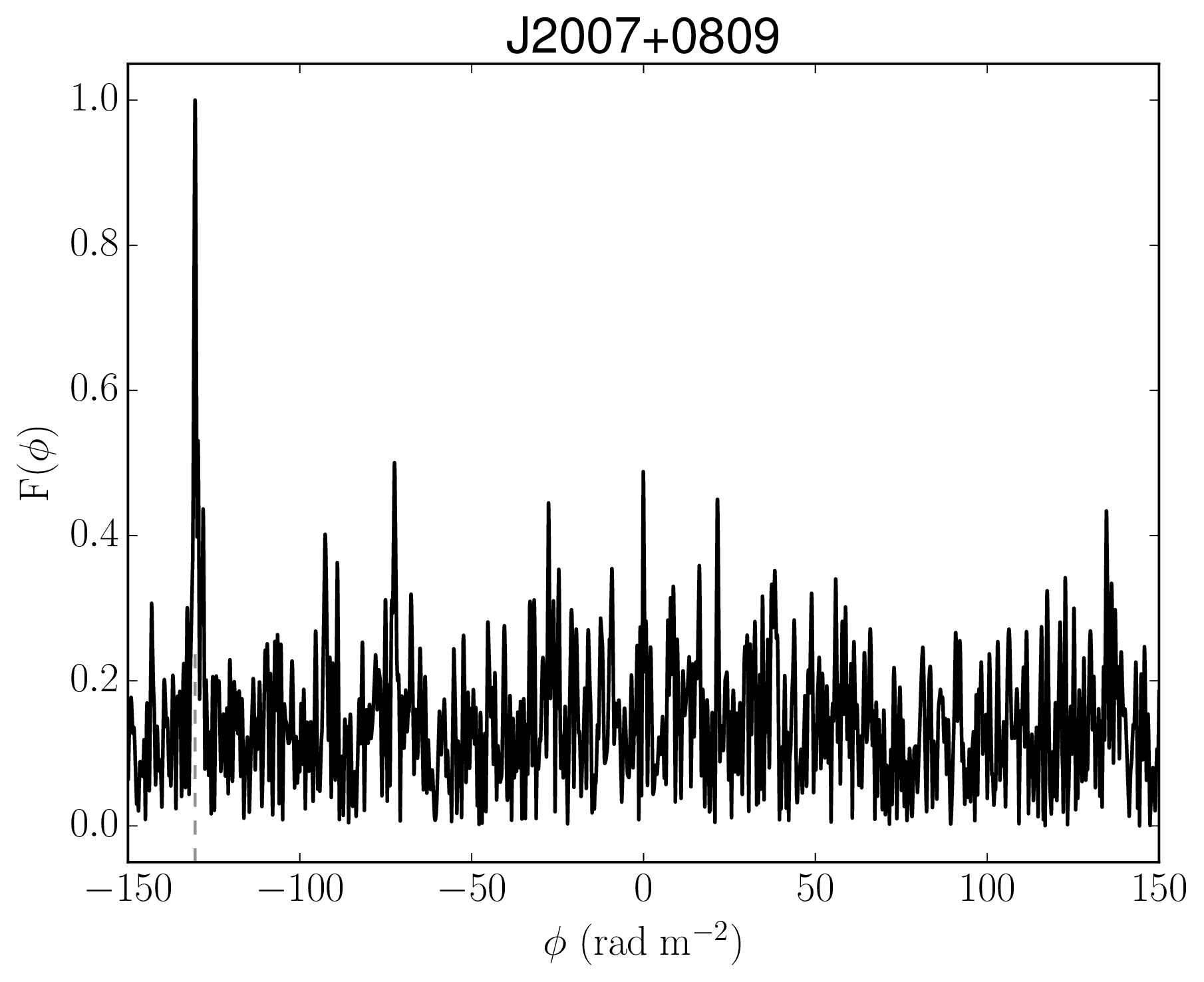}
\includegraphics[width=0.246\columnwidth]{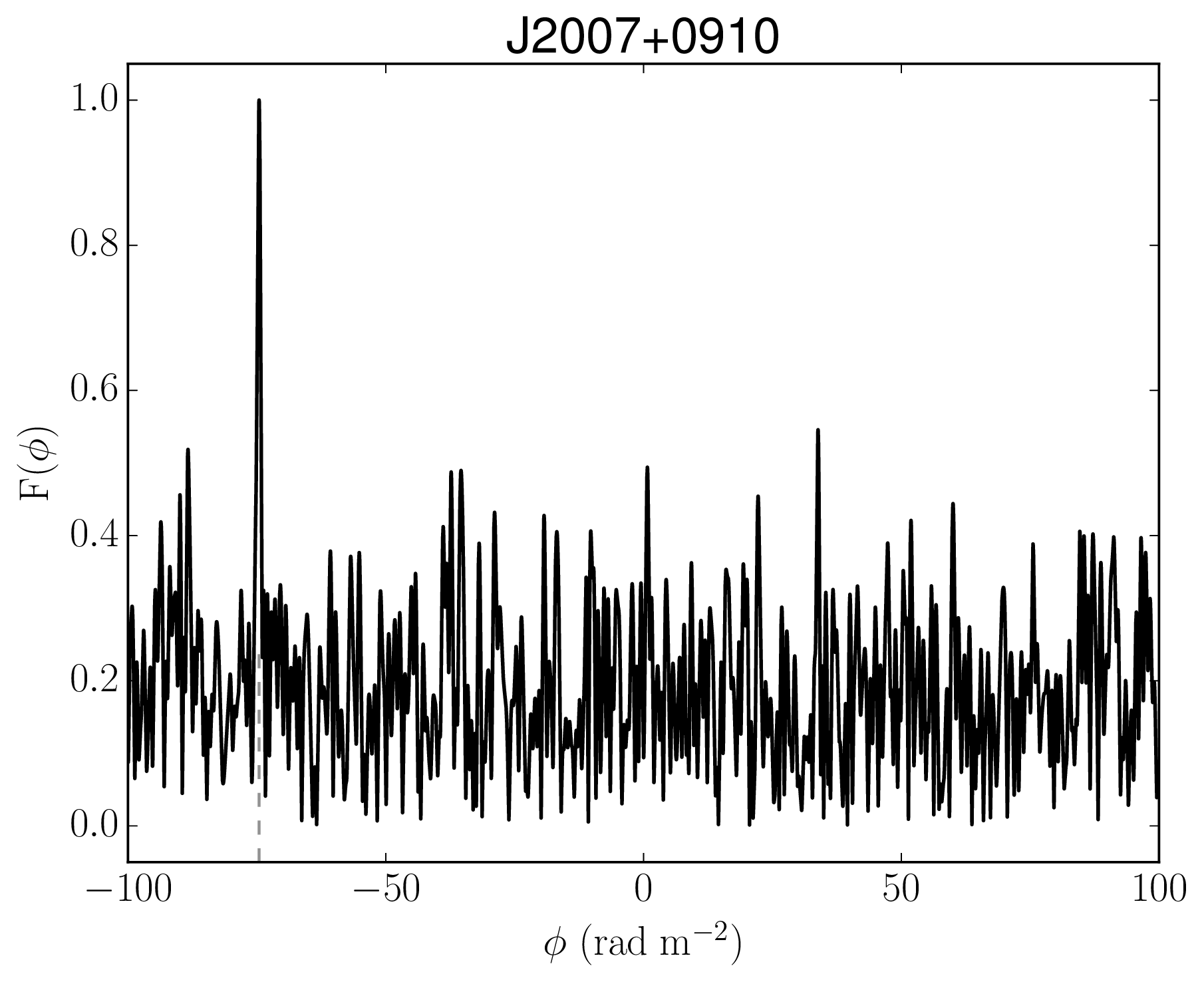}

\includegraphics[width=0.246\columnwidth]{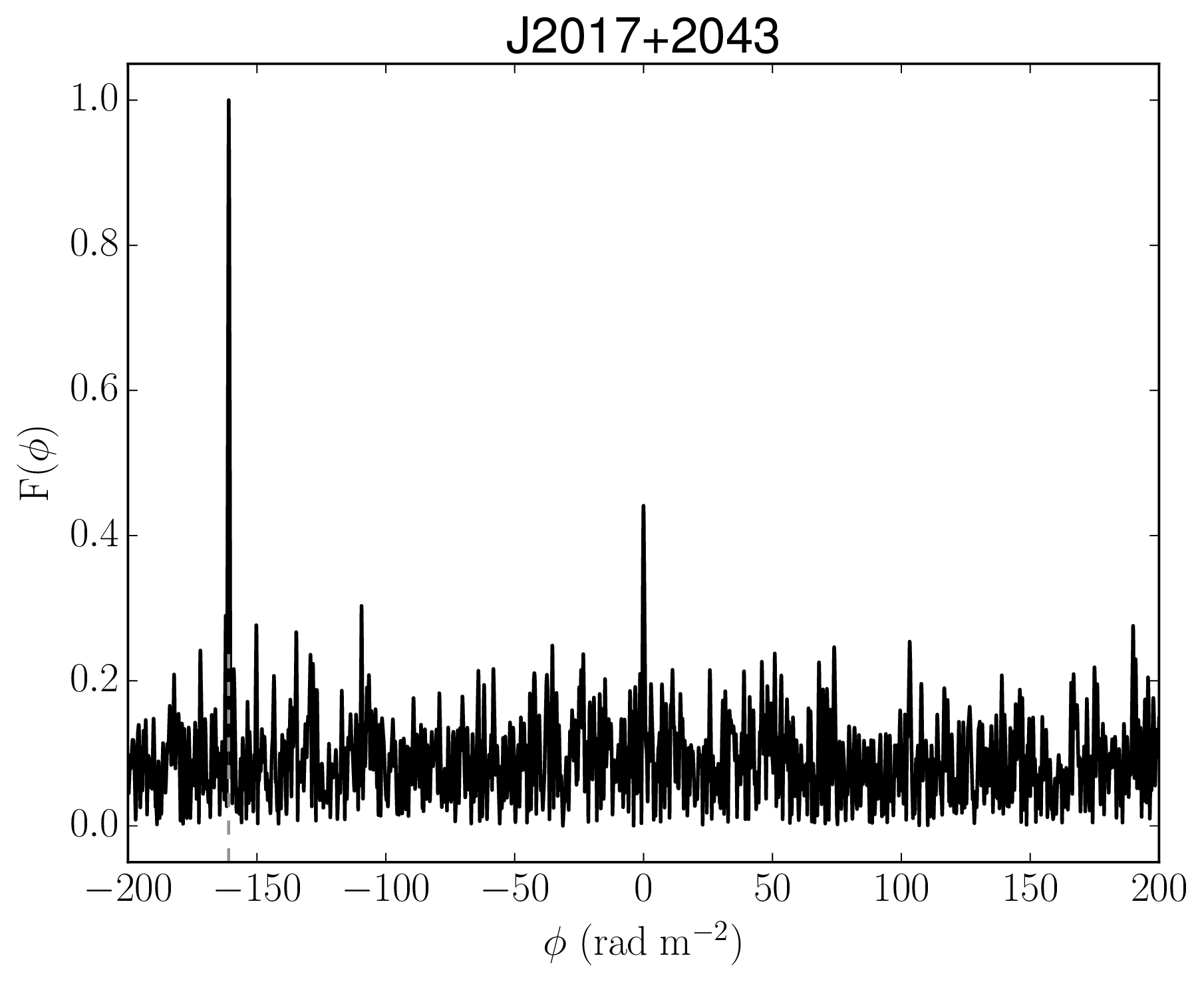}
\includegraphics[width=0.246\columnwidth]{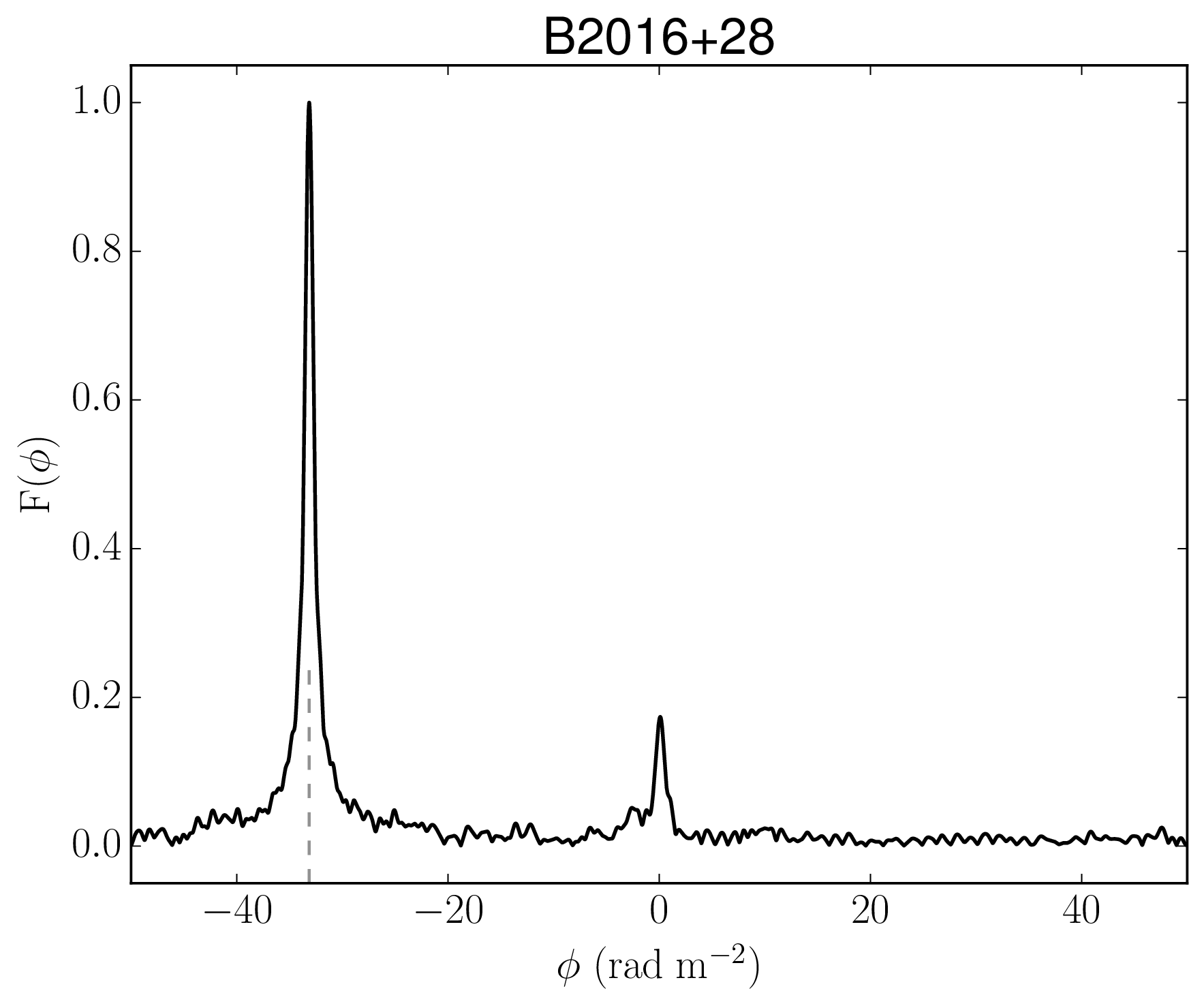}
\includegraphics[width=0.246\columnwidth]{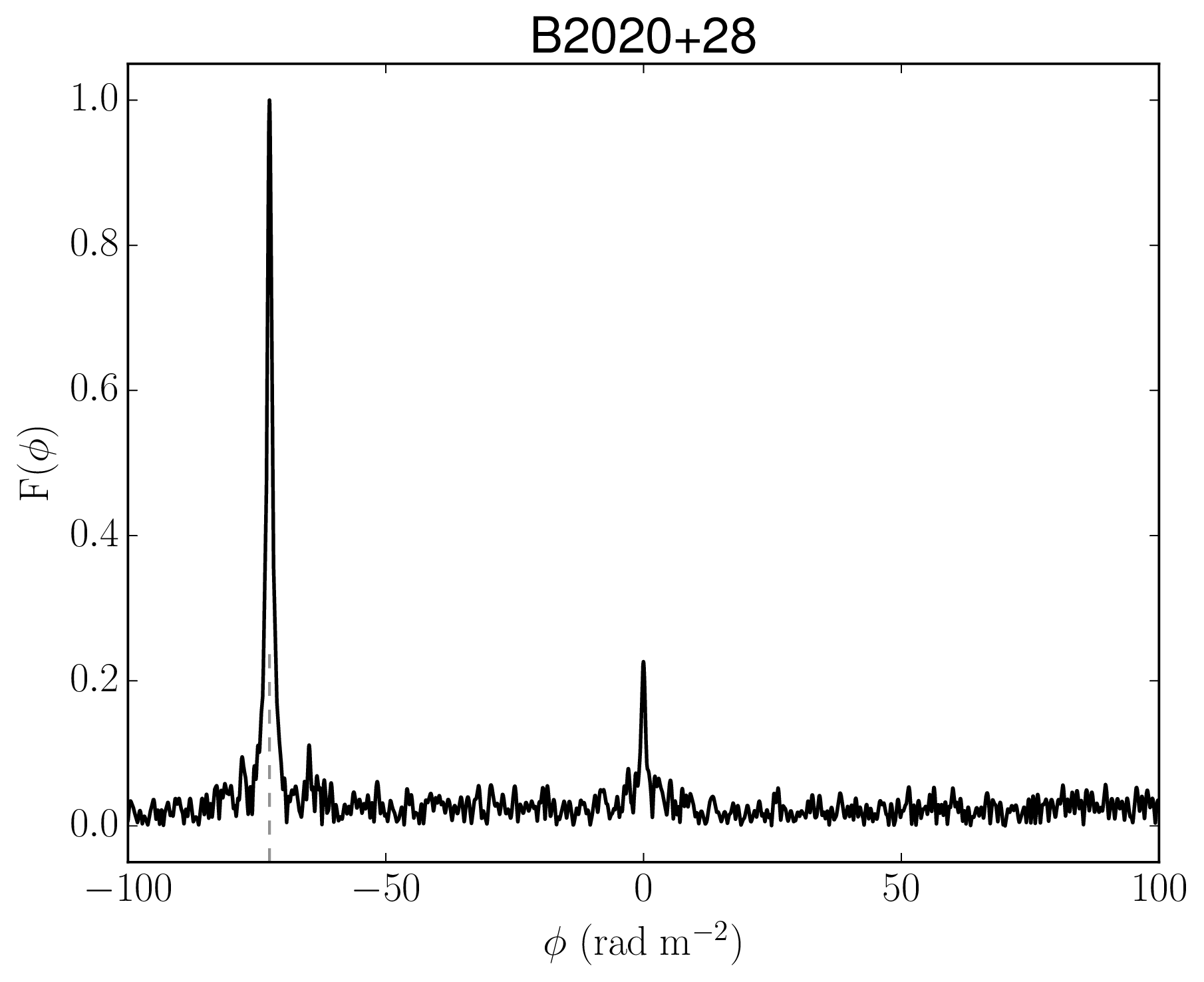}
\includegraphics[width=0.246\columnwidth]{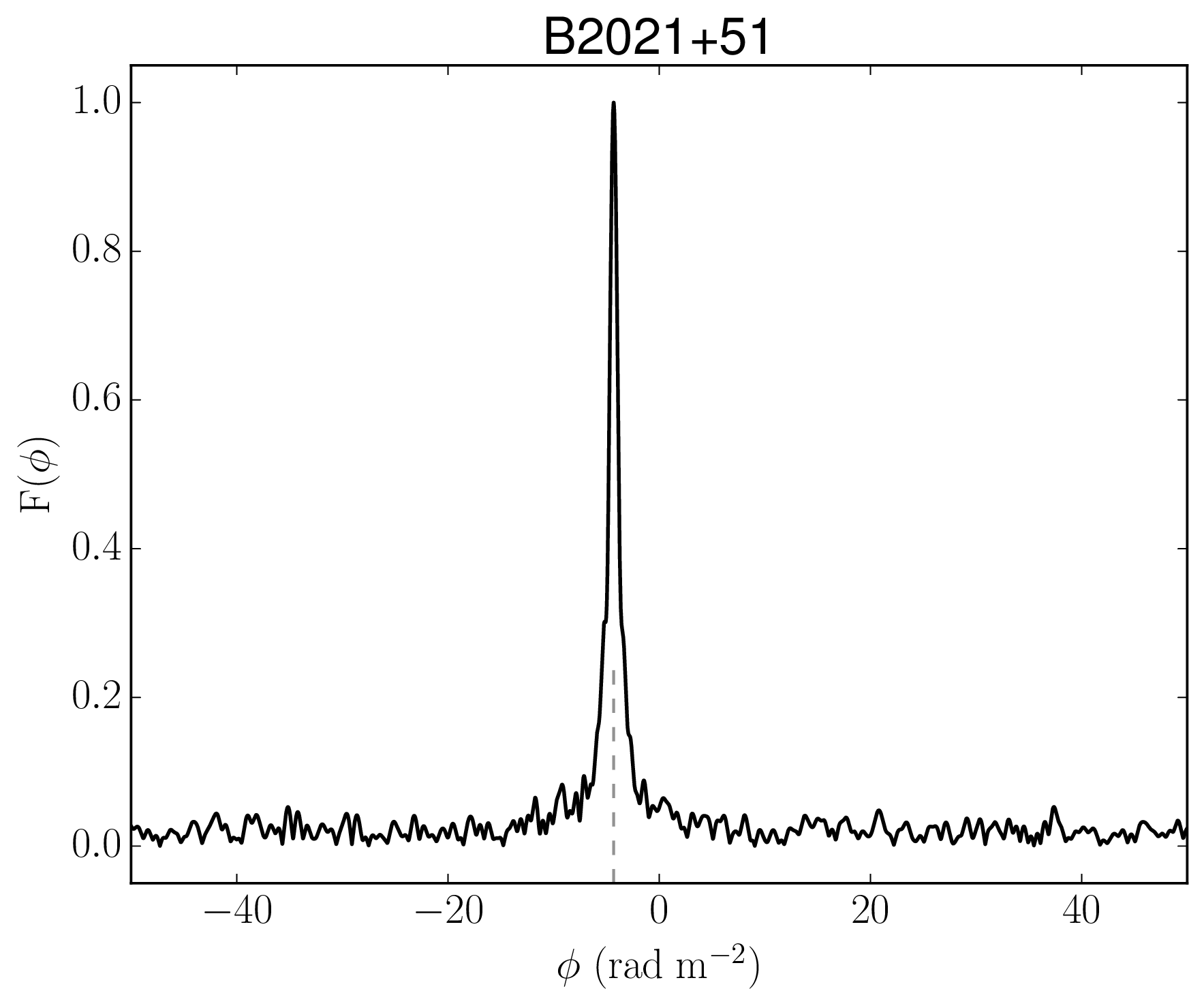}

\includegraphics[width=0.246\columnwidth]{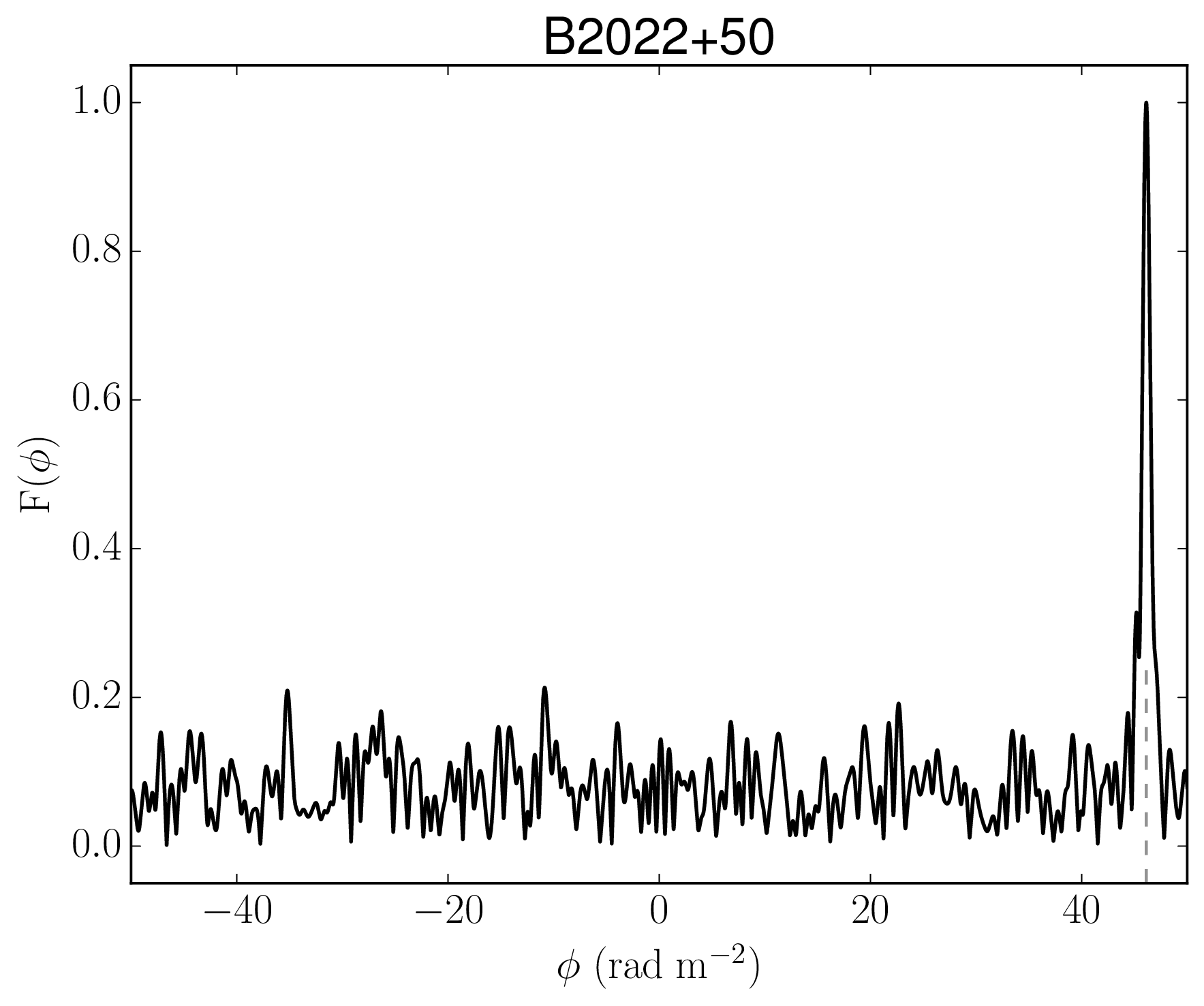}
\includegraphics[width=0.246\columnwidth]{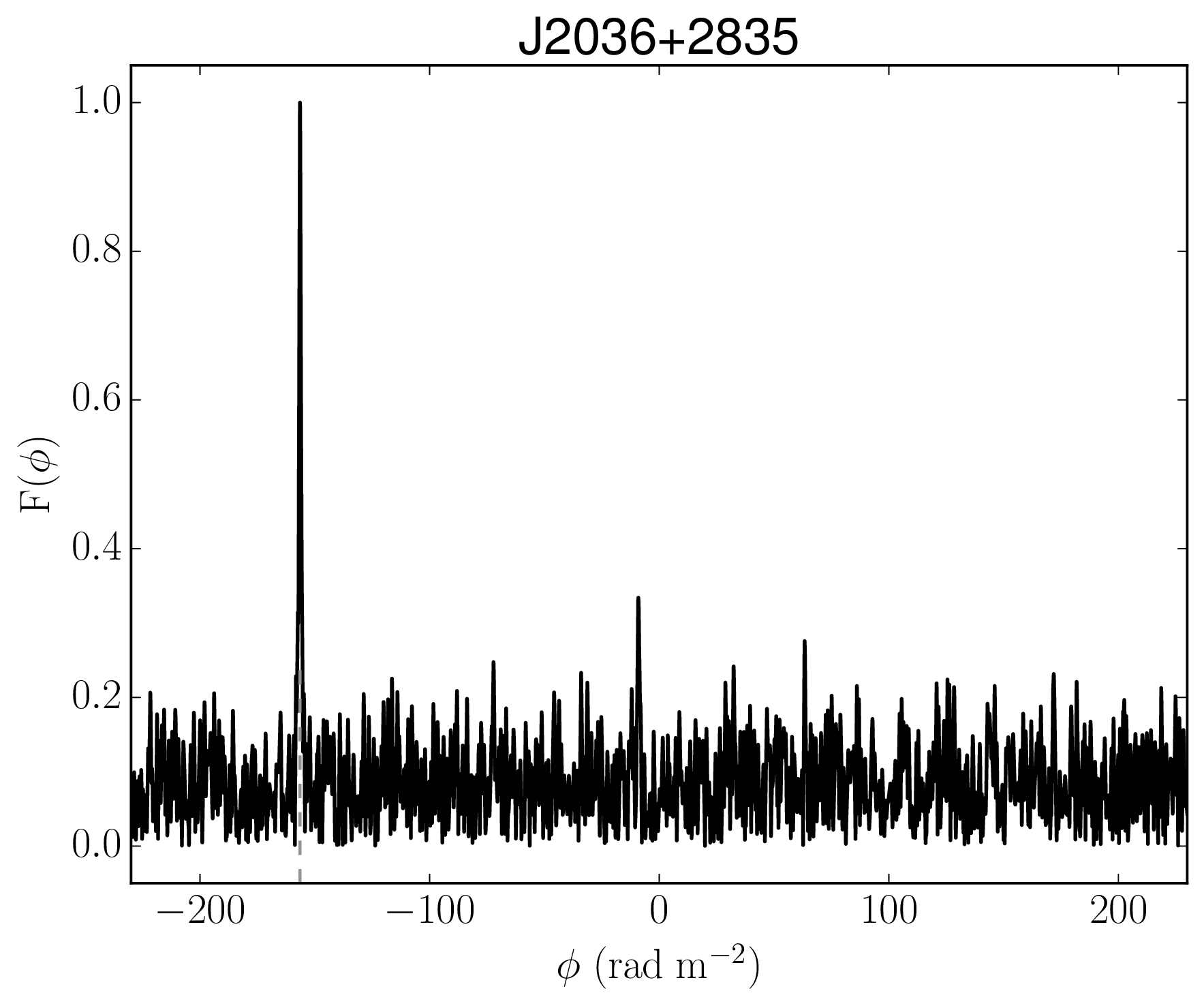}
\includegraphics[width=0.246\columnwidth]{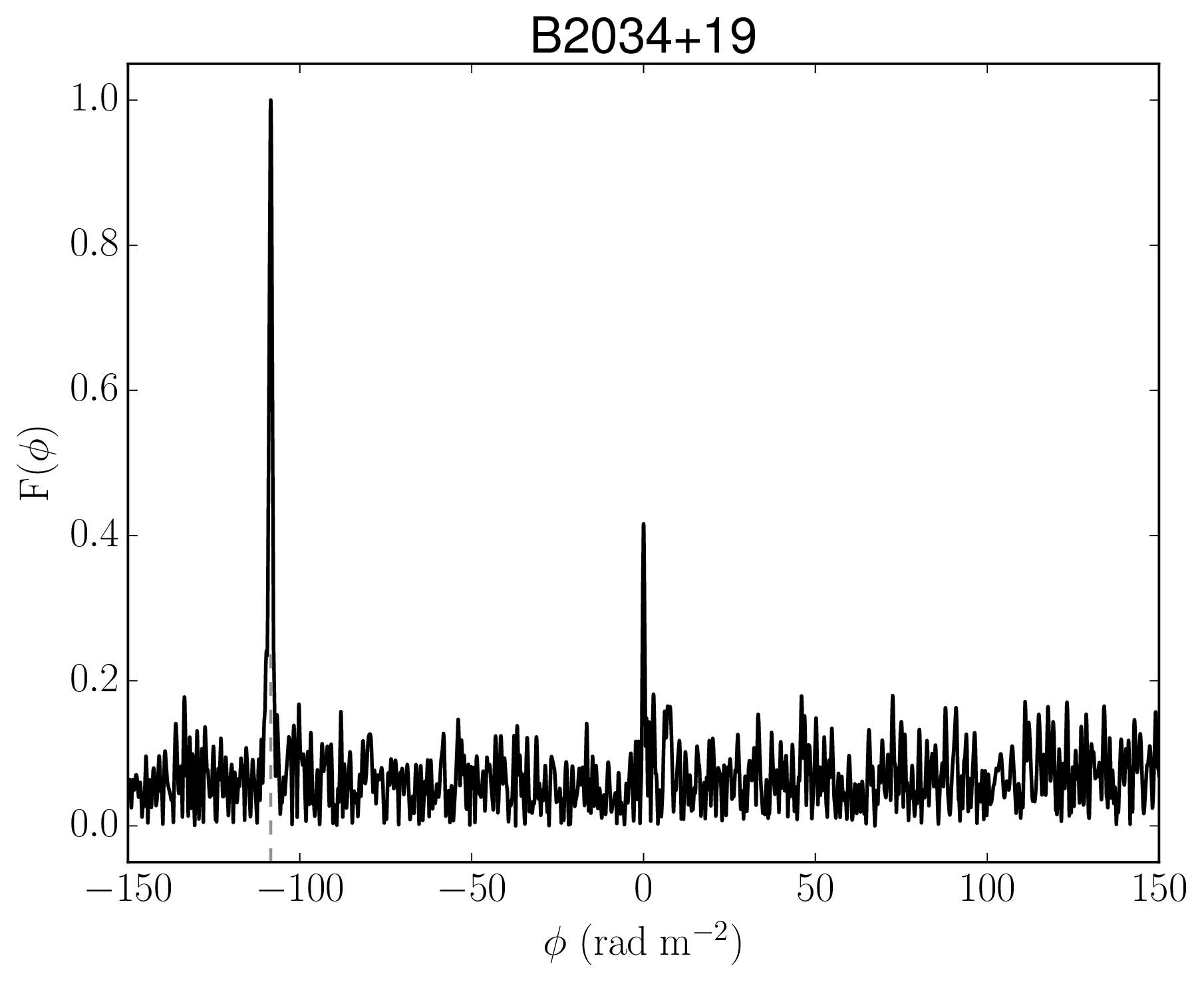}
\includegraphics[width=0.246\columnwidth]{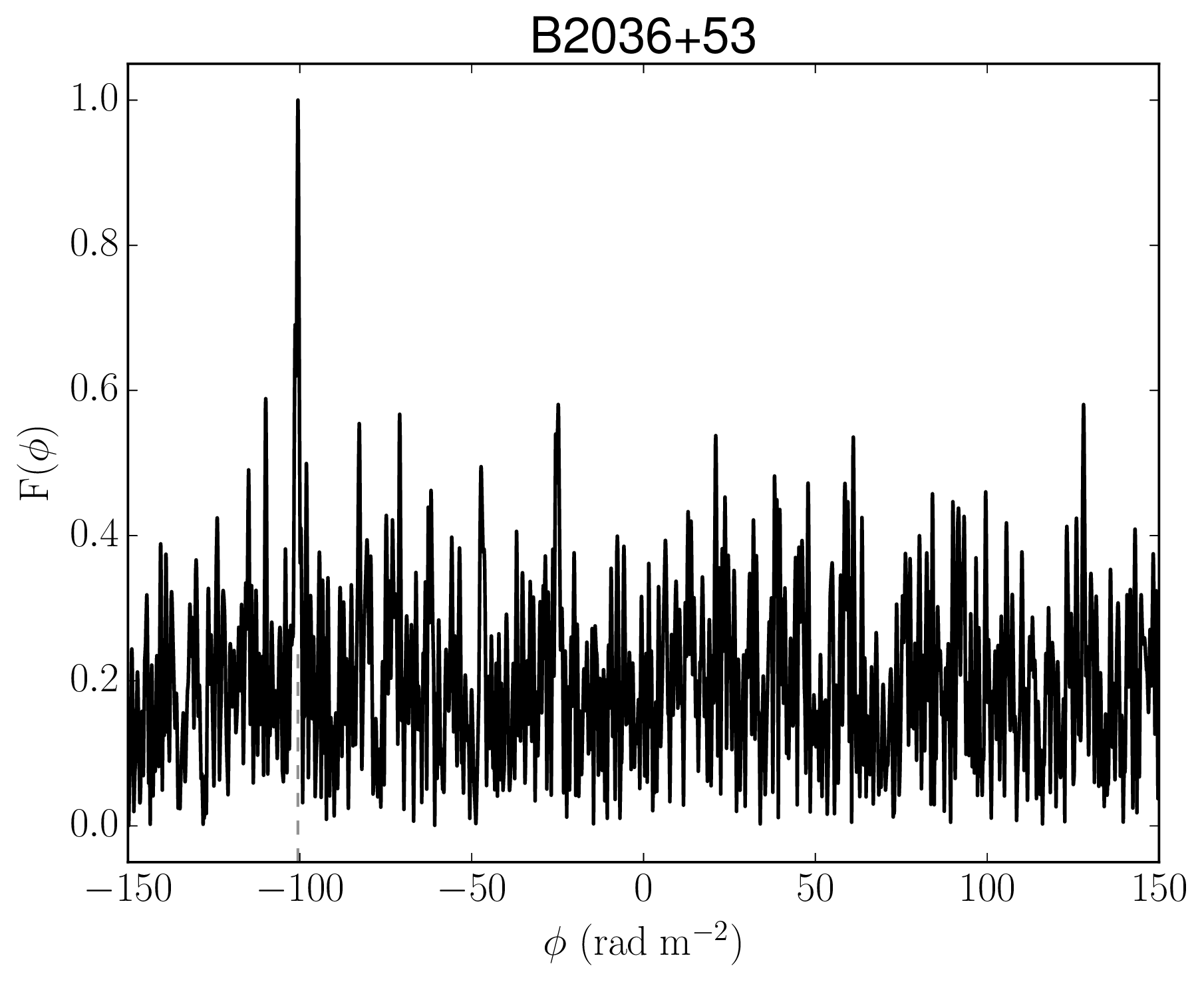}

\includegraphics[width=0.246\columnwidth]{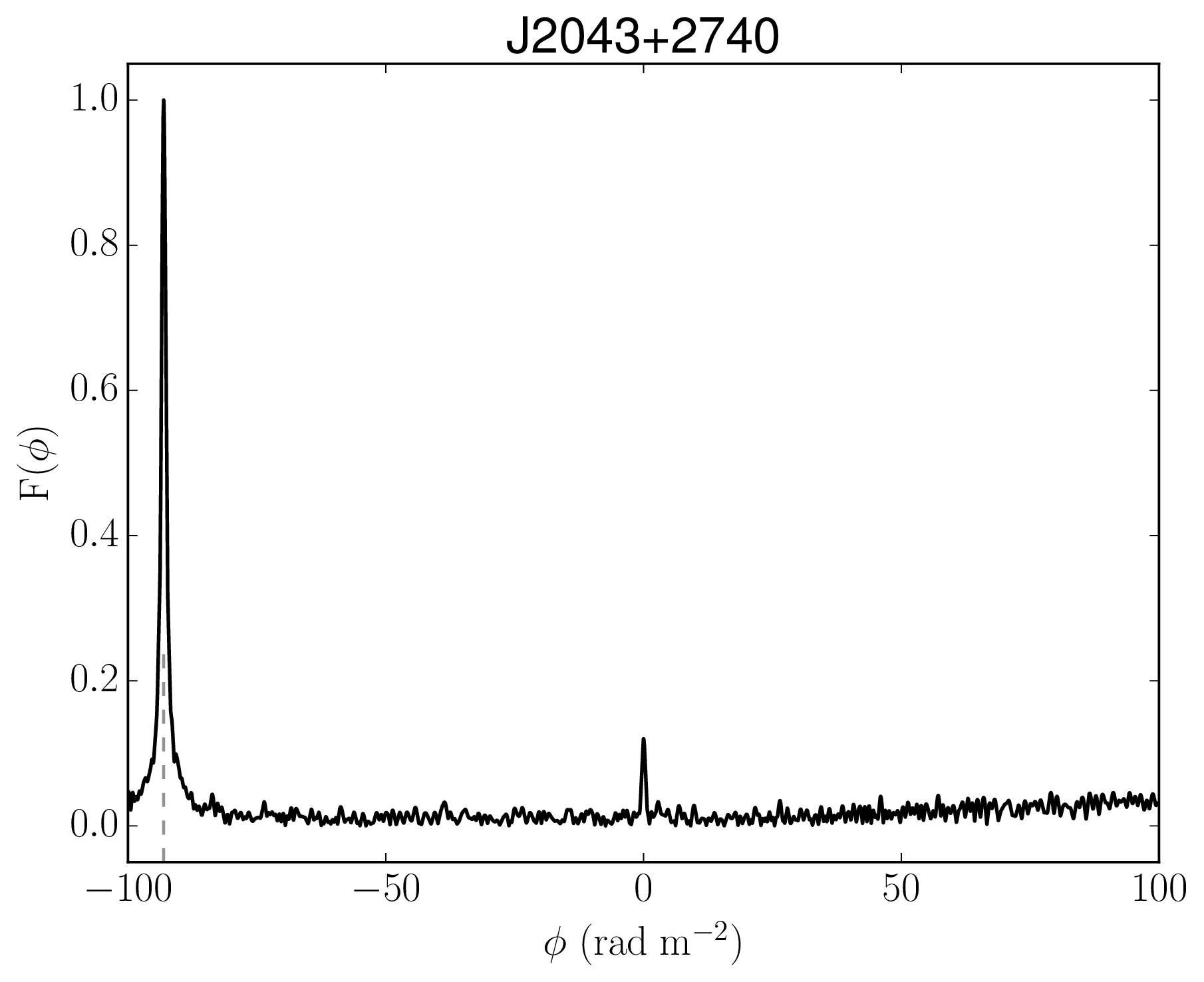}
\includegraphics[width=0.246\columnwidth]{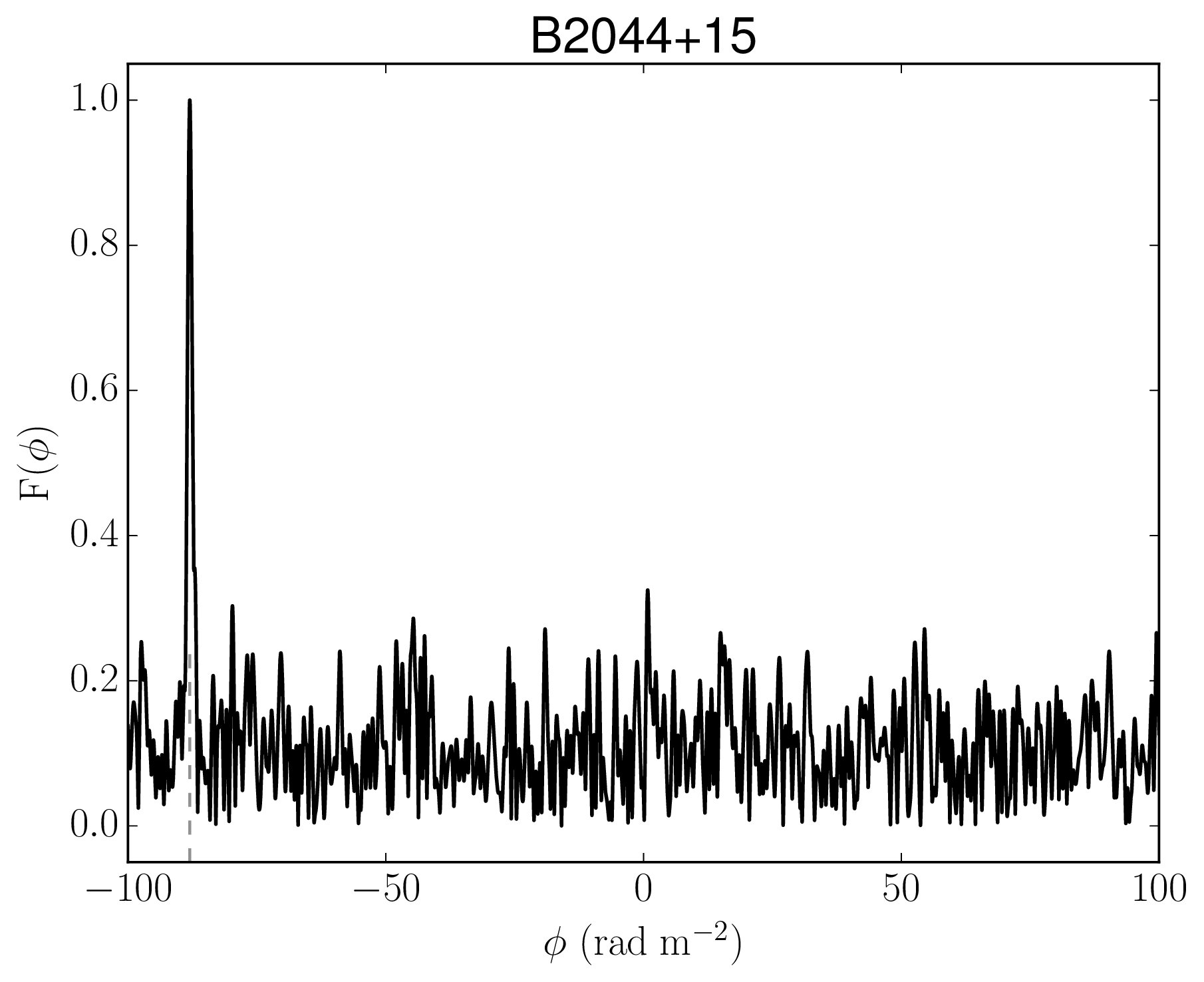}
\includegraphics[width=0.246\columnwidth]{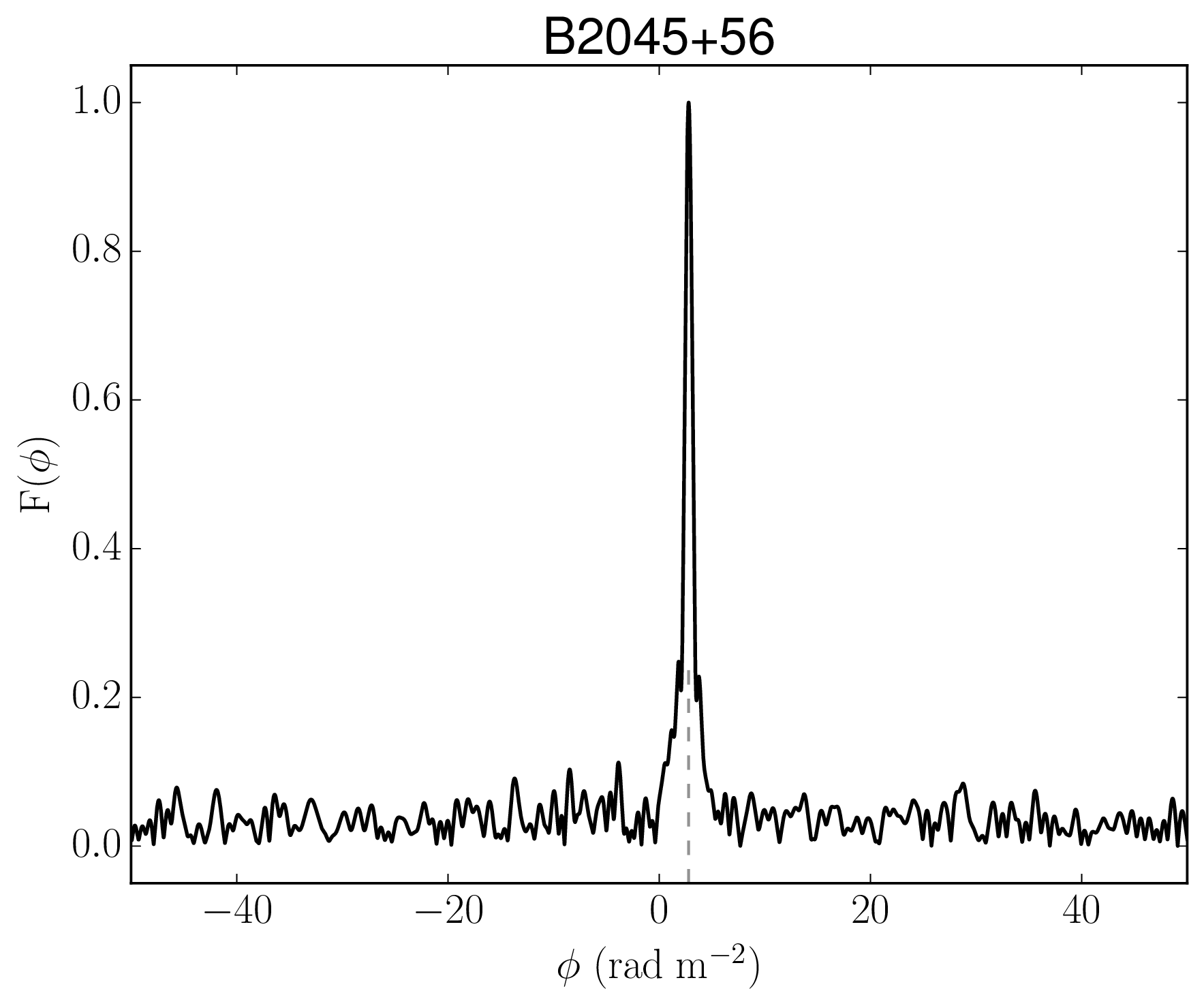}
\includegraphics[width=0.246\columnwidth]{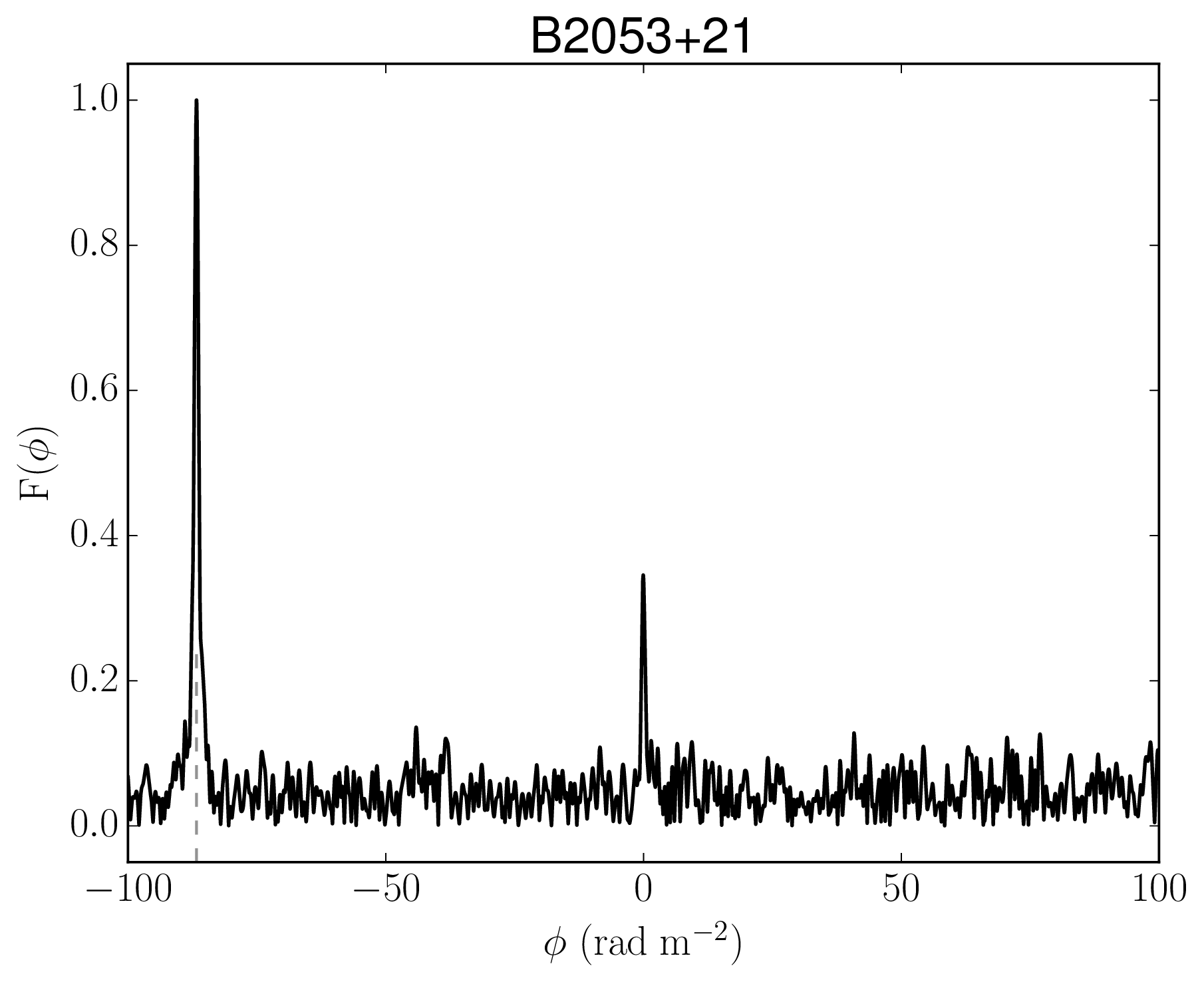}

 \caption{Continued from the previous page.}
% \label{fig:FDFs}
\end{figure*}

\begin{figure*}
 \centering
 %\ContinuedFloat
%\setcounter{subfigure}{2}    

\includegraphics[width=0.246\columnwidth]{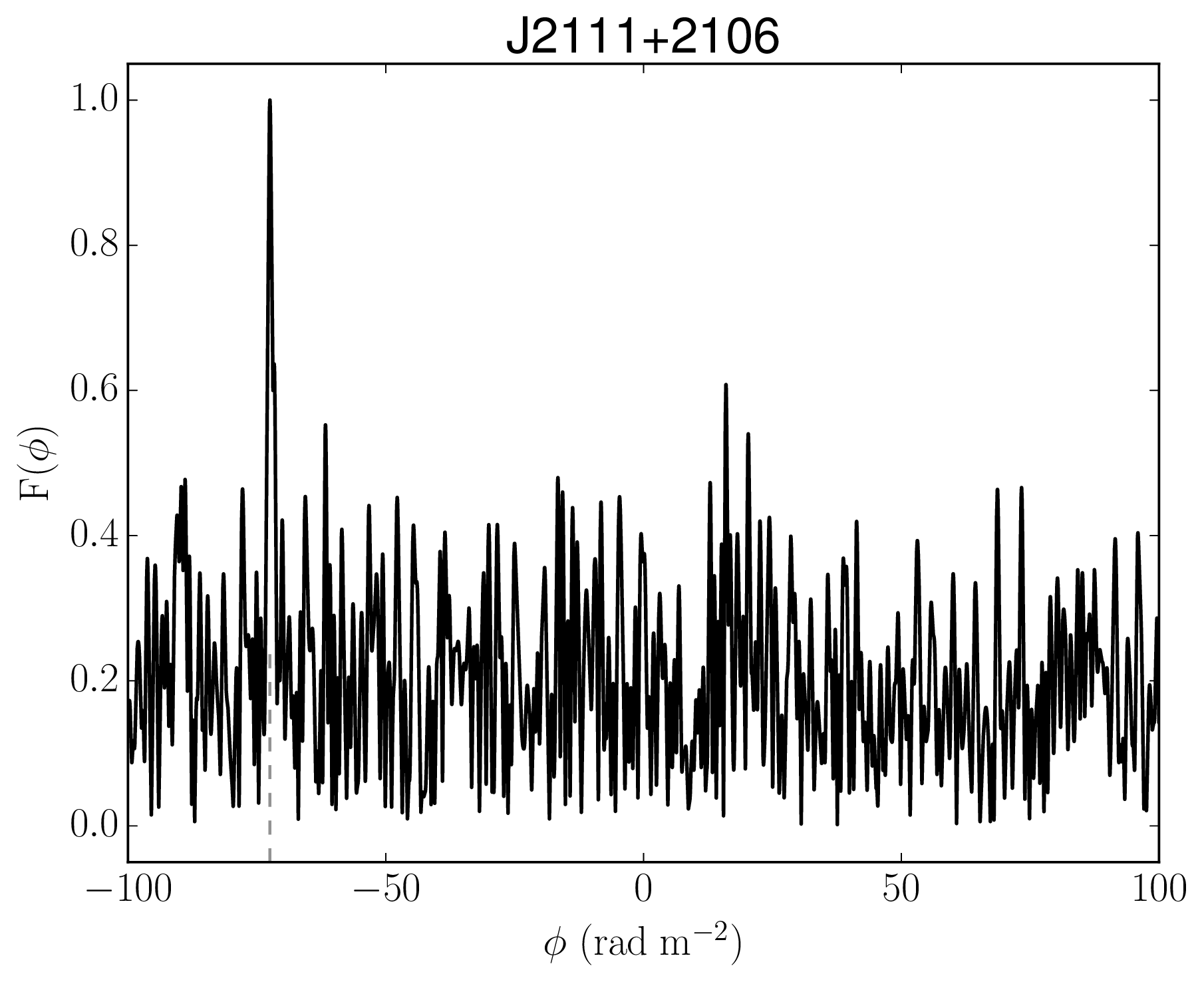}
\includegraphics[width=0.246\columnwidth]{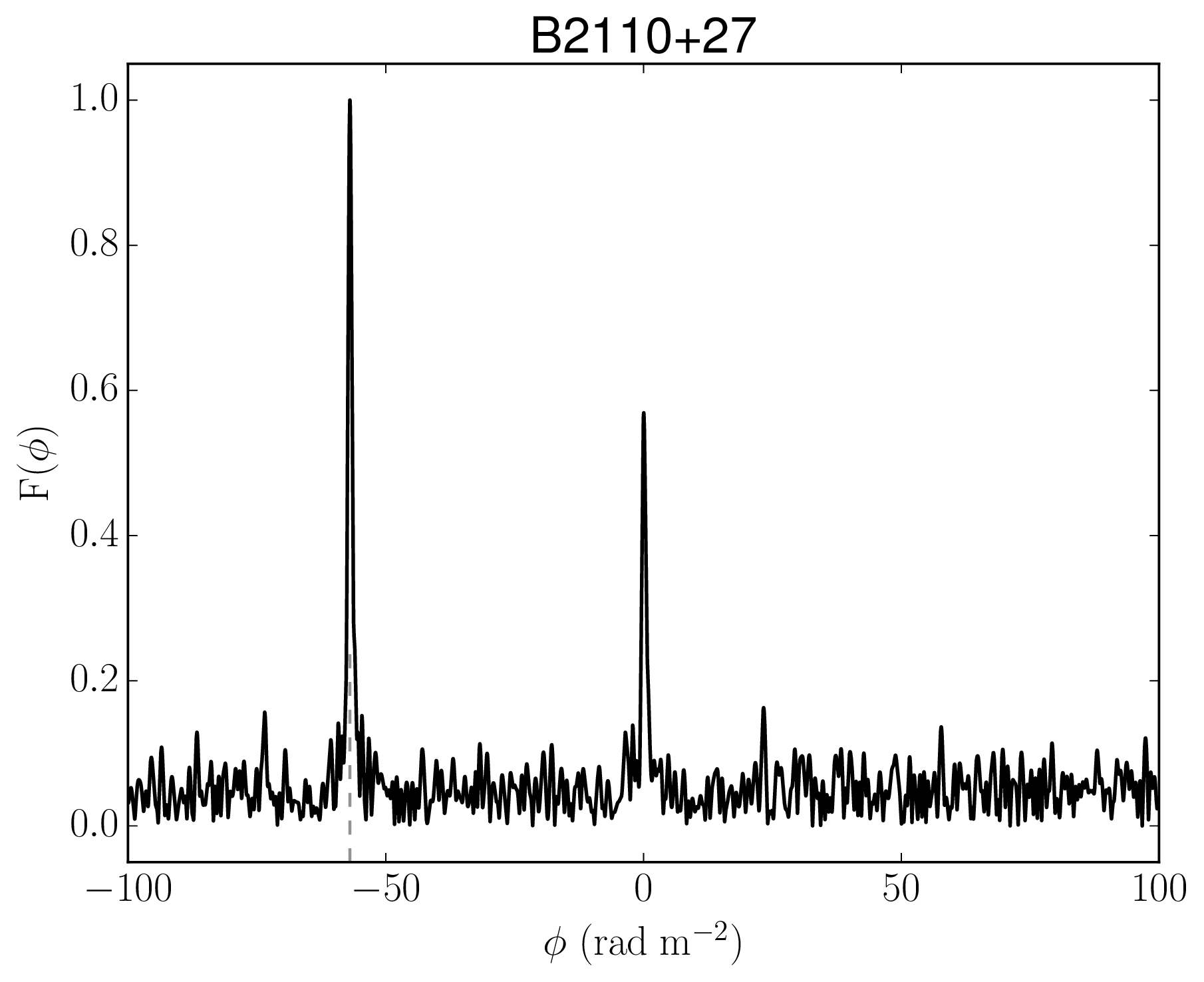}
\includegraphics[width=0.246\columnwidth]{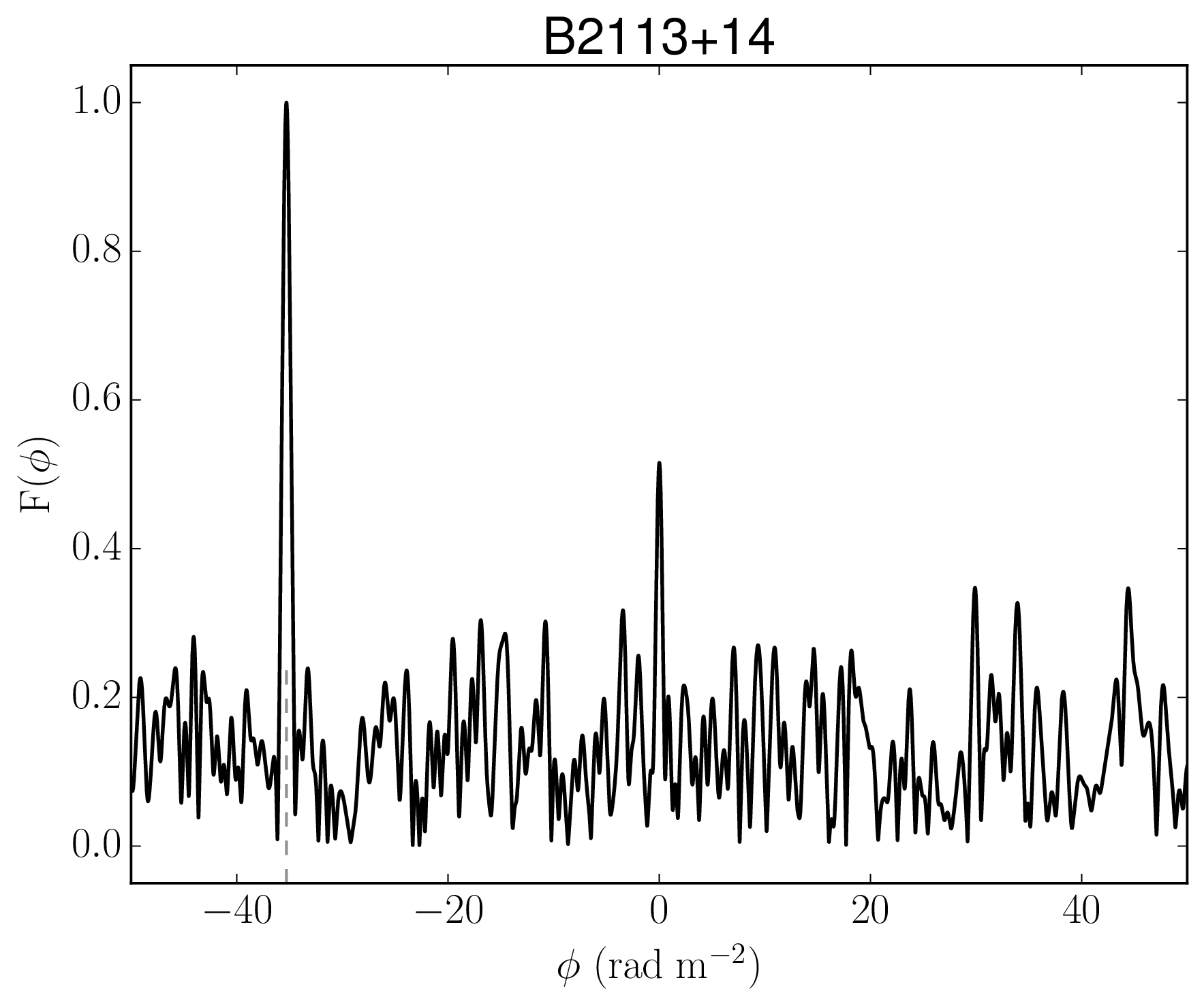}
\includegraphics[width=0.246\columnwidth]{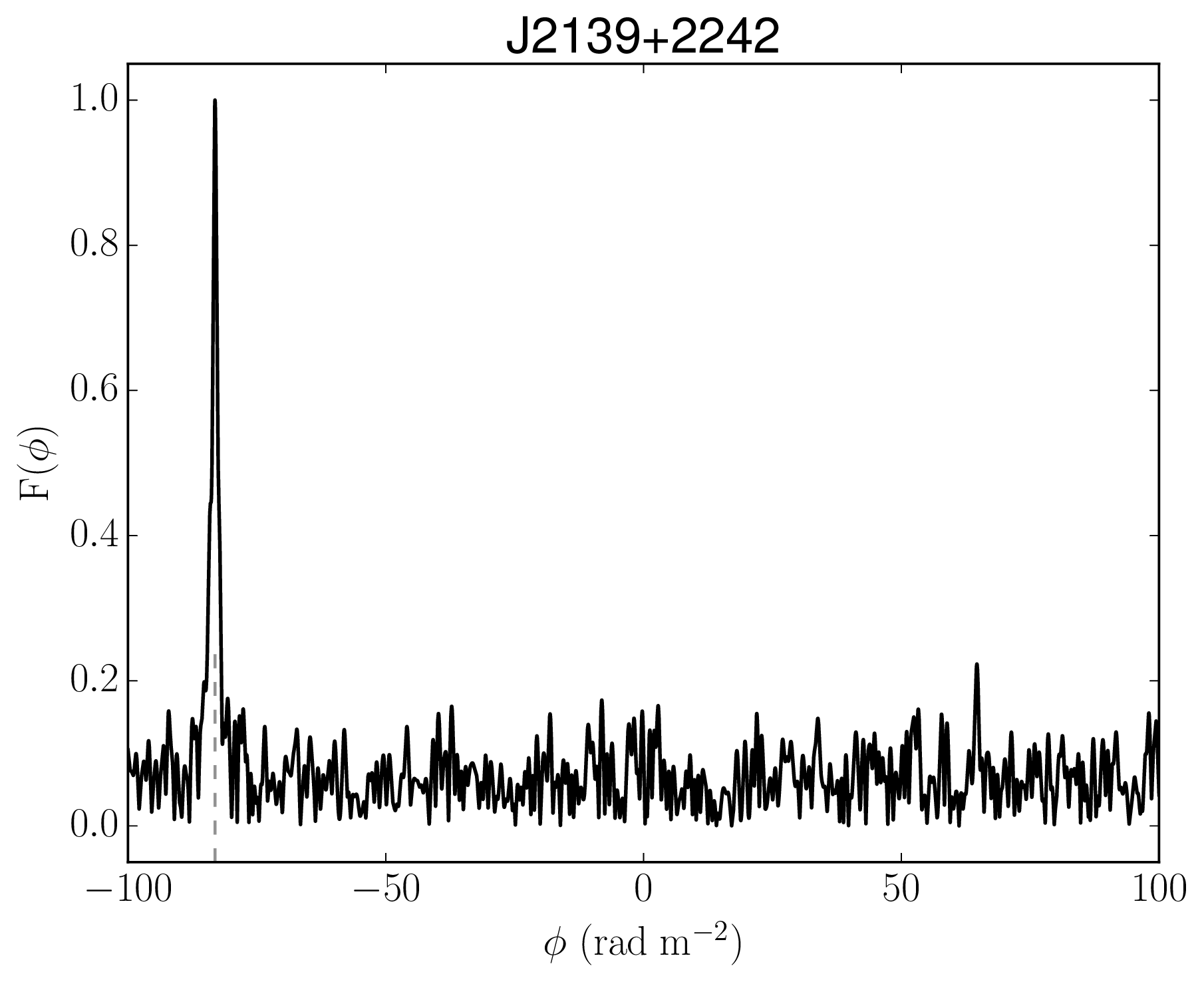}

\includegraphics[width=0.246\columnwidth]{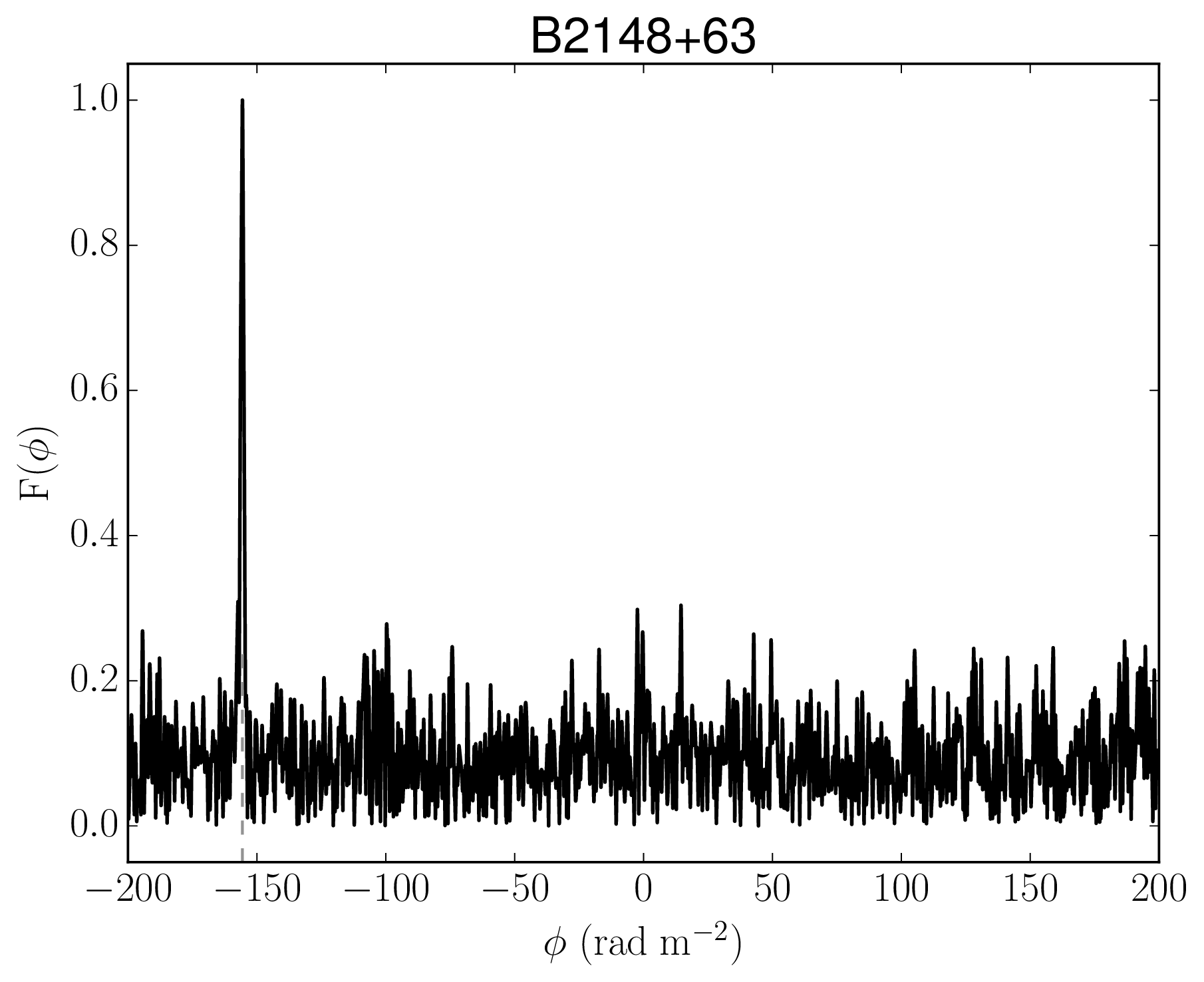}
\includegraphics[width=0.246\columnwidth]{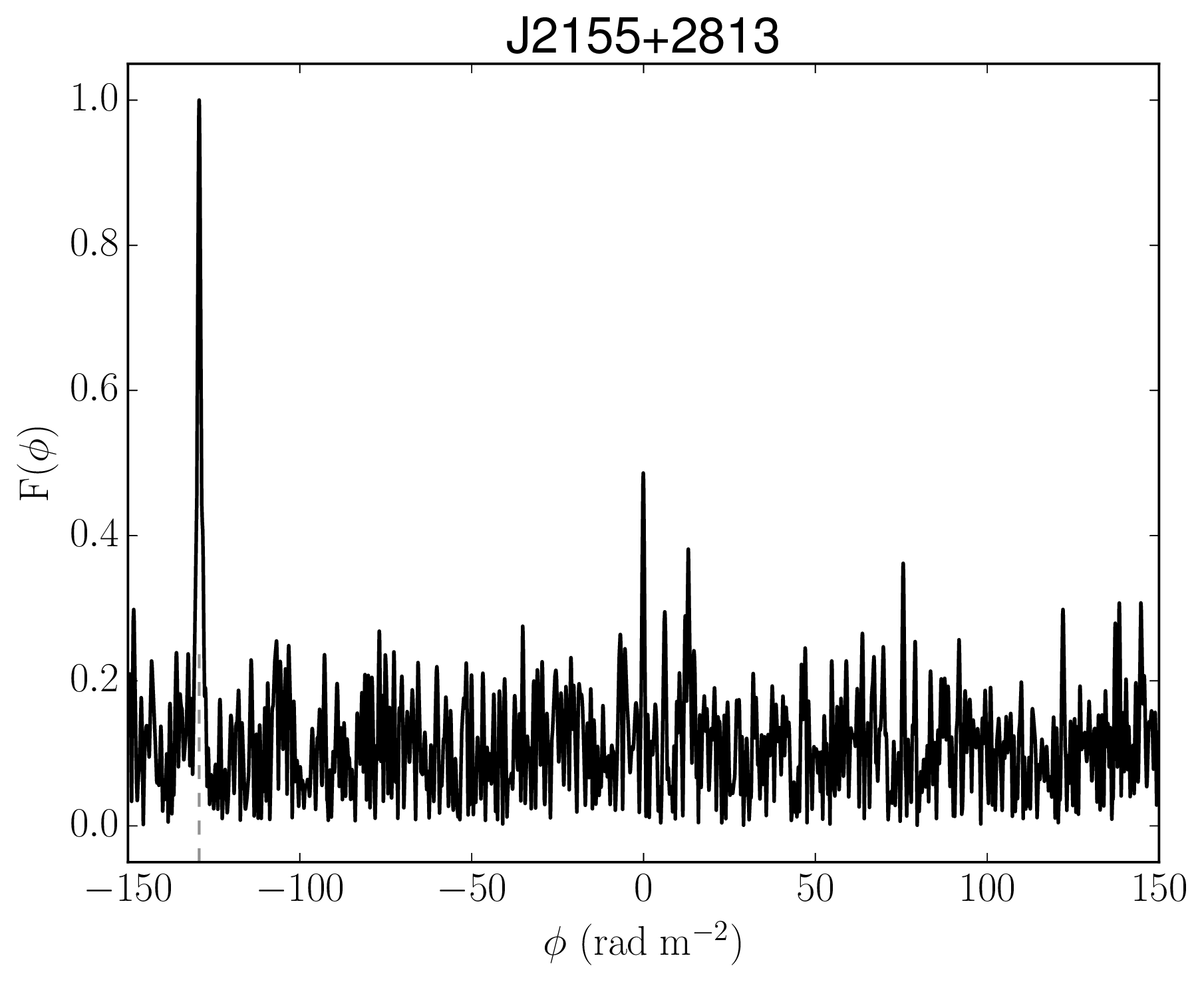}
\includegraphics[width=0.246\columnwidth]{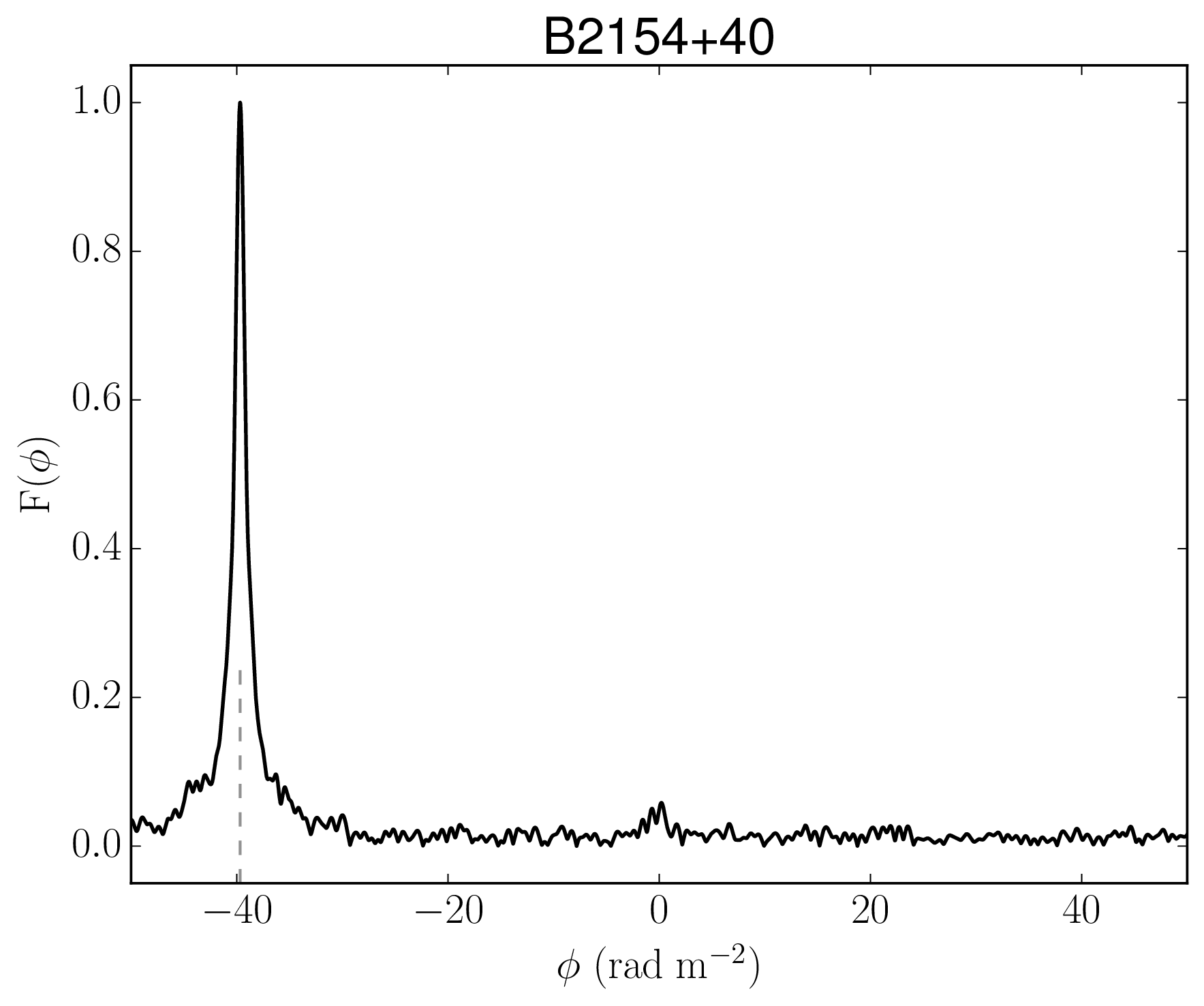}
\includegraphics[width=0.246\columnwidth]{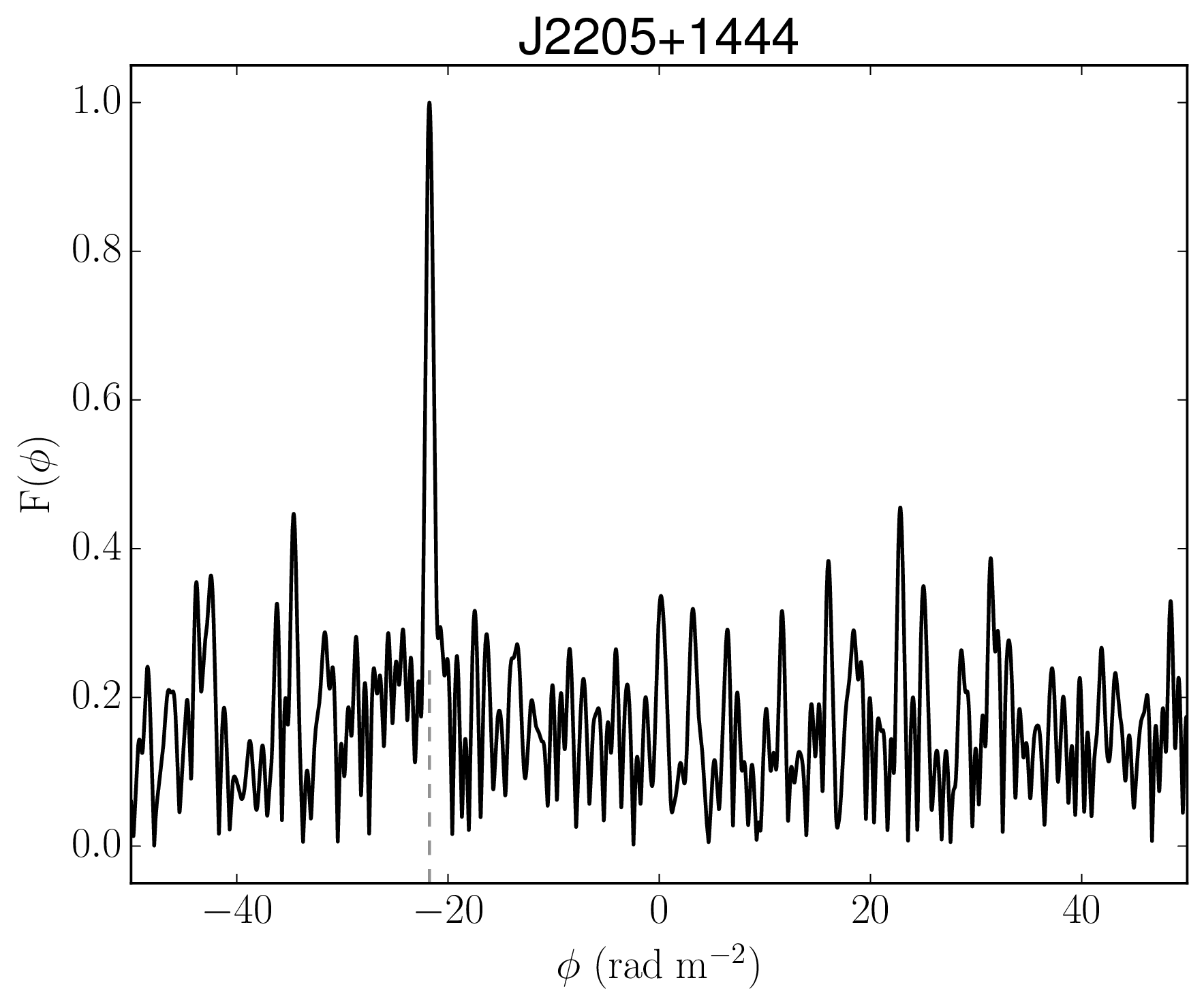}

\includegraphics[width=0.246\columnwidth]{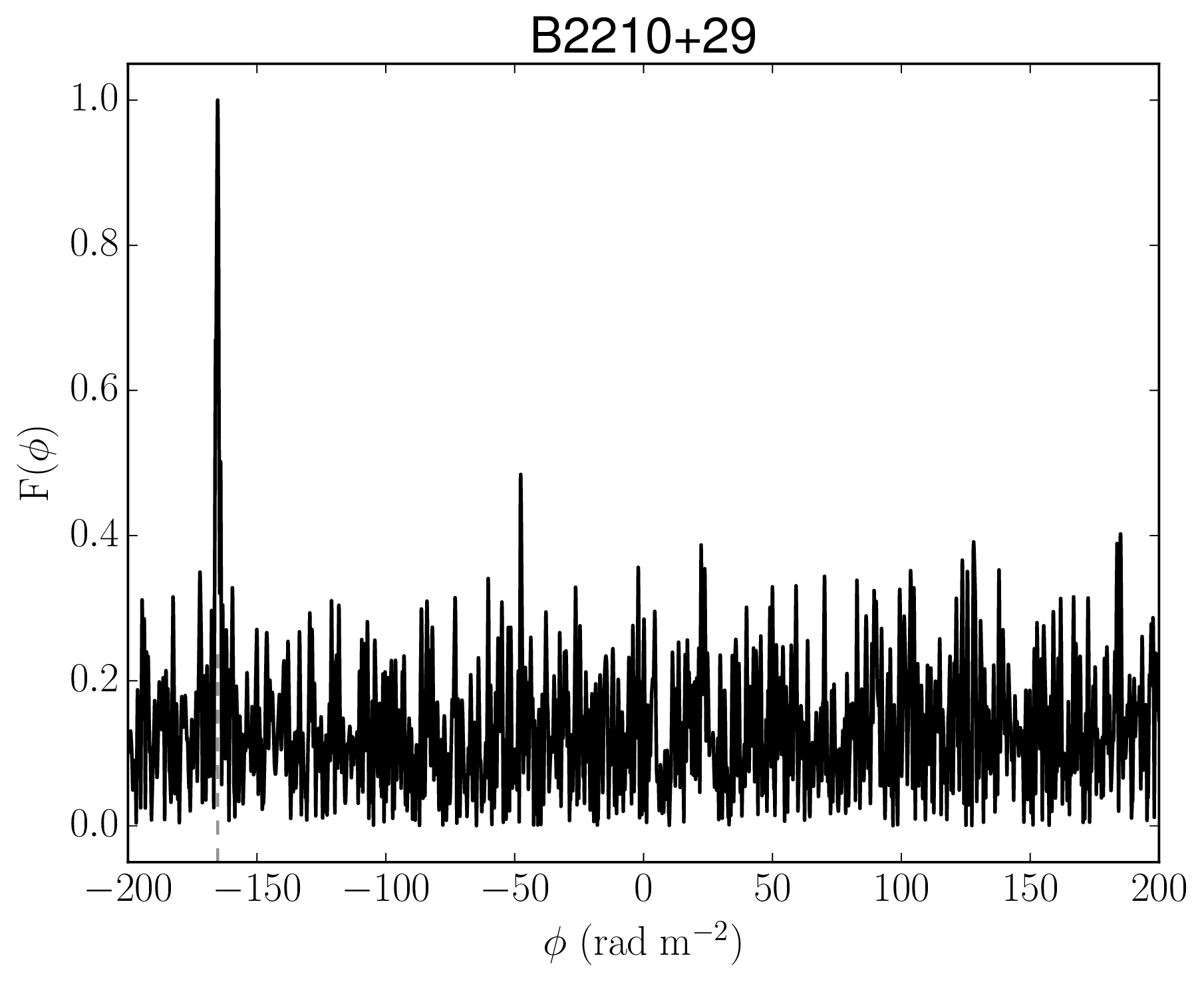}
\includegraphics[width=0.246\columnwidth]{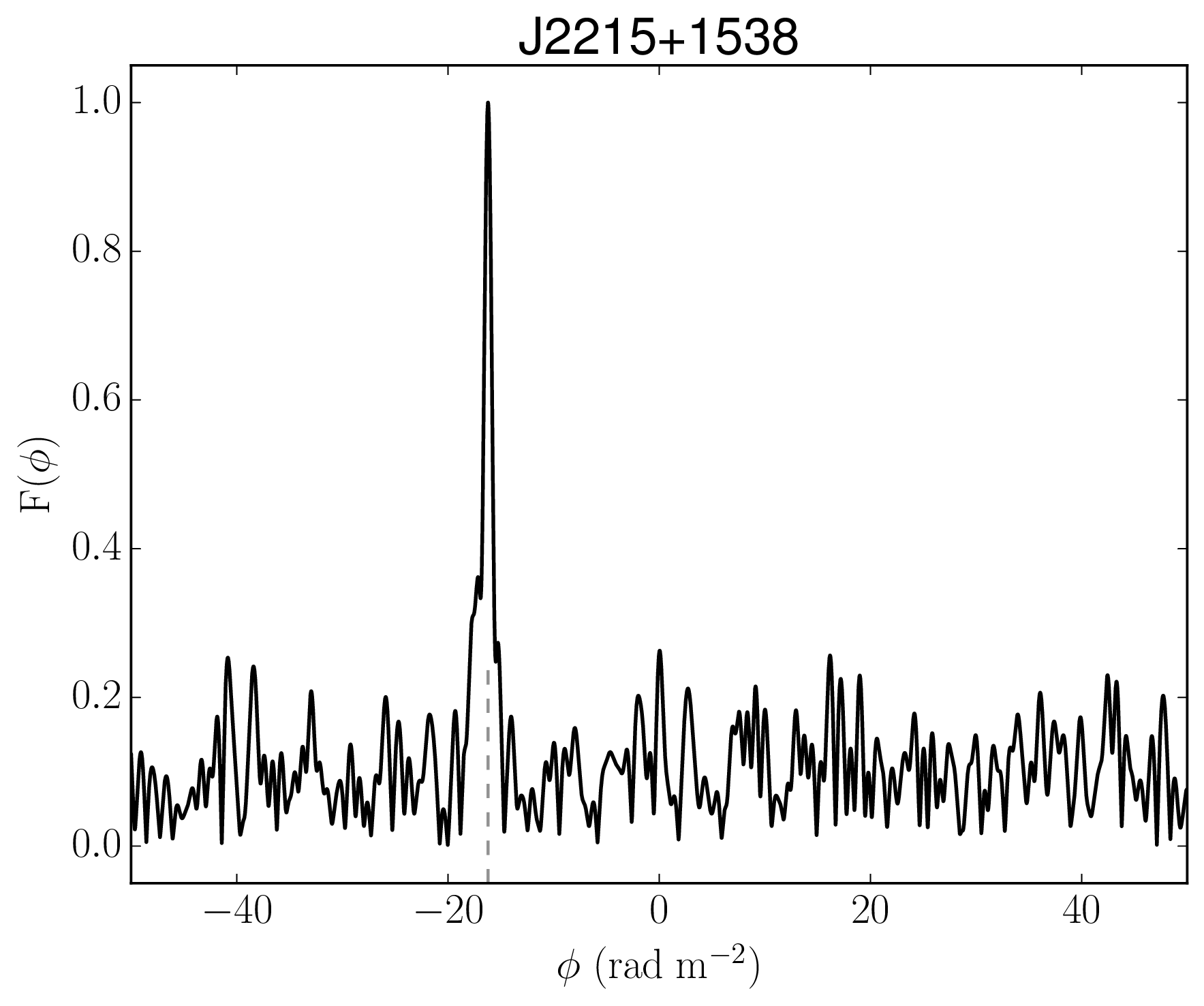}
\includegraphics[width=0.246\columnwidth]{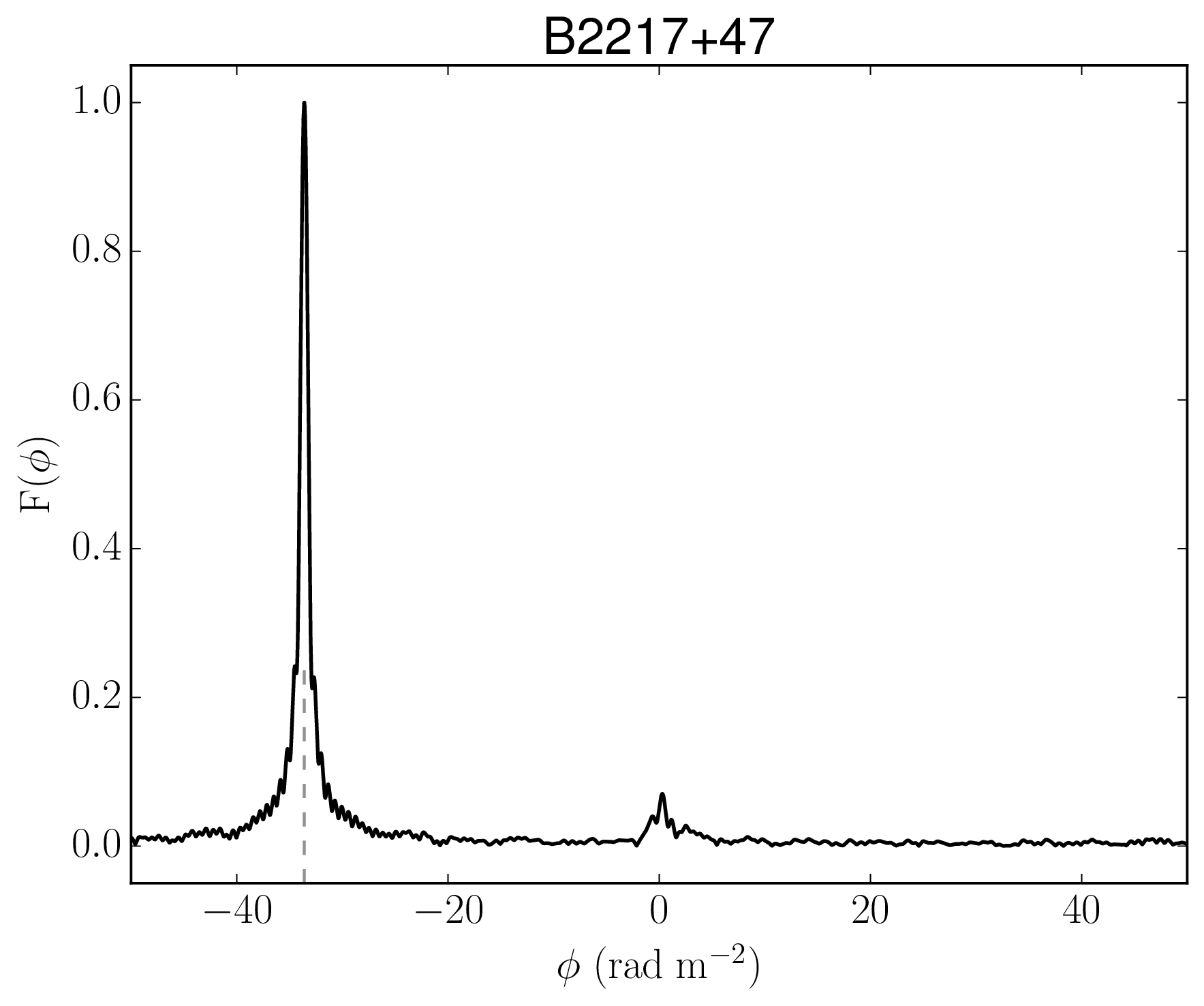}
\includegraphics[width=0.246\columnwidth]{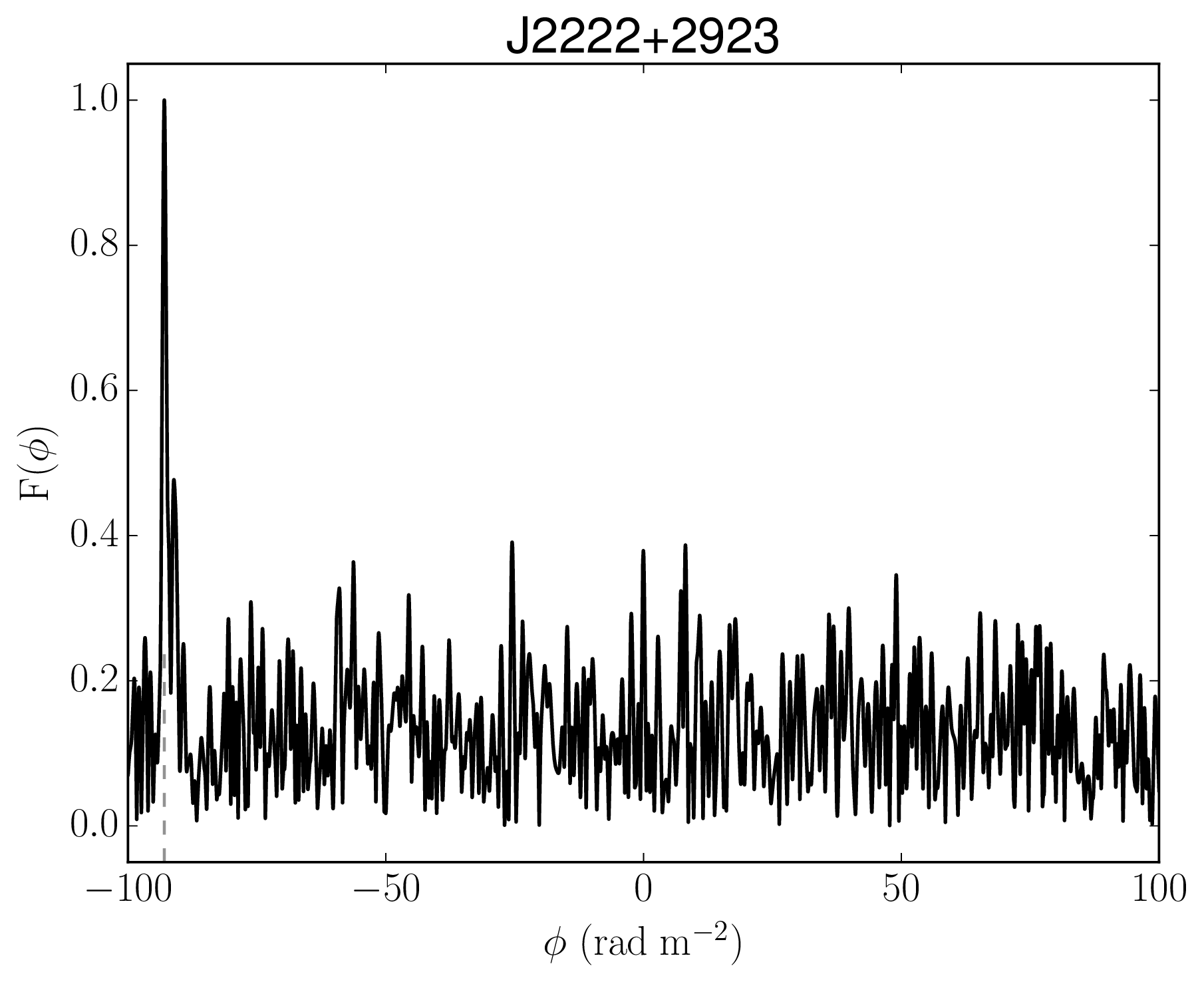}

\includegraphics[width=0.246\columnwidth]{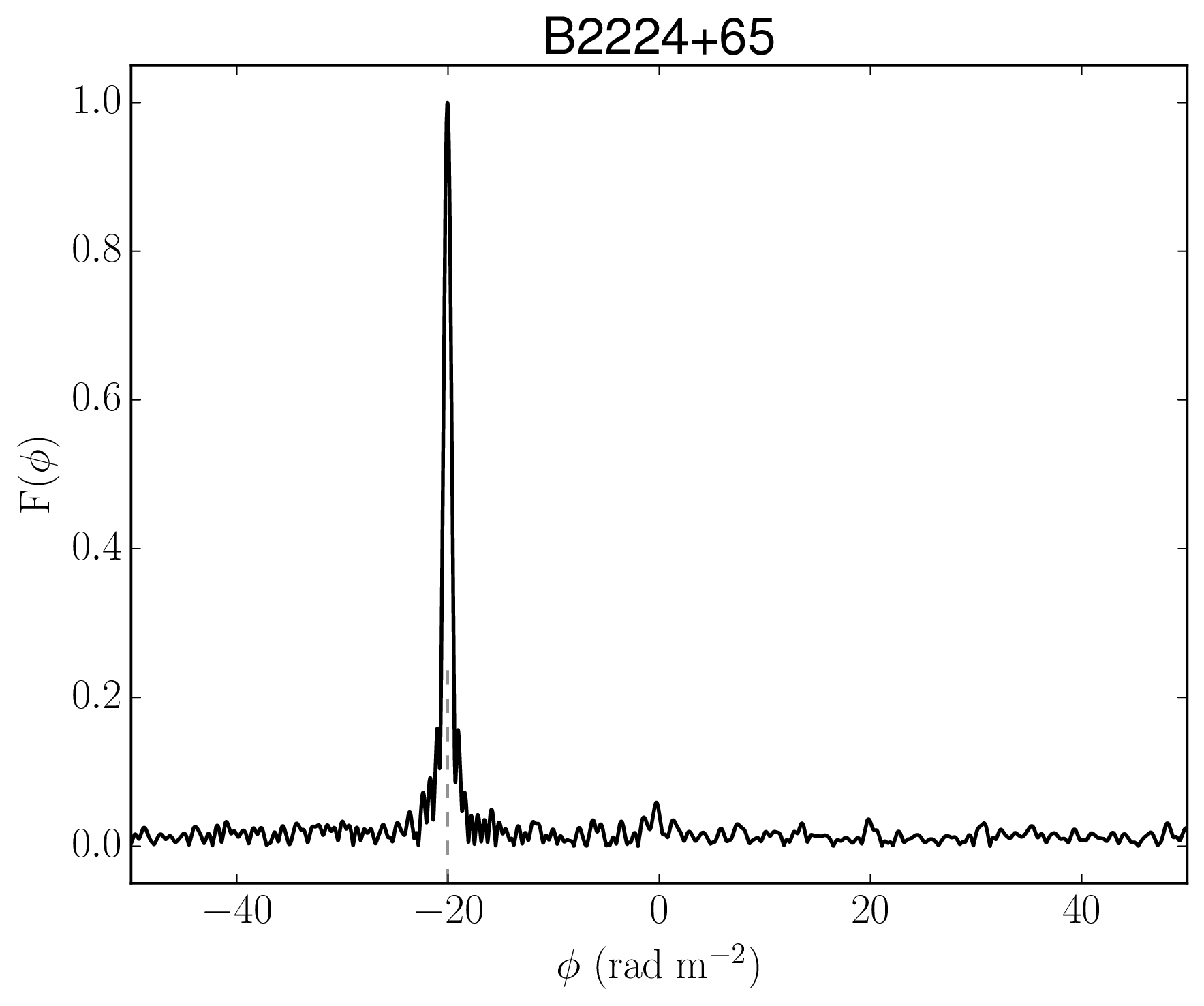}
\includegraphics[width=0.246\columnwidth]{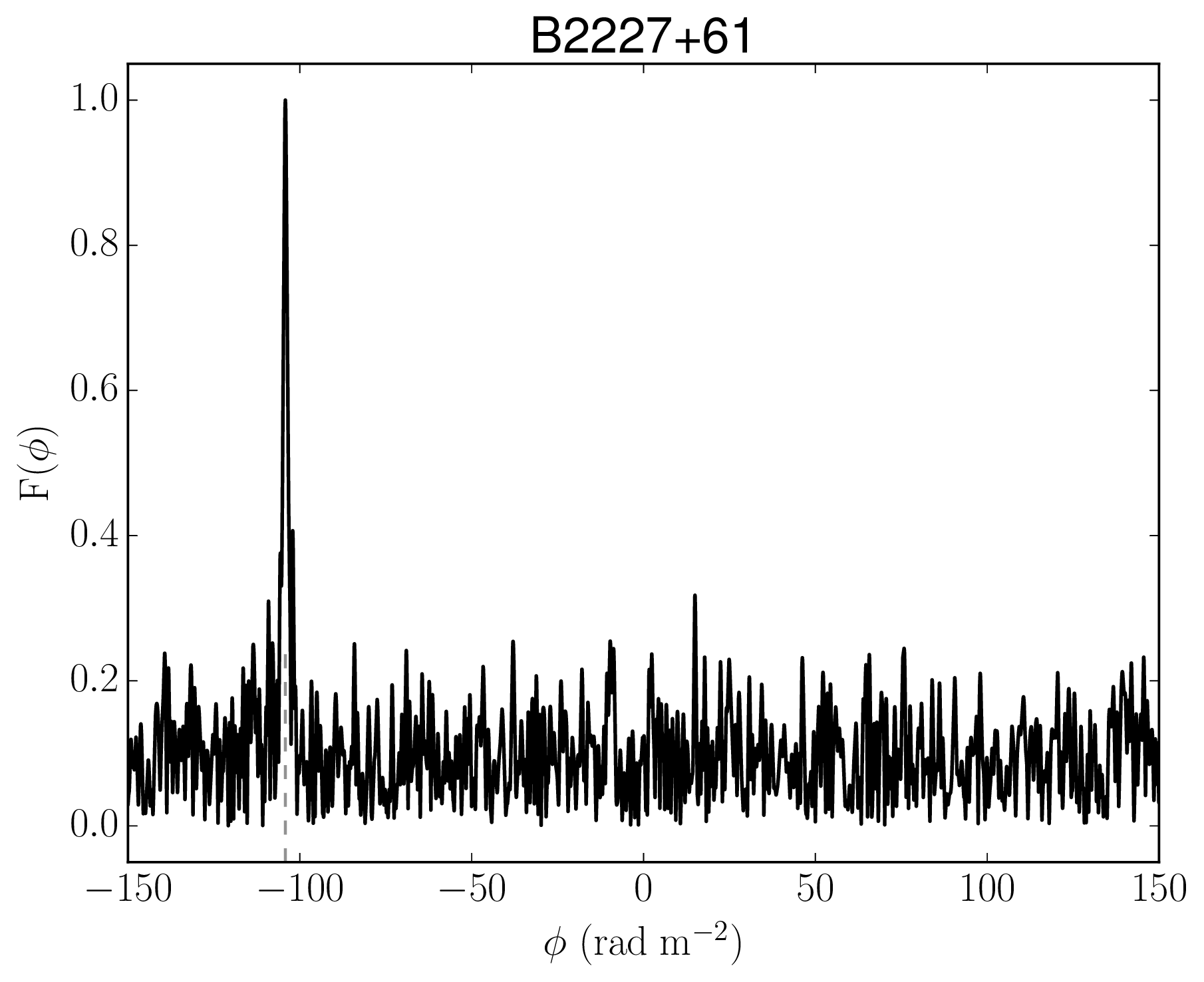}
\includegraphics[width=0.246\columnwidth]{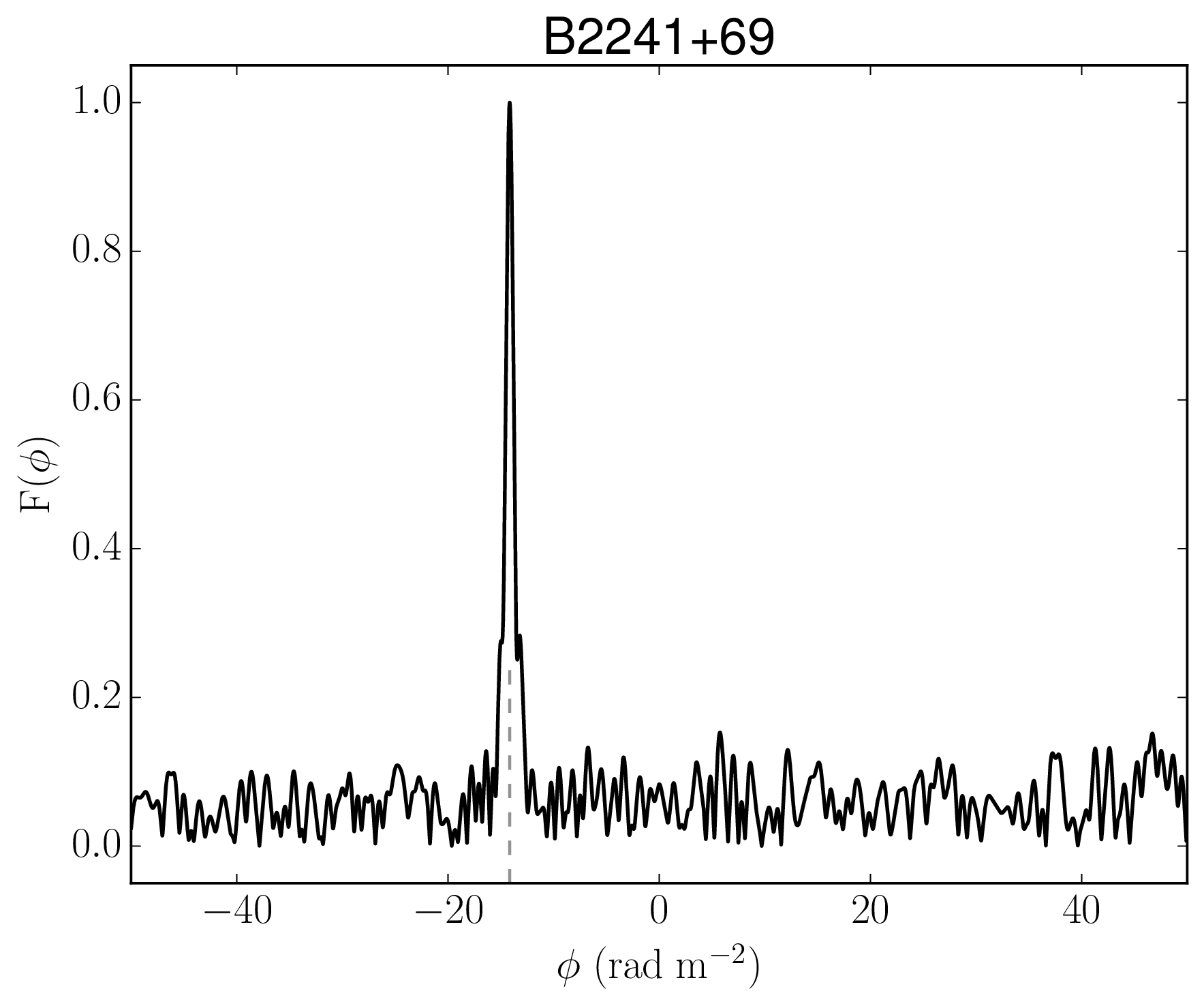}
\includegraphics[width=0.246\columnwidth]{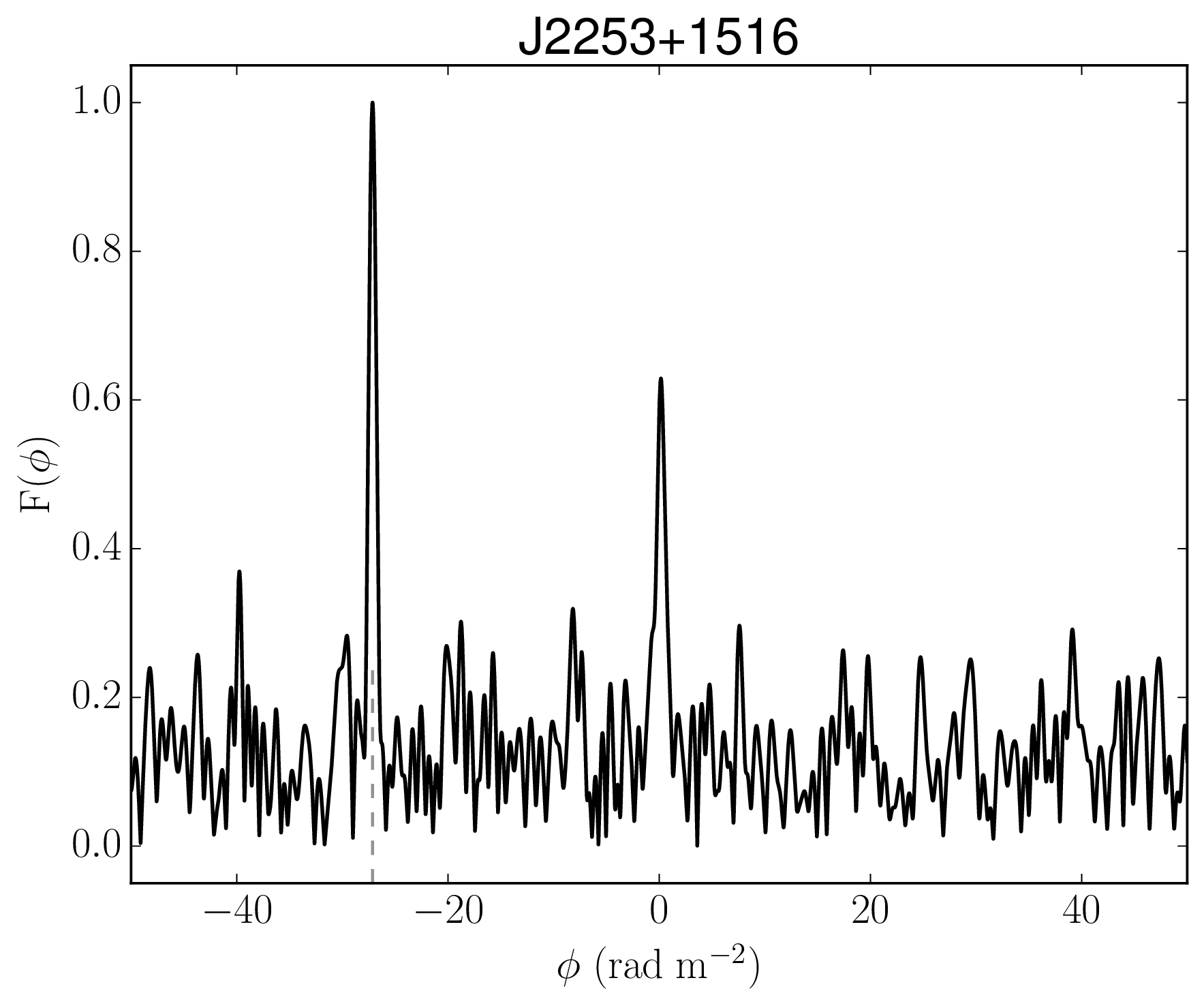}

\includegraphics[width=0.246\columnwidth]{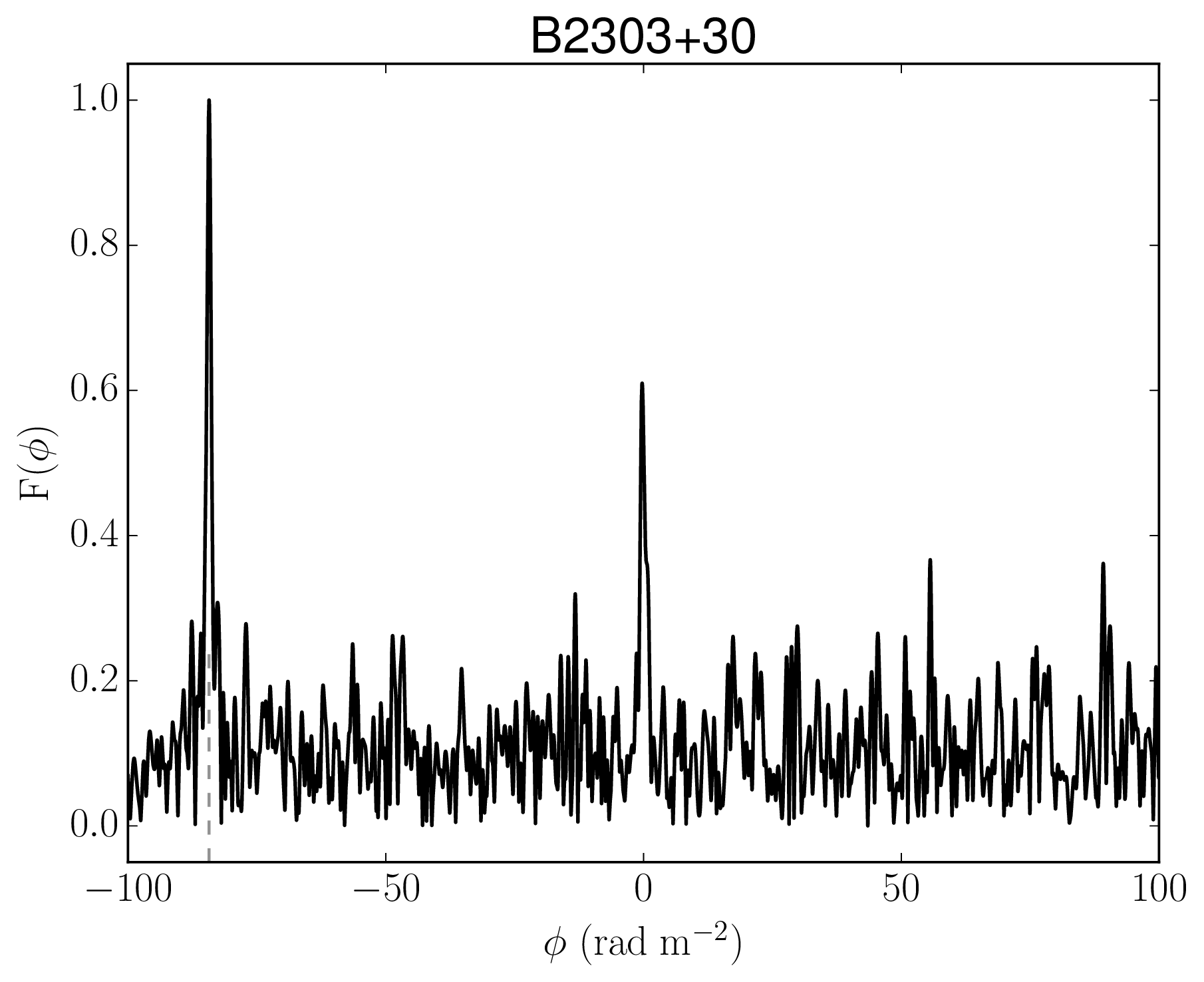}
\includegraphics[width=0.246\columnwidth]{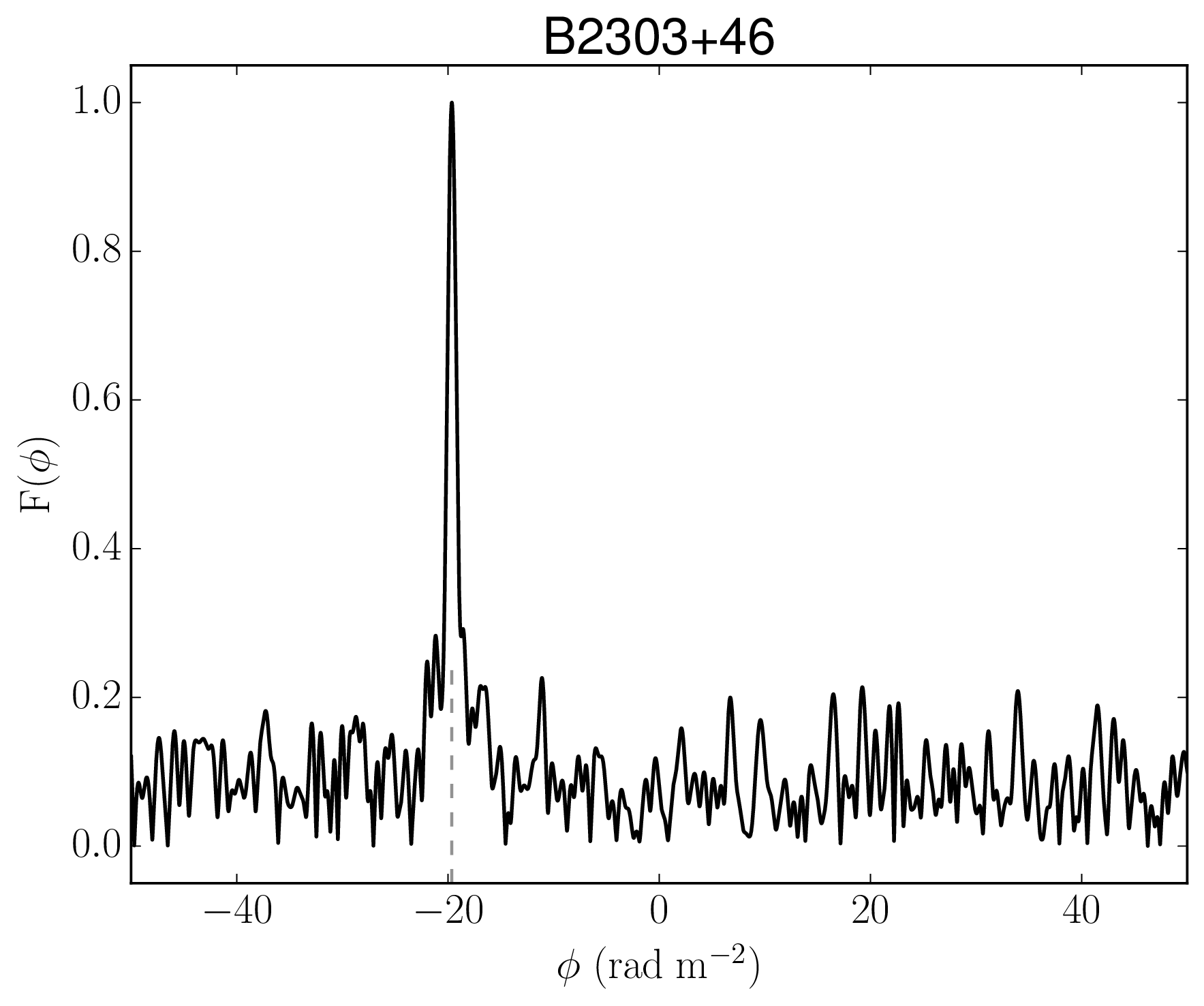}
\includegraphics[width=0.246\columnwidth]{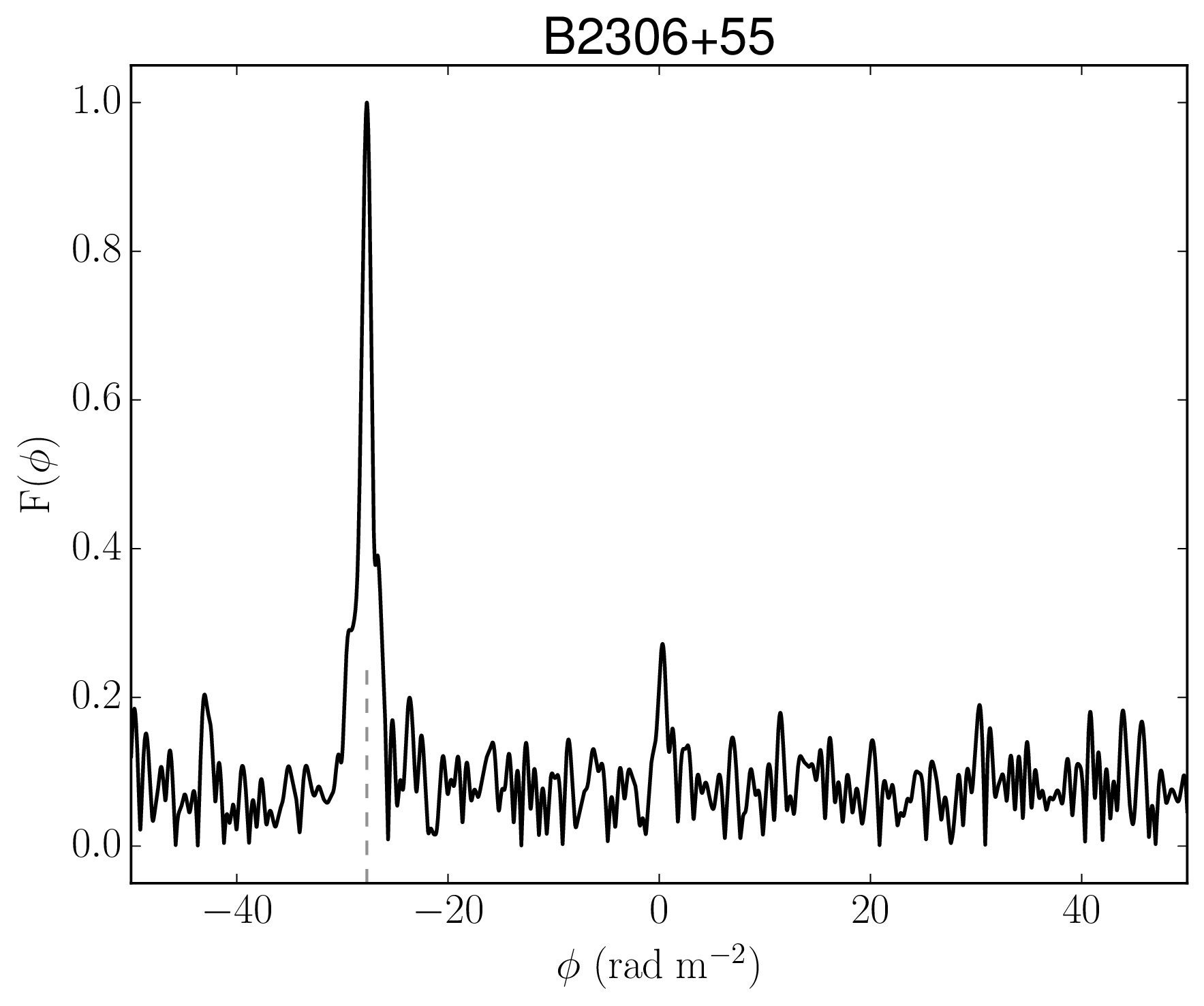}
\includegraphics[width=0.246\columnwidth]{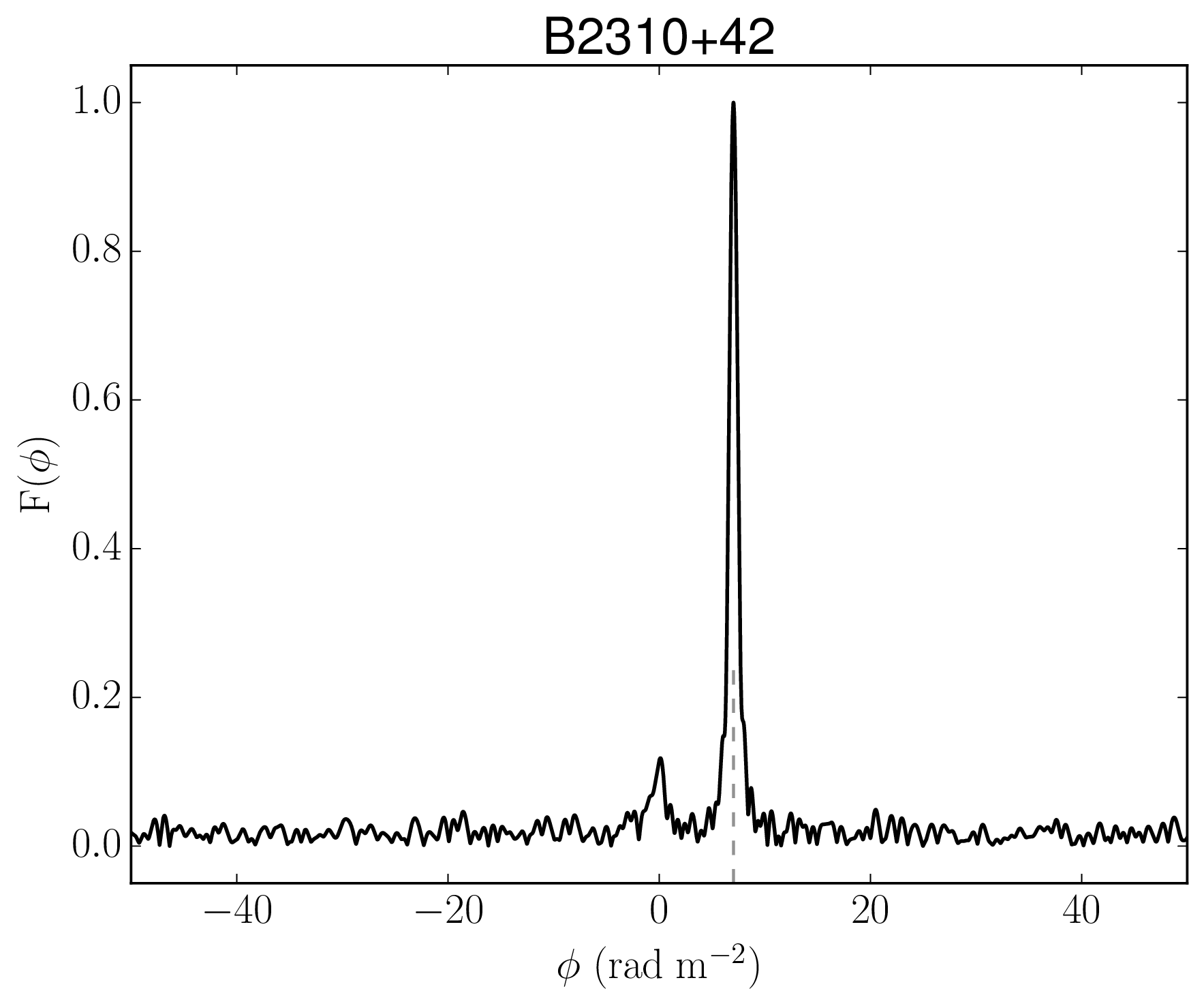}

\includegraphics[width=0.246\columnwidth]{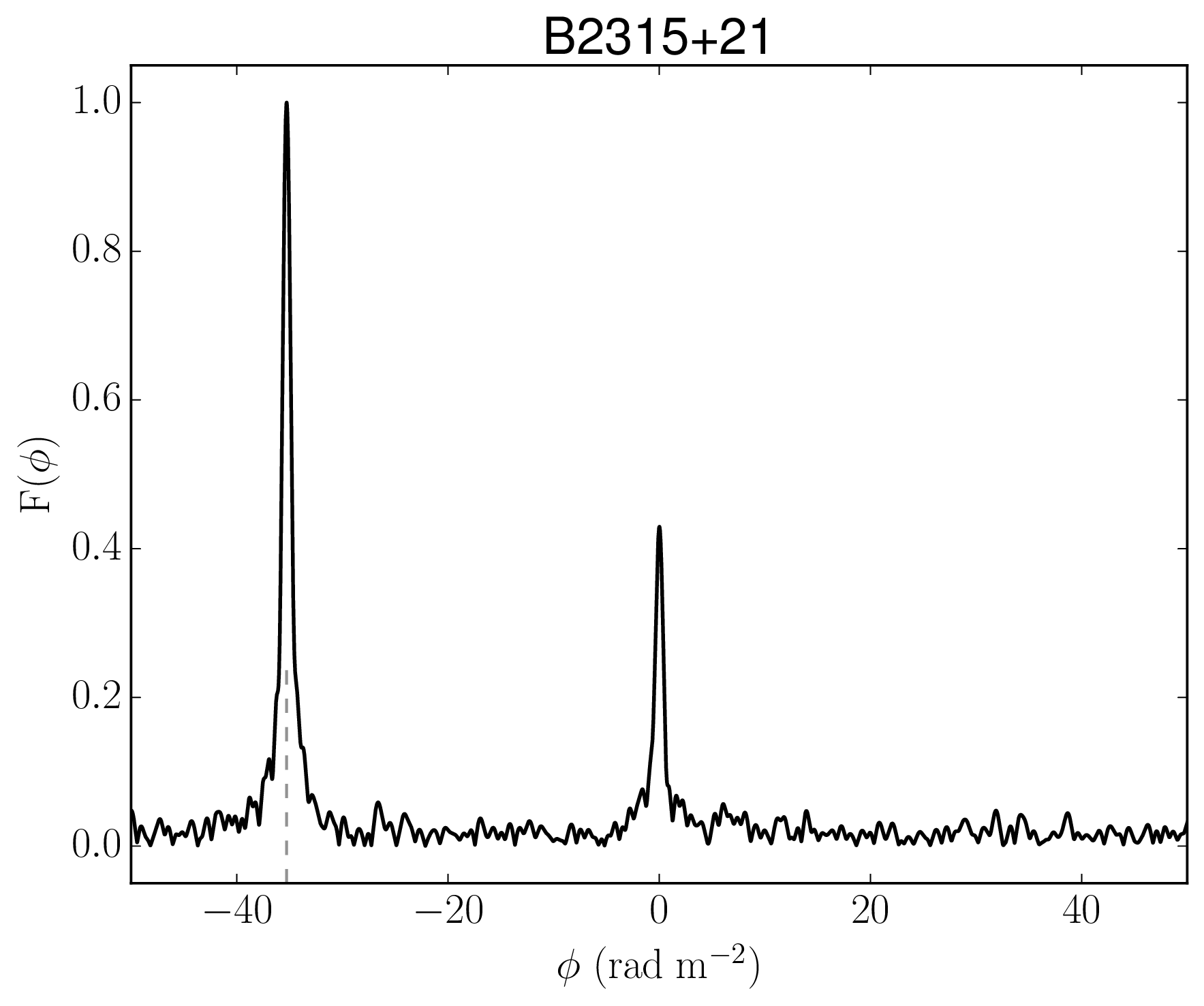}

 \caption{Continued from the previous page.}
% \label{fig:FDFs}
\end{figure*}

\begin{figure*}
 \centering
 %\ContinuedFloat
%\setcounter{subfigure}{2}    

\includegraphics[width=0.246\columnwidth]{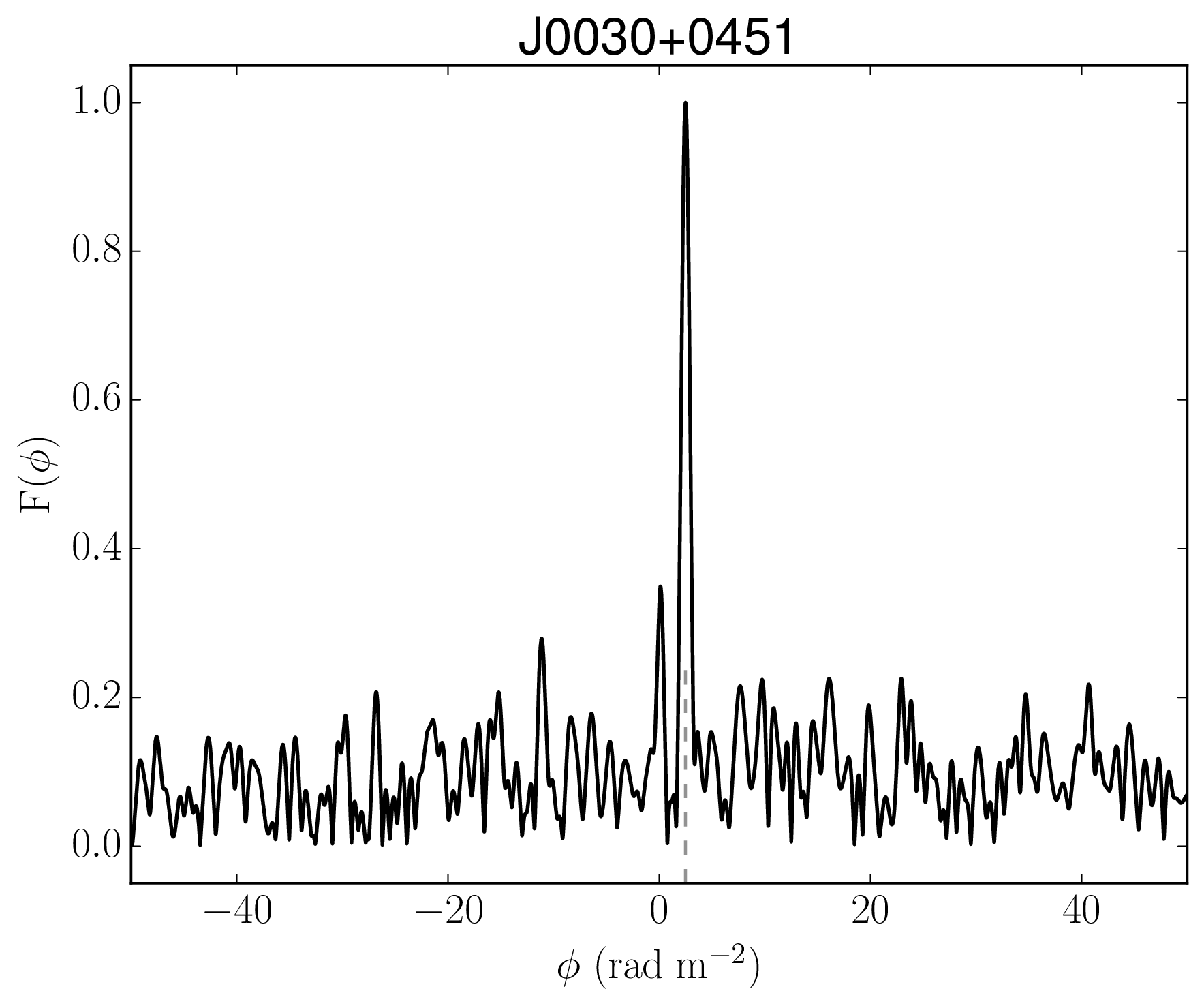}
\includegraphics[width=0.246\columnwidth]{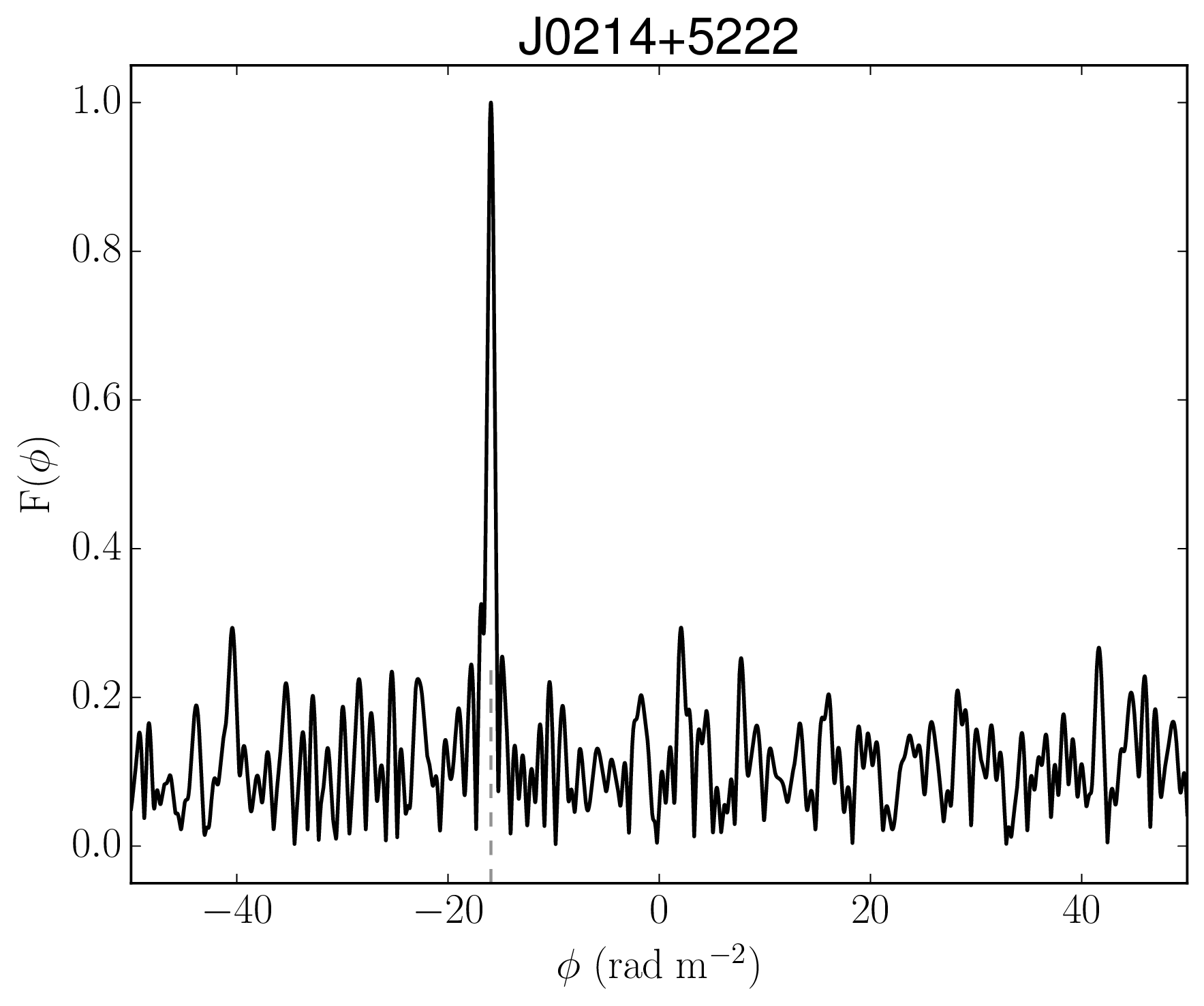}
\includegraphics[width=0.246\columnwidth]{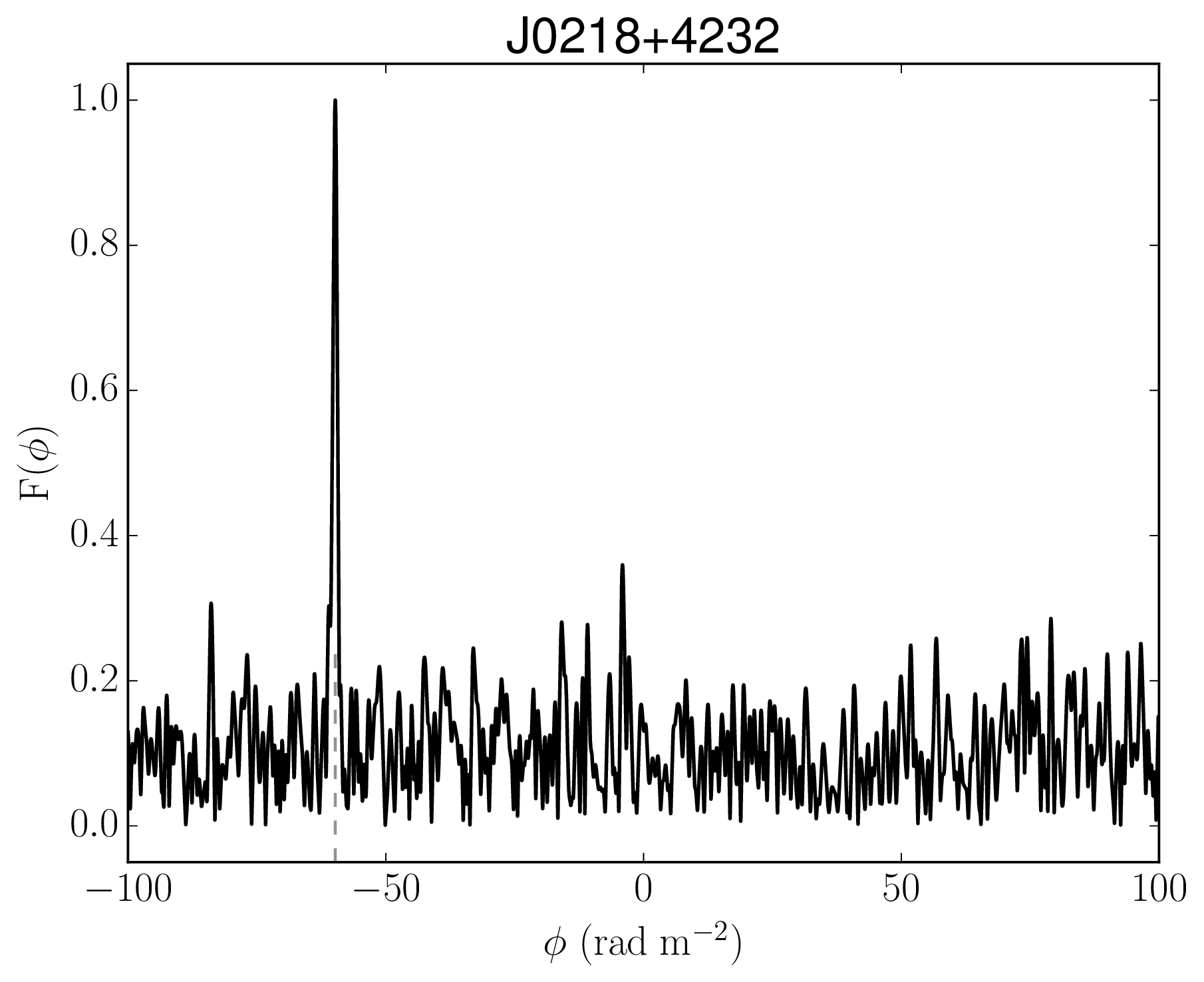}
\includegraphics[width=0.246\columnwidth]{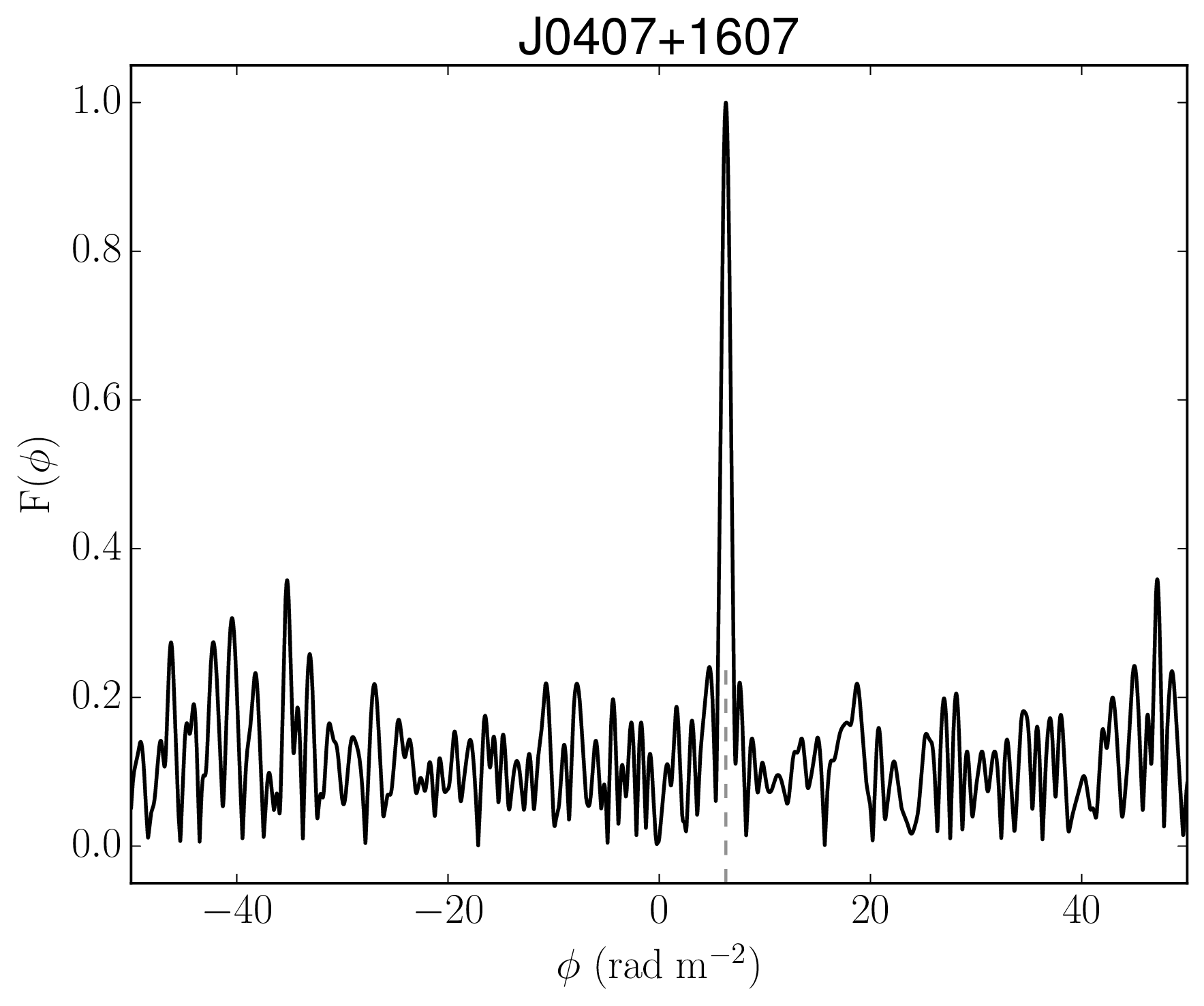}

\includegraphics[width=0.246\columnwidth]{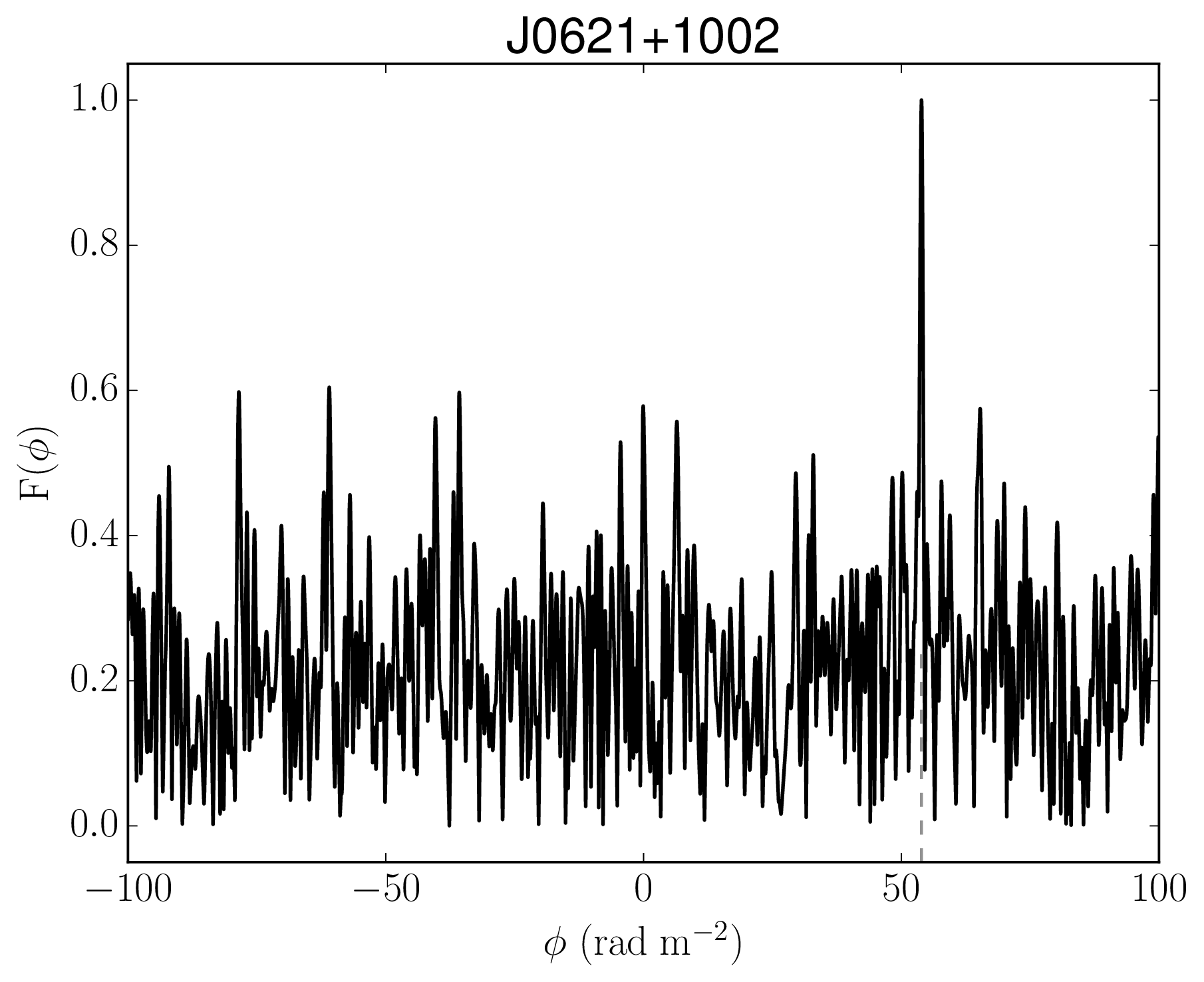}
\includegraphics[width=0.246\columnwidth]{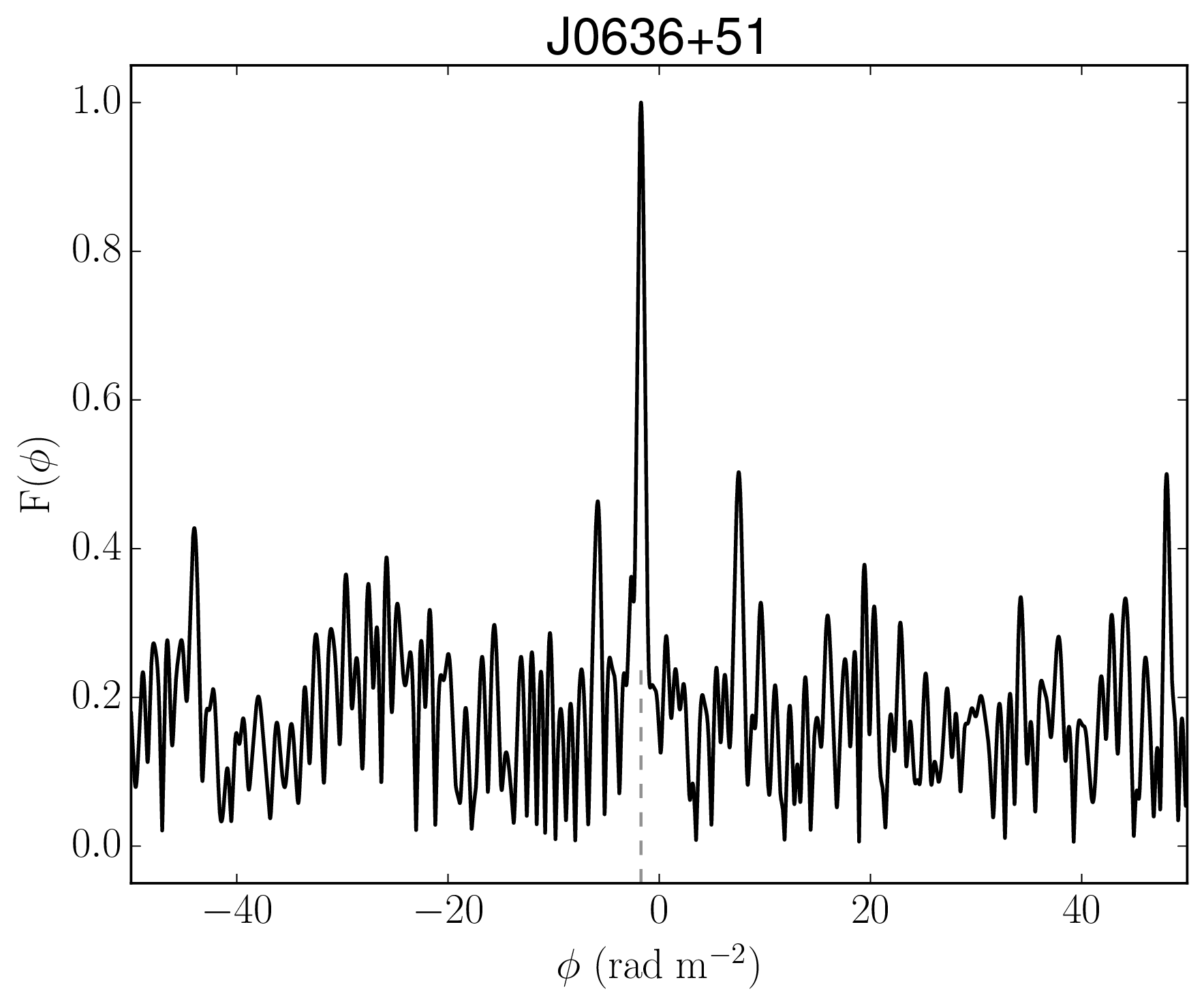}
\includegraphics[width=0.246\columnwidth]{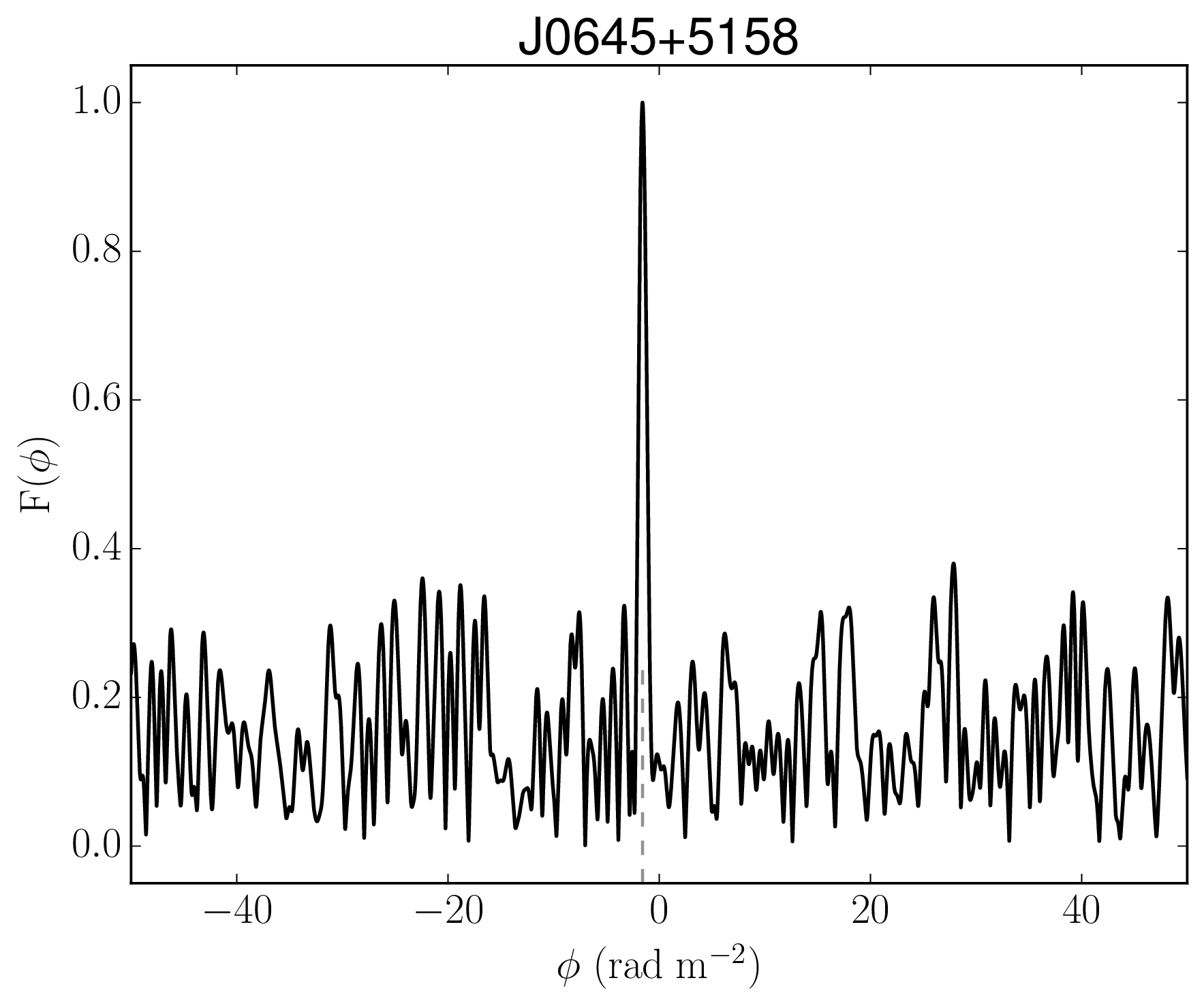}
\includegraphics[width=0.246\columnwidth]{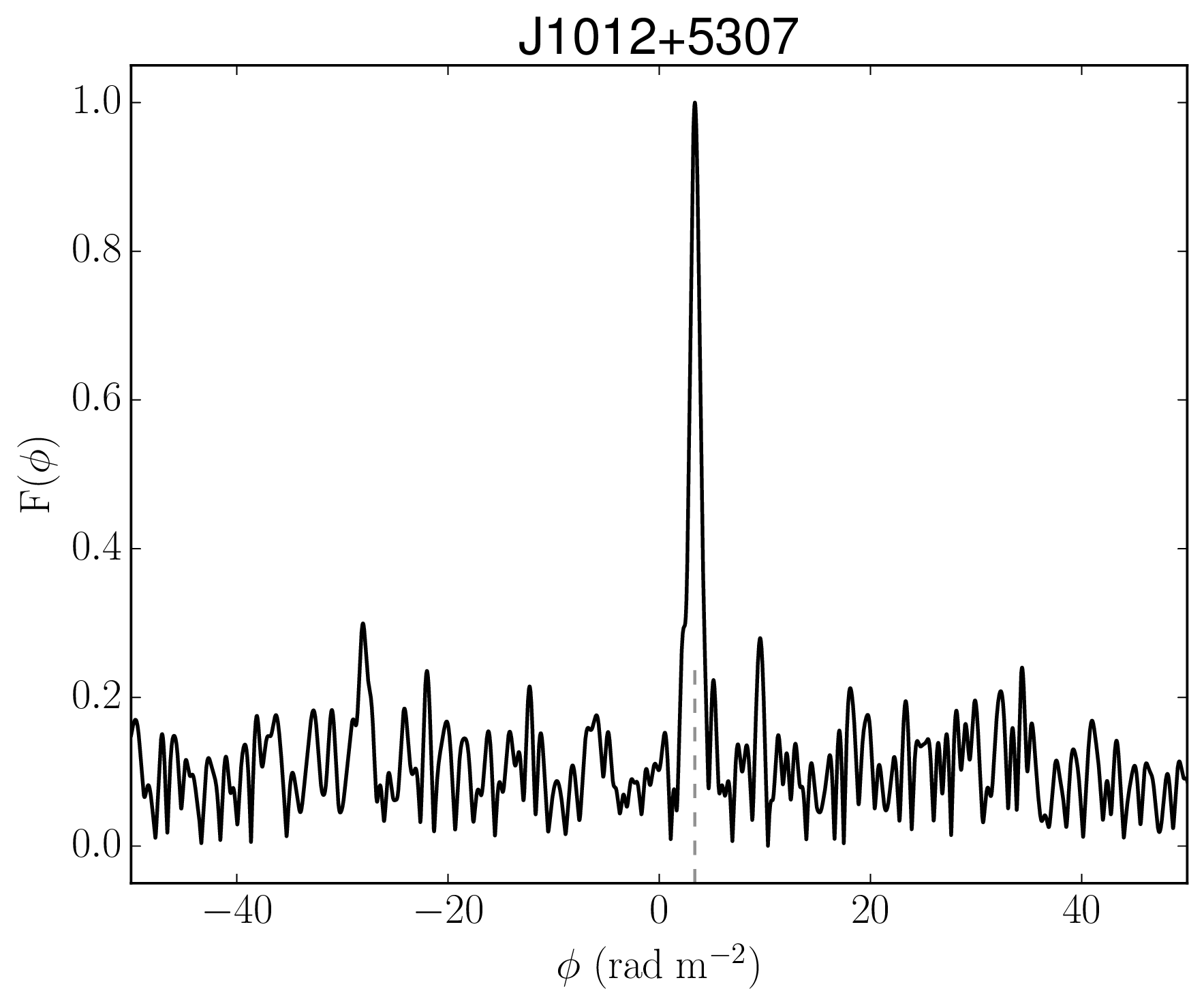}

\includegraphics[width=0.246\columnwidth]{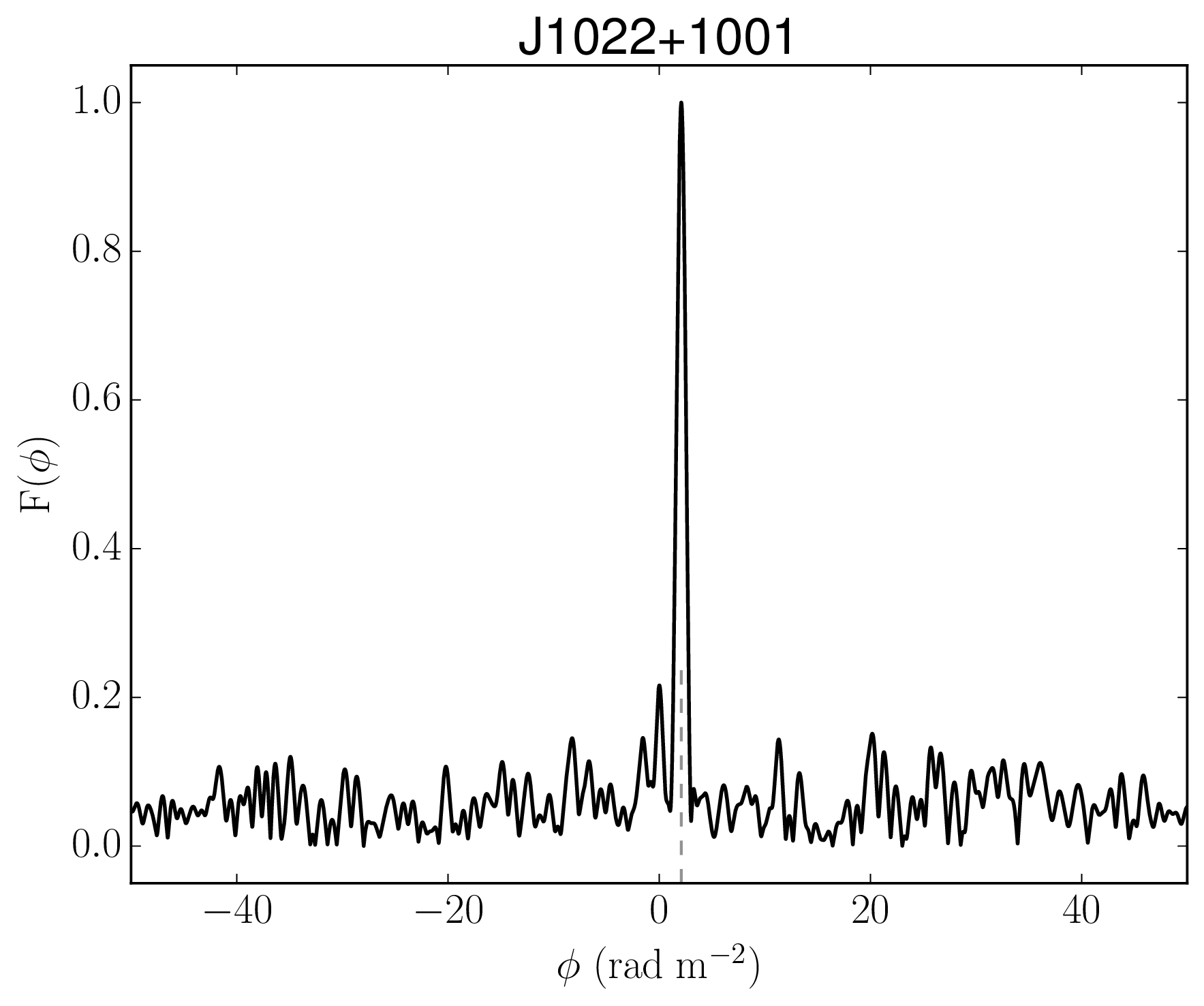}
\includegraphics[width=0.246\columnwidth]{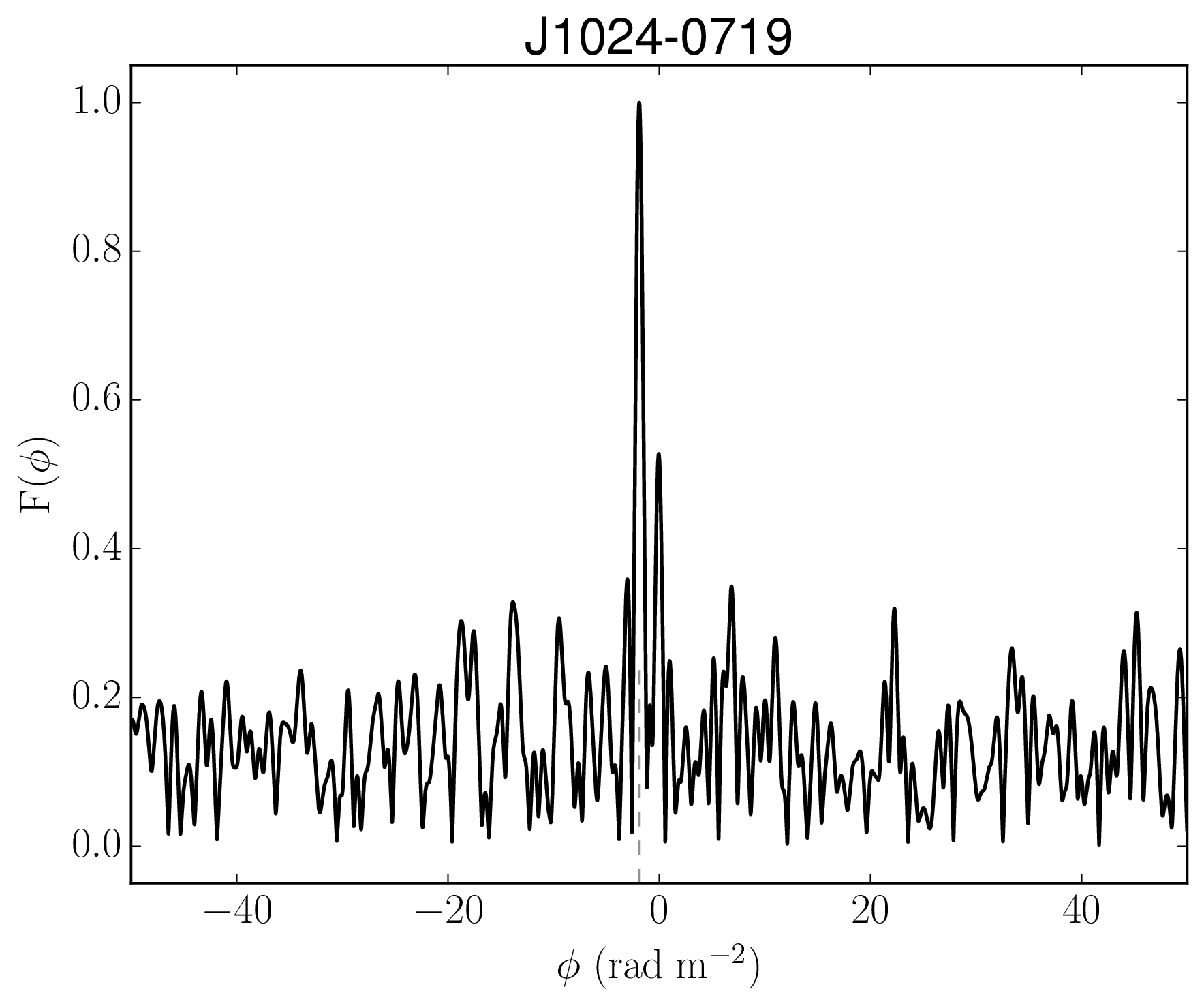}
\includegraphics[width=0.246\columnwidth]{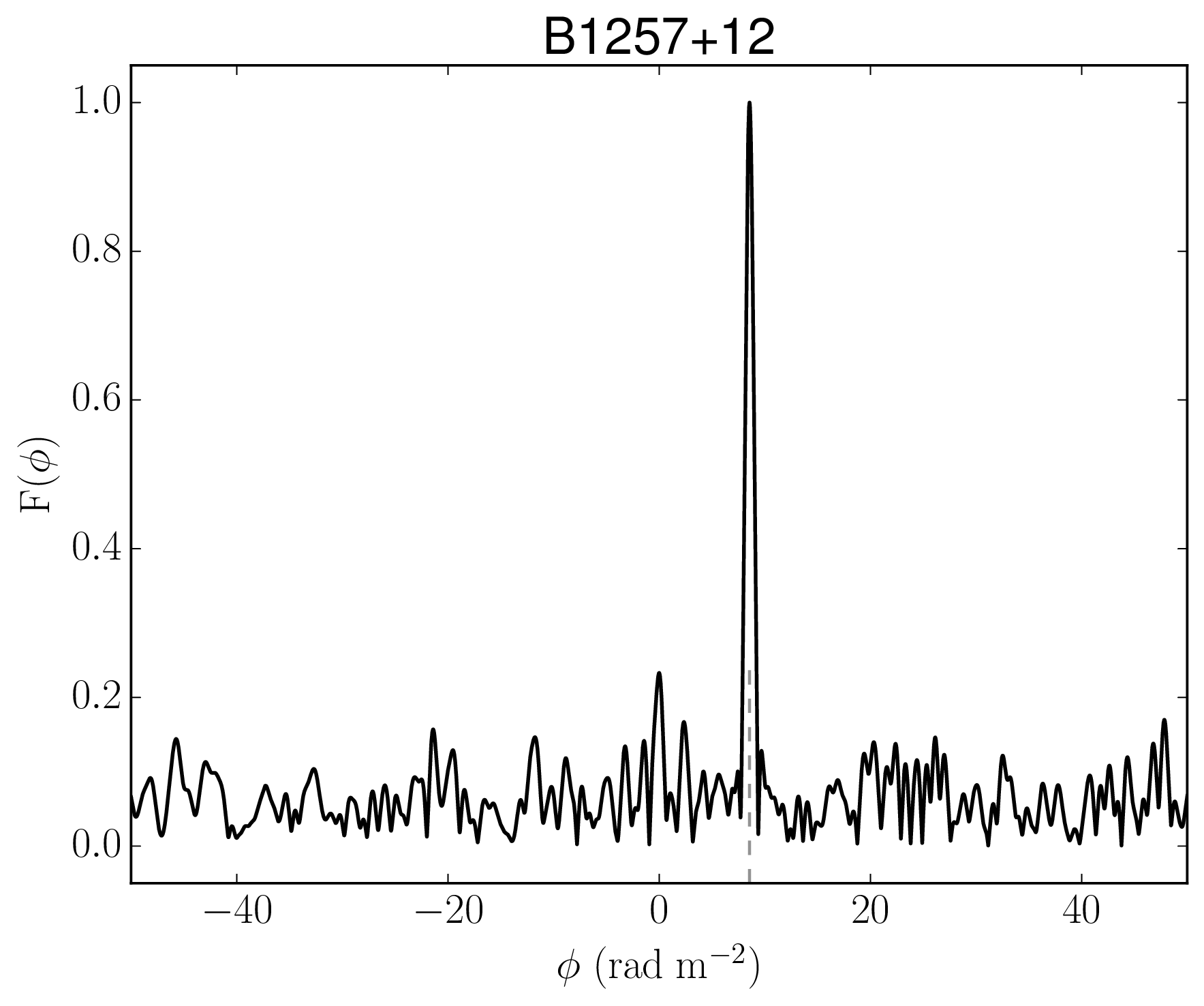}
\includegraphics[width=0.246\columnwidth]{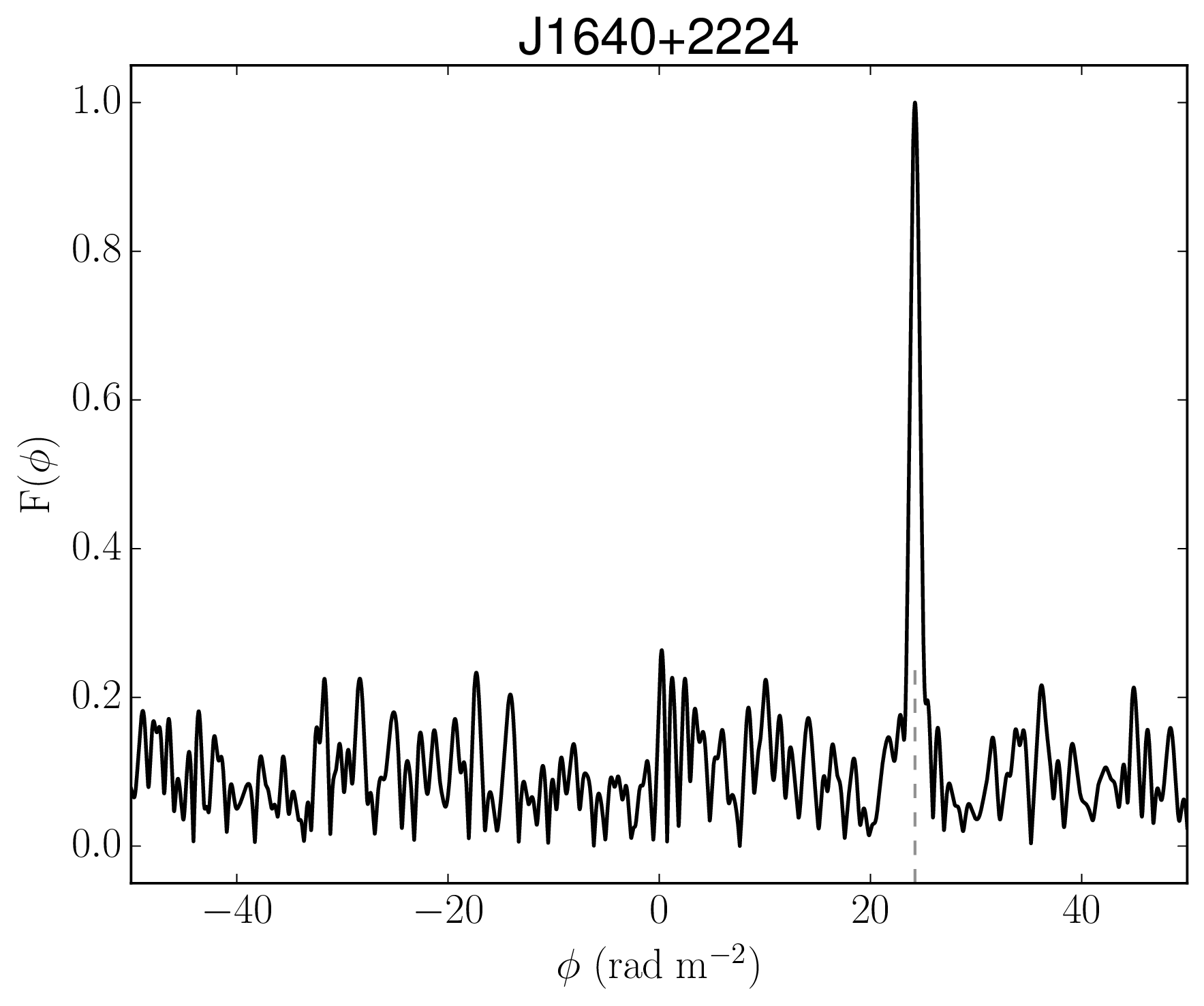}

\includegraphics[width=0.246\columnwidth]{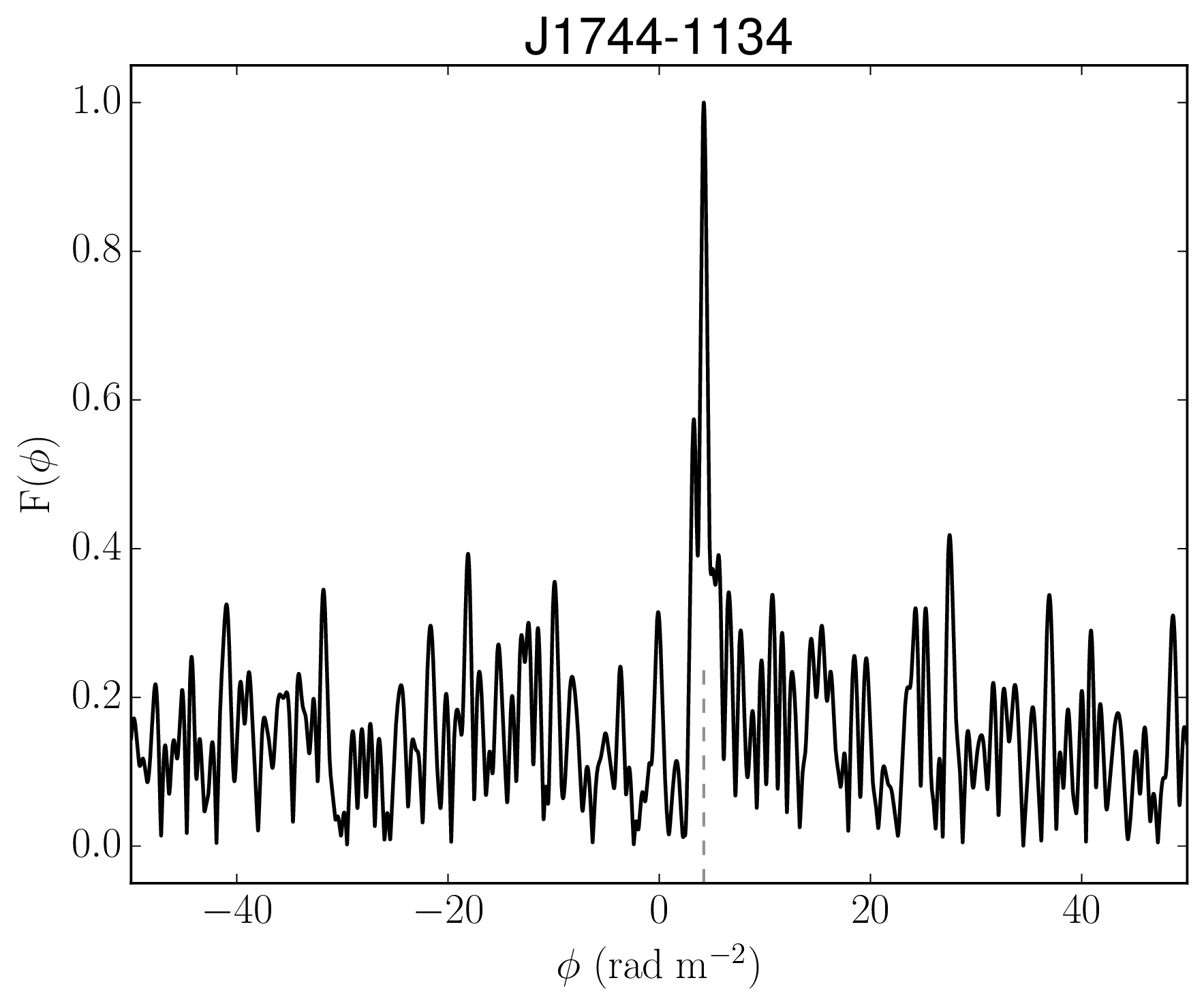}
\includegraphics[width=0.246\columnwidth]{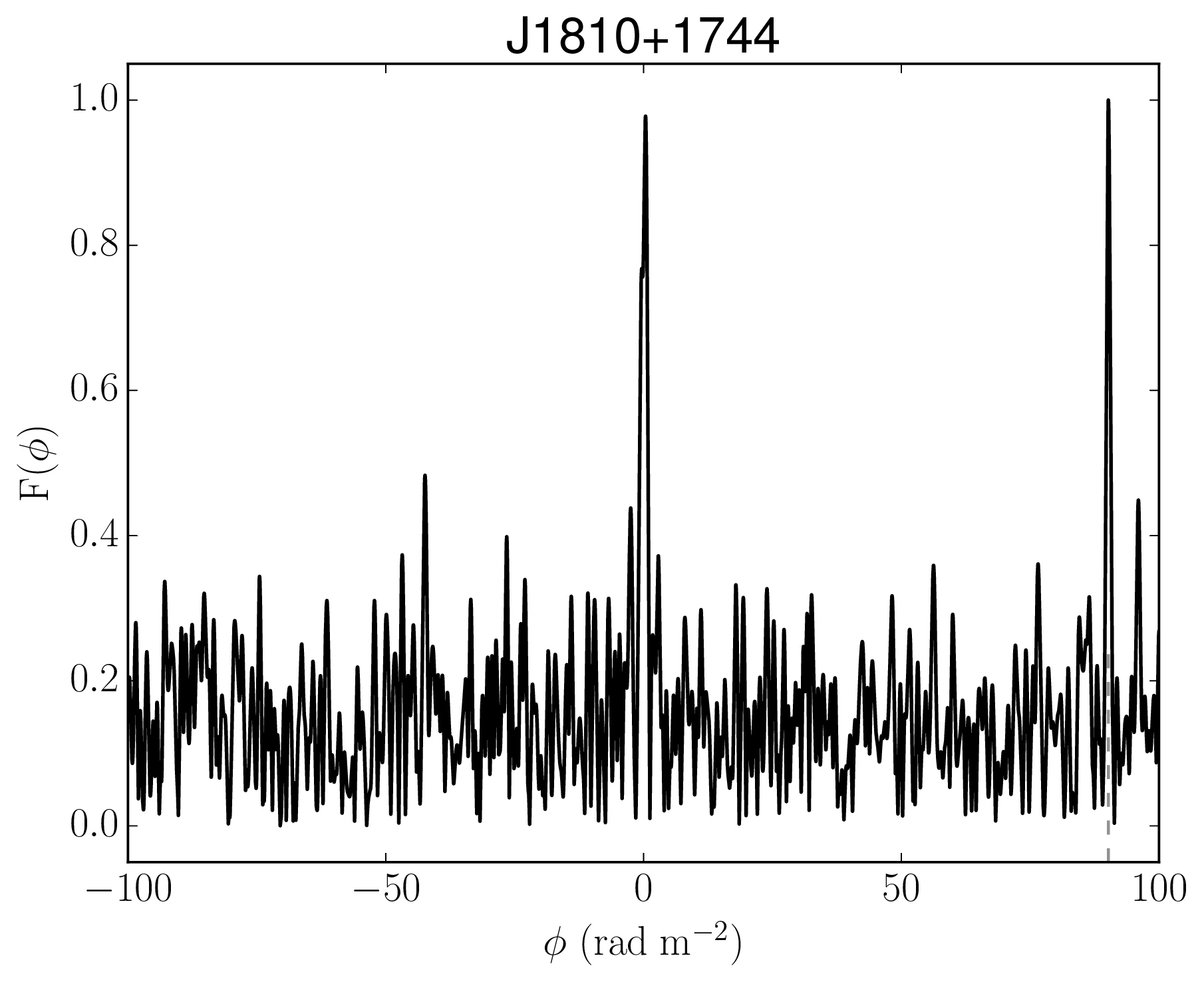}
\includegraphics[width=0.246\columnwidth]{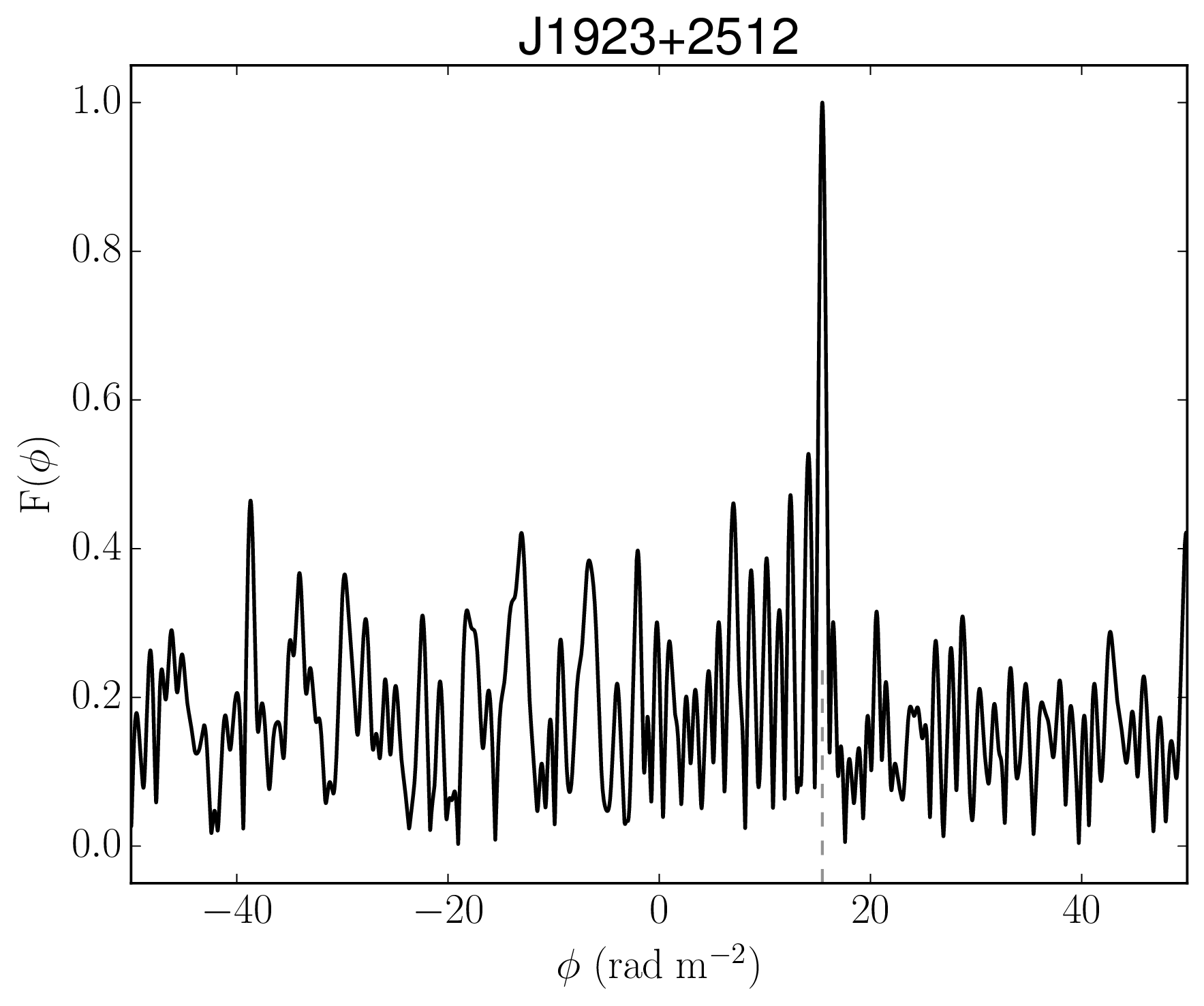}
\includegraphics[width=0.246\columnwidth]{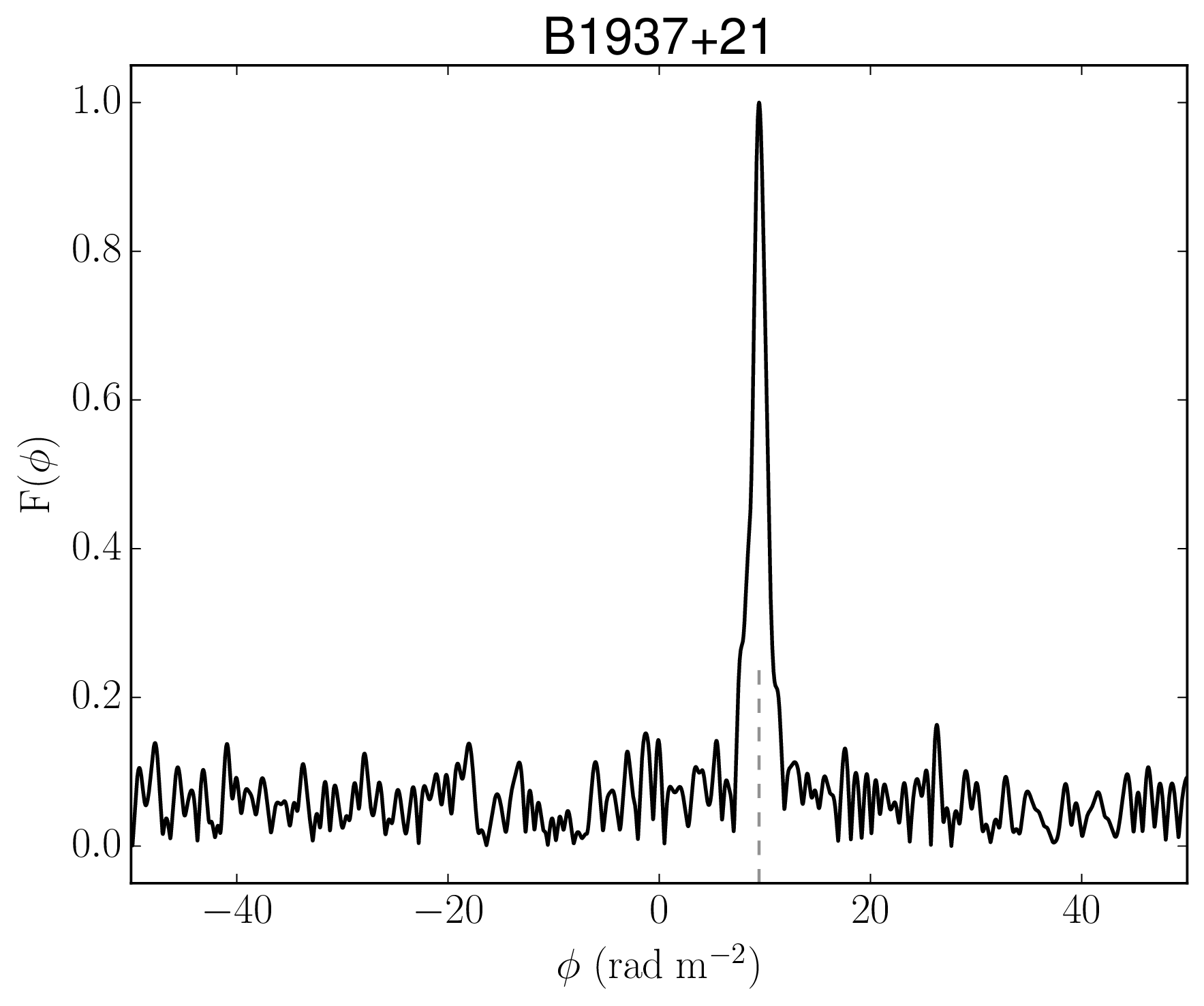}

\includegraphics[width=0.246\columnwidth]{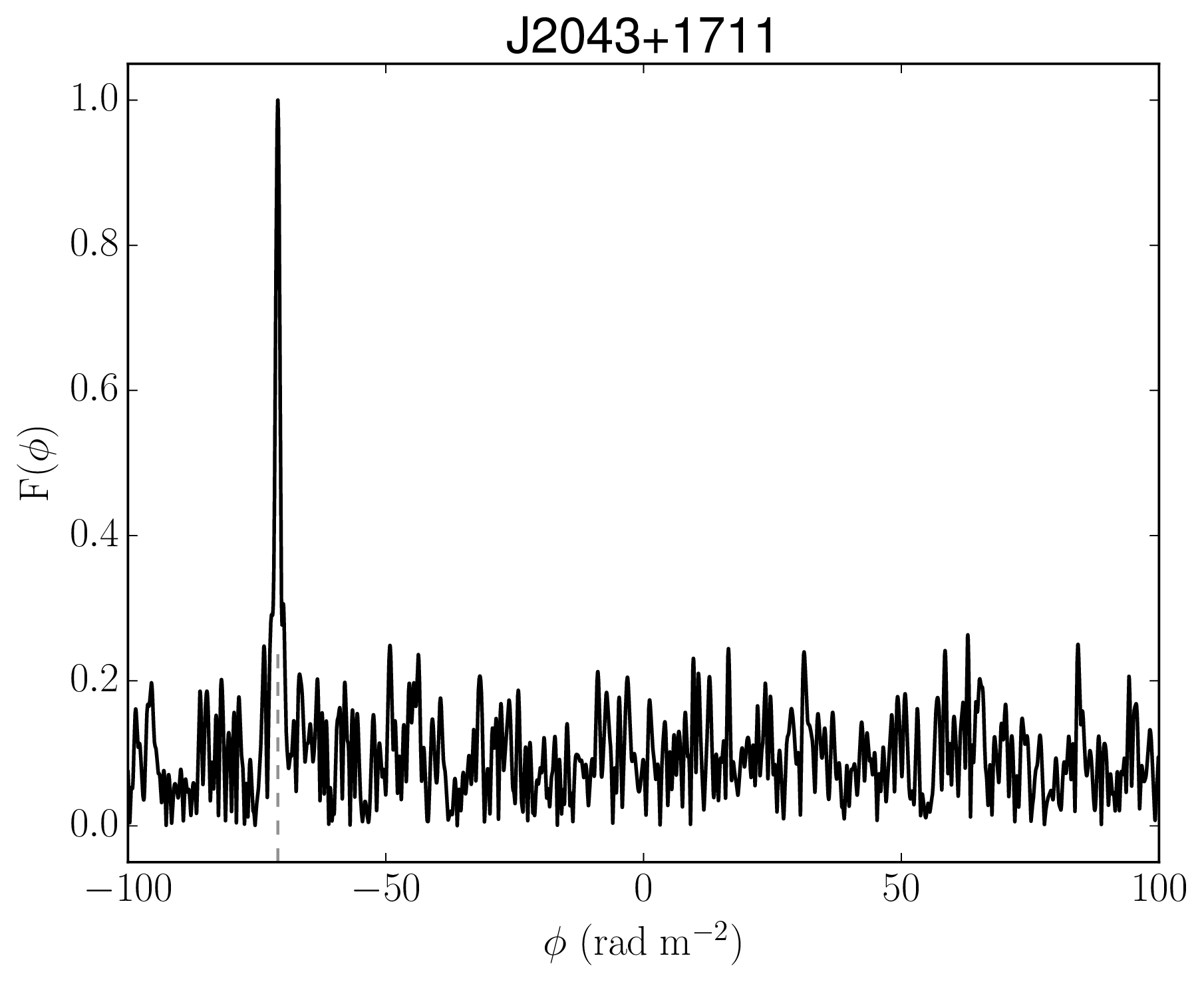}
\includegraphics[width=0.246\columnwidth]{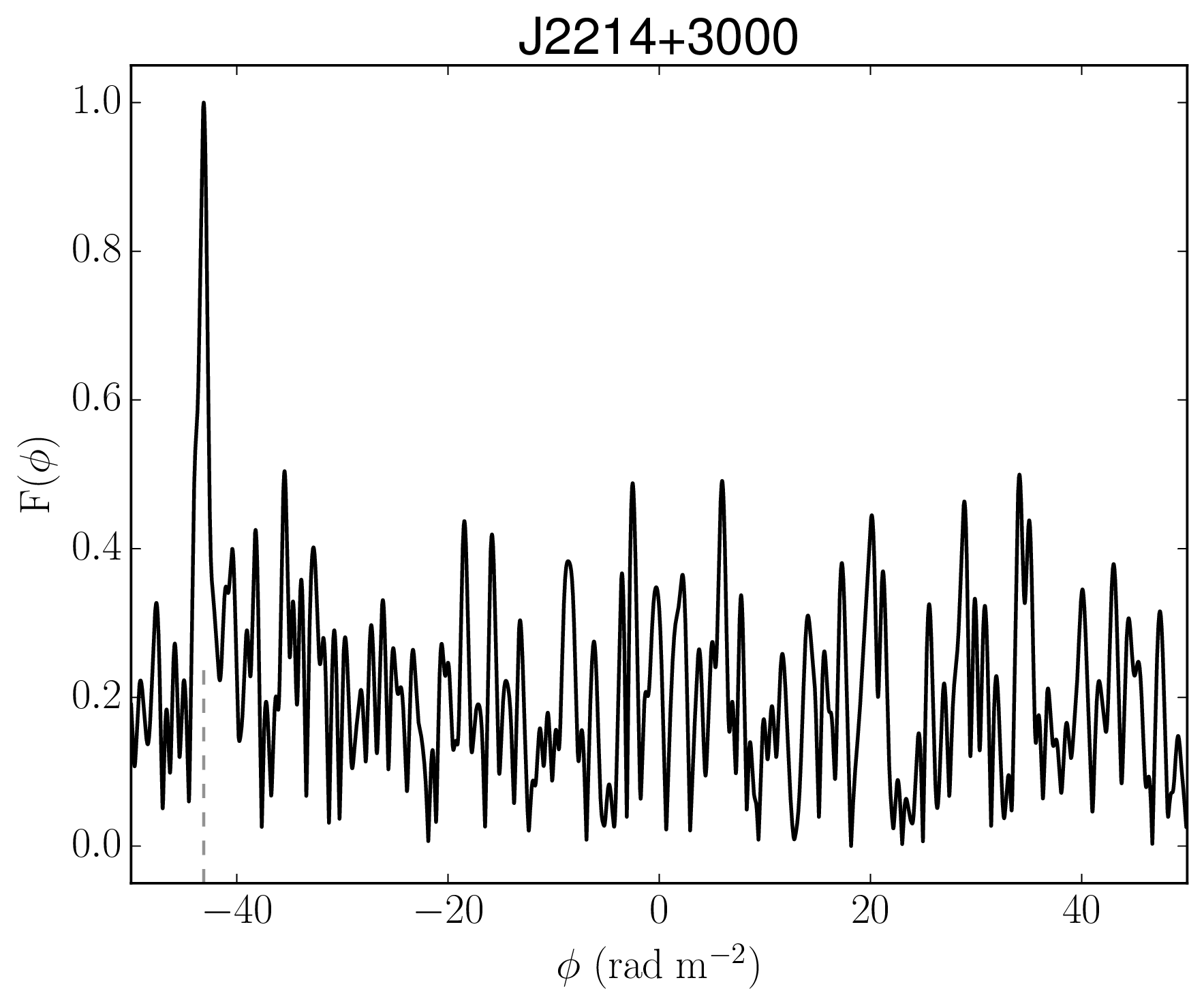}
\includegraphics[width=0.246\columnwidth]{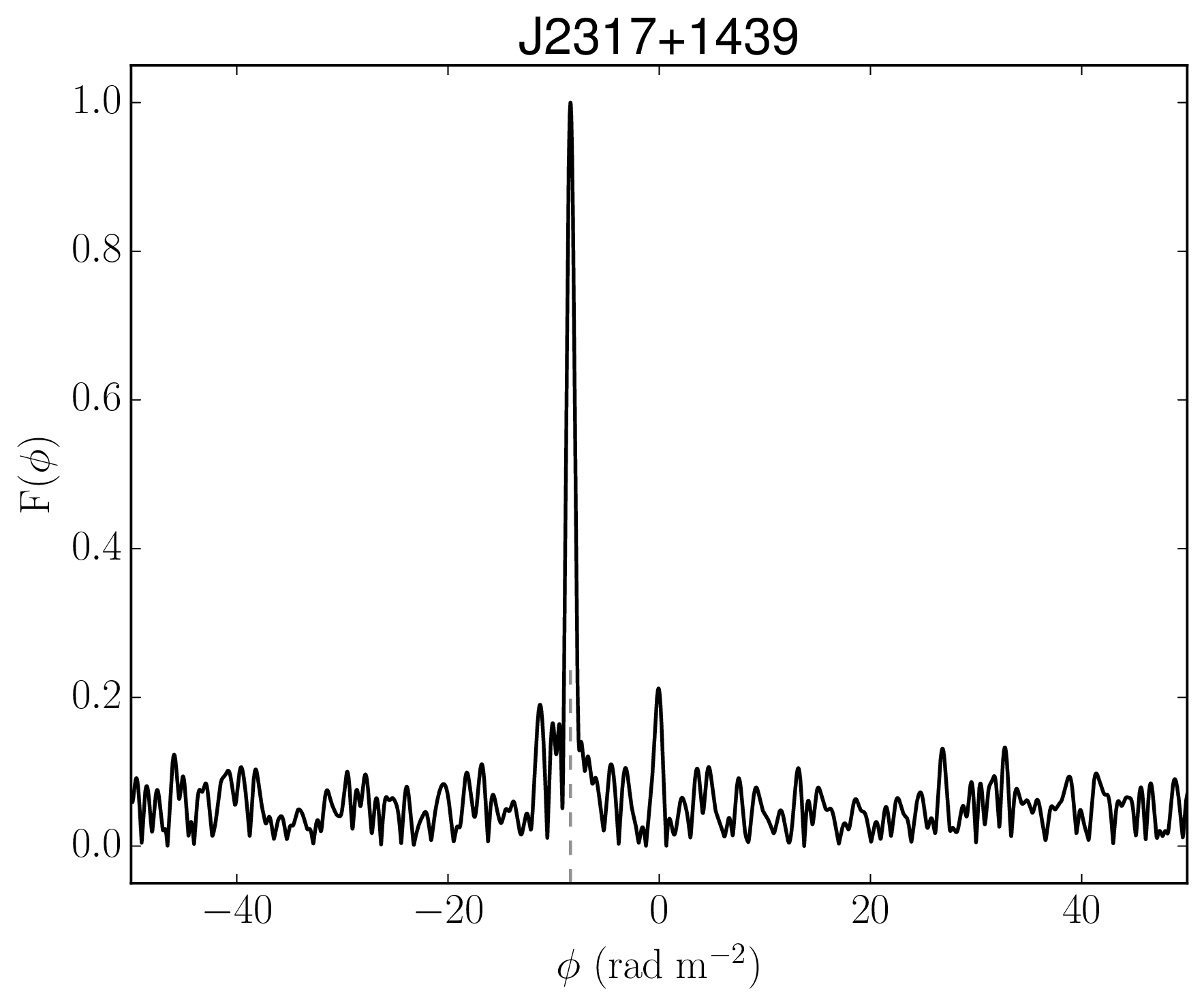}

 \caption{Continued from the previous page.}
 \label{fig:FDFs_MSPs}
\end{figure*}

%%--Start of multi-page subfigure
%\begin{figure}[thbp]
%\centering
%\subfloat[one]{\includegraphics[width=0.75\textwidth]{figure}}
%\subfloat[two]{\includegraphics[width=0.75\textwidth]{figure}}
%\caption[Many]{load of figures}
%\end{figure}
% 
%\begin{figure}[thbp]
%\ContinuedFloat
%\centering
%\subfloat[two]{\includegraphics[width=0.75\textwidth]{figure}}
%\caption[]{load of figures}
%\end{figure}
%%--------------------------------
\label{fig:FDFs}

%%%%%%%%%%%%%%%%% APPENDICES END %%%%%%%%%%%%%%%%%%%%%%%% SEPARATE FILE RMs_mnras_v2.tex %%%%%%%%%%%%%%%%%%
%%%%%%%%%%%%%%%%%%%%%%%%%%%%%%%%%%%%%%%%%%%%%%%%%%

% Don't change these lines
\bsp	% typesetting comment
\label{lastpage}
\end{document}